\documentclass[dvips,12pt]{book}
\usepackage{layout}
\usepackage[latin1]{inputenc}
\usepackage{amsmath}
\usepackage{amssymb}
\usepackage{graphicx}
\usepackage{fancyhdr}
\usepackage[small]{caption}
\usepackage{lscape}
\usepackage{color}
\definecolor{chaptergrey}{rgb}{0.0,0.0,0.0}
\usepackage[nogrey,times]{quotchap}
\usepackage{shapepar}
\usepackage{deluxetable}
\usepackage[times]{quotchap}

\newcommand{\dechms}[4]{$#1^{\rm h}#2^{\rm m}#3\mbox{$^{\rm s}\mskip-7.6mu.\,$}#4$}
\newcommand{\decdms}[4]{$#1^{\circ}#2'#3\mbox{$''\mskip-7.6mu.\,$}#4$}
\newcommand{\inthms}[3]{$#1^{\rm h}#2^{\rm m}#3^{\rm s}$}
\newcommand{\intdms}[3]{$#1^{\circ}#2'#3''$}
\newcommand{\msec}[2]{$#1\mbox{$''\mskip-7.6mu.\,$}#2$}
\newcommand{\mmsec}[2]{$#1\mbox{$^s\mskip-7.6mu.\,$}#2$}
\newcommand{\Av}{A_V}
\newcommand{\HII}{\mbox{H\,{\sc ii}}}
\newcommand{\kmps}{km s$^{-1}$}
\newcommand{\Lsun}{L$_{\odot}$}
\newcommand{\Msun}{M$_{\odot}$}
\newcommand{\Rsun}{R$_{\odot}$}
\newcommand{\aips}{$\cal AIPS$}
\newcommand{\cook}{{\em CookBook}}
\newcommand{\aap}{A\&A}
\newcommand{\apj}{ApJ}
\newcommand{\aj}{AJ}
\newcommand{\apjs}{ApJS}
\newcommand{\pasp}{PASP}
\newcommand{\mnras}{MNRAS}
\newcommand{\araa}{ARAA}

\begin{document}

\numberwithin{equation}{chapter}

\pagestyle{empty}
\thispagestyle{empty}
\begin{flushleft}
\phantom{hola}
\vfill

\selectfont

\begin{tabular*}{3cm}[l]{ll}
\textbf{{\Huge U  N  A  M ~~~~~~~~~~~~~~~}} &
\includegraphics[height=3cm]{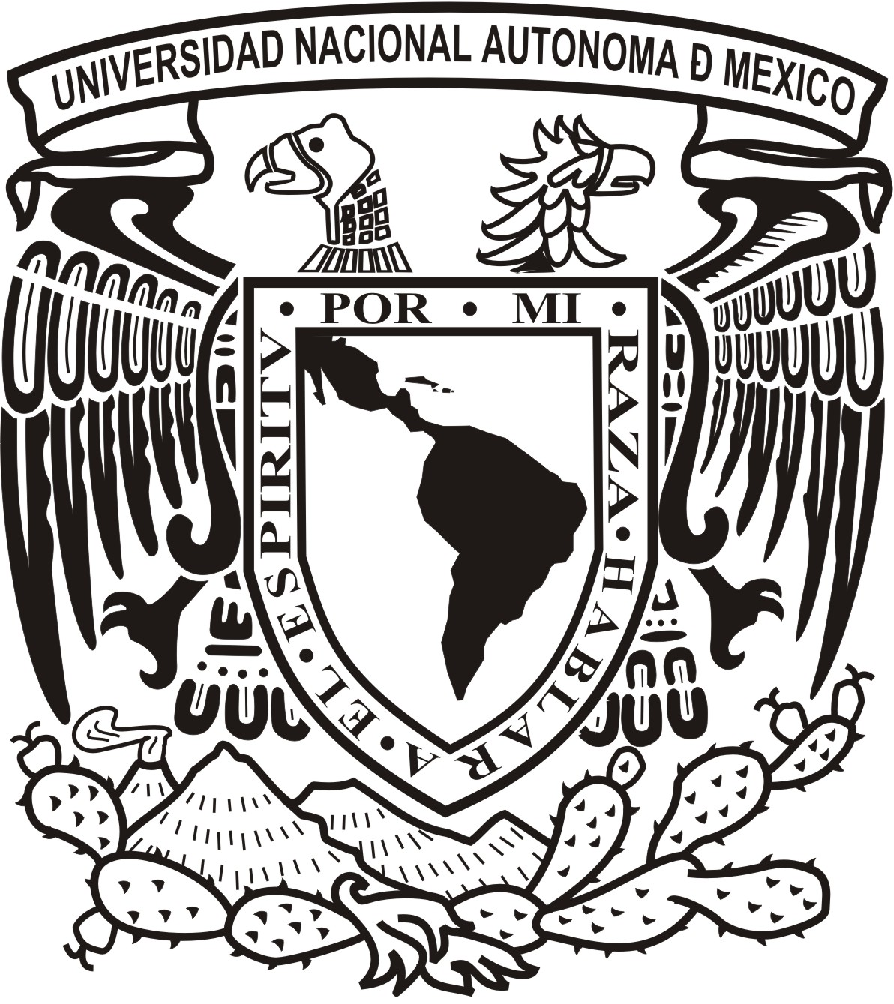}\\
\end{tabular*}

\vspace{.5cm}

{\small T E S I S ~ P A R A ~ O B T E N E R ~ E L ~ G R A D O ~ A C A D É M I C O}\\
{\small D E ~ D O C T O R ~ E N ~ C I E N C I A S ~ ( A S T R O N O M Í A )}\\

\vspace{5cm}

\textbf{\LARGE Midiendo Regiones de Formación}\\
\vspace{.1cm}
\textbf{\LARGE Estelar Cercanas con el VLBA:}\\
\vspace{.2cm}
\textbf{\LARGE de la Distancia a la Dinámica}\\

\vspace{.5cm} 

\textbf {\large R O S A ~ M . ~ T O R R E S}\\ 

\vspace{1.5cm}

{\small Directores}\\
{\small Dr. Laurent Loinard, CRyA - UNAM}\\
{\small Dra. Amy J. Mioduszewski, DSOC - NRAO}\\

\vspace{.5cm}

\textsf{\small U N I V E R S I D A D ~ N A C I O N A L ~ A U T Ó N O M A ~ D E ~ M É X I C O}\\
{\small C e n t r o ~ d e ~ R a d i o a s t r o n o m í a ~ y ~ A s t r o f í s i c a}\\
{\small N a t i o n a l ~ R a d i o ~ A s t r o n o m y ~ O b s e r v a t o r y}\\

\vspace{.5cm}

{\small Morelia, Michoacán, México 2009}\\

\vfill
\end{flushleft}
\cleardoublepage

\thispagestyle{empty}
\begin{center}
\phantom{hola}
\vfill

\selectfont

\textbf{\Huge U  N  A  M}

\vspace{.5cm}

{\small T H E S I S ~ F O R ~ T H E ~ P H D . ~ I N ~ A S T R O N O M Y}\\

\vspace{2.5cm}

\textbf{\LARGE Measuring Nearby Star Forming}\\
\vspace{.2cm}
\textbf{\LARGE Regions with the VLBA:}\\
\vspace{.2cm}
\textbf{\LARGE from the Distance to the Dynamics}\\

\vspace{.5cm} 

{\large R O S A ~ M . ~ T O R R E S}\\ 

\vspace{2.5cm}

{\small Advisors}\\

\vspace{.3cm}

{\small Dr.\ Laurent Loinard, Centro de Radioastronomía y Astrofísica, UNAM}\\
{\small Apartado Postal 72--3 (Xangari), 58089 Morelia, Mich., México}\\

\vspace{.3cm}

{\small Dr.\ Amy J.\ Mioduszewski, Domenici Science Operations Center, NRAO}\\
{1003 Lopezville Road, 87801 Socorro, NM, USA}\\

\vspace{1cm}

\includegraphics[height=2cm]{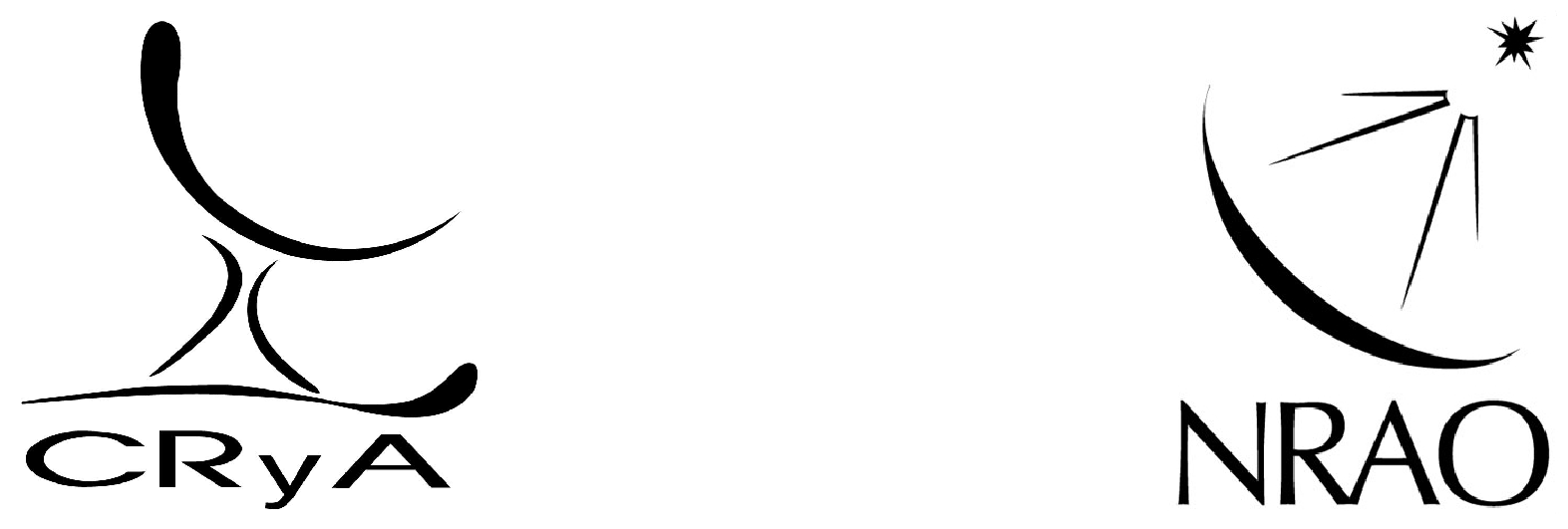}\\

\vspace{.3cm}

{\small UNIVERSIDAD NACIONAL AUTÓNOMA DE MÉXICO}\\
{\small Centro de Radioastronomía y Astrofísica}\\
{\small National Radio Astronomy Observatory}\\

\vspace{.3cm}

{\small September 2009}\\

\vfill
\end{center}
\cleardoublepage

\phantom{hola}

\noindent
\textbf{\LARGE Agradecimientos}

\vspace{1cm}

\noindent
Mi más profunda gratitud a Laurent Loinard por haberme propuesto el
tema de esta tesis, por confiar en que no echaría a perder el proyecto
y por el esfuerzo dedicado durante los últimos años.

\smallskip

\noindent
Agradezco a los directores de la tesis, Laurent Loinard y Amy
Mioduszewski, y a los sinodales Luis Felipe Rodríguez, Paola D'Alessio
y Mark Reid por su atenta y rápida leída de este manuscrito y por los
valiosos comentarios que lo han enriquecido.

\smallskip

\noindent
El trabajo de presentado aquí no habría sido posible sin el apoyo,
dedicación y paciencia de Amy Mioduszewski, quien durante los últimos
tres años, me explicó pacientemente todo lo que ahora sé sobre VLBI,
además revisó cuidadosamente cada uno de los pasos durante la
reducción de datos y me ha motivado a seguir adelante.

\smallskip

\noindent
Agradezco a la UNAM y al CRyA por las facilidades brindadas durante la
maestría y doctorado, a DGAPA-UNAM por el complemento de beca que me
otorgó en maestría, al CONACyT por las becas de maestría y doctorado,
y al NRAO por hacer posible las estancias de trabajo en Socorro. De
igual manera agradezco al Posgrado en Astronomía de la UNAM, al CRyA y
a los proyectos CONACyT y PAPIIT por el apoyo económico proporcionado
para asistir a escuelas y congresos durante estos cuatro años.

\smallskip

\noindent
Doy las gracias a mis profesores de la maestría quienes apoyaron mi
formación académica, a los investigadores que aunque no me dieron
clases siempre estuvieron dispuestos a ayudarme con mis dudas, al
personal administrativo y de cómputo del CRyA por su ayuda brindada
día con día y al personal del Posgrado en Astronomía por el tiempo
invertido en trámites de la maestría y doctorado.

\smallskip

\noindent
De la misma manera agradezco a todo el personal del NRAO, quienes de
alguna u otra manera ayudaron a que mis estancias en Socorro fueran
buenas. En especial agradezco a Kumar, Emmanuel, Mary, Lorant, Vivek,
Urvashi, Sanjay, Veronica, Maurilio, Adam, Nissim, Esteban, Wei-Hao,
Masaya, Aya, Malcolm y Brigette, por el buen café después del lunch,
por las BBQ los fines de semana y por llevarme al Bosque del Apache.

\smallskip

\noindent
Gracias a mis compañeros del posgrado. A Ramiro y Daniel por contestar
todas mis preguntas, por apostar a que sí pasaba el de admisión y los
generales. A Gaby, Edgar y Alfonso por las desveladas, por el trabajo
en equipo, por la compañía y por todo lo que nos divertimos en la
maestría. A Roberto por sus consejos y por buena gente. A todos los
que se fueron agregando en el camino (Alfredo, Alma, Álvaro, Aurora,
Bernardo, Charly, Chente, Chuy, el Hippie, Karla, Eréndira) por su
compañía, por las fiestas, porque me caen bien. A Gabiota y Roberta
por no cuestionarme, por quedarse conmigo y apoyarme en las decisiones
que tomé, por soñar juntos y por las tardes gastadas planeando lo que
haremos cuando seamos doctores.

\smallskip

\noindent
A Roberto Vázquez por descubrirme, por creer en mi, por fracasar
tratando de acercarme a la astronomía óptica y por su invaluable
amistad. A Yolanda por permitirme desarrollar en la parte de
divulgación. A Ramiro por ser mi ejemplo a seguir desde los
propedéuticos y por ser tan chido el tiempo que compartimos oficina. A
Yolanda, Luis Felipe, Paola y Javier por ayudarnos a salir adelante
estos últimos meses. A mis astrónomas favoritas: Amy, Yolanda y Paola,
por todo su cariño, por acercarme a sus familias y porque las quiero
un montón. A Gusana por seguir al otro lado de la línea. A Paty por
derecho de antigüedad. A Nena porque siempre está cuando la
necesito. A Moni por las noches de estrellas. A Gabiota, porque estos
años hubieran sido imposibles sin su amistad.

\smallskip

\noindent
A Mamá, Abuelita, Abuelito, Lore y César porque los adoro y me aceptan
con todas mis virtudes y todos mis defectos. A mis tres amores: Kike,
Tella y Mimí, por superar juntos todo lo que vivimos y porque además
de hermanos son mis grandes amigos. Agradezco de manera especial a mi
familia por su amor incondicional (y condicional a veces), por no
dejar que desaparezca la esperanza, por enseñarme que la felicidad se
consigue con pasión, mucho trabajo y tenacidad. Les agradezco a cada
uno de ellos por darme su ejemplo, bueno o malo, que finalmente me ha
enseñado cómo \textit{quiero} y cómo \textit{no quiero} ser.

\smallskip

\noindent
Y a Rami porque es lo mejor que me pasó en el posgrado. Por
reencontrarme después de tres largos años y darle un rumbo a mi
vida. Por quedarse conmigo, por los planes para el postdoc, por
hacerme sonreír todos los días, por nuestras vidas juntos y porque lo
amo.

\begin{flushright}
\textit{Rosy}\\
\end{flushright}

\begin{flushleft}
Morelia, Michoacán, México\\
Septiembre del 2009
\end{flushleft}

\cleardoublepage

\thispagestyle{empty}
\phantom{hola}
\begin{center}
\vfill

{\Large \textit{To Ramiro}}

\vfill
\end{center}
\cleardoublepage

\renewcommand{\thepage}{\roman{page}}
\pagestyle{fancy}


\tableofcontents

\phantom{hola}

\begin{center}
\textbf{\huge Resumen}
\addcontentsline{toc}{chapter}{Resumen}
\markright{Resumen}
\markboth{Resumen}{Resumen}
\end{center}

\vspace{1cm}

\noindent
{\large \bf Introducción}

\bigskip

\noindent
El problema de la determinación de distancias en el Universo siempre
ha jugado un papel central en la astronomía. A través del tiempo se
han ido desarrollando diferentes métodos para estimar la distancia a
los cuerpos celestes. La gran mayoría de estos métodos son indirectos
y, para objetos fuera del Sistema Solar, solamente hay una técnica
disponible que es cien por ciento directa, sin suposiciones y
puramente geométrica: la paralaje trigonométrica. A la fecha, muchas
de las mediciones de la paralaje trigonométrica han sido basadas en
observaciones ópticas o del cercano infrarrojo, en las cuales la
precisión astrométrica está limitada a 1 milisegundo de arco. Como
consecuencia, la determinación de la paralaje en el óptico e
infrarrojo cercano está limitada a la Vecindad Solar ($d$ $\lesssim$
500 pc).

\bigskip

\noindent
De hecho, a finales de los 90s nuestro conocimiento de la distribución
de las estrellas en la Vecindad Solar ha sido dramáticamente mejorado
gracias al satélite Hipparcos, que midió la paralaje trigonométrica a
miles de estrellas dentro de algunos cientos de parcecs alrededor del
Sol con una precisión de $\sim$ 5--10\% (Perryman et al.\ 1997; van
Leeuwen 2007). Trabajando a longitudes de onda del óptico, Hipparcos
representó una mejora para estrellas brillantes aisladas y fue
comparativamente pobre para aquellas en regiones de formación
estelar. Esto se debe a que las estrellas jóvenes se encuentran
embebidas en su nube madre, lo cuál las hace ópticamente débiles o
tienden a estar rodeadas de nebulosidades. Las barras de error típicas
de Hipparcos para la distancia a estrellas cercanas de pre-secuencia
principal por lo regular exceden 100 pc (resultando en precisiones
típicas del 30\% o peores para la distancia). Consecuentemente,
mientras el satélite Hipparcos nos permitió penetrar en el
entendimiento de la distribución local de estrellas de secuencia
principal, no mejoró mucho nuestro conocimiento sobre la distancia a
regiones de formación estelar cercanas (Bertout et al.\ 1999).

\bigskip

\noindent
Entonces en el 2005, cuando empecé a trabajar en los datos que
presentamos en éste manuscrito, las mejores estimaciones para la
distancia a regiones de formación estelar cercanas seguían siendo
basadas en métodos indirectos. Podría decirse que la situación más
favorable era para el complejo de Tauro, al cuál se le estimaba una
distancia promedio de $140\pm10$ pc (Kenyon et al.\ 1994). Sin
embargo, aún no estaba claro qué tan profundo era el complejo y si
existían gradientes significantes en la distancia a lo largo de la
región. La situación para Ophiuchus era un poco más
incierta. Tradicionalmente se ha estimado una distancia de $165\pm20$
pc (Chini 1981) y recientemente se había sugerido que está más cerca,
a 120--125 pc (de Geus et al.\ 1989; Knude \& Hog 1998). Sin embargo
muchos autores aún prefieren el antiguo valor o uno intermedio de 140
pc (e.g.\ Mamajek 2007). La precisión en la distancia para otros
sitios de formación estelar cercanos menos estudiados (Perseo,
Serpens, Monoceros, Orion, etc.) era más o menos similar o peor, no
mejor que un 20\%.

\bigskip

\noindent
Había entonces una necesidad urgente de mejorar el estado
insatisfactorio del asunto. Primero, porque las grandes incertidumbres
en la distancia limitan la determinación precisa de las propiedades de
las estrellas jóvenes como la masa o la luminosidad. Esto a su vez
limita fuertemente la comparación de modelos teóricos de evolución de
estrellas de pre-secuencia principal con los datos
observacionales. Adicionalmente, la determinación de la estructura
interna y la cinemática de las regiones de formación estelar cercanas
(las cuales pueden ser obtenidas a partir de mediciones astrométricas
precisas) proveen información valiosa sobre la historia de la
formación estelar dentro de cada región y sobre los procesos de
formación estelar mismos. Finalmente, los sitios de formación estelar
no están distribuidos aleatoriamente dentro de la Vía Láctea. En
particular, muchas de las regiones de formación estelar en la Vecindad
Solar parecen estar aproximadamente alineadas en una estructura
toroidal llamada el Cinturón de Gould. Entender el origen de dicha
estructura es relevante en el contexto de Astronomía Galáctica y
requiere una determinación precisa de su geometría. La información
necesaria se puede obtener a partir de la determinación precisa de la
distancia y la profundidad de cada una de las regiones que forman el
Cinturón de Gould.

\bigskip

\noindent
Debido a que las observaciones en el óptico e infrarrojo cercano son
afectadas por el obscurecimiento por polvo, uno debe moverse a
longitudes de onda más grandes para progresar en la determinación de
la paralaje para estrellas jóvenes. Recientemente, la calidad de la
astrometría proveniente de datos del mediano y lejano infrarrojo, así
como de observaciones submilimétricas, sigue siendo pobre. El dominio
del radio (particularmente a 1 cm $\lesssim$ $\lambda$ $\lesssim$ 10
cm) provee, por mucho, la mejor posibilidad porque los grandes
interferómetros de radio pueden proporcionar astrometría
extremadamente precisa (mejor que una décima de milisegundo de arco).

\vspace{1cm}

\noindent
{\large \bf Metas}

\bigskip

\noindent
Esta tesis es parte de una gran cadena de esfuerzos para determinar la
distancia y estructura a todas las regiones de formación estelar
alrededor de algunos cientos de parcecs del Sol usando observaciones
de radio interferometría. Específicamente, las metas principales de
esta tesis pueden resumirse como sigue:

\begin{enumerate}

\item
 Encontrar la distancia promedio a las dos regiones de formación
 estelar cercanas de baja masa más estudiadas (Tauro y Ophiuchus) con
 precisiones de uno o dos órdenes de magnitud mejores que los valores
 presentes. La estrategia básica para lograr ésta meta fue obtener,
 durante el curso de uno o dos años, imágenes multi época con
 radio interferómetros de algunas estrellas jóvenes en cada región. La
 posición de cada estrella a cada época y la paralaje de cada estrella
 puede ser medida muy precisamente a través de una calibración de
 datos cuidadosa. Una vez que conocemos la distancia a cada una de las
 estrellas individuales, simplemente tomamos la media para estimar la
 distancia promedio de cada región.

\item
 Explorar la estructura y la dinámica de estas regiones de formación
 estelar. En particular, la determinación de la distancia a algunas
 estrellas dentro de cada complejo nos permite estimar la profundidad
 y obtener una aproximación de la forma de cada
 región. Adicionalmente, los movimientos propios (que son medidos
 simultáneamente con la paralaje trigonométrica) combinados con
 mediciones de velocidades radiales tomados de la literatura, proveen
 algunas pistas de la cinemática interna de cada región, la cual puede
 ser analizada en términos de modelos dinámicos.

\item
 Estudiar las estrellas por sí mismas. En particular, a través de la
 combinación de temperatura efectiva y luminosidad bolométrica de la
 literatura (reescaladas a las distancias propias de las fuentes),
 hemos sido capaces de refinar la localización en el diagrama HR de
 las estrellas jóvenes estudiadas en ésta tesis. Esto, sucesivamente,
 nos permite refinar la determinación de las edades y poner
 restricciones en los modelos evolutivos de pre-secuencia
 principal. También algunas estrellas en nuestra muestra han resultado
 ser sistemas binarios muy cercanos. En algunos de estos casos
 nuestras observaciones nos permiten refinar la determinación de las
 órbitas y de las masas de los componentes. Esto provee fuertes
 restricciones sobre las propiedades intrínsecas de las estrellas.

\end{enumerate}

\vspace{1cm}

\noindent
{\large \bf Observaciones}

\bigskip

\noindent
Nuestra muestra contiene un total de siete objetos estelares jóvenes:
cinco en el complejo de Tauro (T~Tau~Sb, HDE~283572, Hubble~4,
HP~Tau/G2 y V773~Tau~A), y dos en la región de Ophiuchus (S1 y
DoAr~21). Seis de estos siete objetos son estrellas de pre-secuencia
principal de baja a intermedia masa (M $\lesssim$ 3 \Msun). La séptima
fuente (S1 en Ophiuchus) es una estrella joven de secuencia principal
tipo B. Todos los siete objetos han sido previamente bien
identificados como fuentes brillantes de radio no térmicas, detectados
usando interferómetros de línea de base muy larga (VLBI por sus siglas
en inglés).

\bigskip

\noindent
Para cada fuente, hemos obtenido series de observaciones en el
continuo a 3.6 cm (8.42 GHz) usando el {\it arreglo de linea de base
muy larga} (VLBA por sus siglas en inglés). El número de observaciones
en cada serie, así como la frecuencia con la cuál fueron obtenidas,
fue ajustado de acuerdo a cada fuente. Para todas las observaciones
usamos referencia de fase, método a través del cual se combina la
observación del objetivo científico en cuestión y un calibrador
compacto localizado a unos pocos grados de la fuente. La resolución
angular típica de nuestras imágenes finales fue de 1 a 2 milisegundos
de arco, y la precisión astrométrica típica de cada observación fue de
0.05 a 0.1 milisegundos de arco para las fuentes en Tauro, y de 0.2 a
0.6 milisegundos de arco en Ophiuchus. La pobre precisión en las
observaciones de Ophiuchus se debe a la baja declinación de esta
región ($\delta = -24^\circ$ contra $\delta = +20^\circ$ para Tauro) y
a la binaridad no modelada de las fuentes (ver abajo).

\newpage

\noindent
{\large \bf Resultados principales}

\bigskip

\noindent
{\bf Distancias}

\bigskip

\noindent
Para T~Tau~Sb fueron obtenidas 12 observaciones que resultaron en una
paralaje de $6.82\pm0.03$ milisegundos de arco, correspondiente a una
distancia de $d=146.7\pm0.6$ pc (Loinard et al.\ 2005, 2007). Para
Hubble~4 y HDE~283572 fueron suficientes series de seis observaciones
para precisar las paralajes en $7.53\pm0.03$ milisegundos de arco
($d=132.8\pm0.5$ pc) y $7.78\pm0.04$ milisegundos de arco
($d=128.5\pm0.6$ pc), respectivamente (Torres et al.\ 2007). Para
HP~Tau/G2 se obtuvieron nueve observaciones que dieron la paralaje de
$6.20\pm0.03$ milisegundos de arco, correspondiente a $d=161.2\pm0.9$
pc (Torres et al.\ 2009a). Finalmente, para V773~Tau~A se hicieron 19
observaciones que nos permitieron medir una paralaje de $7.57\pm0.20$
milisegundos de arco, correspondiente a $d=132.0\pm3.5$ pc (Torres et
al.\ 2009b).

\bigskip

\noindent
Dos de estos cinco objetos tienen paralaje trigonométrica medida por
Hipparcos (Bertout et al.\ 1999): T~Tau~Sb con $5.66\pm1.58$
milisegundos de arco y HDE~283572 con $7.81\pm1.30$ milisegundos de
arco. Nuestros resultados son consistentes con estos valores, pero uno
o dos órdenes de magnitud más precisos. Además, la paralaje
trigonométrica de Hubble~4 y HDE~283572 ha sido estimada por Bertout
\& Genova (2006) usando un método de punto convergente modificado. Sus
resultados ($8.12\pm1.50$ milisegundos de arco para Hubble~4 y
$7.64\pm1.05$ milisegundos de arco para HDE~283572) también son
consistentes con los nuestros, pero con un orden de magnitud menos
precisos que los nuestros. Anteriormente ha sido determinada la
distancia a V773~Tau~A ($d=148.4\pm5.3$ pc; Lestrade et al.\ 1999)
usando técnicas de VLBI. Nuestros resultados son marginalmente
consistentes con los anteriores y argumentamos que la discrepancia se
debe al hecho de que Lestrade et al.\ (1999) no modeló la binaridad de
la fuente en su análisis (se sabe que dicha fuente es una binaria
espectroscópica con un periodo orbital de 51 días y semieje mayor de
unos cuantos milisegundos de arco). Adicionalmente obtuvimos una
estimación independiente de la distancia a V773~Tau~A modelando la
órbita física de la binaria (usando una combinación de mediciones en
el óptico para velocidades radiales, observaciones del interferómetro
Keck y nuestros propios datos obtenidos con el VLBA --ver abajo). La
distancia obtenida con éste método alternativo es de $134.5\pm3.2$ pc,
consistente con el valor obtenido a partir de nuestras mediciones de
paralaje, pero nuevamente marginalmente consistente con el valor a
partir de VLBI por Lestrade et al.\ (1999). La distancia promedio
resultante para el complejo de Tauro es de alrededor de 139 pc,
consistente con determinaciones previas. Sin embargo, el complejo
parece tener 30 pc de profundidad, lo cuál discutiremos en un momento.

\bigskip

\noindent
Para S1 y DoAr~21 se obtuvieron seis y siete observaciones
respectivamente. Las paralajes resultantes son $8.55\pm0.50$ y
$8.20\pm0.37$ milisegundos de arco respectivamente. Las cuáles
corresponden a $d=116.9^{+7.2}_{-6.4}$ pc para S1 y
$d=121.9^{+5.8}_{-5.3}$ pc para DoAr~21 (Loinard et al.\ 2008). Note
que la incertidumbre en la distancias para ambas estrellas en
Ophiuchus son significativamente más grandes que aquellas obtenidas
para las estrellas en Tauro. Como se mencionó anteriormente, esto es
una consecuencia de la baja declinación en Ophiuchus relativa a Tauro,
y debido a la binaridad no modelada en las dos fuentes de
Ophiuchus. La media de estas dos paralajes es $8.33\pm0.30$
milisegundos de arco, correspondiente a una distancia de
$120.0^{+4.5}_{-4.2}$ pc. Debido a que S1 y DoAr~21 son miembros del
núcleo de Ophiuchus, dicha distancia debe representar una buena
estimación de la distancia a ésta región de formación estelar. Lo
anterior es consistente con varias de las recientes determinaciones
(p.\ ej.\ Knude \& Hog 1999; de Geus et al.\ 1989; Lombardi et al.\
2008).

\vspace{1cm}

\noindent
{\bf Multiplicidad}

\bigskip

\noindent
Para todas las fuentes de nuestra muestra, las trayectorias observadas
sobre el plano del cielo pueden ser descritas como la combinación del
movimiento elíptico de la Tierra alrededor del Sol y del movimiento
propio de la estrella. Si la fuente es soltera, esperaríamos un
movimiento propio lineal y uniforme. Sin embargo, si la fuente es
miembro de un sistema múltiple, el movimiento propio observado será la
combinación del movimiento propio uniforme del baricentro del sistema
y el movimiento orbital acelerado. Tres de nuestras fuentes (Hubble~4,
HDE~283572 y DoAr~21) anteriormente habían sido reportadas como
solteras, pero las otras cuatro como sistemas múltiples. En uno de
estos cuatro casos (HP~Tau/G2), el periodo orbital esperado es de
algunos miles de años, por lo tanto, la correspondiente aceleración es
muy pequeña como para ser detectada con nuestras observaciones. A
consecuencia de esto, esperábamos un movimiento propio uniforme de la
fuente, como si se tratara de una fuente soltera. Por otro lado, el
periodo orbital de T~Tau~Sb es de unas cuantas décadas, pequeño pero
no insignificante comparado con el lapso de tiempo de nuestras
observaciones. En este caso pensamos que era suficiente incluir un
término de movimiento acelerado en el ajuste astrométrico. Finalmente,
en los dos últimos casos (V773~Tau~A en Tauro y S1 en Ophiuchus)
esperábamos un periodo orbital más corto que el lapso de tiempo de
nuestras observaciones. En tales situaciones uno debería, en
principio, caracterizar por completo el movimiento orbital antes de
poder realizar un ajuste astrométrico completo. Esto fue lo que
hicimos para V773~Tau~A, pero no para S1 porque suponíamos que S1 era
mucho más masiva que su compañera, y como consecuencia, esperábamos
una amplitud pequeña de su movimiento reflejo.

\bigskip

\noindent
En resumen, para Hubble~4, HDE~283572 y DoAr~21 esperábamos
movimientos propios uniformes porque creíamos que las tres fuentes
eran solteras. Vale la pena mencionar que el movimiento propio de
HP~Tau/G2 también debería ser uniforme a nuestro nivel de
precisión. Para S1 esperábamos el movimiento propio uniforme pero con
un pequeño residuo periódico debido a la compañera. Finalmente, para
T~Tau~Sb esperábamos obtener un buen ajuste con un movimiento propio
uniformemente acelerado, mientras que para V773~Tau~A estábamos
conscientes de que se requería una solución orbital completa.

\bigskip

\noindent
Cuatro de nuestras fuentes se encontraron de a cuerdo a nuestras
expectativas:

\begin{itemize}

\item[-]
 Encontramos a V773~Tau~A como una fuente de radio doble resuelta. La
 órbita física del sistema se construyó a partir de la combinación de
 nuestros datos del VLBA, mediciones en el óptico de las velocidades
 radiales y datos del interferómetro Keck (Torres et al.\ 2009b). Este
 ajuste nos permitió refinar la determinación de las masas de las dos
 estrellas en el sistema y deducir la posición del baricentro del
 sistema a cada época. Realizando un nuevo ajuste astrométrico para
 éstas posiciones anteriores (asumiendo movimiento propio uniforme del
 baricentro) nos dio la misma distancia obtenida con el método
 anterior.

\item[-]
 Para describir una adecuada trayectoria de T~Tau~Sb, asumimos un
 ajuste con movimiento propio uniformemente acelerado. En este caso,
 nuestras observaciones no fueron suficientes para determinar la
 órbita del sistema, pero proveen información consistente y
 complementaria con datos existentes del cercano infrarrojo.

\item[-]
 Como se esperaba, el movimiento propio de HDE~283572 se encontró
 lineal y uniforme.

\item[-]
 Se pudo obtener un ajuste adecuado para S1 asumiendo el movimiento
 propio como uniforme, pero se encontraron residuos aproximadamente
 periódicos (en ascensión recta y en declinación) con un periodo de
 alrededor de 0.7 años. Esto es del orden de magnitud correcto si se
 interpreta como el movimiento reflejo de S1 debido a su compañera
 (Loinard et al.\ 2008, Richichi et al.\ 1994).
 
\end{itemize}

\bigskip

\noindent
Las otras tres fuentes no se encontraron como se esperaba. Para
Hubble~4 y HP~Tau/G2 (donde se esperaban movimientos propios
uniformes), encontramos residuos significantes en declinación (pero no
en ascensión recta). Hasta el momento no es claro si los residuos son
consecuencia de una compa-ñera que no vemos, debido a estructura en las
magnetósferas de las estrellas, o debido a errores residuales de
calibración en la fase. Finalmente, DoAr~21 se encontró muy parecida a
S1: con movimiento propio uniforme que provee un buen ajuste, pero con
residuos aproximadamente periódicos (en este caso de alrededor de 1.2
años), y la fuente se encontró doble por lo menos en una de nuestras
imágenes. Concluimos que DoAr~21 probablemente pertenece a un sistema
binario.

\bigskip

\noindent
Entonces, por lo menos 4 de las 7 fuentes en nuestra muestra (57\%)
son sistemas binarios muy cercanos con separaciones entre unos pocos
milisegundos de arco y unas cuantas decenas de milisegundos de
arco. Esto representa una fracción de binaridad mucho más grande que
aquella para estrellas de secuencia principal con el mismo rango de
separación. Nosotros argumentamos que probablemente hay un fuerte
efecto de selección. Los sistemas considerados en ésta tesis fueron
seleccionados porque eran conocidos previamente como emisores no
térmicos con técnicas de VLBI. Entonces, la alta tasa de binaridad
probablemente nos indica que las binarias muy cercanas tienen mayor
emisión en radio que las binarias con separaciones más grandes o que
las estrellas solteras. Esta idea se refuerza con las observaciones de
V773~Tau~A donde confirmamos que el flujo emitido depende fuertemente
de la separación entre las dos estrellas (el flujo incrementa cerca
del periastro y es muy débil al apoastro). Se cree que la emisión no
térmica se crea durante los eventos de reconexión en las magnetósferas
activas de las estrellas. Los eventos de reconexión {\em entre} las
estrellas (además de los producidos dentro de las magnetósferas
individuales) de un sistema binario muy cercano, podrían explicar
naturalmente el flujo de radio tan alto en V773~Tau~A.

\bigskip

\noindent
{\bf Estructura tridimensional}

\bigskip

\noindent
La extensión espacial total de Tauro en el cielo es de aproximadamente
$10^\circ$, correspondiente a un tamaño físico de alrededor de 25
pc. Nuestras observaciones muestran que la profundidad del complejo es
de un tamaño similar debido a que HP~Tau/G2 está aproximadamente 30 pc
más lejos que Hubble~4, HDE~283572 o V773~Tau~A. Esto tiene una
consecuencia importante: aunque la distancia promedio a la asociación
de Tauro fuera determinada con una precisión infinita, se estarían
cometiendo errores del rango de 10--20\% si uno usara
indiscriminadamente dicha distancia promedio para cualquier fuente en
Tauro. Para reducir ésta sistemática fuente de error, necesitaríamos
establecer la estructura tridimensional de toda la asociación de
Tauro, lo cuál se lograría con más observaciones similares a las que
presentamos aquí, y así nos acercaríamos más a la meta. Sin embargo,
las observaciones de las cinco estrellas presentadas en esta tesis nos
dan algunas pistas de cómo debería de ser la estructura tridimensional
de Tauro. Hubble~4, HDE~283572 y V773~Tau~A, que se encontraron a
aproximadamente 130 pc y con una cinemática parecida (Torres et al.\
2007, 2009b), están localizadas en la misma porción de Tauro, cerca de
Lynds 1495. T~Tau~Sb está localizado en la parte Sur de Tauro, cerca
de Lynds 1551, su velocidad tangencial es claramente diferente a la
que presentan Hubble~4, HDE~283572 y V773~Tau~A, y además parece estar
un poco más lejos de nosotros. Finalmente, HP~Tau/G2 está localizada
cerca del borde Este (Galáctico)de Tauro y es la más lejana de las
cinco fuentes consideradas aquí. Aunque se necesitan observaciones
adicionales para dar una conclusión definitiva, nuestros datos
sugieren que la región alrededor de Lynds 1495 corresponde al lado más
cercano del complejo de Tauro, a aproximadamente 130 pc, mientras que
el lado Este de Tauro corresponde a la parte más lejana a 160 pc. La
región alrededor de Lynds 1551 y T~Tau~Sb parecen estar a una
distancia intermedia de alrededor 147 pc.

\bigskip

\noindent
Es bien sabido que Tauro presenta estructura filamentaria. Los dos
filamentos principales están aproximadamente paralelos uno del otro y
tienen un eje de proporción de 7:1. Nuestras observaciones sugieren
que estas estructuras filamentarias están orientadas casi a lo largo
de la línea de visión, p.\ ej.\ aproximadamente a lo largo del eje
centro--anticentro Galáctico. Esta orientación peculiar podría
explicar la baja eficiencia de formación estelar en Tauro comparada
con otras regiones de formación estelar cercanas (Ballesteros-Paredes
et al.\ 2009).

\bigskip

\noindent
Ophiuchus está compuesto de un núcleo compacto de sólo 2 pc de tamaño,
y estructuras filamentarias (llamadas ``streamers'') que se extienden
(en proyec-ción) alrededor de 10 pc. El núcleo de Ophiuchus es
suficientemente compacto que no esperamos resolver estructuras a lo
largo de la línea de visión, y nuestras observaciones muestran que
dicho núcleo se encuentra a una distancia de 120 pc. Podría haber
gradientes en distancia de algunos cuantos parcecs a lo largo de los
streamers. De hecho, hemos notado que Schaefer et al.\ (2008)
determinó la órbita física del sistema binario Haro 1-14C y dedujo una
distancia de $111\pm19$ pc, consistente con nuestra
determinación. Haro 1-14C está localizado en el streamer norte
(asociado con las nubes obscuras L1709/L1704), entonces el resultado
de Schaefer et al.\ (2008) sugiere que ese streamer está un poco más
cerca de nosotros que el núcleo. De hecho, esto es consistente con
resultados recientes de Lombardi et al.\ (2008). Por otro lado, usando
el sistema VLBI japonés (VERA), Imai et al.\ (2007) determinaron la
paralaje a la protoestrella IRAS~16293--2422 que se encuentra embebida
en el streamer Sur de Ophiuchus (en L1689N). Ellos obtuvieron una
distancia de $178^{+18}_{-37}$ pc, la cuál es más consistente con el
antiguo valor de 165 pc usado para Ophiuchus. Aún incluyendo los
streamers, es improbable que Ophiuchus tenga 60 pc de profundidad
debido a que Ophiuchus tiene apenas unos 10 pc de tamaño en
proyección. Entonces, si los resultados de Imai et al.\ (2007) son
confirmados, estos indicarían la existencia de varias regiones de
formación estelar a lo largo de la línea de visión que no están
relacionadas. Se necesitan más observaciones, algunas de las cuales ya
han sido recolectadas y parcialmente analizadas, para llegar a un
acuerdo.

\newpage

\noindent
{\large \bf Perspectivas}

\bigskip

\noindent
Los resultados presentados en ésta tesis nos han permitido refinar la
determinación de la distancia a las dos más importantes regiones de
formación estelar cercanas y comenzar a examinar la estructura
tridimensional de Tauro. Además nos proporcionan un gran número de
temas que podrían ser abordados con nuevos datos y sugiere algunos de
los siguientes estudios.

\bigskip

\begin{itemize}

\item[-]
 Como hemos mencionado, el radio flujo de V773~Tau~A depende de la
 fase de la órbita, siendo más alto al periastro y más débil al
 apoastro. En muchas de nuestras observaciones, la fuente se resuelve
 en dos componentes asociadas con las dos estrellas del sistema. No
 obstante, nuestras observaciones han mostrado que cerca del periastro
 la posición de las radio fuentes están significativamente desplazadas
 de la posición de las estrellas asociadas. Esto nos da evidencia
 adicional de que la emisión no térmica en el sistema es afectada por
 la presencia de la compañera. En una propuesta recientemente
 aceptada, solicitamos tiempo para observar V773~Tau~A cerca del
 periastro con el High Sensitivity Array (arreglo interferométrico
 compuesto por el VLBA, el radiotelescopio de Green Bank, el VLA y el
 disco de Arecibo en Puerto Rico). Estas observaciones nos permitirán
 examinar la evolución espacial de la binaria cuando está cerca del
 periastro en un periodo de seis horas. Esto debería decirnos mucho a
 cerca del origen de la variabilidad de la fuente.

\item[-]
 Para decidir si están o no relacionadas varias regiones de formación
 estelar existentes en Ophiuchus a lo largo de la linea de visión,
 hemos estado observando algunas nuevas fuentes en la dirección del
 complejo: VSSG~14 en la subregión Oph--B y ROX~39 en la región entre
 L1686 y L1689.

\item[-]
 Otras regiones de formación estelar (p.\ ej.\ Serpens o Perseo) han
 estado siendo estudiadas con mucho detalle a muchas otras longitudes
 de onda, pero no tienen una buena determinación de la distancia. Se
 sabe que existen algunas fuentes no térmicas en dichas regiones,
 entonces observaciones multiépoca con el VLBA permitirían mejorar
 significativamente la determinación de sus distancias. Ya hemos
 obtenido algunas de las observaciones para Serpens, mientras que para
 Perseo serán parte de una propuesta futura.

\end{itemize}
\cleardoublepage

\phantom{hola}

\begin{center}
\textbf{\huge Summary}
\addcontentsline{toc}{chapter}{Summary}
\markright{Summary}
\markboth{Summary}{Summary}
\end{center}

\vspace{1cm}

\noindent
{\large \bf Introduction}

\bigskip

\noindent
The problem of the determination of distances in the Universe has
always played a central role in astronomy, and many different methods
have been devised to obtain distance estimates to celestial sources.
The vast majority of these methods are indirect, and --for objects
outside of the Solar System-- there is only one direct,
assumption-free, purely geometrical technique available: the
determination of the trigonometric parallax. Most parallax
measurements to date have been based on optical or near-infrared
observations, whose astrometric accuracy is limited to about 1
milli-arcsecond. As a consequence, optical and near-infrared parallax
determinations are largely limited to the Solar Neighborhood ($d$
$\lesssim$ 500 pc).

\bigskip

\noindent
Indeed, our knowledge of the distribution of stars in the Solar
Neighborhood dramatically improved at the end of the 1990s thanks to
the Hipparcos satellite mission. Hipparcos measured the
trigonometric parallax of thousands of stars within several hundred
parsecs of the Sun with a precision better than $\sim$ 5--10\%
(Perryman et al.\ 1997; van Leeuwen 2007). Working at optical
wavelengths, it performed best for optically bright isolated stars,
and did comparatively poorly in star-forming regions. This is because
young stars tend to be surrounded by nebulosities, and --being still
enshrouded in their parental cloud-- they are usually optically rather
faint. Indeed, the typical Hipparcos error bars for the distances to
individual nearby pre-main sequence stars usually exceeded 100 pc
(resulting in typical distance accuracies of 30\% or
worse). Consequently, the Hipparcos satellite, while allowing a
breakthrough in our understanding of the local distribution of main
sequence stars, did little to improve our knowledge of the distance to
nearby star-forming regions (Bertout et al.\ 1999).

\bigskip

\noindent
Thus, in 2005 when I started working on the data that will be
presented in this manuscript, the best estimates for the distances to
the nearest star-forming regions remained those based on indirect
methods. Arguably the most favorable situation was that of the Taurus
complex which was assumed to be at an average distance of 140 $\pm$ 10
pc (Kenyon et al.\ 1994). It remained unclear, however, how deep the
complex was, and if significant distance gradients existed across the
region. The situation for Ophiuchus was a bit more
uncertain. Traditionally assumed to be at 165 $\pm$ 20 pc (Chini 1981),
it had recently been suggested to be somewhat nearer: at 120--125 pc
(de Geus et al.\ 1989; Knude \& Hog 1998). Many authors, however,
still preferred the older value, or an intermediate one of about 140
pc (e.g.\ Mamajek 2007). The accuracy for other, less studied, nearby
sites of star formation (Perseus, Serpens, Monoceros, Orion, etc.)
was similar or worse, and it is fair to say that the distances to
nearby star-forming regions were, on average, not known to better than
20\%.

\bigskip

\noindent
There was an urgent need to improve this unsatisfactory state of
affair. First because the large distance uncertainties precluded
accurate determinations of such basic properties of young stars as
their mass or luminosity. This, in turn, strongly limited the
precision with which theoretical pre-main sequence evolutionary models
could be constrained by observational data. In addition, the
determination of the internal structure and kinematics of star-forming
regions (which can be obtained from accurate astrometric measurements)
provides valuable information on the history of star-formation within
each regions, and on the star-forming process itself. Finally, sites
of star-formation are not randomly distributed within the Milky
Way. In particular, most of the star-forming regions in the Solar
Neighborhood appear to delineate a roughly toroidal structure called
the Gould Belt. Understanding the origin of such a structure is
relevant in the context of Galactic Astronomy, and requires an
accurate determination of its exact geometry. Precise determinations
of the distance and depth of each of the regions forming the Gould
Belt would directly provide the required constraints.

\bigskip

\noindent
Since optical and near-infrared observations are affected by dust
obscuration, one must turn to longer wavelengths to make progress in
the determination of the parallax of young stars. Currently, the
astrometry quality provided by mid- and far-infrared data as well as
by sub-millimeter observations remains poor. The radio domain
(particularly at 1 cm $\lesssim$ $\lambda$ $\lesssim$ 10 cm) provides,
by far, the best prospect because large radio-interferometers can
deliver extremely accurate astrometry (better than a tenth of a
milli-arcsecond).

\newpage

\noindent
{\large \bf Goals}

\bigskip

\noindent
This thesis inserts itself in a large ongoing effort to determine the
distance and structure of all star-forming regions within several
hundred parsecs of the Sun using radio-interferometric
observations. Specifically, the main goals of the thesis can be
summarized as follows:

\begin{enumerate}

\item
 Find the mean distance to the two best-studied nearby regions of
 low-mass star-formation (Taurus and Ophiuchus) with accuracies (a few
 percent or better) one to two orders of magnitude better than the
 present values. The basic strategy to achieve this goal was to obtain
 multi-epoch radio-interferometric images of several young stars in
 each region over the course of one to two years. Through careful data
 calibration, the position of each star at each epoch, and the
 parallax of each star could be measured very accurately. Once the
 distances to the individual stars were known, we simply took their
 weighted mean to estimate the average distance of each region.

\item
 Explore the structure and dynamics of these star forming regions. In
 particular, the determination of the distance to several stars within
 each complex allowed us to estimate the depth and to obtain a rough
 approximation of the shape of each region. In addition, the proper
 motions (which are measured simultaneously with the trigonometric
 parallaxes) combined with radial velocity measurements taken from the
 literature provide some hints on the internal kinematics of each
 region, which can be analyzed in terms of dynamical models.

\item
 Study the stars themselves. In particular, we were able to refine the
 location in the HR diagram of the young stars studied in this thesis
 by combining effective temperature and bolometric luminosities
 (re-scaled to the proper distances) taken from the literature. This,
 in turns, allowed us to refine age determinations, and to put
 constraints on pre-main sequence evolutionary models. Also, several
 stars in our sample turned out to be tight binary systems. In some of
 these cases, our observations allowed us to refine the determination
 of their orbital paths, and of the mass of the individual components.
 This provided further constrains on the intrinsic properties of the
 stars.

\end{enumerate}

\newpage

\noindent
{\large \bf Observations}

\bigskip

\noindent
Our sample contains a total of seven young stellar objects: five in
the Taurus complex (T~Tau~Sb, HDE~283572, Hubble~4, HP~Tau/G2, and
V773~Tau~A), and two in the Ophiuchus region (S1 and DoAr~21). Six of
these seven objects are low- to intermediate-mass (M $\lesssim$ 3
\Msun) pre-main sequence stars. The seventh source (S1 in Ophiuchus)
is a young main sequence B star. All 7 objects were previously known
to be fairly bright non-thermal radio sources, detectable using Very
Long Baseline Interferometers.

\bigskip

\noindent
For each source, we obtained a series of 3.6 cm (8.42 GHz) continuum
observations using the {\it Very Long Baseline Array} (VLBA) of the
National Radio Astronomy Observatory (NRAO). The number of
observations in each series as well as the cadence at which they were
obtained were adjusted to each source. Phase-referencing --whereby
observations of the scientific target and a compact calibrator located
within a few degrees of the source are intertwined-- was used for all
the observations. The typical angular resolution of our final images
is 1 to 2 milli-arcsecond, and the typical astrometric accuracy of
each observation is 0.05 to 0.1 milli-arcseconds for the sources in
Taurus, and 0.2 to 0.6 milli-arcseconds in Ophiuchus. The poorer
performance in Ophiuchus is related to the lower declination of this
region ($\delta = -24^\circ$ against $\delta = +20^\circ$ for Taurus),
and to the unmodeled binarity of the sources (see below).

\vspace{1cm}

\noindent
{\large \bf Main results}

\bigskip

\noindent
{\bf Distances}

\bigskip

\noindent
For T~Tau~Sb, 12 observations were obtained, with a resulting parallax
of $6.82\pm0.03$ mas, corresponding to a distance $d=146.7\pm0.6$ pc
(Loinard et al.\ 2005, 2007). For Hubble~4 and HDE~283572, series of
six observations were sufficient to constrain the parallaxes to
$7.53\pm0.03$ mas ($d=132.8\pm0.5$ pc) and $7.78\pm0.04$ mas
($d=128.5\pm0.6$ pc), respectively (Torres et al.\ 2007). Nine
observations of HP~Tau/G2 were obtained, and yielded a parallax of
$6.20\pm0.03$ mas, corresponding to $d=161.2\pm0.9$ pc (Torres et al.\
2009a). Finally, for V773~Tau~A, 19 observations were obtained, and
allowed us to measure a parallax of $7.57\pm0.20$ mas, corresponding
to $d=132.0\pm3.5$ pc (Torres et al.\ 2009b).

\bigskip

\noindent
Two of these five objects have measured Hipparcos parallaxes (Bertout
et al.\ 1999): T~Tau~Sb with $5.66\pm1.58$ mas, and HDE~283572 with
$7.81\pm1.30$ mas. Our results are consistent with these values, but
one to two orders of magnitude more accurate. Also, the parallax of
both Hubble~4 and HDE~283572 were estimated by Bertout \& Genova
(2006) using a modified convergent point method. Their results
($8.12\pm1.50$ mas for Hubble~4 and $7.64\pm1.05$ mas for HDE~283572)
are also consistent with our results, but again more than one order of
magnitude less accurate. The distance to V773~Tau~A had been obtained
using Very Long Baseline Interferometer (VLBI) observations before
($d=148.4\pm5.3$ pc; Lestrade et al.\ 1999). Our result is only
marginally consistent with this earlier figure, and we argue that the
discrepancy is due to the fact that Lestrade et al.\ (1999) did not
model the binarity of the source in their analysis (this source is a
know spectroscopic binary with an orbital period of 51 days and a
semi-major axis of a few mas). Indeed, we obtained a mostly
independent estimate of the distance to V773~Tau~A by modeling the
physical orbit of the binary (using a combination of optical radial
velocity measurements, Keck Interferometer observations, and our own
VLBA data --see below). The distance obtained by this alternative
method is $134.5\pm3.2$ pc, in excellent agreement with the value
obtained from our parallax measurement, but again only marginally
consistent with the older VLBI value. The resulting mean distance to
the Taurus complex is about 139 pc, in excellent agreement with
previous determination. The complex, however, appears to be at least
30 pc deep, as will be discussed in a moment.

\bigskip

\noindent
Six observations of S1 and seven of DoAr~21 were obtained. The
resulting parallaxes are $8.55\pm0.50$ mas and $8.20\pm0.37$ mas,
respectively. This corresponds to $d=116.9^{+7.2}_{-6.4}$ pc for S1
and $d=121.9^{+5.8}_{-5.3}$ pc for DoAr~21 (Loinard et al.\
2008). Note that the uncertainties on the distances to the stars in
Ophiuchus are significantly larger than those for the stars in
Taurus. As mentioned earlier, this is a consequence of the lower
declination of Ophiuchus relative to Taurus, and of the unmodeled
binarity of both sources in Ophiuchus. The weighted mean of these two
parallaxes is $8.33\pm0.30$ mas, corresponding to a distance
$120.0^{+4.5}_{-4.2}$ pc. Since both S1 and DoAr~21 are members of the
Ophiuchus core, this figure must represent a good estimate of the
distance to this star-forming region. Note that it is in excellent
agreement with several recent determination (e.g.\ Knude \& Hog 1999;
de Geus et al.\ 1989; Lombardi et al.\ 2008).

\newpage

\noindent
{\bf Multiplicity}

\bigskip

\noindent
The observed trajectories on the plane of the sky for all the stars in
our sample can be described as combinations of a parallactic ellipse
and a proper motion. If the source is single, the proper motion is
expected to be linear and uniform. If the source is a member of a
multiple system, however, the observed proper motion will be the
combination of the uniform proper motion of the barycenter of the
system and the accelerated orbital motion. Three of our sources
(Hubble~4, HDE~283572, and DoAr~21) were believed to be single, but
the other four were previously known to be members of multiple
systems. In one case (HP~Tau/G2), the orbital period is expected to be
several thousand years, and the corresponding acceleration is too
small to be detectable with our observations. As a consequence, the
proper motion of the source is expected to be uniform as if it were
single. For T~Tau~Sb, on the other hand, the orbital period is a few
decades, small but not negligible compared with our observing time
span. In this situation, it is sufficient to include a uniform
acceleration term in the astrometric fit. Finally, in the last two
cases (V773~Tau~A in Taurus, and S1 in Ophiuchus), the orbital period
is expected to be shorter than the time spans covered by our
observations. In such situations, one should in principle fully
characterize the orbital motion before a full astrometric fit can be
performed. This was done in V773~Tau~A, but not in S1 because S1
itself is expected to be much more massive than its companion. As a
consequence, the amplitude of its reflex motion is anticipated to be
small.

\bigskip

\noindent
In summary, we expected to proper motions of Hubble~4, HDE~283572,
and DoAr~21 to be uniform because all three of these source were
believed to be single. We anticipated that the proper motion of
HP~Tau/G2 would also be uniform at our level of precision. For S1, we
expected the proper motion to be uniform but with a small periodic
residual due to its known companion. Finally, for T~Tau~Sb we expected
to obtain a good fit with a uniformly accelerated proper motion,
whereas we were aware that a complete orbital solution would have to
be sought for V773~Tau~A.

\bigskip

\noindent
Four of the sources were found to behave according to our
expectations:

\begin{itemize}

\item[-]
 V773~Tau~A was found to be a resolved double radio source. The
 physical orbit of the system was constrained by combining optical
 radial velocity measurements, Keck Interferometer data, and our own
 VLBA data (Torres et al.\ 2009b). This fit allowed us to refine the
 determination of the masses of the two stars in the system, and to
 deduce the position of the barycenter of the system at each epoch. An
 astrometric fit to these positions (assuming a uniform proper motion
 for the barycenter) yielded the distance quoted earlier.

\item[-]
 A fit assuming a uniformly accelerated proper motion appeared to
 provide an adequate description of the trajectory of T~Tau~Sb. In
 this case, our observations are not sufficient to constrain the orbit
 of the system, but they provide information consistent with, and
 complementary of, existing near infrared data.

\item[-]
 As expected, the proper motion of HDE~283572 was found to be linear
 and uniform.

\item[-]
 Assuming the proper motion of S1 to be uniform appears to provide an
 adequate fit, but with roughly periodic residuals (in both right
 ascension and declination) with a period of about 0.7 years. This is
 of the correct order of magnitude to be interpreted as the reflex
 motion of S1 due to its known companion (Loinard et al.\ 2008,
 Richichi et al.\ 1994).
 
\end{itemize}

\bigskip

\noindent
The other three sources did not behave as expected. For both Hubble~4
and HP~Tau/G2 (where the proper motions were expected to be uniform)
we find significant residuals in declination (but not in right
ascension). It is unclear at the moment whether these residuals are
the consequence of an unseen companion, of structure in the
magnetospheres of the stars, or of residual phase errors in our
calibration. Finally, DoAr~21 was found to behave much like S1: a
uniform proper motion provides a good fit but with roughly periodic
residuals (with a period of about 1.2 years in this case), and the
source was found to be double in at least one of our images. We
conclude that DoAr~21 is likely to belong to a binary system.

\bigskip

\noindent
Thus, at least 4 of the 7 sources in our sample (57\%) turn out to be
tight binary systems with separations between a few and a few tens of
milli-arcseconds. This represents a binarity fraction much larger than
that of main sequence stars for the same separation range. We argue
that a strong selection effect is likely to be at work. The systems
considered in this thesis were selected because they were known to be
non-thermal emitters previously detected with VLBI techniques. The
high binary rate may, therefore, indicate that tight binaries are more
likely to emit non-thermal radio emission than looser binaries or
single stars. This idea is reinforced by the observations of
V773~Tau~A where we confirm that the emitted flux depends strongly on
the separation between the two stars (the flux is strongest near the
periastron of the system, and weakest near apoastron). Non-thermal
emission is believed to be created during reconnection events in the
active magnetospheres of the stars. Reconnection events {\em between}
the stars (in addition to those within the individual magnetospheres)
in tight binary systems might naturally explain the higher radio flux
of V773~Tau~A.

\bigskip

\noindent
{\bf Three dimensional structure}

\bigskip

\noindent
The total spatial extent of Taurus on the sky is about $10^\circ$,
corresponding to a physical size of about 25 pc. Our observations show
that the depth of the complex is similar since HP~Tau/G2 is about 30
pc farther than Hubble~4, HDE~283572 or V773~Tau~A. This has an
important consequence: even if the mean distance of the Taurus
association were known to infinite accuracy, one would still make
errors as large as 10--20\% by using the mean distance
indiscriminately for all sources in Taurus. To reduce this systematic
source of error, one needs to establish the three-dimensional
structure of the Taurus association, and observations similar to those
presented here currently represent the most promising avenue toward
that goal. Indeed, the observations of the five stars presented here
already provide some hints of what the three-dimensional structure of
Taurus might be. Hubble~4, HDE~283572, and V773~Tau~A which were found
to be at about 130 pc and to share a similar kinematics (Torres et
al.\ 2007, 2009b), are also located in the same portion of Taurus,
near Lynds 1495. T~Tau~Sb is located in the southern part of Taurus
near Lynds 1551, its tangential velocity is clearly different from
that of Hubble~4, HDE~283572, and V773~Tau~A, and it appears to be
somewhat farther from us. Finally, HP~Tau/G2 is located near the
(Galactic) eastern edge of Taurus, and is the farthest of the five
sources considered here. Although additional observations are needed
to draw definite conclusions, our data, therefore, suggest that the
region around Lynds 1495 corresponds to the near side of the Taurus
complex at about 130 pc, while the eastern side of Taurus corresponds
to the far side at 160 pc. The region around Lynds 1551 and T~Tau~Sb
appears to be at an intermediate distance of about 147 pc.

\bigskip

\noindent
Taurus has long been known to present a filamentary structure. The two
main filaments are roughly parallel to one another, and have an axis
ratio of about 7:1. Our observations suggest that these filaments are
oriented nearly along the line of sight, i.e.\ roughly along the
Galactic center--anticenter axis. This peculiar orientation might
indeed explain the low star-forming efficiency of Taurus compared with
other nearby star-forming regions (Ballesteros-Paredes et al.\ 2009).

\bigskip

\noindent
Ophiuchus is composed of a compact core, only about 2 pc across, and
filamentary structures (called ``streamers'') extending (in
projection) to about 10 pc. The Ophiuchus core is sufficiently compact
that we do not expect to resolve any structure along the line of
sight, and our observations show that it is at a distance of 120
pc. There could potentially be distance gradients of several parsecs
across the streamers. We note, however, that Schaefer et al.\ (2008)
determined the physical orbit of the binary system Haro 1-14C, and
deduced a distance of $111\pm19$ pc, in good agreement with our
determination. Haro 1-14C is located in the northern streamer
(associated with the darks clouds L1709/L1704), so the result of
Schaefer et al.\ (2008) suggests that streamer is, if anything,
somewhat closer that the core.  This is, indeed, in agreement with
recent results of Lombardi et al.\ (2008).  On the other hand, Imai et
al.\ (2007) used the Japanese VLBI system (VERA) to determine the
parallax to the very young protostar IRAS~16293--2422 deeply embedded
in the southern Ophiuchus streamer (in L1689N). They obtain a distance
of $178^{+18}_{-37}$ pc, which would be more consistent with the older
value of 165 pc. Even including the streamers, Ophiuchus is only 10 pc
across in projection, so it is unlikely to be 60 pc deep. Thus, if the
results of Imai et al.\ (2007) are confirmed, they would indicate the
existence of several unrelated star-forming regions along the line of
sight. More observations --some of which have already been collected
and partially analyzed-- will be necessary to settle this issue.

\newpage


\noindent
{\large \bf Perspectives}

\bigskip

\noindent
The results presented in this thesis have allowed us to refine the
determination of the distance to two important regions of nearby
star-formation, and to start examining the three-dimensional structure
of Taurus. They also raised a number of issues that could be tackled
with new data, and suggested several follow-up studies.

\bigskip

\begin{itemize}

\item[-]
 As we mentioned, the radio flux of V773~Tau~A depends on the orbit
 phase, being highest at periastron and weakest at apoastron. In most
 of our observations, it is resolved into two components associated
 with the two stars in the system. Our observations have shown,
 however, that near periastron, the position of the radio source is
 significantly displaced from the position of the associated
 star. This is additional evidence that the non-thermal emission in
 the system is affected by the presence of the companion. In a
 recently accepted proposal, we requested time to observe V773~Tau~A
 near periastron with the High Sensitivity Array (a composite VLBI
 array comprised of the VLBA, the Green Bank Telescope, and the VLA
 plus the Arecibo dish in Puerto Rico). These observations will allow
 us to examine the spatial evolution of this interacting binary when
 it is near periastron over a six hour period. This ought to shed
 light on the origin of the variability of the source.

\item[-]
 To decide whether or not several unrelated star-forming regions exist
 along the line of sight to Ophiuchus, we have also been observing
 several new sources in the direction of the complex: VSSG~14 in the
 Oph--B sub-region, and ROX~39 in the region between L1686 and L1689.
 
\item[-]
 Several other nearby star-forming regions (e.g.\ Serpens or Perseus)
 have been studied in detail at many wavelengths but have poorly
 determined distances. Non-thermal sources are known to exist in these
 regions, so multi-epoch VLBA observations would allow significant
 improvements in the determination of their distances. The
 corresponding observations of Serpens, are indeed already being
 obtained, whereas Perseus will be the subject of forthcoming
 proposals.

\end{itemize}
\cleardoublepage

\pagestyle{empty}
\thispagestyle{empty}
\begin{center}
\phantom{hola}
\vfill

\selectfont

\textbf{\LARGE Measuring Nearby Star Forming}\\
\vspace{.2cm}
\textbf{\LARGE Regions with the VLBA:}\\
\vspace{.2cm}
\textbf{\LARGE from the Distance to the Dynamics}\\

\vspace{10cm}

\vfill
\end{center}
\cleardoublepage

\pagestyle{fancy}

\renewcommand{\thepage}{\arabic{page}}
\setcounter{page}{1}

\chapter{Introduction}\label{chap-introduction}

\begin{quote}
\noindent
This thesis will describe how multi-epoch VLBA observations can be
used to measure the distance to nearby young stars to extremely high
accuracy. This first chapter is intended as a broad introduction to
the subject. We start with a very brief description of astrometry
(Sect.\ \ref{c1-astrometry}), and explain why high accuracy astrometry
of young stars is important (Sect.\ \ref{c1-motivation}). We then move
on to describe the two star-forming regions on which we will
concentrate: Taurus and Ophiuchus (Sect.\ \ref{c1-taurus} and
\ref{c1-ophiuchus}). In Sect.\ \ref{c1-parallax}, we explain why VLBI
instruments are ideal to obtain high accuracy astrometric observations
of young stars. This leads us to state the specific goals of this
thesis (Sect.\ \ref{c1-goals}). Finally, in Sect.\ \ref{c1-sources},
we describe the main characteristics of the young stars that will be
studied in the rest of the manuscript.
\end{quote}

\section{Astrometry}\label{c1-astrometry}

\begin{figure*}[!t]
\centerline{\includegraphics[height=.8\textwidth,angle=-90]{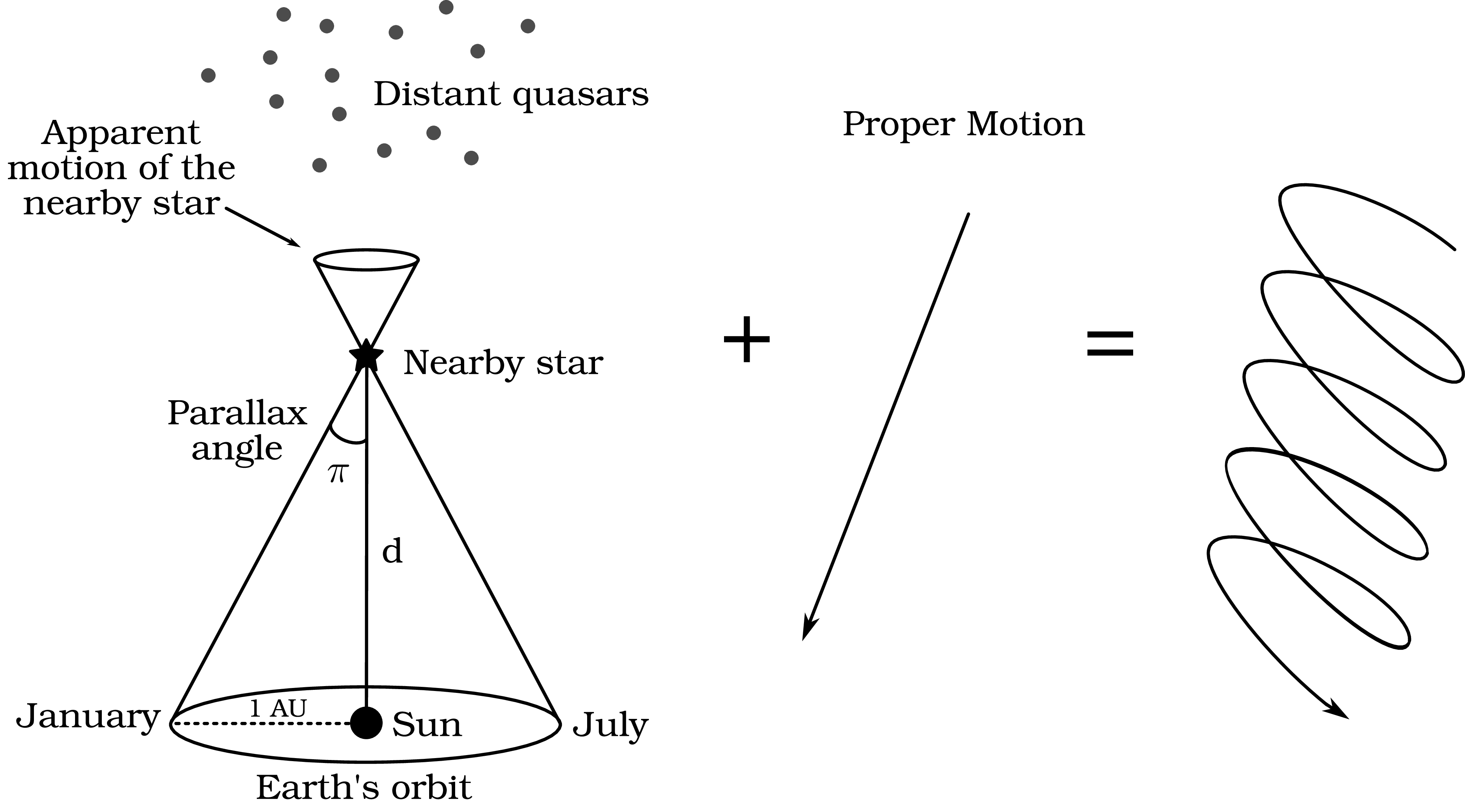}}
\caption{\footnotesize{Parallax and proper
motion.}\label{fig-parpm}}\end{figure*}

Astrometry is the branch of astronomy concerned with making precise
measurements of the positions of celestial bodies, and of their
movements on the celestial sphere. For objects outside of the Solar
System, the observed displacements will result from two
contributions. First, the true relative motion between the source and
the Sun: the {\em proper motion}. In most cases, the proper motion can
be assumed to be linear and uniform, but in some specific situations
(for instance if a star is a member of a multiple system), the proper
motion can be non-uniform.  The second contribution is the apparent
change in the position of the source due to the annual rotation of the
Earth about the Sun. The semi-major axis of the elliptical path
resulting from this effect is called the {\em trigonometric parallax}
($\pi$), and provides a direct estimate of the distance $d$ to the
object (see Fig.\ \ref{fig-parpm}). The value in milli-arcsecond (mas)
of the trigonometric parallax as a function of distance are indicated
in Tab.\ \ref{tab-dpimu}, together with the value of the proper motion
in mas yr$^{-1}$ for an object with a rather modest transverse speed
of 10 km s$^{-1}$.

\begin{deluxetable}{llllllllllllll}
\tabletypesize{\scriptsize}
\tablecolumns{14}
\tablewidth{0pc}
\tablecaption{\footnotesize{Parallax and proper motions as a function of the distance.}
\label{tab-dpimu}}
\tablehead{
\colhead{Distance [pc]} &
\colhead{1}        &
\colhead{2}        &
\colhead{5}        &
\colhead{10}       &
\colhead{20}       &
\colhead{50}       &
\colhead{100}      &
\colhead{200}      &
\colhead{500}      &
\colhead{1000}     &
\colhead{2000}     &
\colhead{5000}     &
\colhead{10000}    }
\startdata
$\pi$ [mas]                                   &	1000 & 500  & 200 & 100 & 50  & 20 & 10 & 5  & 2 & 1 & 0.5 & 0.2 & 0.1 \\%
$\mu$ [mas yr$^{-1}$] \tablenotemark{\dagger} &	2000 & 1000 & 400 & 200 & 100 & 40 & 20 & 10 & 4 & 2 & 1   & 0.4 & 0.2 \\%
\enddata
\tablenotetext{\dagger}{Proper motions calculated for $v=10$ km s$^{-1}$.}
\end{deluxetable}

\section{Motivation}\label{c1-motivation}

Astrometric observations of young stellar objects can provide a wealth
of important information on their properties. First and foremost, an
accurate trigonometric parallax measurement is a pre-requisite to the
derivation, from observational data, of their most important
characteristics (luminosity, age, mass, etc.). Unfortunately, even in
the current post-Hipparcos era, the distance to even the nearest
star-forming regions (Taurus, Ophiuchus, Perseus, etc.) is rarely
known to better than 20 to 30\% (e.g.\ Knude \& Hog 1998, Bertout et
al.\ 1999). At this level of accuracy, the luminosity of any given
star cannot be assessed to better than 50\%. As a consequence, the
accuracy with which young stars can be positioned on an HR diagram is
rather limited, and the comparison between observations and detailed
theoretical models can only be approximate. This unsatisfactory state
of affairs is largely the result of the fact that young stars are
still embedded in their opaque parental cloud. They are, therefore,
dim in the visible bands that were observed by Hipparcos.

\begin{deluxetable}{lcl}
\tabletypesize{\scriptsize}
\tablecolumns{3}
\tablewidth{0pc}
\tablecaption{\footnotesize{Distances to Taurus and Ophiuchus star-forming regions.}
\label{tab-distancias}}
\tablehead{
\colhead{Region       }&
\colhead{Distance [pc]}&
\colhead{Author       }}
\startdata
Taurus & $135\pm10$        & Racine (1968)            \\%
       & $140\pm20$        & Elias (1978)             \\%
       & $140\pm10$        & Kenyon et al.\ (1994)    \\%
       & $139^{+10}_{-9}$  & Bertout et al.\ (1999)   \\%
       & $140^{+40}_{-12}$ & Bertout \& Genova (2006) \\%
\\[-0.2cm]
\hline
\\[-0.2cm]
Ophiuchus & $145\pm10$ & Racine (1968)	       \\%
          & $160\pm20$ & Elias (1978)           \\%
          & $165\pm20$ & Chini (1981)	       \\%
          & $125\pm25$ & de Geus et al.\ (1989) \\%
          & $120-150$  & Knude \& Hog (1998)    \\%
          & $135\pm8$  & Mamajek (2007)         \\%
          & $119\pm6$  & Lombardi et al.\ (2008)\\%
\enddata
\end{deluxetable}

\medskip

\noindent
The proper motions of young stars are also important because they can
be used to characterize the overall dynamics of star-forming regions
and the orbital paths of young multiple systems. This last point is
particularly important because the mass of the individual stars which
can be obtained from Kepler's law provides strong constraints for
pre-main sequence evolutionary models (e.g.\ Hillenbrand \& White
2004).

\medskip

\noindent
Much of what we know about the formation of stars has been derived
from the observation and modeling of a few nearby regions (Taurus,
Ophiuchus, Perseus, Orion, Serpens, etc.). Thus, a significant
improvement in the determination of the distance to these few regions
would represent a major step forward. In this thesis, we will
concentrate on the two nearest regions of low-mass star formation:
Taurus and Ophiuchus.

\section{Taurus}\label{c1-taurus}

The Taurus star-forming region has been intensively studied, starting
with the historical observations of T~Tauri and its surroundings by
Hind in the 1850s. It contains a few 10$^{4}$ \Msun\ in molecular gas
(e.g.\ Goldsmith et al.\ 2008), mostly distributed along three
filamentary structures. At smaller scales, these filaments break up
into dense clumps and cores with individual masses of 1--100
\Msun. These dense regions are associated with optically visible dark
nebulae, cataloged by Barnard or Lynds.

\medskip

\noindent
The Taurus region contains a large number of pre-main sequence stars
(nearly 400 according to the recent review by Kenyon et al.\
2008). Numerous surveys conducted in the last several decades at
optical, infrared, radio, and (more recently) X-ray frequencies have
revealed the existence of three populations of young stars (e.g.\ Lada
1987). The embedded (optically invisible) population is believed to be
the youngest and is only found in association with dense molecular
gas. The two somewhat older populations (the Classical and Weak Line T
Tauri stars --WTTS and CTTS) are more dispersed across the cloud, but
still follow closely the contours of the molecular gas.

\medskip

\noindent
Most young stars in Taurus are grouped into small clusters of a few
tens of objects (e.g.\ Gómez et al.\ 1993). The most prominent such
groups in the central portion of Taurus are associated with the dark
clouds L1495, B18, and B22. Another prominent group located to the
south of the complex (and containing the famous young star T~Tauri) is
associated with the dark cloud L1551 (see Kenyon et al.\ 2008 for a
recent review).

\begin{figure*}[!t]
\centerline{\includegraphics[height=1\textwidth,angle=-90]{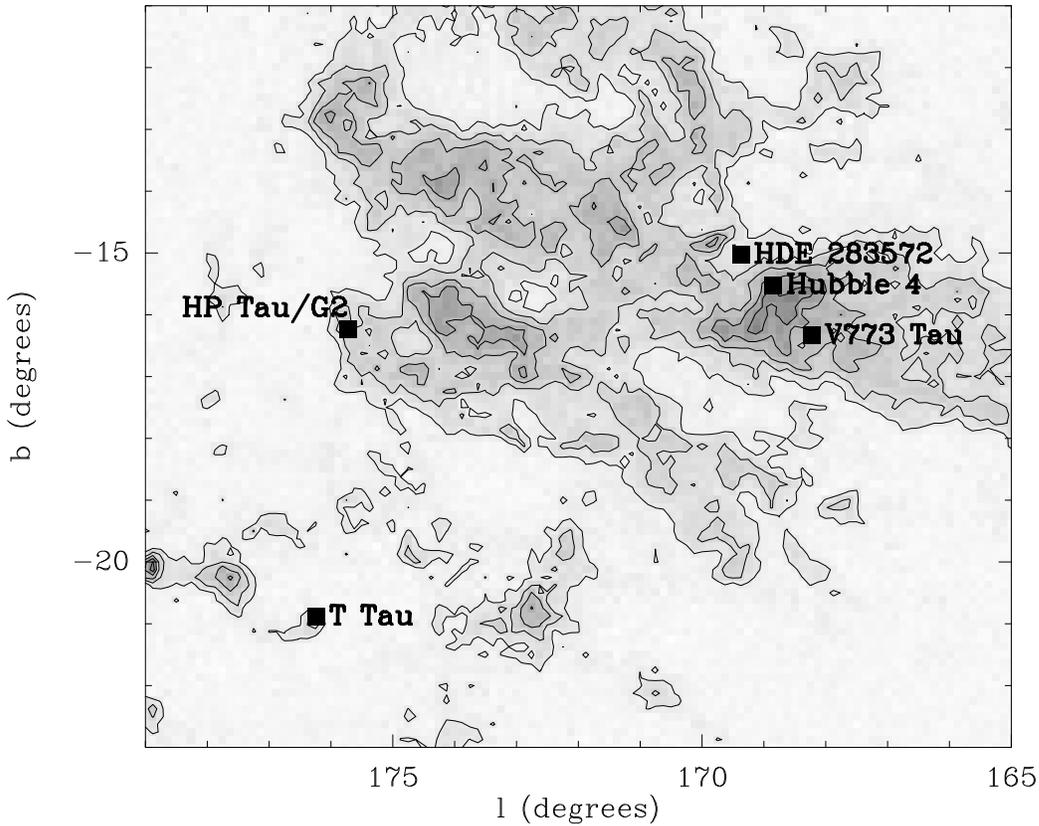}}
\caption
{\footnotesize{$^{12}$CO map of Taurus from Dame et al.\ (2001). The
positions of the five stars studied in this thesis are shown as black
squares.}\label{fig-taurus}}\end{figure*}

\medskip

\noindent
Many pre-main sequence stars in Taurus are in binary or multiple
systems (see Kenyon et al.\ 2008). Indeed, the fraction of binaries
among pre-main sequence stars in Taurus is similar to or slightly
larger than the corresponding figure for field main sequence
stars. Most of these binaries are pairs of either two WTTS, or two
CTTS, but there are a number of mixed pairs, containing two stars in
seemingly different evolutionary stages. We shall see an example of
such a system in this thesis (V773~Tau).

\medskip

\noindent
The mean distance to Taurus has been determined by a number of authors
using various methods. In the last few decades, there has been a
consensus for a value of about 140 pc (see Tab.\
\ref{tab-distancias}).  However, the extent of Taurus on the plane of
the sky is about 25 pc, so it likely to also be about 10--30 pc
deep. Moreover, given its filamentary structure, the Taurus complex is
unlikely to be ``round'', and systematic distance differences may well
exist within the complex.

\section{Ophiuchus}\label{c1-ophiuchus}

\begin{figure*}[!b]
\centerline{\includegraphics[height=.5\textwidth,angle=0]{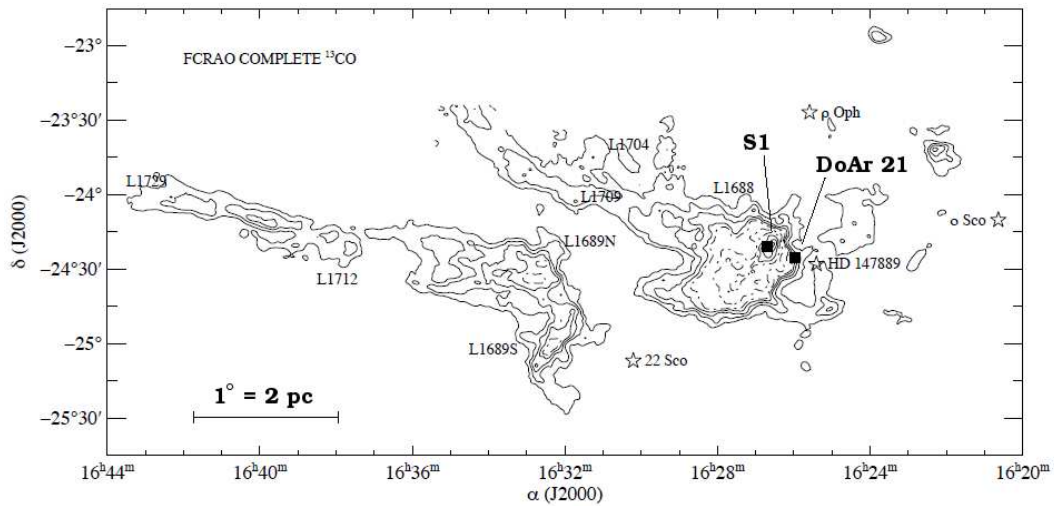}}
\caption
{\footnotesize{$^{13}$CO map of Ophiuchus from Ridge et al.\ (2006).
The main dark clouds are indicated, and the positions of the two stars
studied in this thesis are shown as black
squares.}\label{fig-ophiuchus}}\end{figure*}

Ophiuchus is one of the most active regions of star formation within a
few hundred parsecs of the Sun (e.g.\ Lada \& Lada 2003). It comprises
a main core associated with the dark cloud L1688, and lower density
filaments (called ``streamers'') associated with the clouds L1704 and
L1689 (Fig.\ \ref{fig-ophiuchus}). The core contains over 300 young
stellar objects (Wilking et al.\ 2008). The youngest, more embedded
population has a mean age of about 0.3 Myr whereas the more diffusely
distributed population is a few million years old. The star formation
in Ophiuchus is somewhat more clustered than in Taurus, and more
massive stars are being formed in Ophiuchus than in Taurus. Indeed,
one the stars that will be included in this thesis (S1) is a B star of
about 6 \Msun.

\medskip

\noindent
The Ophiuchus cloud has played an important role in the development of
our understanding of star formation, and remains an important
benchmark for this field of research. As for Taurus, detailed
multi-wavelengths observations are available (see the review by
Wilking et al.\ 2008).

\medskip

\noindent
Traditionally assumed to be at 165 $\pm$ 20 pc (Chini 1981), Ophiuchus
has recently been suggested to be somewhat closer. For example, de
Geus et al.\ (1989) found a mean photometric distance of 125 $\pm$ 25
pc. Knude \& Hog (1998), who examined the reddening of stars in the
direction of Ophiuchus as a function of their Hipparcos distances,
also found a clear extinction jump at 120 pc. Using a similar method,
Lombardi et al.\ (2008) also find a distance of about 120 pc for the
Ophiuchus core. Finally, Mamajek (2007) identified reflection nebulae
within 5$^\circ$ of the center of Ophiuchus, and obtained the
trigonometric parallax of the illuminating stars from the Hipparcos
catalog. From the average of these Hipparcos parallaxes, he obtains a
mean distance to Ophiuchus of 135 $\pm$ 8 pc. While the core of
Ophiuchus is fairly compact (about 1$^\circ$, or 2 pc), the total
extent of the streamers on the plane of the sky is about 5 degrees
($\equiv$ 10 pc; Fig.\ \ref{fig-ophiuchus}). It is therefore possible
that distance gradients might exist across the streamers.

\section{VLBI parallax measurements}\label{c1-parallax}

Since observations of young stars in the visible range are limited by
the effect of dust extinction, one must turn to a more favorable
wavelength regime in order to obtain high quality astrometric
data. Radio observations, particularly using large interferometers is
currently the best prospect because the interstellar medium is largely
transparent at these wavelengths, and because the astrometry delivered
by radio-interferometers is extremely accurate and calibrated against
fixed distant quasars. Of course, only those young stars associated
with radio sources are potential targets. Moreover,
radio-interferometers effectively filter out any emission more
extended than a certain limiting angular size, so only compact sources
will be detectable.

\medskip 

\noindent
Low-mass young stars often generate non-thermal continuum emission
produced by the interaction of free electrons with the intense
magnetic fields that tend to exist near their surfaces (e.g.\
Feigelson \& Montmerle 1999; see Chap.\ 2). Since the magnetic field
strength decreases quickly with the distance to the stellar surface
(as $r^{-3}$ in the magnetic dipole approximation), the emission is
strongly concentrated to the inner few stellar radii. If the magnetic
field intensity and the electron energy are sufficient, the resulting
compact radio emission can be detected with Very Long Baseline
Interferometers (VLBI --e.g.\ André et al.\ 1992). The relatively
recent possibility of accurately calibrating the phase of VLBI
observations of faint, compact radio sources using nearby quasars
makes it possible to measure the absolute position of these objects
(or, more precisely, the angular offset between them and the
calibrating quasar) to better than a tenth of a milli-arcsecond. This
level of precision is sufficient to constrain the trigonometric
parallax of sources within a few hundred parsecs of the Sun (in
particular of nearby young stars) with a precision better than a few
percent using multi-epoch VLBI observations (see Tab.\
\ref{tab-dpimu}).

\section{Goals}\label{c1-goals}

Taking advantage of the very accurate astrometry attainable with the
\textit{Very Long Baseline Array} (VLBA) of the National Radio Astronomy
Observatory (NRAO), we initiated a large project aimed at accurately
measuring the trigonometric parallax of a significant sample of
magnetically active young stars in nearby star-forming regions
(Taurus, Ophiuchus, Perseus, Cepheus, and Serpens) using the VLBA with
accuracies one to two orders of magnitude better than the present
values.

\medskip

\noindent
The specific goal of this thesis is to use multi-epoch
radio-interferometric VLBA observations of seven young stellar objects
in Taurus and Ophiuchus in order to measure their displacement over
the celestial sphere, and deduce their trigonometric parallax and
proper motion with a level of precision that currently cannot be
attained at any other wavelength.

\section{Sources}\label{c1-sources}

We chose from the literature a list of seven young stellar objects:
five in the Taurus complex (T~Tau, Hubble~4, HDE~283572, HP~Tau/G2 and
V773~Tau~A; Fig.\ \ref{fig-taurus}), and two in the Ophiuchus complex
(S1 and DoAr~21; Fig.\ \ref{fig-ophiuchus}). Those sources are low- to
intermediate-mass (M $\lesssim$ 3 \Msun) pre-main sequence stars (with
the exception of S1 which is a main sequence B star). All seven
sources were previously known to be non-thermal radio emitters, and
had been detected with VLBI techniques in the past.

\medskip

\noindent
\textbf{T~Tau}---
T~Tau was initially identified as a single optical star, with unusual
variability and peculiar emission lines (Barnard 1895, Joy 1945, and
references therein). Early infrared observations then revealed the
existence of a heavily obscured companion (hereafter T~Tau~S) located
about \msec{0}{7} to the south of the visible star (Dyck et al.\ 1982) and
most likely gravitationally bound to it (Ghez et al.\ 1991). Recently,
this infrared companion was itself found to contain two sources
(T~Tau~Sa and T~Tau~Sb; Koresko 2000 and Kohler et al.\ 2000) in rapid
relative motion (Duchêne et al.\ 2002; Furlan et al.\ 2003). Thus,
T~Tau is now acknowledged to be at least a triple stellar system. At
radio wavelengths, T~Tau has long been known to be a double source
(Schwartz et al.\ 1986). The northern radio component is associated
with the optical star and mostly traces the base of its thermal jet
(e.g., Johnston et al.\ 2003), whereas the southern radio source is
related to the infrared companion and is thought to be the
superposition of a compact component of magnetic origin and an
extended halo, presumably related to stellar winds (Johnston et al.\
2003; Loinard et al.\ 2003).

\medskip

\noindent
The non-thermal mechanisms at the origin of the compact radio emission
in T~Tau~S require the presence of an underlying, magnetically active
star (Skinner 1993). Specifically, the emission is expected to be
either gyrosynchrotron radiation associated with reconnection flares
in the stellar magnetosphere and at the star-disk interface or
coherently amplified cyclotron emission from magnetized accretion
funnels connecting the disk to the star (Dulk 1985; Feigelson \&
Montmerle 1999; Smith et al.\ 2003). In all cases, the emission is
produced within less than about 10 stellar radii (roughly 30 \Rsun) of
the star itself. Indeed, 3.6 cm VLBA observations revealed the
existence, near the expected position of T~Tau~Sb, of a source with an
angular size less than about 15 \Rsun\ (Smith et al.\ 2003).  Because
it is so small, any structural changes in this compact radio component
would occur on such small scales that the effects on the astrometry
would be very limited. Thus, observations focusing on it should
accurately trace the path of the underlying pre-main sequence star.

\medskip

\noindent
\textbf{Hubble~4 and HDE~283572}---
Hubble~4 (V1023~Tau, HBC~374, IRAS 0415+2813) is a K7 naked T~Tauri
star with an effective temperature of 4060 K (Briceño et al.\
2002). It has long been known to have a particularly active
magnetosphere that produces non-thermal radio emission characterized
by significant variability, large circular polarization and a nearly
flat spectral index (Skinner 1993). It was detected in VLBI
experiments, with a flux of a few mJy by Phillips et al.\ (1991), and
is also an X-ray source (Güdel et al.\ 2007). The superficial magnetic
field of Hubble~4 has been estimated to be about 2.5 kG using
Zeeman-sensitive Ti I lines (Johns-Krull et al.\ 2004). HDE~283572
(V987~Tau, HIP~20388), on the other hand, is a somewhat hotter
($T_{\mbox{eff}} = $ 5770 K --Kenyon \& Hartmann 1995) G5 naked
T~Tauri star. Early observations with the Einstein satellite showed
that it has a fairly bright X-ray counterpart (Walter et al.\
1987). It was initially detected as a radio source by O'Neal et al.\
(1990), and in VLBI observations by Phillips et al.\ (1991) with a
flux of about 1 mJy.

\medskip

\noindent
\textbf{HP~Tau/G2}---
The well-known variable star HP~Tau (IRAS 04328+2248) was discovered
by Cohen \& Kuhi (1979) to be surrounded by a small group of young
stars (called HP~Tau/G1, HP~Tau/G2, and HP~Tau/G3). HP~Tau/G1 is
located about 20$''$ north of HP~Tau, whereas HP~Tau/G2 and HP~Tau/G3
are about 15$''$ to its south-east (see the finding charts in Fig.\ 22
of Cohen \& Kuhi 1979). HP~Tau/G2 and HP~Tau/G3 are believed to form a
gravitationally bound system with a separation of about 10$''$.
Recently, HP~Tau/G3 was itself found to be a tight binary (Richichi et
al.\ 1994), so the HP~Tau/G2 - HP~Tau/G3 system appears to be a
hierarchical triple system. HP~Tau/G2 is a weak-line T Tauri star of
spectral type G0, with an effective temperature of 6030 K (Briceño et
al.\ 2002). It is somewhat obscured ($\Av$ $\sim$ 2.1 mag) and has a
bolometric luminosity of 6.5 \Lsun\ (Briceño et al.\ 2002; Kenyon \&
Hartmann 1995). This corresponds to an age of about 10.5 Myr and a
mass of 1.58 \Msun\ (Brice\~no et al.\ 2002). The first radio
detection of HP~Tau/G2 was reported by Bieging et al.\ (1984) who
found a 5 GHz flux of 5--7 mJy. A few years later, however, the flux
had fallen to only about 0.3 mJy (Cohen \& Bieging 1986). Such strong
variability is suggestive of non-thermal processes (e.g.\ Feigelson \&
Montmerle 1999). The successful detection of HP~Tau/G2 in VLBI
experiments (at levels of 1 to 3 mJy) by Phillips et al.\ (1991)
confirmed the non-thermal origin of the radio emission.

\medskip

\noindent
\textbf{V773~Tau}---
The young stellar system V773~Tau (HD~283447, HBC~367, HIP~19762, IRAS
04111+2804) is located in the surroundings of the dark cloud Lynds
1495 in Taurus. V773~Tau has been known to be a multiple system since
it was almost simultaneously found to be a double-line spectroscopic
binary (V773~Tau~A) with an orbital period of about 51 days (Welty
1995), and to have a companion at about 150 mas (V773~Tau~B; Ghez et
al.\ 1993, Leinert et al.\ 1993). More recently, Duchêne et al.\
(2003) identified a fourth component in the system (V773~Tau~C),
showing that V773~Tau is (at least) a quadruple system. This fourth
component belongs to the still poorly understood class of ``infrared
companions'' (young stellar sources, fairly bright in the infrared but
invisible at optical wavelengths, which have been discovered around a
small number of T~Tauri stars). It is interesting to point out that
the four (almost certainly coeval) stars in V773~Tau belong to three
different spectral classes: both members of the spectroscopic binary
are WTTS, the optical companion is a CTTS, and the fourth member is
(as mentioned above) an infrared companion. Such a variety of
apparently distinct evolutionary stages in a single system likely
reflects the effect of binarity on the evolution of young stars.

\medskip

\noindent
The relative orbital motion between the spectroscopic binary and the
two companions has been monitored by Duchêne et al.\ (2003), and the
orbit between the two members of the spectroscopic binary has recently
been investigated in detail by Boden et al.\ (2007) who combined
(optical) Radial Velocity measurements, Keck Interferometer data, and
radio VLBI images. Using these data, Boden et al.\ (2007) constructed
a preliminary physical orbit for the spectroscopic binary system,
which yields masses of 1.54 and 1.33 \Msun\ for the primary and the
secondary, respectively.  The distance to the system obtained from
these data is 136.2 $\pm$ 3.7 pc. There is also a direct trigonometric
parallax measurement based on multi-epoch VLBI observations for this
source (Lestrade et al.\ 1999). This VLBI-based distance measurement
($d=148.4^{+5.7}_{-5.3}$ pc) is roughly consistent (at the 2--3
$\sigma$ level) with the value obtained from the orbital fit.

\medskip

\noindent
V773~Tau~A has long been known to be a strong radio source (Kutner et
al.\ 1986). Indeed, it was the strongest source in the 5 GHz VLA
survey of WTTS in the Taurus-Auriga molecular cloud complex by O'Neal
et al.\ (1990). The two companions V773~Tau~B and C, on the other
hand, are not known to be radio sources. From detailed multi-frequency
observations, Feigelson et al.\ (1994) concluded that the radiation
was most likely of non-thermal origin. This was confirmed by Phillips
et al.\ (1991) who obtained VLBI observations, and resolved the radio
emission into a clear double source, most likely corresponding to the
two components of the spectroscopic binary. More recently, Massi et
al.\ (2002, 2006) showed that the radio emission exhibits periodic
variations with a period corresponding to the 51 day orbital period of
the spectroscopic binary. This variability is due to an increase in
the flaring activity near periastron and likely reflect interactions
between the magnetospheres of the two stars when they get close to one
another. Finally, Boden et al.\ (2007) and Massi et al.\ (2008) also
resolved the radio emission from V773~Tau~A into two components, which
they associate with the two stars in the spectroscopic binary.

\medskip

\noindent
\textbf{S1 and DoAr~21}---
The star S1 (IRAS 16235-2416, ROX~14, YLW~36) of spectral type B4, and
$M\sim6$ \Msun, is among the brightest red and near-infrared objects
in Ophiuchus (Grasdalen et al.\ 1973). It is also the brightest
far-infrared member of the cluster (Fazio et al.\ 1976), a very bright
X-ray source (ROX~14 --Montmerle et al.\ 1983), and the brightest
steady radio stellar object in Ophiuchus (Leous et al.\ 1991). S1 is
fairly heavily obscured ($A_V\sim10$), and there is clear evidence for
an interaction between S1 and the dense gas associated with Oph-A
sub-region, and traced by DCO$^+$ emission (Loren et al.\
1990). Moreover, the age of the \HII\ region excited by S1 is
estimated to be about 5,000 yr (André et al.\ 1988). DoAr~21
(V2246~Oph, Haro 1-6, HBC~637, ROX~8, YLW~26) is a somewhat less
massive star ($\sim2.2$ \Msun) of spectral type K1. Like S1, it is
fairly obscured ($A_V\sim6-7$), and probably younger than 10$^6$
yr. It is associated with a bright X-ray source (ROX~8 --Montmerle et
al.\ 1983), and with a strongly variable radio source (Feigelson \&
Montmerle 1985). Although it has long been classified as a naked
T~Tauri star (e.g.\ André et al.\ 1990), it was recently found to show
a substantial infrared excess at 25 $\mu$m (Jensen et al.\ 2009)
suggestive of a circumstellar disk. As mentioned above, both S1 and
DoAr~21 are fairly strong radio sources. Indeed, both have been
detected at 6 cm in previous VLBI experiments: S1 with a flux density
of 6--9 mJy (André et al.\ 1991), and DoAr~21 with a flux density of
nearly 10 mJy (Phillips et al.\ 1991).

\section{Conclusions}\label{c1-conclusions}

In this chapter, we justified why VLBI parallax measurements of young
stars in nearby star-forming regions are important. We also described
briefly the structure of the two regions (Taurus and Ophiuchus) on
which we will focus in this thesis, and presented the main
characteristics of the seven young stars that were observed as part of
this work. In the next chapter, we will examine in more detail the
processes at work to generate the non-thermal radiation emitted by
young stars.

\chapter{Emission Processes of Radio Waves}\label{chap-emission}

\begin{quote}
\noindent
In Chapter \ref{chap-introduction}, we mentioned that all seven
sources presented in this thesis produce non-thermal radio
emission. This kind of emission in young stellar objects is due to
electrons spiraling in a magnetic field, as will be described in this
chapter. Our treatment is based primarily on the book by Rybicki \&
Lightman (1986), and the articles of Dulk (1985), Robinson \& Melrose
(1984), and Petrosian \& McTiernan (1983).
\end{quote}

\section{Electromagnetic waves}\label{c2-electromagnetic}

An electromagnetic wave is a transversal wave composed of an electric
and magnetic field oscillating together. The fields are oriented
perpendicular to each other, and the wave travels in a direction
perpendicular to both fields. Electromagnetic waves can be
characterized by any of three properties: wavelength $\lambda$,
frequency $\nu$, and energy $E$ of the individual photons. The
relationships between wavelength, frequency, and energy are:
\begin{equation}\begin{split}\label{ec-relacion}
\lambda &= \frac{c}{\nu}=\frac{hc}{E}
\end{split}\end{equation}
where $c$ is the speed of light, and $h$ is the Planck constant.

\medskip

\noindent
The polarization is a property of electromagnetic waves that describes
the orientation of their electric fields, that may be oriented in a
single direction (linear polarization) or rotate (circular and
elliptical polarization). For circular polarization the electric field
vector describes a helix along the direction of wave propagation, and
may be referred to as right or left, depending on the direction in
which the electric field vector rotates.

\medskip

\noindent
In the most general sense, an electromagnetic wave is generated by
accelerating charges. In the next sections we will describe the
relevant processes at radio wavelengths.

\section{Radiation at radio wavelengths}\label{c2-radiation}

Much of the radiation emitted as radio waves by astrophysical objects
is due to individual electrons accelerated by collisions with ions or
by spiraling in a magnetic field. In this case the radiation is the
result of a random process of collisions and it is incoherent.

\medskip

\noindent
In some cases there could exist an efficient process where the energy
of the electrons is converted into some natural wave mode of the
plasma (e.g.\ electron-cyclotron or Langmuir waves). These waves are
in the radio-frequency domain because the characteristic frequencies
of plasma are typically $\lesssim 10^{10}$ Hz, and they are:\\
the electron plasma frequency
\begin{equation}\begin{split}\label{ec-plasma}
\nu_{\rm p} &= {\left[\frac{n_e e^{2}}{\pi m_{e}}\right]}^{1/2}
\simeq 9000 \ n_{e}^{1/2}
\end{split}\end{equation}
and the electron-cyclotron frequency
\begin{equation}\begin{split}\label{ec-ciclotron}
\nu_{\rm B} &= \frac{eB}{2\pi m_{e}c}\simeq 2.8\times 10^{6} \ B
\end{split}\end{equation}
where $n_{e}$ is the electron density, $e$ the electric charge,
$m_{e}$ the electron mass, $B$ the magnetic field, and $c$ is the
speed of light.

\medskip

\noindent
It is possible to have resonances between particles and waves with
these characteristic frequencies (Eqs.\ \ref{ec-plasma} and
\ref{ec-ciclotron}). These can rapidly extract any free energy that
might exist in the electron distribution. Plasmas with free energy can
only exist when the electron-electron and electron-ion collision
frequencies are neither as high as the resonance frequencies nor so
high as to restore the plasma quickly to equilibrium and thus quench
the instabilities. This is the major reason that amplified radiation
(maser) is generally confined to radio frequencies. The resistive
instabilities could amplify a particular wave mode and lead to
coherent emission as in the case of an electron-cyclotron maser.

\section{Radiative transfer}\label{c2-radiative}

The equation of radiative transfer is usually written in terms of the
specific intensity $I_{\nu}$:
\begin{equation}\begin{split}\label{ec-transferencia1}
\dfrac{dI_{\nu}}{d\tau_{\nu}} =-I_{\nu}+S_{\nu}
\end{split}\end{equation}
where $\tau_{\nu}$ is the optical depth and $S_{\nu}$ is the source
function defined as the ratio of the emission coefficient to the
absorption coefficient:
\begin{equation}\begin{split}\label{ec-fnfuente1}
S_{\nu} &= \frac{\eta_{\nu}}{\kappa_{\nu}} .
\end{split}\end{equation}

\medskip

\noindent
From Eq.\ \ref{ec-transferencia1} we see that if $I_{\nu}>S_{\nu}$
then $\dfrac{dI_{\nu}}{d\tau_{\nu}}<0$, and $I_{\nu}$ tends to
decrease. If $I_{\nu}<S_{\nu}$ then $I_\nu$ tends to increase. Thus
the source function is the quantity that the specific intensity tries
to approach, and does approach if given sufficient optical depth. In
this respect the transfer equation describes a ``relaxation'' process.

\medskip

\noindent
A black body is an idealized object that absorbs all electromagnetic
radiation that falls on it. No electromagnetic radiation passes
through it and none is reflected. Because no electromagnetic radiation
is reflected or transmitted, the object appears black when it is
cold. The thermal radiation from a black body is called blackbody
radiation.

\medskip

\noindent
Two important properties of $I_{\nu}$ in that case, are that (1) it
depends only on the temperature, and (2) it is isotropic. Therefore we
have the relation
\begin{equation}\begin{split}\label{ec-blackbody}
I_{\nu}(T) \equiv B_{\nu}(T)
\end{split}\end{equation}
where the function $B_{\nu}(T)$ is called the Planck function.

\medskip

\noindent
If we consider that the material of the object is emitting at
temperature $T$, so that its emission depends solely on its
temperature and internal properties, then we have the relation
\begin{equation}\begin{split}\label{ec-fnfuente2}
S_{\nu} &= B_{\nu}(T)
\end{split}\end{equation}
and the emissivity $\eta_{\nu}$ can be related to the absorption
coefficient $\kappa_{\nu}$ and temperature by the Kirchoff law:
\begin{equation}\begin{split}\label{ec-kirchoff}
\eta_{\nu} &= \kappa_{\nu}B(T)
\end{split}\end{equation}
the transfer equation for thermal radiation is, then
\begin{equation}\begin{split}\label{ec-transferencia2}
\dfrac{dI_{\nu}}{d\tau_{\nu}} &=-I_{\nu}+B_{\nu}(T) .
\end{split}\end{equation}

\medskip

\noindent
Note that for blackbody radiation $I_{\nu}=B_{\nu}$, whereas for
thermal radiation $S_{\nu}=B_{\nu}$. Thermal radiation becomes
blackbody radiation only for optically thick media.

\medskip

\noindent
The Planck function represents the emitted power per unit area of
emitting surface, per unit solid angle, and per unit frequency. It can
be expressed as a function of frequency, and is written as
\begin{equation}\begin{split}\label{ec-planck}
B_{\nu}(T) &= \frac{2h\nu^{3}/c^{2}}{{\rm e}^{(h\nu/kT)-1}} .
\end{split}\end{equation}

\medskip

\noindent
Eq.\ \ref{ec-planck} is called the Planck law, and describes the
spectral radiance of electromagnetic radiation at all wavelengths from
a black body at temperature $T$.

\medskip

\noindent
For $h \nu \ll k T$ the exponential in Eq.\ \ref{ec-planck} can be
expanded and we have the Rayleigh--Jeans law:
\begin{equation}\begin{split}\label{ec-rayleigh}
I_{\nu}^{\rm RJ}(T) &= \frac{2\nu^{2}}{c^{2}}kT
\end{split}\end{equation}
that applies at low frequencies. In particular, it almost always
applies in the radio region.

\medskip

\noindent
One way to characterize brightness at a certain frequency is to give
the temperature of the blackbody having the same brightness at that
frequency. That is, for any value $I_{\nu}$ we define the brightness
temperature $T_{\rm b}(\nu)$ by
\begin{equation}\begin{split}\label{ec-brightness}
I_{\nu} &= B_{\nu}(T_{\rm b}) .
\end{split}\end{equation}

\medskip

\noindent
The effective temperature $T_{\rm eff}$ of a source is derived from the
total amount of flux, integrated over all frequencies, radiated at the
source. We obtain $T_{\rm eff}$ by equating the actual flux $F$ to the
flux of a blackbody at temperature $T_{\rm eff}$:
\begin{equation}\begin{split}\label{ec-effective}
F &= \int{cos\theta \ I_{\nu} \ d\nu d\Omega}\equiv\sigma \ T_{\rm eff}^{4} .
\end{split}\end{equation}

\medskip

\noindent
In radio astronomy, where the Rayleigh--Jeans law is applicable, it is
convenient to change variables from $I_{\nu}$ to brightness temperature
$T_{\rm b}$. So that for $h \nu \ll k T$ we have
\begin{equation}\begin{split}\label{ec-intensidad}
I_\nu &= \frac{2\nu^{2}}{c^{2}}k \ T_{\rm b} .
\end{split}\end{equation}

\medskip

\noindent
It is also convenient to replace the source function $S_{\nu}$ by the
effective temperature $T_{\rm eff}$ of the radiating electrons, using
the Kirchoff law:
\begin{equation}\begin{split}\label{ec-fnfuente3}
S_{\nu} &= \frac{2\nu^{2}}{c^{2}}k \ T_{\rm eff} .
\end{split}\end{equation}

\medskip

\noindent
$I_{\nu}$ can be written as a combination of the orthogonal
polarizations as
\begin{equation}\begin{split}\label{ec-intensidadtotal}
I_{\nu}^{\rm tot} &= I_{\nu}^{\rm p1}+I_{\nu}^{\rm p2}
\end{split}\end{equation}
in this case, there will be no factor of 2 in the equations for
$I_{\nu}$ and $S_{\nu}$.

\medskip

\begin{figure*}[!t]
\centerline{\includegraphics[height=.3\textwidth,angle=0]{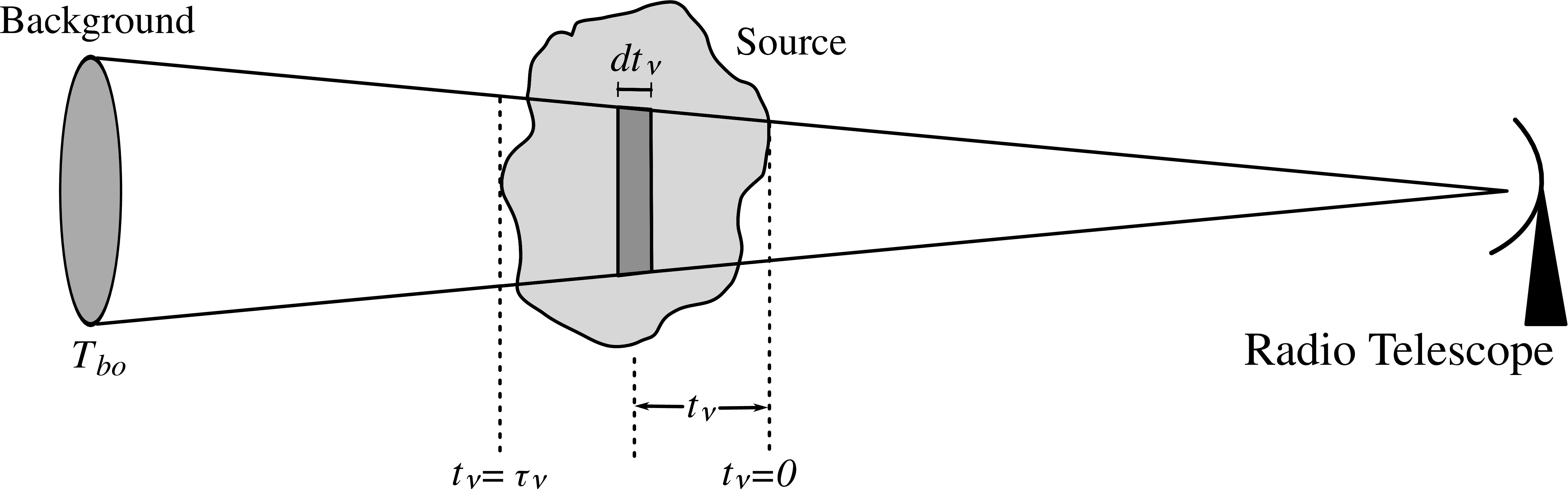}}
\caption
{\footnotesize{A source of optical depth $\tau_{\nu}$ located in front
of a background of brightness temperature $T_{\rm
bo}$.}\label{fig-geometria}}\end{figure*}

\noindent
Eqs.\ \ref{ec-intensidad} and \ref{ec-fnfuente3} allow us to
write Eq.\ \ref{ec-transferencia1} for radiative transfer in the
form:
\begin{equation}\begin{split}\label{ec-transferencia3}
\dfrac{dT_{\rm b}}{d\tau_{\nu}} &= -T_{\rm b}+T_{\rm eff}
\end{split}\end{equation}
or, if we use the geometry of Fig.\ \ref{fig-geometria}, we can write:
\begin{equation}\begin{split}\label{ec-transferencia4}
T_{\rm b}=\int_{0}^{\tau_\nu}{T_{\rm eff} \ {\rm e}^{-t_\nu} \ dt_\nu}
+T_{\rm bo} \ {\rm e}^{-\tau_\nu}
\end{split}\end{equation}

\medskip

\noindent
In the special case of an isolated source with constant $T_{\rm eff}$,
Eq.\ \ref{ec-transferencia4} reduces to
\begin{equation}\begin{split}\label{ec-tb1}
T_{\rm b} &= T_{\rm eff}[1 - e^{-\tau_\nu}]
\end{split}\end{equation}
\begin{equation}\label{ec-tb2}
T_{\rm b}= \left\{ 
\begin{array}{l@{\quad}l@{\quad}l}
T_{\rm eff} &
{\rm if} & \tau_{\nu}\gg1 \\
T_{\rm eff} \ \tau_\nu=\dfrac{c^{2}}{\nu^{2}}\dfrac{\eta_{\nu} L}{k} &
{\rm if} & \tau_{\nu}\ll1
\end{array}\right .
\end{equation}
where $L$ is the size of the source along the line of sight.

\medskip

\noindent
For incoherent radiation, Eq.\ \ref{ec-tb2} shows that the radiation
cannot attain a value of $T_{\rm b}$ higher than $T_{\rm eff}$, where
$T_{\rm eff}$ is related to the average energy of the emitting
particles:
\begin{equation}\begin{split}\label{ec-energia}
\left\langle E \right\rangle &= kT_{\rm eff}
\end{split}\end{equation}
$T_{\rm eff}$ and $T_{\rm b}$ of incoherent emission are usually
limited to about $10^{9}$ to $10^{10}$ K. Higher values imply a
coherent mechanism, such as maser or plasma radiation.

\medskip

\noindent
The flux density $J_{\nu}$ of a radio source is related to the
brightness temperature by
\begin{equation}\begin{split}\label{ec-flujo}
J_{\nu} &= \frac{\nu^{2}}{c^{2}}k\int{T_{\rm b} \ d\Omega} ,
\end{split}\end{equation}
where $d\Omega$ is a differential solid angle and the integral is over
the projected area of the source.

\medskip

\noindent
The circular polarization reflects the direction of the magnetic field
in the source. The degree of circular polarization $r_{\rm c}$ is
given by
\begin{equation}\begin{split}\label{ec-polarizacion}
r_{\rm c} &= \frac{T_{\rm b-}-T_{\rm b+}}{T_{\rm b-}+T_{\rm b+}}
\end{split}\end{equation}
where $T_b$ is the brightness temperature, and the two signs, ($+$)
and ($-$), correspond to two different directions of rotation of
electric field. The ($-$) sign is related to the right-hand-side
circular polarization of electromagnetic wave, when the direction of
the electric field rotation coincides with the direction of an
electron gyration in the magnetic field. In optics, such a wave is
usually referred to as the extraordinary wave. The ($+$) sign is
related to the left-hand-side circular polarization of electromagnetic
wave, that in optics is usually referred to as the ordinary wave.

\medskip

\noindent
Eq.\ \ref{ec-polarizacion} will be valid when we have no significant
variations in the magnetic field strength, the effective temperature
or the angle between the line of sight and the magnetic field across
the region.

\begin{figure*}[!b]
\centerline{\includegraphics[height=.5\textwidth,angle=0]{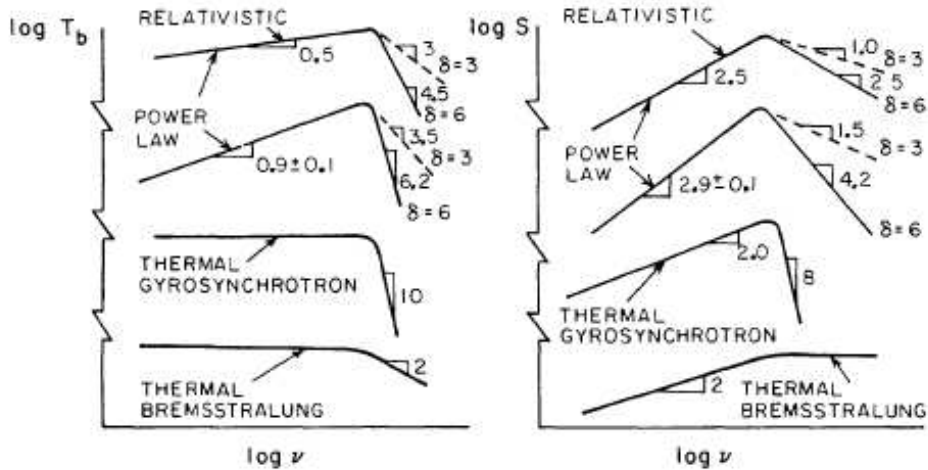}}
\caption
{\footnotesize{Spectra of brightness temperature and flux density for
gyrosynchrotron radiation from Dulk (1985).}\label{fig-espectro}}
\end{figure*}

\medskip

\noindent
Finally, a very useful quantity is the frequency of peak flux density
$\nu_{\rm peak}$ in a spectrum that changes from a positive to a
negative slope (see Fig.\ \ref{fig-espectro}). The peak occurs at that
frequency where
\begin{equation}\begin{split}\label{ec-peak}
\tau_{\nu} &= \kappa_{\nu}L\approx 1 .
\end{split}\end{equation}

\medskip

\noindent
For gyrosynchrotron emission, $\nu_{\rm peak}$ depends very strongly
on the magnetic field strength and the average electron energy, but
very weakly on electron numbers or path lengths.

\section{Continuous radiation from radio sources}\label{c2-continuous}

Radio sources can be classified into two categories: those which
radiate by thermal mechanisms and the others, which radiate by
non-thermal processes. In principle many different radiation
mechanisms could be responsible for non-thermal emission, but in
practice one single mechanism seems to dominate: synchrotron emission.

\medskip

\noindent
With the exception of thermal line emission of atoms and molecules,
and thermal emission from solid bodies, radio emission always arises
from free electrons, and since free electrons can exchange energy by
arbitrary amounts, no definite energy jumps will occur: thus we are
dealing with a continuous spectrum.

\section{Acceleration due to particle gyration}\label{c2-acceleration}

\begin{figure*}[!t]
\centerline{\includegraphics[height=.4\textwidth,angle=0]{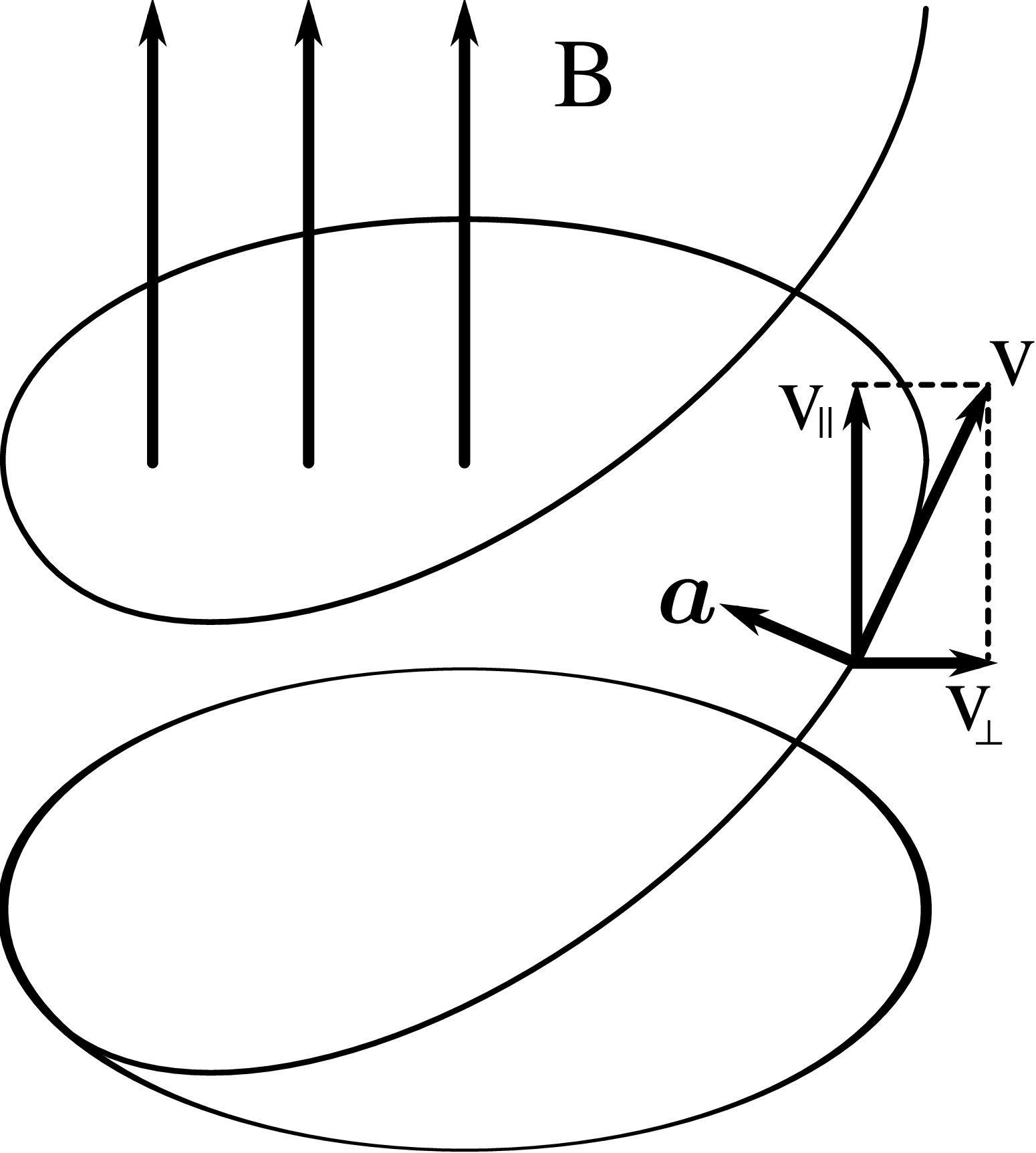}}
\caption
{\footnotesize{Helical motion of a particle in a uniform magnetic
field.}\label{fig-sincrotron}}\end{figure*}

When a plasma contains a magnetic field, accelerations due to particle
collisions can often be negligible in comparison with those due to
gyration around the field lines (see Fig.\
\ref{fig-sincrotron}). There are three possible cases of emission:
\textit{cyclotron} for non-relativistic particles (Lorentz
factor $\gamma=$1), \textit{gyrosynchrotron} for mildly relativistic
particles ($\gamma \lesssim$ 2 or 3), and \textit{synchrotron} for
highly relativistic particles ($\gamma\gg$1).

\medskip

\noindent
In the case of cyclotron emission, \textit{thermal electron}
distributions are of most interest because the average energy of
electrons is low, leading to frequent collisions and generally to
\textit{Maxwellian} distributions. Emission is concentrated at low harmonics
($s\lesssim10$) and at the fundamental frequency $\omega=\Omega_{\rm
e}$, where $\Omega_{\rm e}$ is the electron cyclotron
frequency. Radiation at the fundamental frequency ($s=1$) is directed
mainly along the magnetic field, and radiation at low harmonics is
mainly at moderate angles.

\medskip

\noindent
In the opposite limit of synchrotron radiation, electron collisions
are rare, a \textit{non-Maxwellian tail} is generally dominant, and
the energy distribution of the electrons is usually well described by
a \textit{power-law}. Emission is distributed over a broad continuum
at high harmonics near $s\approx(\gamma \ sin\theta)^{3}$, i.e.\ near
frequency $\omega\approx\Omega_{\rm e}\gamma^{2} \ sin\theta$. Emission
is directed very strongly in the direction of the instantaneous
electron motion, which leads, for approximately isotropic
distributions, to a peak of radiation perpendicular to the field.

\medskip

\noindent
In the intermediate case of gyrosynchrotron radiation, both
\textit{thermal} and \textit{power-law} distributions are of interest,
emission at harmonics $10\lesssim s\lesssim100$ is of major
importance, and the emission from approximately isotropic electrons
has a broad maximum perpendicular to the field.

\section{Gyrosynchrotron emission from thermal electrons}\label{c2-thermal}

For temperatures in the range from $10^7$ to $10^9$ K, emission and
absorption are usually important in the range of harmonics from 10 to
100. Thus, one cannot use only the leading terms of a power-series
expansion of the Bessel functions (as for non-relativistic electrons)
or the Airy integral approximation (as for relativistic
electrons). Analytical expressions had been derived for the radiation,
and here we will give a simple expression from Dulk et al.\ (1979)
that is valid in the ranges $10^{8}\lesssim T\lesssim10^{9}$ K and
$10\lesssim s\lesssim100$:
\begin{equation}\begin{split}\label{ec-potencia-termica}
\frac{\kappa_{\nu}B}{N}\approx 50 \ T^{7}
\left[sin\theta\right]^{6}\left[\frac{B}{\nu}\right]^{10}
\end{split}\end{equation}
where $\kappa_{\nu}$ is the absorption coefficient, $N$ is the number
of electrons per cubic centimeter, and $\theta$ is the angle between
the line of sight and magnetic field.

\medskip

\noindent
The source function, the degree of circular polarization for small
optical depth, and the frequency of maximum flux density for a
homogeneous source are given by:
\begin{equation}\begin{split}\label{ec-absorcion-termica}
S_{\nu}&=\frac{\eta_{\nu}}{\kappa_{\nu}}\approx 1.2\times10^{-24} \
T\left[\frac{B\nu}{\nu_{\rm B}}\right]^{2}
\end{split}\end{equation}
\begin{equation}\begin{split}\label{ec-polarizacion-termica}
r_{\rm c} \approx 13.1 \ T^{-0.138}10^{0.231cos\theta-0.219cos^{2}\theta}
\left[\frac{\nu}{\nu_{\rm B}}\right]^{-0.782+0.545cos\theta}
\ \ \ (\tau_{\nu}\ll 1)
\end{split}\end{equation}
\begin{equation}\label{ec-pico-termica}
\nu_{\rm peak}\approx \left\{
\begin{array}{l@{\quad}l@{\quad}}
1.4\left[\frac{N \ L}{B}\right]^{0.1}\left[sin\theta\right]^{0.6}T^{0.7} \ B &
(10^{8}<T<10^{9} \ {\rm K})\\[0.5cm]
475\left[\frac{N \ L}{B}\right]^{0.05}\left[sin\theta\right]^{0.6}T^{0.5} \ B &
(10^{7}<T<10^{8} \ {\rm K})
\end{array}\right .
\end{equation}
%

\section{Gyrosynchrotron emission from power-law electrons}\label{c2-power}

We now consider an electron population that is isotropic in pitch
angle and has a power law energy distribution:
\begin{equation}\begin{split}\label{ec-1-leypotencias}
n(E) &= KE^{-\delta}
\end{split}\end{equation}
where $K$ is related to N, the number of electrons per cubic
centimeter with $E>E_{0}$, by
\begin{equation}\begin{split}\label{ec-2-leypotencias}
K &= (\delta-1) \ E^{\delta-1}_{0} \ N .
\end{split}\end{equation}

\medskip

\noindent
Electrons with energy less than 50 to 100 keV contribute very little
to the radiation, and for normalization we assume $E_{0}=10~{\rm
keV}=1.6\times 10^{-8}~{\rm erg}$.

\medskip

\noindent
Empirical expressions (from Dulk \& Marsh 1982) for the quantities,
valid over the range $2\lesssim\delta\lesssim 7$, $\theta\gtrsim
20^{\circ}$, and $10\lesssim\nu / \nu_{\rm B}\lesssim 100$ is given
by:

\begin{equation}\begin{split}\label{ec-potencia-leypotencias}
\frac{\kappa_{\nu}B}{N}\approx1.4\times10^{-9}10^{-0.22\delta}
\left[sin\theta\right]^{-0.9+0.72\delta}
\left[\frac{\nu}{\nu_{\rm B}}\right]^{-1.30-0.98\delta}
\end{split}\end{equation}

\medskip

\noindent
The emissivity, degree of circular polarization, effective
temperature, and the frequency of maximum flux density are given by:
\begin{equation}\begin{split}\label{ec-absorcion-leypotencias}
\frac{\eta_{\nu}}{BN}\approx 3.3\times10^{-24}10^{-0.52\delta}
\left[sin\theta\right]^{-0.43+0.65\delta}
\left[\frac{\nu}{\nu_{\rm B}}\right]^{1.22-0.90\delta}
\end{split}\end{equation}
\begin{equation}\begin{split}\label{ec-polarizacion-leypotencias}
r_{c} \approx 1.26 \ 10^{0.035\delta}10^{-0.071cos\theta}
\left[\frac{\nu}{\nu_{\rm B}}\right]^{0.782+0.545cos\theta}
~~~ (\tau_{\nu}\ll 1)
\end{split}\end{equation}
\begin{equation}\begin{split}\label{ec-temperatura-leypotencias}
T_{\rm eff} \approx 2.2\times10^{9}10^{-0.31\delta}\left[sin\theta\right]
^{-0.36-0.06\delta}\left[\frac{\nu}{\nu_{\rm B}}\right]^{0.50+0.085\delta}
\end{split}\end{equation}
\begin{equation}\begin{split}\label{ec-pico-leypotencias}
\nu_{\rm peak} \approx 2.72\times10^{3}10^{0.27\delta}\left[sin\theta\right]
^{0.41+0.03\delta}\left[N \ L\right]^{0.32-0.03\delta}
\times B^{0.68+0.03\delta}.
\end{split}\end{equation}
%

\section{Conclusions}\label{c2-conclusions}

In this chapter we briefly reviewed the possible cases of emission due
to particle gyration around magnetic fields: cyclotron,
gyrosynchrotron, and synchrotron radiation. In the case (appropriate
for young stars) of mildly relativistic electrons, gyrosynchrotron
emission is produced. Depending on the situation, the electrons can
have a \textit{thermal} or a \textit{power-law} energy distribution.

\chapter{Interferometry}\label{chap-interferometry}

\begin{quote}
\noindent
Having examined the processes that generate non-thermal radiation in
young stars, we now move on to describe the instruments used to detect
this kind of emission. Treatment is based mostly on the books by Wilson
et al.\ (2009) and Thompson et al.\ (2001).
\end{quote}

\section{The interferometer}\label{c3-interferometer}
The angular resolution of a radio telescope is
$\delta\sim\frac{\lambda}{D}$ where $\lambda$ is the wavelength of the
radiation received, and $D$ is the diameter of the telescope. In order
to improve this angular resolution, for a given wavelength, the
diameter $D$ of the telescope must be increased. Obviously, there are
practical limits on how large a telescope can be built. In the radio
domain, this limit is in the range of 100--300 m depending on the
technical and financial means available.

\medskip

\begin{figure*}[!h]
\centerline{\includegraphics[height=.5\textwidth,angle=0]{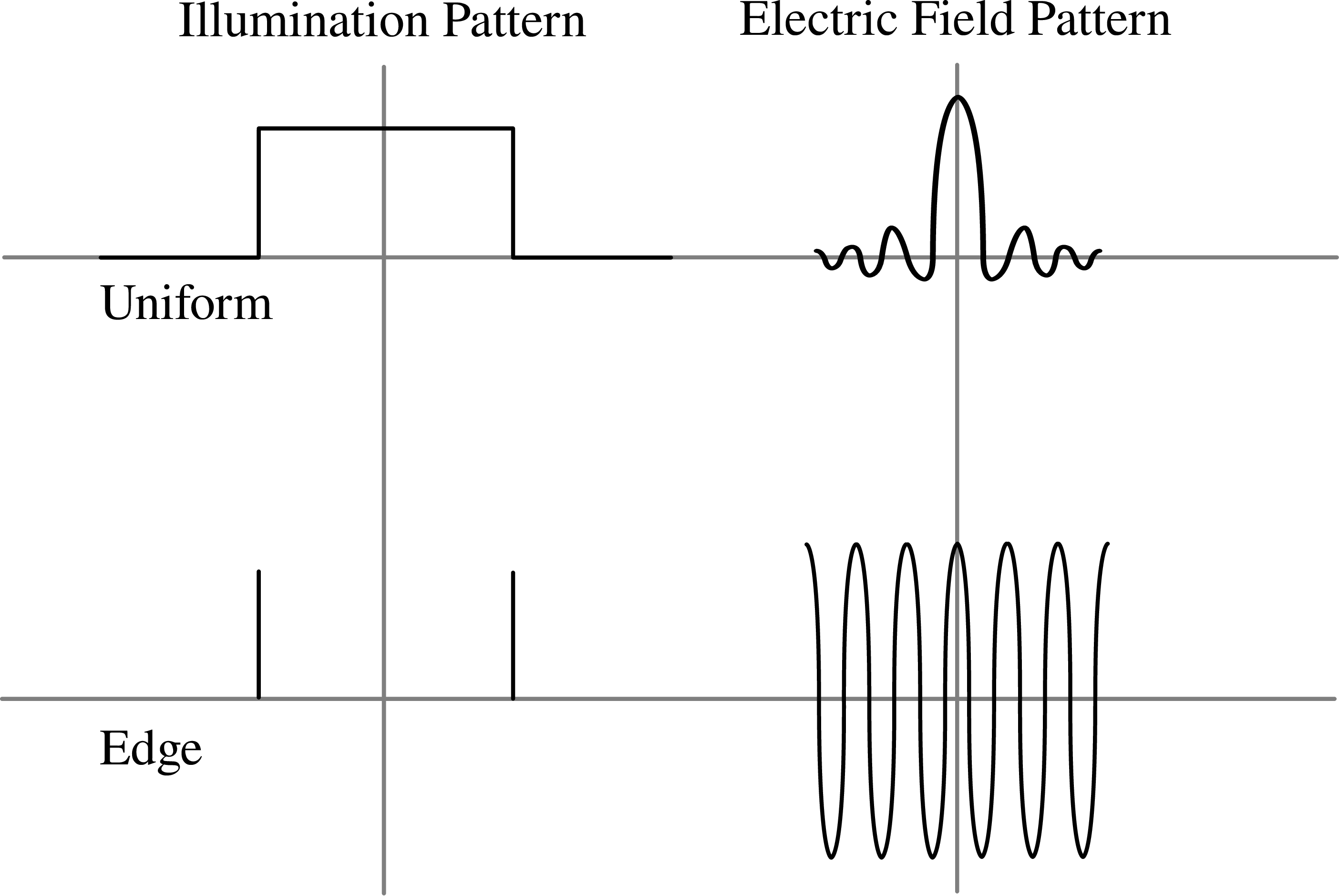}}
\caption
{\footnotesize{Comparison of aperture illumination and resulting
electric field pattern. The upper panel shows a uniformly illuminated
full aperture, the lower one, an instrument in which only the outer
edge is illuminated.}\label{fig-apertura}}\end{figure*}

\noindent
Therefore, radio astronomers have increased the effective resolving
power of their instruments by joining together the outputs of several
small telescopes separated by a maximum distance $D$. The basic
principle behind such an \textit{interferometer} may become clear from
considering Fig.\ \ref{fig-apertura} where we compare the electric
field patterns produced be a filled aperture telescope of uniform
illumination, with that of a reflector in which only the outer edge of
the aperture is illuminated. This figure shows that the resolving
powers of the main beams of these two configurations are
comparable. The main difference is in the side lobe level; in the
second case this level is of the same order as the main beam, whereas
it is much lower in the first situation. An image formed by such a
two-element instrument therefore will differ considerably from that of
a conventional full aperture. It will be corrupted by contributions of
a large number of side lobes. As we shall see, there are techniques to
remove (or at least minimize) those effects.

\section{The mutual coherence function}\label{c3-mutual}

Suppose that we have a plane monochromatic wave propagating through
space. If we know the intensity of the electromagnetic field at a
position $P_{1}$, it is possible to calculate the field intensity at
another position $P_{2}$ for all times. In this case the wave is said
to be fully \textit{coherent}. Now, if we have an arbitrary
polychromatic wave, the field and the time variation at $P_{1}$ has no
relation to the field at $P_{2}$, and this is the case of an
\textit{incoherent} wave.

\medskip

\noindent
A measure of coherence is given by the mutual coherence function of
the (complex) wave field $U(P_{1},t_{1})$ and $U(P_{2},t_{2})$:
\begin{equation}\begin{split}\label{ec-1}
\Gamma(P_{1},P_{2},\tau)
&= {\rm lim}_{T\rightarrow\infty}
\frac{1}{2T}\int^T_{-T}{U(P_{1},t)U^{*}(P_{2},t+\tau) \ dt} \\
&= \left\langle U(P_{1},t)U^{*}(P_{2},t+\tau) \right\rangle
\end{split}\end{equation}
where the brackets are used to indicate time averaging.

\medskip

\noindent
For a plane monochromatic wave field propagating in the $z$ direction,
$\Gamma$ is easily computed. Using a complex representation,
$U(P,t)=U_{0}e^{{\rm i}\left(kz-\omega t\right)}$, where $P=(x,y,z)$,
$k=2\pi/\lambda={\rm const}$, $\omega=2\pi\nu={\rm const}$, we get
\begin{equation}\begin{split}\label{ec-2}
\Gamma(P_{1},P_{2},\tau) &=
\mid{U_{0}}\mid{\rm e}^{{\rm i}\left[k\left(z_{1}-z_{2}\right)
+\omega\tau\right]} ,
\end{split}\end{equation}
where $\tau$ is the time delay. The mutual coherence function of the
travelling monochromatic wave field is thus periodic with a constant
amplitude and a wavelength equal to that of the original wave
field. The coherence function does not propagate; it is a standing
wave with a phase such that, for $\tau=0$, $\Gamma=\Gamma_{max}$ for
$z_{1}=z_{2}$. It is often useful to normalize Eq.\ \ref{ec-1} by
referring it to a wave field of intensity $I$. Thus
\begin{equation}\begin{split}\label{ec-3}
\gamma(P_{1},P_{2},\tau)=
\frac{\Gamma(P_{1},P_{2},\tau)}{\sqrt{I(P_{1})I(P_{2})}} .
\end{split}\end{equation}
For this complex coherence, we always have
\begin{equation}\begin{split}\label{ec-4}
\mid{\gamma(P_{1},P_{2},\tau)}\mid\leqq 1 .
\end{split}\end{equation}
%

\section{The coherence function of extended sources}\label{c3-coherence}

A wave field that is only slightly more complex than a monochromatic
plane wave is formed by the (incoherent) superposition of two such
wave fields with identical wavelengths but propagating in different
directions:
\begin{equation}\begin{split}\label{ec-5}
U_{a} &= U_{0a}~{\rm e}^{{\rm i}(k\textbf{s}_{a}\cdot\textbf{x}-\omega t)},\\
U_{b} &= U_{0b}~{\rm e}^{{\rm i}(k\textbf{s}_{b}\cdot\textbf{x}-\omega t)}.
\end{split}\end{equation}
where $\textbf{s}_{a}$ and $\textbf{s}_{b}$ are unit vectors
describing the propagation direction, and both $k=2\pi/\lambda$ and
$\omega=2\pi\nu$ are assumed to be equal for both waves. The total
wave field is then formed by $U=U_{a}+U_{b}$ and the mutual coherence
function (Eq.\ \ref{ec-1}) is
\begin{equation}\begin{split}\label{ec-6}
\left\langle U(P_{1},t_{1})U^{*}(P_{2},t_{2})\right\rangle =&
\left\langle U_{a}(P_1,t_1)U_{a}^{*}(P_2,t_2)\right\rangle \\
&+ \left\langle U_{b}(P_{1},t_{1})U_{b}^{*}(P_{2},t_{2})\right\rangle \\
&+ \left\langle U_{a}(P_{1},t_{1})U_{b}^{*}(P_{2},t_{2})\right\rangle \\
&+ \left\langle U_{b}(P_{1},t_{1})U_{a}^{*}(P_{2},t_{2})\right\rangle .
\end{split}\end{equation}

\medskip

\noindent
If we now assume the two wave fields $U_{a}$ and $U_{b}$ to be
incoherent, we accept the field strengths $U_{a}$ and $U_{b}$ to be
uncorrelated even when taken at the same point so that
\begin{equation}\begin{split}\label{ec-7}
\left\langle U_{a}(P_{1},t_{1})U_{b}^{*}(P_{2},t_{2})\right\rangle =&
\left\langle U_{b}(P_{1},t_{1})U_{a}^{*}(P_{2},t_{2})\right\rangle\equiv 0 .
\end{split}\end{equation}

\medskip

\noindent
Such incoherence is not possible for strictly monochromatic waves of
identical polarization consisting of a wave train of infinite duration
and length. Eq.\ \ref{ec-7} is true only if the wave is made up of
sections of finite duration between which arbitrary phase jumps
occur. In such a situation, the waves are not strictly monochomatric
but have a finite, although small bandwidth. Substituting Eq.\
\ref{ec-7} into Eq.\ \ref{ec-6} we obtain
\begin{equation}\begin{split}\label{ec-8}
\Gamma(P_{1},P_{2},\tau)
&= \left\langle U(P_{1},t)U^{*}(P_{2},t+\tau)\right\rangle \\
&= \left\langle U_{a}(P_{1},t)U_{a}^{*}(P_{2},t+\tau)\right\rangle
+  \left\langle U_{b}(P_{1},t)U_{b}^{*}(P_{2},t+\tau)\right\rangle
\end{split}\end{equation}
or using Eq.\ \ref{ec-5}
\begin{equation}\begin{split}\label{ec-9}
\Gamma(P_{1},P_{2},\tau)
&= \mid{U_{0a}}\mid^{2}{\rm e}^{{\rm i}(k\textbf{s}_{a}\cdot\textbf{u}+
\omega t)}+\mid{U_{0b}}\mid^{2}{\rm e}^{{\rm i}(k\textbf{s}_{b}\cdot\textbf{u}
+\omega t)}
\end{split}\end{equation}
where $\textbf{u}=\textbf{x}_{1}-\textbf{x}_{2}$.

\medskip

\noindent
Thus only the difference of the two positions $P_{1}$ and $P_{2}$
enter into the problem. For the case of two waves of equal amplitude,
$\mid{U_{0a}}\mid=\mid{U_{0b}}\mid=\mid{U_{0}}\mid$, Eq.\ \ref{ec-9}
can be simplified using the identities $\textbf{s}_{a} =
1/2(\textbf{s}_{a} + \textbf{s}_{b}) + 1/2(\textbf{s}_{a} -
\textbf{s}_{b})$, $\textbf{s}_{b} = 1/2(\textbf{s}_{a} +
\textbf{s}_{b}) - 1/2(\textbf{s}_{a} - \textbf{s}_{b})$, resulting in
\begin{equation}\begin{split}\label{ec-10}
\Gamma(\textbf{u},\tau) &=
2\mid{U_{0}}\mid^{2}cos\left(\frac{k}{2}
(\textbf{s}_{a}-\textbf{s}_{b})\cdot\textbf{u}\right) 
{\rm e}^{{\rm i}\left(\frac{k}{2}
(\textbf{s}_{a}+\textbf{s}_{b})\cdot\textbf{u}+\omega\tau\right)} ,
\end{split}\end{equation}
or, if normalized,
\begin{equation}\begin{split}\label{ec-11}
\gamma(\textbf{u},\tau) &=
cos\left(\frac{k}{2}(\textbf{s}_{a}-\textbf{s}_{b})\cdot\textbf{u}\right) 
{\rm e}^{{\rm i}\left(\frac{k}{2}(\textbf{s}_{a}+\textbf{s}_{b})\cdot\textbf{u}
+\omega\tau\right)} .
\end{split}\end{equation}

\medskip

\noindent
For two waves propagating in directions that differ only slightly,
$\mid{\textbf{s}_{a}-\textbf{s}_{b}}\mid/2$ is a small quantity, while
$({\textbf{s}_{a}+\textbf{s}_{b}})/2$ differs only little from either
$\textbf{s}_{a}$ or $\textbf{s}_{b}$. The normalized coherence
function therefore is similar to that of a single plane wave, but with
an amplitude that varies slowly with position. We will have a complete
loss of coherence for
\begin{equation}\begin{split}\label{ec-12}
\frac{k}{2}(\textbf{s}_{a}-\textbf{s}_{b})
\cdot\textbf{u}=(2n+1)\frac{\pi}{2}, \ \ \ \ \ n=0,1,2,...
\end{split}\end{equation}

\medskip

\noindent
This principle of superposition of simple monochromatic plane waves
can be extended to an arbitrary number of plane waves, and the result
will be a simple generalization of Eq.\ \ref{ec-9} if we assume these
fields to be mutually incoherent. The signals at $P_{1}$ and $P_{2}$
are then the sum of the components $U_{n}(P,t)$,
\begin{equation}\begin{split}\label{ec-13}
U(P,t) &= \sum_{n}U_{n}(P,t) ,
\end{split}\end{equation}
and, if the different waves are incoherent, then $\left\langle U_{m}
(P_{1},t) U_{n}^{*} (P_{2},t + \tau) \right\rangle = 0$ for all $m\neq
n$, while $\left\langle U_{n} (P_{1},t) U_{n}^{*} (P_{2},t + \tau)
\right\rangle = \mid{U_{0n}}\mid^{2} {\rm e}^{\rm i(ks_{n}\cdot u +
\omega\tau)}$, so that
\begin{equation}\begin{split}\label{ec-14}
\Gamma(\textbf{u},\tau) &= \left\langle U(P_{1},t)U^{*}(P_{2},t+\tau)
\right\rangle=\sum_{n}\mid{U_{n}}\mid^{2}{\rm e}^{{\rm i}
(k\textbf{s}_{n}\cdot\textbf{u}+\omega\tau)} .
\end{split}\end{equation}

\medskip

\noindent
Or, if we go to the limit $n\rightarrow\infty$
\begin{equation}\begin{split}\label{ec-15}
\Gamma(\textbf{u},\tau) &= \int\int I(\textbf{s}){\rm e}^{{\rm i}
(k\textbf{s}\cdot\textbf{u}+\omega\tau)} \ d\textbf{s} ,
\end{split}\end{equation}
where
\begin{equation}\begin{split}\label{ec-16}
I(\textbf{s}) &= \int\int U(\textbf{s}+\sigma)
U^{*}(\textbf{s}+\sigma) \ d\sigma
\end{split}\end{equation}
is the total intensity at the position $P$ if the integral is taken
over the angular extent of those positions $\textbf{s}+\sigma$ that
contribute to the radiation field propagating into the direction
$\textbf{u}$. The generalization of Eq.\ \ref{ec-13} is
\begin{equation}\begin{split}\label{ec-17}
U(\textbf{x},t) &= \int\int U(\textbf{s}){\rm e}^{{\rm i}
(k\textbf{s}\cdot\textbf{x}-\omega t)} \ d\textbf{s} .
\end{split}\end{equation}

\medskip

\noindent
Eq.\ \ref{ec-15} is the monochromatic version of the van
Cittert-Zernike theorem. This theorem specifies how the mutual
coherence function of an arbitrary monochromatic wave field, built up
from plane waves, is related to the intensity distribution.

\medskip

\noindent
Provided that $\Gamma(\textbf{u},\tau)$ can be measured and Eq.\
\ref{ec-15} can be solved for $I(\textbf{s})$, we can measure the total
intensity. The possible angular resolution with which $I(\textbf{s})$
can be determined depends on the size of the telescope used. For Eq.\ 
\ref{ec-15} the difference in the positions where measurements are made,
$\mid{\textbf{u}}\mid$ is introduced. Since it is possible to measure
$\Gamma(\textbf{u},\tau)$ for values of $\mid{\textbf{u}}\mid$ much
larger than the largest single telescope diameters possible, the
resolution with which $I(\textbf{s})$ can be measured from the
inversion of Eq.\ \ref{ec-15} is much greater than that which can be
achieved by using single-dish telescopes.

\section{Two-element interferometers}\label{c3-element}

\begin{figure*}[!h]
\centerline{\includegraphics[height=.8\textwidth,angle=0]{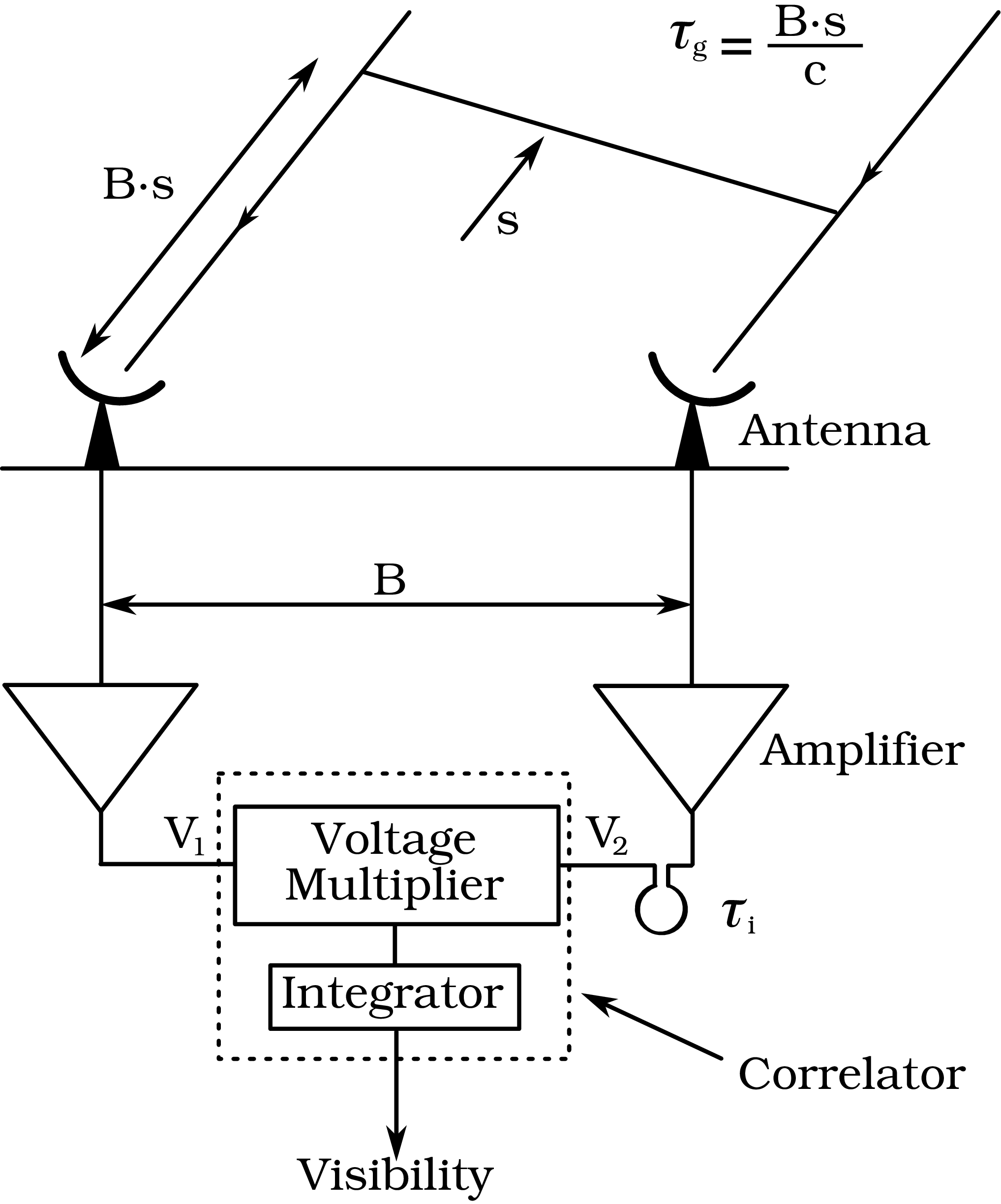}}
\caption
{\footnotesize{Schematic diagram of a two-element
interferometer.}\label{fig-elementos}}\end{figure*}

The simplest way to measure the coherence function is by using a
two-element interferometer. Let two telescopes $T_{1}$ and $T_{2}$ be
separated by the distance $\textbf{B}$, both telescopes being
sensitive only to radiation of the same state of polarization (see
Fig.\ \ref{fig-elementos}).

\medskip

\noindent
An electromagnetic wave induces the voltage $U_{1}\propto E \ {\rm
e}^{{\rm i}\omega t}$ at the output of antenna $T_{1}$ while at
$T_{2}$ we obtain $U_{2}\propto E \ {\rm e}^{{\rm i}\omega(t-\tau)}$,
where $\tau$ is the geometric delay caused by the relative orientation
of the interferometer baseline $\textbf{B}$ and the direction of the
wave propagation. At the correlator, the signals are injected to a
multiplying device followed by a low-pass filter such that the output
is proportional to
\begin{equation}\begin{split}\label{ec-18}
R(\tau) \propto \frac{1}{2}E^{2}{\rm e}^{{\rm i}\omega\tau}
\end{split}\end{equation}

\medskip

\noindent
The output of the correlator and integrator varies periodically with
the delay time, and is the mutual coherence function of the received
wave. If the relative orientation of interferometer baseline and the
wave propagation direction remain fixed, the delay time remains
constant, and so does $R(\tau)$. But if the wave propagation direction
is slowly changing due to the rotation of the earth, the delay time
will vary, and we will measure interference fringes as a function of
time.

\medskip

\noindent
The basic constituents of a two-element interferometer are shown in
Fig.\ \ref{fig-elementos}. If the radio brightness distribution is
given by $I_{\nu}(\textbf{s})$, the power received per bandwidth $d\nu$
from the source element $d\Omega$ is $A(\textbf{s})
I_{\nu}(\textbf{s}) d\Omega d\nu$, where $A(\textbf{s})$ is the
effective collecting area in the direction $\textbf{s}$; we will
assume the same $A(\textbf{s})$ for each of the antennas. The
amplifiers introduce a constant gain factor which we will omit for
simplicity.

\medskip

\noindent
The output of the correlator for radiation from the direction
$\textbf{s}$ is
\begin{equation}\begin{split}\label{ec-19}
r_{12} &= A(\textbf{s})I_{\nu}(\textbf{s}){\rm e}^{{\rm i}\omega\tau}
\ d\textbf{s} \ d\nu
\end{split}\end{equation}
where $\tau$ is the difference between the geometrical $\tau_{g}$ and
instrumental $\tau_{i}$ delays. If $\textbf{B}$ is the baseline vector
between the two antennas
\begin{equation}\begin{split}\label{ec-20}
\tau &= \tau_{g}-\tau_{i}=\frac{1}{c}\textbf{B}\cdot\textbf{s}-\tau_{i}
\end{split}\end{equation}
and the total response is obtained by integrating over the source $S$
\begin{equation}\begin{split}\label{ec-21}
R(\textbf{B}) &= \int\int_{S}A(\textbf{s})I_{\nu}(\textbf{s}){\rm e}^{\left[
{\rm i}2\pi\nu\left(\frac{1}{c}\textbf{B}\cdot\textbf{s}-\tau_{i}\right)
\right]} \ d\textbf{s} \ d\nu .
\end{split}\end{equation}

\medskip

\noindent
This function is closely related to the mutual coherence function of
the source but, due to the power pattern $A(\textbf{s})$ of the
individual antennas, it is not identical to $\Gamma(\textbf{B},
\tau)$. For parabolic antennas it is usually assumed that $A(\textbf{s})=0$
outside the main beam area so that Eq.\ \ref{ec-21} is integrated only
over this region.

\medskip

\noindent
The reduction in the error caused by finite bandwidth can be estimated
by assuming that the expression in Eq.\ \ref{ec-21} is averaged over a
range of frequencies $\Delta\nu=\nu_{1}-\nu_{2}$. Then, for an average
over frequencies, there will be an additional factor
$sin(\Delta\nu\tau)/2\pi\Delta\nu\tau$ in $R(\textbf{B})$. This will
reduce the response if $\Delta\nu$ is large compared to the time delay
$\tau$. For typical bandwidths of 100 MHz, the offset from the zero
delay must be $\ll 10^{-8}~{\rm s}$ and this adjustment of delays is
referred to as a \textit{fringe stopping}. This causes the response of
$R(\textbf{B})$ to loose a component. To recover the full information
in the complex $R(\textbf{B})$, an extra delay of a quarter wavelength
relative to the input of the correlator is inserted, so that the sine
and cosine response in Eq.\ \ref{ec-21} can be measured.

\section{Aperture synthesis}\label{c3-aperture}

Aperture synthesis is a method of solving Eq.\ \ref{ec-21} for
$I_{\nu}(\textbf{s})$ by measuring $R(\textbf{B})$ at suitable values
of $\textbf{B}$. To do this effectively, a convenient coordinate
system has to be introduced for the two vectorial quantities
$\textbf{s}$ and $\textbf{B}$. The image center can be chosen at the
position of zero phase. This geometry can be introduced using a unit
vector $\textbf{s}=\textbf{s}_{0}+\sigma$, where $\textbf{s}_{0}$ is a
conveniently chosen position close to the center of the region
investigated, and $\mid\sigma\mid=1$. Thus, substituting the unit
vector in $R(\textbf{B})$, Eq.\ \ref{ec-21} can be written as
\begin{equation}\begin{split}\label{ec-22}
R(\textbf{B}) &= {\rm e}^{{\rm i}\omega\left(
\frac{1}{c}\textbf{B}\cdot\textbf{s}_{0}-\tau_{i}\right)}
\int\int_{S}A(\sigma)I(\sigma){\rm e}^{{\rm i}
\frac{\omega}{c}\textbf{B}\cdot\sigma} \ d\sigma .
\end{split}\end{equation}
The exponential factor extracted from the integral is describing a
plane wave which defines the phase of $R(\textbf{B})$ for the image
center. The integral is called the visibility function $V$ of the
intensity distribution $I(\sigma)$,
\begin{equation}\begin{split}\label{ec-23}
V(\textbf{B}) &= \int\int_{S}A(\sigma)I(\sigma) \
{\rm e}^{{\rm i}\frac{\omega}{c}\textbf{B}\cdot\sigma} \ d\sigma .
\end{split}\end{equation}
Since all phases are adjusted to produce a zero delay at the image
center the visibility is referred to this position.

\medskip

\noindent
We will choose a coordinate system such that $\omega/2\pi c \
\textbf{B}=(u,v,w)$, and $(\omega\pm\delta\omega)/2\pi
c=(f/c)(1\pm\Delta f/f)$, where $u$,$v$,$w$ are measured in units of
wavelength $\lambda=2\pi c/\omega$ and the direction (0,0,1) is
parallel to $\textbf{s}_{0}$, $u$ points in the local east direction
while $v$ points north; the vector $\sigma=(x,y,z)$ is defined such
that $x$ and $y$ are the direction cosines with respect to the $u$ and
$v$ axes. Then the $xy$ plane represents a projection of the celestial
sphere onto a tangent plane with the tangent point (and origin) at
$\textbf{s}_{0}$. In these coordinates the visibility function $V$
becomes
\begin{equation}\begin{split}\label{ec-24}
V(u,v,w) &= \int_{-\infty}^{\infty}\int_{-\infty}^{\infty}A(x,y)I(x,y)\\
& \times {\rm e}^{{\rm i}2\pi\left(ux,vy,w\sqrt{1-x^2-y^2}\right)}
\frac{dx \ dy}{\sqrt{1-x^2-y^2}} .
\end{split}\end{equation}

\medskip

\noindent
The integration limits have been formally extended to $\pm\infty$ by
demanding that $A(x,y)=0$ for $x^2+y^2 > l^2$; where $l$ is the full
width of the primary telescope beams. Interestingly, Eq.\ \ref{ec-24}
closely resembles a two dimensional Fourier integral; this would be
identical if the term $w\sqrt{1-x^2-y^2}$ could be extracted from
under the integral sign. If only a small region of the sky is to be
mapped, then $\sqrt{1-x^2-y^2}\cong$ const $\cong$ 1 and Eq.\
\ref{ec-24} becomes
\begin{equation}\begin{split}\label{ec-25}
V(u,v,w) {\rm e}^{{\rm i}2\pi\omega} &=
\int_{-\infty}^{\infty}\int_{-\infty}^{\infty}A(x,y)I(x,y)
{\rm e}^{{\rm i}2\pi\left(ux,vy\right)} \ dx \ dy .
\end{split}\end{equation}

\medskip

\noindent
The factor ${\rm e}^{{\rm i}2\pi\omega}$ is the approximate conversion
required to change the observed phase of $V$ to the value that would
be measured with antennas in the $uv$ plane: $V(u,v,w) {\rm e}^{{\rm
i}2\pi\omega} \cong V(u,v,0)$. Substituting this into Eq.\ \ref{ec-25}
and performing the inverse Fourier transform we obtain
\begin{equation}\begin{split}\label{ec-26}
I'(x,y) &= A(x,y)I(x,y)=\int_{-\infty}^{\infty}V(u,v,0)
{\rm e}^{{\rm i}2\pi\left(ux,vy\right)} \ du \ dv .
\end{split}\end{equation}
where $I'(x,y)$ is the intensity $I(x,y)$ as modified by the primary
beam shape $A(x,y)$. One can easily correct $I'(x,y)$ by dividing it,
point for point, by $A(x,y)$. The input $V(u,v)$ is modified by a
linear transfer function, which produces the output image $I'$.

\section{Interferometer sensitivity}\label{c3-sensitivity}

The random noise limit to an interferometer system is calculated
following the method used for a single telescope. The $rms$
fluctuations in antenna temperature are
\begin{equation}\begin{split}\label{ec-27}
\Delta T_{\rm A} &= \frac{M~T_{\rm sys}}{\sqrt{t~\Delta\nu}},
\end{split}\end{equation}
where $M$ is a factor of order unity used to account for extra noise
from analog to digital conversions, digital clipping, etc. Applying
the definition of flux density given in Chapter \ref{chap-emission} in
terms of antenna temperature for a two-element system, we find:
\begin{equation}\begin{split}\label{ec-28}
\Delta S_{\nu} &= 2~k~\frac{T_{\rm sys}~{\rm e}^{\tau}}{A_{\rm e}\sqrt{2}},
\end{split}\end{equation}
where $\tau$ is the atmospheric opacity and $A_{\rm e}$ is the
effective collecting area of a single telescope of diameter $D$. The
extra factor of $\sqrt{2}$ arises from the use of two antennas each of
collecting area $A_{\rm e}$, and the fact that the correlation of two
noisy signals leads to a decrease in the noise by a factor of
$\sqrt{2}$. We denote the system noise corrected for atmospheric
absorption by $T'_{\rm sys}=T_{\rm sys} \ {\rm e}^{\tau}$ in order to
simplify the following equations. For an array of $n$ identical
telescopes, there are $N=n(n-1)/2$ simultaneous pair-wise
correlations. Then the $rms$ variation in flux density is
\begin{equation}\begin{split}\label{ec-29}
\Delta S_{\nu} &= \frac{2~M~k~T'_{\rm sys}}{A_{\rm e}\sqrt{2~N~t~\Delta\nu}}.
\end{split}\end{equation}
This relation can be recast in the form of brightness temperature
fluctuations using the Eq.\ \ref{ec-rayleigh}:
\begin{equation}\begin{split}\label{ec-30}
S &= 2~k~\frac{T_{\rm b}~\Omega_{\rm b}}{\lambda^{2}}.
\end{split}\end{equation}
Then the $rms$ brightness temperature, due to random noise, in
aperture synthesis image is
\begin{equation}\begin{split}\label{ec-31}
\Delta T_{\rm b} &= \frac{2~M~k~\lambda^{2}~T'_{\rm sys}}
{A_{\rm e}~\Omega_{\rm b}{\sqrt{2~N~t~\Delta\nu}}}.
\end{split}\end{equation}

\medskip

\noindent
A few qualitative comments in regard to the last equation should be
made. With shorter wavelengths, the $rms$ temperature fluctuations are
lower. Thus, for the same collecting area and system noise, a
millimeter image is more sensitive than an image made at centimeter
wavelengths. With a larger main beam solid angle, these fluctuations
will also decrease, if the effective collecting area remains the
same. For this reason, smoothing an image will result in a lower $rms$
noise in an image. It is frequently noted that multi-element
interferometers are capable of producing images faster than single
dishes. This is due to the fact that there are $n$ receivers in an
interferometer system.

\medskip

\noindent
For a Gaussian beam, $\Omega_{\rm mb}=1.133~\theta^{2}$, we can then
relate the $rms$ temperature fluctuations to observed properties of a
synthesis image. Thus, inserting numerical values in Eqs.\ \ref{ec-29}
and \ref{ec-31}, we have
\begin{equation}\begin{split}\label{ec-32}
\Delta S_{\nu}&=1.02~\frac{f~T'_{\rm sys}}{A_{\rm e}\sqrt{N~t~\Delta\nu}},\\
\Delta T_{\rm b}&=13.58~\frac{f~T'_{\rm sys}}{A_{\rm e}{\sqrt{N~t~\Delta\nu}}}
~\frac{\lambda^{2}}{\theta^{2}},
\end{split}\end{equation}
where $\lambda$ is expressed in mm, $\theta$ in arcsec, and
$\Delta\nu$ in kHz.

\medskip

\noindent
The limitations of the interferometer systems are two-fold. The
sensitivity in Kelvins of the system is usually worse than for a
single telescope. From Eq.\ \ref{ec-rayleigh}, since the sensitivity
in Jansky is fixed by the antenna collecting area and the receiver
noise, the only parameters which can be varied are the wavelength and
the angular resolution. And as can be easily seen, the increase in
angular resolution is made at the expense of temperature
sensitivity. This is not such a great problem for the high-brightness
radio sources, which radiate by non-thermal processes, such as
synchrotron radiation, but would be for thermal sources, for which the
maximum brightness temperatures is about $2\times10^4$ K for regions
of photo ionized gas surrounding massive stars.

\medskip

\noindent
Compared with single telescopes, interferometers have the great
advantage that uncertainties such as pointing and beam size depend on
electronics and fundamentally on timing. Such timing uncertainties can
be made very small compared to all other uncertainties. In contrast,
the single dish measurements are critically dependent on mechanical
deformations of the telescope. In summary, the single dish results are
easier to obtain, but source positions and sizes on the arc second
scales are difficult to estimate. The interferometer system has a much
greater degree of complexity, but allows one to measure such
details. The single dish system responds to the source irrespective of
the relation of source to beam size; the correlation interferometer
will respond to source structures smaller than the beam corresponding
to the minimum separation between the antennas.

\medskip

\noindent
One of the most important advantages of an interferometer over a single
dish is that with a single dish one needs to subtract an off-source
value. This can be corrupted by short term variations in atmospheric
emission/absorption or things that affect telescope gain such as
pointing. However, the interferometer ``switches'' at the natural
fringe rate, which for VLBI is typically kHz, and this cancels these
effects.

\medskip

\noindent
The method of aperture synthesis is based on sampling the visibility
function $V(u,v,0)$ with separate telescopes distributed in the
$(u,v)$ plane. Many configurations are possible, because all that is
needed is a reasonably dense covering of the $(u,v)$ plane. If one
calculates the $rms$ noise in a synthesis image obtained by simple
Fourier transforming the $(u,v)$ data, one usually finds a corrupted
image with a noise level frequently many times higher than that given
by $\Delta S_{\nu}$ or $\Delta T_{\rm b}$. The reason is that the
phases are affected by atmospheric or instrumental influences, but
another cause of higher $rms$ noise is that the $(u,v)$ plane is
usually incompletely sampled and instrumental effects are present,
such as stripe-like features in the final image. Yet another
systematic effect is the presence of grating rings around more intense
sources. It has been found that these effects can be eliminated by
software techniques.

\section{Very long baseline interferometers}\label{c3-vlbi}

For a given wavelength the angular resolving power of an
interferometer depends only on the length of the interferometer
baselines $B$. But the need to provide a phase-stable links (optical
fibers) between individual antennas and the correlator set limits on
$\mid B \mid$ to $\sim200$ km. Over longer paths, it becomes more
difficult to guarantee the phase stability, since transient
irregularities in the transmission path will have detrimental effects,
so several systems are limited to baselines of a few hundred
kilometers.

\medskip

\noindent
The development of atomic clocks with extremely phase-stable
oscillators opened up the possibility of avoiding altogether the
transmission of a phase-stable local oscillator signal. The
measurements are made independently at the individual antennas of the
interferometer. The data are recorder on storage media together with
precise time marks. These data are correlated later. Currently the
data are recorder on hard disks that are shipped to a central
correlator location. The antenna outputs contain accurate records of
the time variation of the electrical field strengths that the
appropriately time-averaged product obtained by multiplying the
digitized signal gives the mutual correlation function directly.

\medskip

\noindent
Local oscillators with extreme phase stability are needed at each
station for two reasons. The disks which are normally used usually
permit the recording of signals in a band reaching from zero to at
most a hundred MHz; the signal therefore must be mixed down to this
band, and for this a phase-stable local oscillator is needed, since
all phase jumps of the oscillator affect the IF signal. The second use
of the phase stability is to provide the extremely precise time marks
needed to align the signals from two stations. Again phase jumps would
destroy the coherence, that is the correlation of signals from the
source. For the local oscillators, different systems have been used
with varying success, ranging from rubidium clocks, free-running
quartz oscillators and, most successfully, hydrogen masers. With
present day hydrogen maser technologies, it is possible to have
frequency and phase stability that allows measurements for many
minutes. At longer centimeter wavelengths, the maximum time over which
the visibility function can be coherently integrated, that is, where
we can determine the amplitude and phase of the visibility function is
not limited by the best currently commercially available maser
clocks. At wavelengths or 1 cm or shorter, the atmosphere is the
limit.

\medskip

\noindent
Today, in VLBI only digital data recording is used. The media from
different observatories are processed on special purpose digital
correlators that align the signals in time, account for local
oscillator offsets and geometric delays, clock rate offsets and
differential Doppler shifts arising from Earth rotation and then
generate a correlation function for each pair of observation
sites. Amplitudes and phases of these correlation functions are
directly comparable to the complex visibilities of a conventional
connected element interferometer. The delay time between the two
independent telescopes can vary rapidly. In the past, one could not
determine the instrumental phases from measurements of a calibration
source, so one had to use \textit{fringe fitting} to allow the
accumulation of data over much longer times. The correlator delivers
an amplitude as a function of time for a delay range larger than any
uncertainty. In the Fourier transform domain the time axis becomes
frequency (\textit{residual fringe frequency} or \textit{fringe rate})
and then the maximum should appear as a peak in this two-dimensional
distribution. The coordinates of the \textit{peak}, the
\textit{fringe rate} and the \textit{lag} are the required
parameters. For strong sources this maximum can be determined
directly, but for weaker sources more sophisticated techniques are
needed, and will be described in some detail in Chapter
\ref{chap-observations} and Appendix A. Once these parameters are
determined, further reduction procedures are basically identical to
those used for the analysis of conventional synthesis array data. To
remove the remaining errors , one solves and corrects for residual
delays and fringe rate offsets. This is an additional step,
\textit{fringe fitting}, is necessary for VLBI reductions. There
are several reasons for the phase variation; (1) random delays in the
atmospheric propagation properties at the individual sites and (2)
phase changes in the electronics and the independent clocks. For these
reasons, without fringe fitting the correlation will be only fairly
short for a given source.

\section{Conclusions}\label{c3-conclusions}

In this chapter, we have described in some detail how interferometers
work. As we mentioned in Chapter \ref{chap-introduction}, VLBI
instruments are ideal for astrometry work because phase-referencing to
an extragalactic source with accurately-known position, can be
performed. A detailed description of the interferometer that we used
to collect the observations presented in this thesis (the Very Long
Baseline Array) is given in Appendix A. We are now ready to describe
the observations and the calibration made in this thesis.

\chapter{Observations}\label{chap-observations}

\begin{quote}
\noindent
In this chapter we will describe the observations, the data reduction,
and calibration that form the basis of this thesis.
\end{quote}

\section{Summary of observations}\label{c4-summary}

We made use of a total of 67 observations obtained with the VLBA. This
represents about 382 hours of telescope time. As mentioned earlier,
five stars in Taurus and two in Ophiuchus were considered. Tab.\
\ref{tab-sources} lists the sources and the corresponding position of
the phase centers (i.e.\ the center of the synthesized field, rather
than source position).

\begin{deluxetable}{llll}
\tabletypesize{\scriptsize}
\tablecolumns{4}
\tablewidth{0pc}
\tablecaption{\footnotesize{Sources in the thesis.}\label{tab-sources}}
\tablehead{
\colhead{SFR}                     &
\colhead{~~~~~~~~~Source~~~~~~~~~}&
\colhead{$\alpha_{\rm phase}$}     &
\colhead{$\delta_{\rm phase}$}     }
\startdata
Taurus & T~Tau~Sb   \dotfill & \dechms{04}{21}{59}{426} & \decdms{19}{32}{05}{730}\\%
\\[-0.2cm]
       & Hubble~4   \dotfill & \dechms{04}{18}{47}{033} & \decdms{28}{20}{07}{398}\\%
\\[-0.2cm]
       & HDE~283572 \dotfill & \dechms{04}{21}{58}{847} & \decdms{28}{18}{06}{502}\\%
\\[-0.2cm]
       & HP~Tau/G2  \dotfill & \dechms{04}{35}{54}{161} & \decdms{22}{54}{13}{492}\\%
\\[-0.2cm]
       & V773~Tau~A \dotfill & \dechms{04}{14}{12}{922} & \decdms{28}{12}{12}{180}\\%
\\[-0.2cm]
\hline
\\[-0.2cm]
Ophiuchus & S1         \dotfill & \dechms{16}{26}{34}{174} & \decdms{$-$24}{23}{28}{428}\\%
\\[-0.2cm]
          & DoAr~21    \dotfill & \dechms{16}{26}{03}{019} & \decdms{$-$24}{23}{36}{340}\\%
\\[-0.2cm]
\enddata
\end{deluxetable}

\medskip

\noindent
All 67 observations were made at 3.6 cm (8.42 GHz), in
phase-referenced mode. We used a bit rate of 256 Mb/s rather than 128
Mb/s for several reasons. First, we wanted to schedule when the
sources are at high elevation to give us the best chance of successful
phase transfer between the calibrators and the targets --this was
particularly important for the sources in Ophiuchus. Also, we need as
much sensitivity as possible to obtain the astrometric accuracy need
to reach our scientific goals.

\medskip

\noindent
Since our goal was to measure the annual parallax, each source was
observed at least every several months over a few years. In the case
of T~Tau~Sb, we also re-reduced an observation gathered in 1999, and
obtained a single additional pointing in 2008, to help characterize
the orbital path of the system. For V773~Tau~A, we also obtained 13
additional observations spread over about 80 days, which were needed
to help determine the physical orbit of the system. Tab.\
\ref{tab-observaciones} contains a summary of these observations, and
is divided in seven blocks where each block corresponds to one
source. The epoch, mean UT date, and mean Julian Date, are listed in
cols.\ [2], [3] and [4]. The number of antennas, and the total hours
of telescope time used in the source and calibrators, are given
explicitly in cols.\ [5] and [6]. Finally, the noise in each resulting
image is given in col.\ [7].

\medskip

\noindent
Each observation consisted in a series of cycles with two minutes
spent on source and one minute spent on the main phase-referencing
quasar. Because the VLBI astrometric accuracy can be improved if more
than one reference calibrator is observed, during most observations,
secondary quasars were also observed every 24 minutes. Since all seven
stars observed are weak targets, then all of the calibrators are used
to determine the phase correction at the target source. A simple
example of the observing schedule used in the thesis is:

\noindent
\begin{center}
p--T--p--T--p--T--p--T--p--T--p--T--p--T--p--T--p--\textbf{1}--\textbf{2}--\textbf{3}--p--T--p--T--p--T--\\
p--T--p--T--p--T--p--T--p--T--p--\textbf{1}--\textbf{2}--\textbf{3}--p--T--p--T--p--T--p--T--p--T--p--T--\\
p--T--p--T--p--\textbf{1}--\textbf{2}--\textbf{3}--p--T--p--T--p--T--p--T--p--T--p--T--p--T--p--T--p--\textbf{1}...
\end{center}

\noindent
where \textbf{p} is the primary calibrator, \textbf{T} is the target,
and \textbf{1}, \textbf{2}, \textbf{3} are the secondary
calibrators. Note that one out of every nine \textbf{T} observations
is replaced by \textbf{1}--\textbf{2}--\textbf{3} observations.

\medskip

\noindent
Tab.\ \ref{tab-calibradores} contains all primary and secondary
calibrators used for each source in the project. All calibrators are
very compact extragalactic sources whose absolute positions are known
to better than a few milli-arcseconds (see cols.\ [4] and [6] of Tab.\
\ref{tab-calibradores}), and are located in the vicinity of the target
(the separation between target and calibrator is given in col.\ [7] of
Tab.\ \ref{tab-calibradores}). Also, in Fig.\ \ref{fig-configuracion}
is shown the calibration configuration for the astronomical targets
HDE~283572 and Hubble~4.

\medskip

\noindent
The data were edited and calibrated using the Astronomical Image
Processing System (\aips ~--Greisen 2003) following the procedures
described in the next sections.

\section{Basic data reduction}\label{c4-basic}

In this section we explain all the steps that we followed for the
initial data reduction. It follows closely the standard VLBA procedure
for phase-referenced observations (Appendix C, \aips~\cook).

\begin{enumerate}

\item
 Ionospheric corrections were applied with the \textsf{VLBATECR}
 procedure which automatically downloads the needed Global Positioning
 System (GPS) models of the electron content in the Earth atmosphere,
 and correct the dispersive delays caused by the ionosphere with the
 task \textsf{TECOR}. This procedure is recommended for all
 experiments at 8 GHz and lower frequencies, in particular for
 phase-referencing.

\item
 The VLBA correlator must use measurements of the Earth Orientation
 Parameters (EOPs). It uses the best available estimate at the time of
 the correlation, but better measurements usually become available
 several weeks after correlation. It is recommended that all
 phase-referencing experiments be corrected to these newer EOPs
 measurements. The procedure \textsf{VLBAEOPS} automatically downloads
 a file from the US Naval Observatory database and correct the EOPs
 using \textsf{CLCOR}.

\item
 To calibrate visibility amplitudes, we use the procedure
 \textsf{VLBACALA} which runs several tasks in sequence. The task
 \textsf{ACCOR} uses the autocorrelation to correct the sampler
 voltage offsets, and creates a solution table \textsf{SN} which is
 then smoothed with \textsf{SNSMO} in order to remove any outlying
 points. This smoothed \textsf{SN} table is applied to the highest
 calibration table \textsf{CL} using \textsf{CLCAL}, and a new
 \textsf{CL} table is created. The task \textsf{APCAL} is then run on
 the highest \textsf{TY} (system temperature) and \textsf{GC} (gain
 curve) tables to apply the amplitude calibration, and a new
 \textsf{SN} table is created. Finally, \textsf{CLCAL} is run to apply
 the last \textsf{SN} table to the \textsf{CL} table created before.

\item
 The RCP and LCP feeds on alt-az antennas will rotate in position
 angle with respect to the source during the course of the
 observation. Since this rotation is a simple geometric effect, it can
 be corrected by adjusting the phases without looking at the
 data. This correction is important in phase-referencing experiments,
 because the parallactic angle difference between calibrator and
 target is different at different stations. This leads to an extra
 phase error which must be corrected. We used the procedure
 \textsf{VLBAPANG} that uses \textsf{TACOP} to copy the highest
 \textsf{CL} table and then runs \textsf{CLCOR} to correct the
 parallactic angles.

\item
 Instrumental delay residuals are caused by the passage of the signal
 through the electronics of the VLBA baseband converters. We applied
 the procedure \textsf{VLBAMPCL} that uses the fringes on a strong
 source to compute the delays and phase residuals for each antenna and
 IF. This procedure runs \textsf{FRING} to find the corrections and
 then \textsf{CLCAL} to apply them.

\item
 At this point we remove global frequency- and time-dependent phase
 errors using \textsf{FRING}. For phase-referencing experiments, one
 could run \textsf{CALIB} instead to \textsf{FRING}, but fringe
 fitting is recommended because it solves for rates while
 \textsf{CALIB} does not. \textsf{FRINGE} produces a new \textsf{SN}
 table which is then smoothed with \textsf{SNSMO} to remove outlying
 points. Finally, the \textsf{SN} table is used to run \textsf{CLCAL}
 once for each source. In this last step, we used the \textsf{2PT}
 linear vector interpolation method. Note that at this point,
 \textsf{FRINGE} assumed the calibrator is a point source at the phase
 center.

\item
 In this last step, we apply the calibration to the visibility data
 and make single-source data set using \textsf{SPLIT}. A set of images
 were produced using \textsf{IMAGR}. As an example, in Fig.\
 \ref{fig-reduce} we shown the images produced for target, primary and
 secondary calibrators in the first epoch of HDE~283572 (22/Sep/04;
 see Tab.\ \ref{tab-observaciones}).

\end{enumerate}

\section{Self-calibration}\label{c4-self}

Even after global fringe fitting, transferring the phases from the
calibrator to the target, the phases on the target can still vary with
time. Most of these variations are due to inadequate removal of
antenna-based atmospheric phases, but some variations also can be
caused by an inadequate model of the source structure during fringe
fitting (e.g.\ it is unlikely that the calibrator is a perfect point
source). If the calibrator is sufficiently strong, it is possible to
reduce these phase fluctuations by looping through cycles of Fourier
transform imaging and deconvolution, combined with phase
self-calibration in a time interval shorter than the time scale for
the variations that one want to remove (Cornwell 1995; Walker 1995;
Cornwell \& Fomalont 1999). The resulting images can be deconvolved to
rid them of substantial sidelobes arising from relatively sparse
sampling of the $(u,v)$ plane (Cornwell, Braun, \& Briggs 1999).

\medskip

\noindent
The first step is to use a point-source model to make the first image
of the main calibrator by using the task \textsf{IMAGR} (see second
map in Fig.\ \ref{fig-reduce}). It is a good idea to use the task
\textsf{CCMRG} in the resulting image to reduce the number of
components in the model. This improves the speed of the calibration
and make the first negative component be a real negative rather than a
minor correction to previous positive components. The resulting clean
image can then be used as a model of the source in the following
iteration of \textsf{CALIB} to correct only phases. \textsf{CALIB}
compares the input $(u,v)$ data set with the predictions of the source
model in order to compute a set of antenna-based phase corrections as
a function of time which would bring the data into better agreement
with the current model. \textsf{CALIB} produces a new improved data
set, and will catalog the gain corrections as an \textsf{SN} extension
to the input $(u,v)$ data file. If the corrections are believable, we
run \textsf{IMAGR} to produce a new clean image. The next step is to
use the new clean image as a model in \textsf{CALIB} to produce a new
improved data set. The whole process is repeated a few times, and
results in an improved image. During these first iterations, we only
seek improved phase corrections. During a second set of
self-calibration iterations, we also determine improved amplitude
corrections. When the new iterations provide little improvement in the
image quality, we stop and the last image is kept as our best model
for the calibrator.

\medskip

\noindent
A different phase calibrator model was produced for each epoch for
each source to account for possible small changes in the main
calibrator structure from epoch to epoch. In Fig.\ \ref{fig-model} are
shown the main calibrator models used at all six epochs for
HDE~283572. Using an image model for the calibrator rather than
assuming a point source improved the position accuracy by a few tens
of $\mu$as.

\section{Re-fringe fit main phase calibrator}\label{c4-re-fringe}

In the images obtained after the self-calibration iterations, the main
phase calibrator for almost all epochs was found to be slightly
extended or with some source structure (see Fig.\ \ref{fig-model}). To
take this into account, we followed the extra steps listed below.

\begin{enumerate}

\item
 The global fringe fitting (\textsf{FRINGE}, \textsf{SNSMO} and
 \textsf{CLCAL} in step 6 of Sect.\ \ref{c4-basic}) was repeated using
 the image of the main phase calibrator (obtained in Sect.\
 \ref{c4-self}) as a model instead of assuming it to be a point
 source.

\item
 After phases have been corrected with \textsf{FRING}, an improvement
 of the {\it a priori} antenna amplitude gains (step 3 in Sect.\
 \ref{c4-basic}) can be obtained using the observations of the main
 phase calibrator. To achieve that, we ran \textsf{CALIB} instead to
 \textsf{FRING} to produce a new \textsf{SN} table which was then
 smoothed with \textsf{SNSMO}. The corrections were transferred to the
 target and secondary calibrators using \textsf{CLCAL}.

\item
 This last step is the same as step 7 of Sect.\ \ref{c4-basic}: apply
 the calibration to the visibility data and make single-source data
 set using \textsf{SPLIT}. The single-source data sets are ready for
 imaging. In Fig.\ \ref{fig-redo} are shown the images produced in
 this step for the target, primary and secondary calibrators in the
 first epoch of HDE~283572 (22/Sep/04; see Tab.\
 \ref{tab-observaciones}). Compare Fig.\ \ref{fig-reduce} with Fig.\
 \ref{fig-redo} and note that in the last one, the peak flux is
 highest, and the r.m.s.\ is lowest.

\end{enumerate}

\section{Phase referencing using more than one calibrator}\label{c4-phase}

\begin{figure*}[!b]
\centerline{
\includegraphics[height=.8\textwidth,angle=-90]{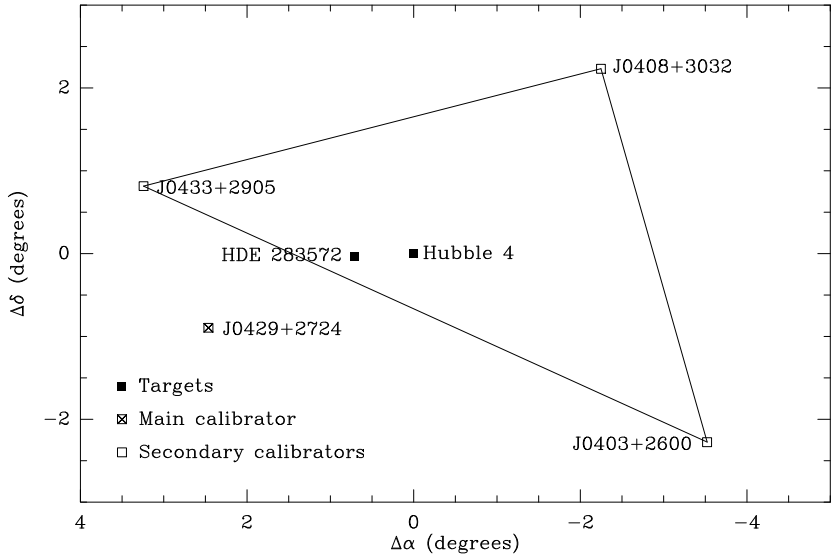}}
\caption{\footnotesize{
Relative position of the astronomical targets HDE~283572 and Hubble~4,
the main calibrator J0429+2724, and the secondary calibrators
J0433+2905,J0408+3032, J0403+2600.}
\label{fig-configuracion}}\end{figure*}

The astrometry precision of VLBA observations depends critically on
the quality of the phase calibration. Systematic errors, unremoved by
the standard calibration procedures described above, usually dominate
the phase calibration error budget, and limit the astrometric
precision achieved to several times the value expected theoretically
(Fomalont 1999, Pradel et al.\ 2006). At the frequency of the present
observations, the main sources of systematic errors are inaccuracies
in the troposphere model used, as well as clock, antenna and \textit{a
priori} source position errors. These effects combine to produce a
systematic phase difference between the calibrator and the target,
causing position shifts. One effective strategy to measure and correct
these systematic errors consists of including observations of more
than one phase calibrator chosen to surround the target (Fomalont,
2005). This allows phase gradients around the source due to errors in
the troposphere model or related to uncertainties in the cataloged
position of the calibrators, to be measured and corrected. This
strategy was applied to most of our observations using several
secondary calibrators (e.g.\ see the calibrator configuration for our
observations of HDE~283572 and Hubble~4 in Fig.\
\ref{fig-configuracion}), and resulted in significant improvements in
the final phase calibration and image quality.

\medskip

\noindent
The task \textsf{ATMCA} in \aips~ has been written to combine the data
from several calibrators to improve the image quality and position
accuracy of the target source. This task only requires calibrator data
in the vicinity of the target. In Fig.\ \ref{fig-configuracion} the
phase differences between each pair of calibrators are sufficient to
determine the two-dimensional phase gradient, from which an estimate
of the phase at the target position can be obtained. This
configuration gives a robust solution. Thus, our observing schedules
included observation of secondary calibrators every 24 minutes (see
Sect.\ \ref{c4-summary}).

\begin{figure*}[!b]
\centerline{
\includegraphics[height=.9\textwidth,angle=0]{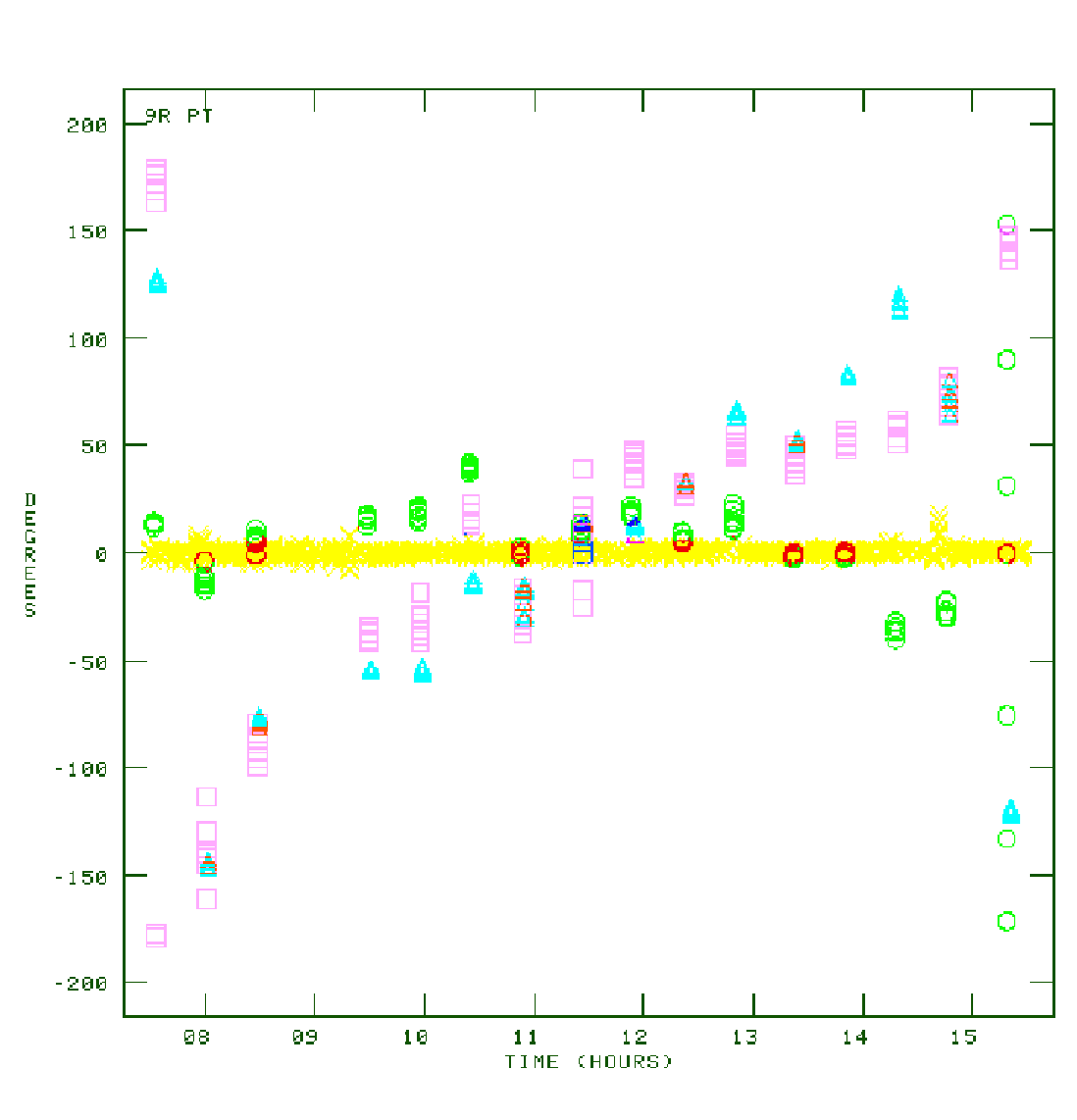}}
\caption{\footnotesize{
The plot shows the residual phases for the calibrators that was input
to \textsf{ATMCA} for HDE~283572 at the first epoch (22/Sep/04; see
Tab.\ \ref{tab-observaciones}). The yellow crosses are for main phase
calibrator J0429+2724, the green circles, pink squares, and light blue
triangles are for the secondary calibrators J0433+2905, J0408+3032,
and J0403+2600, respectively.}
\label{fig-phases}}\end{figure*}

\medskip

\noindent
Several steps have to be taken before running \textsf{ATMCA}. Since
most calibrators are not point sources, self-calibration loops for
each secondary calibrator are necessary. From the resulting
self-calibrated images, we measure the offsets between the observed
and the cataloged calibrator positions (i.e.\ the offset between the
observed position of the calibrators, and their respective phase
centers). The task \textsf{CLCOR} is then used to correct for these
offsets, and ``move'' the calibrators to the phase
centers. \textsf{SPLAT} was used to obtain data sets averaged in
frequency. Finally, we run \textsf{CALIB} on all of the calibrators to
produce an \textsf{SN} table with the residual phases for all
calibrators. Fig.\ \ref{fig-phases} is an example of the residual
phases (in antenna 9=PT with all IFs and Stokes RR) for all four
calibrators used in HDE~283572 at the first epoch (22/Sep/04). Note
the residuals for secondary calibrators at the begin and at the end of
observation.

\medskip

\noindent
The input to \textsf{ATMCA} is that \textsf{SN} table containing the
time variable residual antenna-based phases, sampled in the direction
to the calibrators. The phase origin is defined by the main
calibrator. The goal of \textsf{ATMCA} is to determine the phase
gradient in the sky which is consistent with the phases sampled by the
secondary calibrators, and then to apply this gradient to the phase of
all sources. The images produced after \textsf{ATMCA} for HDE~283572,
primary and secondary calibrators at the first epoch (22/Sep/04) is
shown in Fig.\ \ref{fig-atmca}.

\section{Improvements}\label{c4-improvements}

In order to give an estimate of the magnitude of the improvements
following the calibration method presented in this chapter, we include
images (Fig.\ \ref{fig-hde-mejoras} and \ref{fig-cal-mejoras}) for
target and phase calibrator after each step produced for HDE~283572 at
the first epoch (22/Sep/04; see Tab.\ \ref{tab-observaciones}). Also,
in Tab.\ \ref{tab-mejoras} we give intensities and positions after
each step, to compare the improvements in calibration.


\begin{deluxetable}{lllllcc}
\tabletypesize{\scriptsize}
\tablecolumns{7}
\tablewidth{0pc}
\tablecaption{\footnotesize{Observations.}\label{tab-observaciones}}
\tablehead{
\colhead{Source}                                    &
\colhead{Epoch}                                     &
\multicolumn{2}{c}{Mean Date}                       &
\colhead{Antennas\tablenotemark{a}}                 &
\colhead{$t_{\rm obs}$\tablenotemark{c}}            &
\colhead{rms}                                       \\
\cline{3-4}                                         \\[-0.2cm]
\multicolumn{1}{c}{}                                &
\multicolumn{1}{c}{}                                &
\multicolumn{1}{c}{~~~~~~~~~~~~~~~UT~~~~~~~~~~~~~~~}&
\multicolumn{1}{c}{JD}                              &
\multicolumn{1}{c}{}                                &
\multicolumn{1}{c}{[hr]}                            &
\multicolumn{1}{c}{[mJy/beam]}                      }
\startdata
\\[-0.2cm]
T~Tau~Sb & 1  & 2003 Sep 24~~11:33\dotfill & 2452906.981 & VLBA    & 5  & 0.074 \\
         & 2  & 2003 Nov 18~~08:02\dotfill & 2452961.834 & VLBA    & 5  & 0.066 \\
         & 3  & 2004 Jan 15~~04:09\dotfill & 2453019.672 & VLBA-MK & 5  & 0.072 \\
         & 4  & 2004 Mar 26~~23:26\dotfill & 2453091.476 & VLBA    & 5  & 0.070 \\
         & 5  & 2004 May 13~~20:17\dotfill & 2453139.345 & VLBA-PT & 5  & 0.111 \\
         & 6  & 2004 Jul 08~~16:37\dotfill & 2453195.192 & VLBA    & 5  & 0.064 \\
         & 7  & 2004 Sep 16~~11:59\dotfill & 2453264.999 & VLBA    & 6  & 0.070 \\
         & 8  & 2004 Nov 09~~08:27\dotfill & 2453318.852 & VLBA-HN & 6  & 0.104 \\
         & 9  & 2004 Dec 28~~05:14\dotfill & 2453367.718 & VLBA    & 6  & 0.070 \\
         & 10 & 2005 Feb 24~~01:26\dotfill & 2453425.559 & VLBA    & 6  & 0.080 \\
         & 11 & 2005 May 09~~20:32\dotfill & 2453500.355 & VLBA-OV & 6  & 0.106 \\
         & 12 & 2005 Jul 08~~16:36\dotfill & 2453560.191 & VLBA    & 6  & 0.075 \\
         & 13 & 2008 May 29~~19:09\dotfill & 2454616.298 & VLBA    & 5  & 0.080 \\
         & 14\tablenotemark{b}
              & 1999 Dec 15~~06:04\dotfill & 2451527.753 & VLBA    & 11 & 0.069 \\
\\[-0.1cm]\hline\\[-0.1cm]
Hubble~4 & 1  & 2004 Sep 19~~11:47\dotfill & 2453267.991 & VLBA & 9 & 0.054 \\
         & 2  & 2005 Jan 04~~04:46\dotfill & 2453374.699 & VLBA & 9 & 0.073 \\
         & 3  & 2005 Mar 25~~23:28\dotfill & 2453455.478 & VLBA & 9 & 0.114 \\
         & 4  & 2005 Jul 04~~16:51\dotfill & 2453556.202 & VLBA & 9 & 0.058 \\
         & 5  & 2005 Sep 18~~11:52\dotfill & 2453631.994 & VLBA & 9 & 0.053 \\
         & 6  & 2005 Dec 28~~05:15\dotfill & 2453732.719 & VLBA & 9 & 0.051 \\
\\[-0.1cm]\hline\\[-0.1cm]
HDE~283572 & 1  & 2004 Sep 22~~11:35\dotfill & 2453270.983 & VLBA    & 9 & 0.081 \\
           & 2  & 2005 Jan 06~~04:39\dotfill & 2453376.693 & VLBA    & 9 & 0.058 \\
           & 3  & 2005 Mar 30~~23:08\dotfill & 2453460.464 & VLBA    & 9 & 0.065 \\
           & 4  & 2005 Jun 23~~17:34\dotfill & 2453545.232 & VLBA    & 9 & 0.080 \\
           & 5  & 2005 Sep 23~~11:32\dotfill & 2453636.981 & VLBA-NL & 9 & 0.062 \\
           & 6  & 2005 Dec 24~~05:31\dotfill & 2453728.729 & VLBA    & 9 & 0.047 \\
\\[-0.1cm]\hline\\[-0.1cm]
HP~Tau/G2 & 1  & 2005 Sep 07~~12:36\dotfill & 2453621.024 & VLBA    & 6 & 0.06 \\
          & 2  & 2005 Nov 16~~08:01\dotfill & 2453690.833 & VLBA    & 6 & 0.07 \\
          & 3  & 2006 Jan 23~~03:33\dotfill & 2453758.648 & VLBA    & 6 & 0.07 \\
          & 4  & 2006 Mar 31~~23:06\dotfill & 2453826.462 & VLBA    & 6 & 0.07 \\
          & 5  & 2006 Jun 10~~18:27\dotfill & 2453897.268 & VLBA    & 6 & 0.08 \\
          & 6  & 2006 Sep 08~~12:33\dotfill & 2453987.022 & VLBA    & 6 & 0.06 \\
          & 7  & 2007 Jun 04~~18:56\dotfill & 2454256.289 & VLBA    & 9 & 0.07 \\
          & 8  & 2007 Sep 03~~12:53\dotfill & 2454347.037 & VLBA-HN & 9 & 0.06 \\
          & 9  & 2007 Dec 04~~06:51\dotfill & 2454438.785 & VLBA-SC & 9 & 0.05 \\
\\[-0.1cm]\hline\\[-0.1cm]
V773~Tau~A & 1  & 2005 Sep 08~~12:01\dotfill & 2453622.001 & VLBA       & 2 & 0.145 \\
           & 2  & 2005 Nov 15~~07:31\dotfill & 2453689.813 & VLBA       & 2 & 0.117 \\
           & 3  & 2006 Jan 21~~03:11\dotfill & 2453756.632 & VLBA       & 2 & 0.109 \\
           & 4  & 2006 Apr 01~~22:31\dotfill & 2453827.438 & VLBA-PT    & 2 & 0.238 \\
           & 5  & 2006 Jun 12~~17:48\dotfill & 2453899.242 & VLBA       & 2 & 0.102 \\
           & 6  & 2006 Sep 05~~12:14\dotfill & 2453984.010 & VLBA       & 2 & 0.159 \\
           & 7  & 2007 Aug 23~~13:06\dotfill & 2454336.046 & VLBA       & 5 & 0.207 \\
           & 8  & 2007 Aug 29~~12:42\dotfill & 2454342.029 & VLBA       & 5 & 0.148 \\
           & 9  & 2007 Sep 05~~12:15\dotfill & 2454349.010 & VLBA-HN    & 5 & 0.117 \\
           & 10 & 2007 Sep 11~~11:51\dotfill & 2454354.994 & VLBA-HN-SC & 5 & 0.110 \\
           & 11 & 2007 Sep 16~~11:32\dotfill & 2454359.980 & VLBA-HN-SC & 5 & 0.122 \\
           & 12 & 2007 Sep 21~~11:12\dotfill & 2454364.966 & VLBA-SC-PT & 5 & 0.106 \\
           & 13 & 2007 Sep 27~~10:48\dotfill & 2454370.950 & VLBA-SC    & 5 & 0.135 \\
           & 14 & 2007 Oct 03~~10:25\dotfill & 2454376.934 & VLBA-SC-NL & 5 & 0.148 \\
           & 15 & 2007 Oct 09~~10:01\dotfill & 2454382.917 & VLBA-SC-PT & 5 & 0.256 \\
           & 16 & 2007 Oct 17~~09:30\dotfill & 2454390.896 & VLBA-SC    & 5 & 0.405 \\
           & 17 & 2007 Oct 23~~09:06\dotfill & 2454396.879 & VLBA-SC-KP & 5 & 0.267 \\
           & 18 & 2007 Oct 27~~08:50\dotfill & 2454400.868 & VLBA-SC    & 5 & 0.159 \\
           & 19 & 2007 Nov 17~~07:28\dotfill & 2454421.811 & VLBA-SC-FD & 5 & 0.139 \\
\\[-0.1cm]\hline\\[-0.1cm]
S1 & 1  & 2005 Jun 24~~05:32\dotfill & 2453545.730 & VLBA       & 4 & 0.311 \\
   & 2  & 2005 Sep 15~~00:05\dotfill & 2453628.504 & VLBA-LA    & 4 & 0.199 \\
   & 3  & 2005 Dec 17~~17:56\dotfill & 2453722.247 & VLBA       & 4 & 0.196 \\
   & 4  & 2006 Mar 15~~12:10\dotfill & 2453810.007 & VLBA-KP    & 4 & 0.135 \\
   & 5  & 2006 Jun 03~~06:56\dotfill & 2453889.789 & VLBA-KP-HN & 4 & 0.184 \\
   & 6  & 2006 Aug 22~~01:41\dotfill & 2453969.570 & VLBA       & 4 & 0.209 \\
\\[-0.1cm]\hline\\[-0.1cm]
DoAr~21 & 1  & 2005 Sep 08~~00:33\dotfill & 2453621.523 & VLBA    & 4 & 0.144  \\
        & 2  & 2005 Nov 16~~19:58\dotfill & 2453691.332 & VLBA    & 4 & 0.185  \\
        & 3  & 2006 Jan 08~~16:30\dotfill & 2453744.187 & VLBA    & 4 & 0.054  \\
        & 4  & 2006 Jan 19~~15:46\dotfill & 2453755.157 & VLBA    & 4 & 0.063  \\
        & 5  & 2006 Mar 28~~11:19\dotfill & 2453822.971 & VLBA-MK & 4 & 0.056  \\
        & 6  & 2006 Jun 04~~06:52\dotfill & 2453890.786 & VLBA-HN & 4 & 0.073  \\
        & 7  & 2006 Aug 24~~01:33\dotfill & 2453971.565 & VLBA    & 4 & 0.043  \\
\enddata
\tablenotetext{a}{VLBA antennas: MK--Mauna Kea, PT--Pie Town,
HN--Hancock, OV--Owens Valley, NL--North Liberty, SC--St. Croix,
KP--Kitt Peak, FD--Fort Davis, BR--Brewster, LA--Los Alamos.}
\tablenotetext{b}{Data from VLBA archive, project BB112, 99-Dec-15.}
\tablenotetext{c}{This time include time spent in source and calibrators.}
\end{deluxetable}

\begin{deluxetable}{llllccc}
\tabletypesize{\scriptsize}
\tablecolumns{7}
\tablewidth{0pc}
\tablecaption{\footnotesize{Calibrators.}\label{tab-calibradores}}
\tablehead{
\colhead{Source}&
\colhead{Calibrator}&
\colhead{$\alpha_{\rm J2000.0}$ \tablenotemark{a}}&
\colhead{$\sigma_\alpha$ \tablenotemark{a}}&
\colhead{$\delta_{\rm J2000.0}$ \tablenotemark{a}}&
\colhead{$\sigma_\delta$ \tablenotemark{a}}&
\colhead{Sep.}\\
\multicolumn{1}{c}{}&
\multicolumn{1}{c}{[J2000]}&
\multicolumn{1}{c}{[\inthms{~}{~}{~}~]}&
\multicolumn{1}{c}{[mas]}&
\multicolumn{1}{c}{[\intdms{~}{~}{~}~]}&
\multicolumn{1}{c}{[mas]}&
\multicolumn{1}{c}{[deg]}}
\startdata
\\[-0.3cm]
T~Tau~Sb & J0428+1732 \tablenotemark{p} & 04:28:35.633683 & 0.49 & +17:32:23.58810 & 0.77 &  2.54 \\%
         & J0412+1856 \tablenotemark{s} & 04:12:45.944191 & 0.58 & +18:56:37.07668 & 0.71 &  2.26 \\%
         & J0431+1731 \tablenotemark{s} & 04:31:57.379257 & 0.73 & +17:31:35.77540 & 1.23 &  3.10 \\%
         & J0426+2327 \tablenotemark{s} & 04:26:55.734792 & 0.32 & +23:27:39.63378 & 0.59 &  4.09 \\%
\\[-0.2cm]\hline\\[-0.2cm]
Hubble~4 & J0429+2724 \tablenotemark{p} & 04:29:52.960767 & 0.23 & +27:24:37.87633 & 0.41 &  2.62 \\%
         & J0408+3032 \tablenotemark{s} & 04:08:20.377570 & 0.36 & +30:32:30.48991 & 0.58 &  3.17 \\%
         & J0433+2905 \tablenotemark{s} & 04:33:37.829861 & 0.23 & +29:05:55.47708 & 0.40 &  3.34 \\%
         & J0403+2600 \tablenotemark{s} & 04:03:05.586060 & 0.23 & +26:00:01.50283 & 0.40 &  4.20 \\%
\\[-0.2cm]\hline\\[-0.2cm]
HDE~283572 & J0429+2724 \tablenotemark{p} & 04:29:52.960767 & 0.23 & +27:24:37.87633 & 0.41 & 1.96 \\%
           & J0408+3032 \tablenotemark{s} & 04:08:20.377570 & 0.36 & +30:32:30.48991 & 0.58 & 3.72 \\%
           & J0433+2905 \tablenotemark{s} & 04:33:37.829861 & 0.23 & +29:05:55.47708 & 0.40 & 2.68 \\%
           & J0403+2600 \tablenotemark{s} & 04:03:05.586060 & 0.23 & +26:00:01.50283 & 0.40 & 4.79 \\%
\\[-0.2cm]\hline\\[-0.2cm]
HP~Tau/G2 & J0426+2327 \tablenotemark{p} & 04:26:55.734792 & 0.32 & +23:27:39.63378 & 0.59 & 2.14 \\%
          & J0435+2532 \tablenotemark{s} & 04:35:34.582945 & 1.23 & +25:32:59.69695 & 2.12 & 2.65 \\%
          & J0449+1754 \tablenotemark{s} & 04:49:12.511544 & 0.56 & +17:54:31.59559 & 0.69 & 5.89 \\%
\\[-0.2cm]\hline\\[-0.2cm]
V773~Tau~A & J0408+3032 \tablenotemark{p} & 04:08:20.377570 & 0.36 & +30:32:30.48991 & 0.58 & 2.67 \\%
           & J0403+2600 \tablenotemark{s} & 04:03:05.586060 & 0.23 & +26:00:01.50283 & 0.40 & 3.31 \\%
           & J0429+2724 \tablenotemark{s} & 04:29:52.960767 & 0.23 & +27:24:37.87633 & 0.41 & 3.55 \\%
           & J0356+2903 \tablenotemark{s} & 03:56:08.461936 & 0.68 & +29:03:42.32059 & 1.46 & 4.06 \\%
\\[-0.2cm]\hline\\[-0.2cm]
S1 & J1625-2527 \tablenotemark{p} & 16:25:46.891640 & 0.22 & -25:27:38.32687 & 0.40 & 1.08 \\%
   & J1617-1941 \tablenotemark{s} & 16:17:27.093081 & 0.96 & -19:41:32.01350 & 1.79 & 5.15 \\%
   & J1644-1804 \tablenotemark{s} & 16:44:35.746823 & 2.37 & -18:04:32.45913 & 5.36 & 7.58 \\%
   & J1626-2951 \tablenotemark{s} & 16:26:06.020836 & 0.24 & -29:51:26.97118 & 0.41 & 5.47 \\%
\\[-0.2cm]\hline\\[-0.2cm]
DoAr~21 & J1625-2527 \tablenotemark{p} & 16:25:46.891640 & 0.22 & -25:27:38.32687 & 0.40 & 1.07 \\%
        & J1617-1941 \tablenotemark{s} & 16:17:27.093081 & 0.96 & -19:41:32.01350 & 1.79 & 5.11 \\%
        & J1644-1804 \tablenotemark{s} & 16:44:35.746823 & 2.37 & -18:04:32.45913 & 5.36 & 7.65 \\%
        & J1626-2951 \tablenotemark{s} & 16:26:06.020836 & 0.24 & -29:51:26.97118 & 0.41 & 5.46 \\%
\enddata
\tablenotetext{a}{Taken from L. Petrov, solution 2008a$\_$astro (unpublished,
available at\\ \textsf{http://vlbi.gsfc.nasa.gov/solutions/2008a$\_$astro})}
\tablenotetext{p}{Main phase calibrator.}
\tablenotetext{s}{Secondary calibrator.}
\end{deluxetable}

\begin{figure*}[!h]
\centering
\begin{tabular}{cc}
\includegraphics[width=.45\textwidth,angle=0]{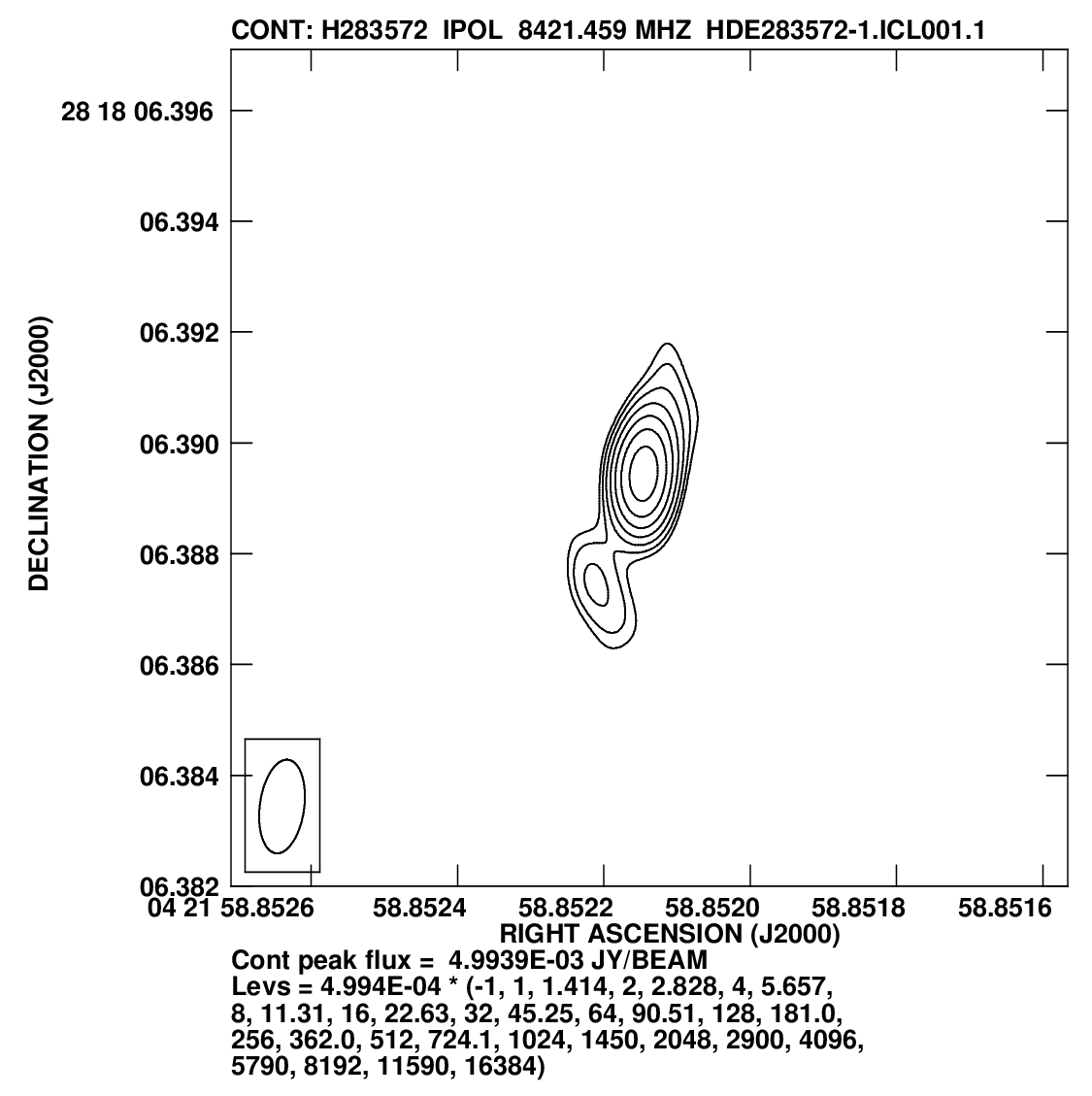}&
\includegraphics[width=.45\textwidth,angle=0]{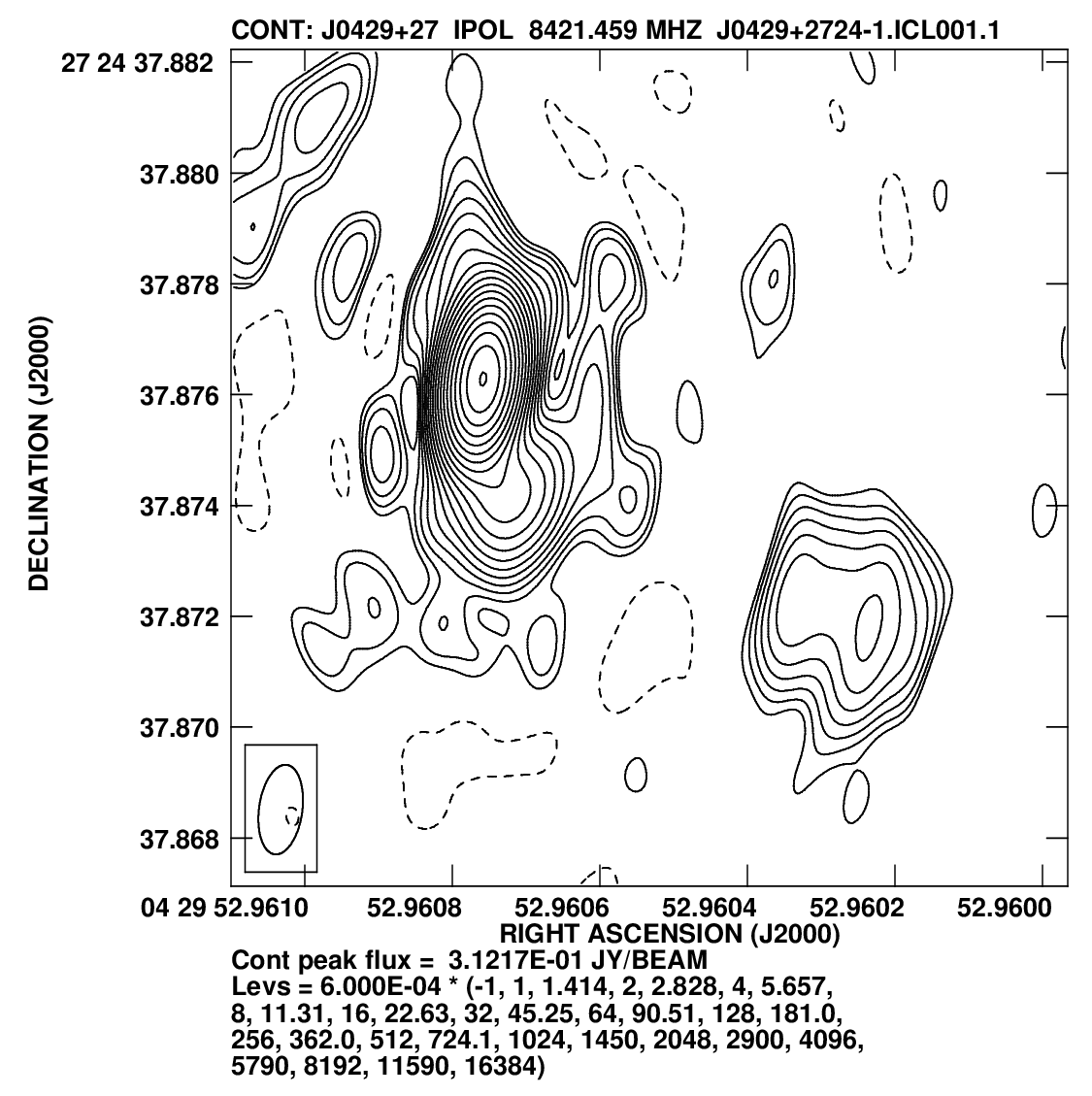}\\
\includegraphics[width=.45\textwidth,angle=0]{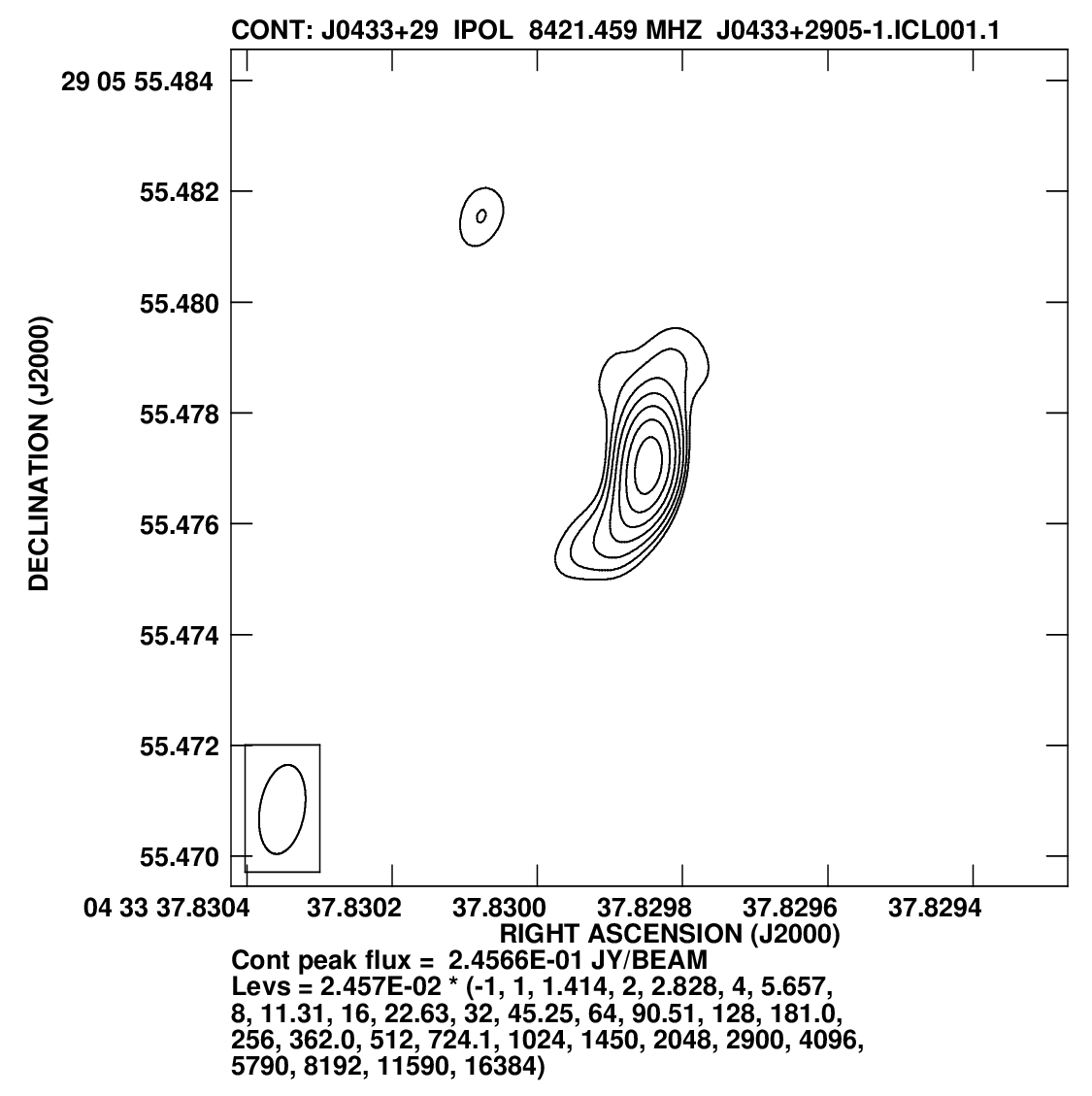}&
\includegraphics[width=.45\textwidth,angle=0]{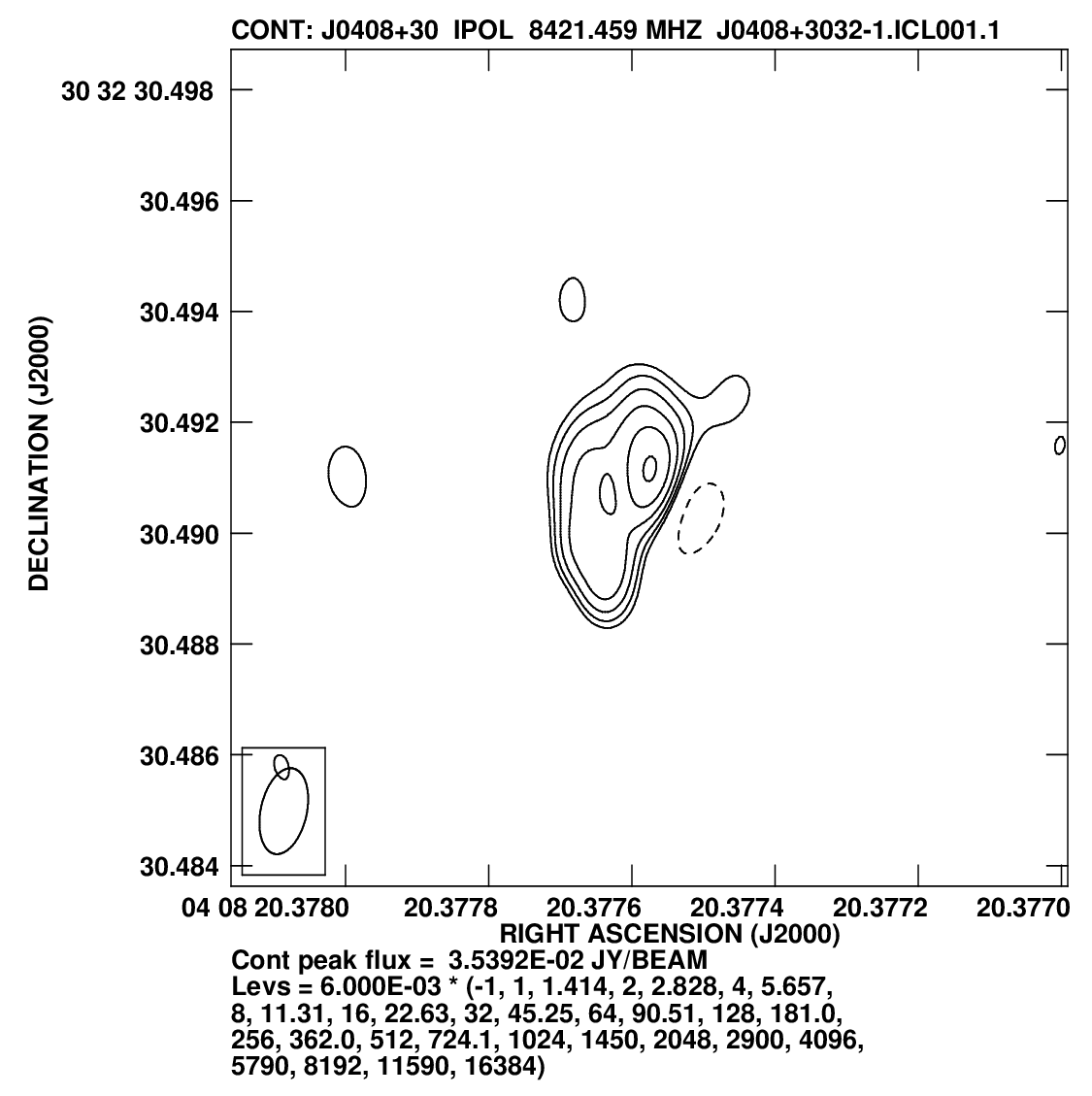}\\
\includegraphics[width=.45\textwidth,angle=0]{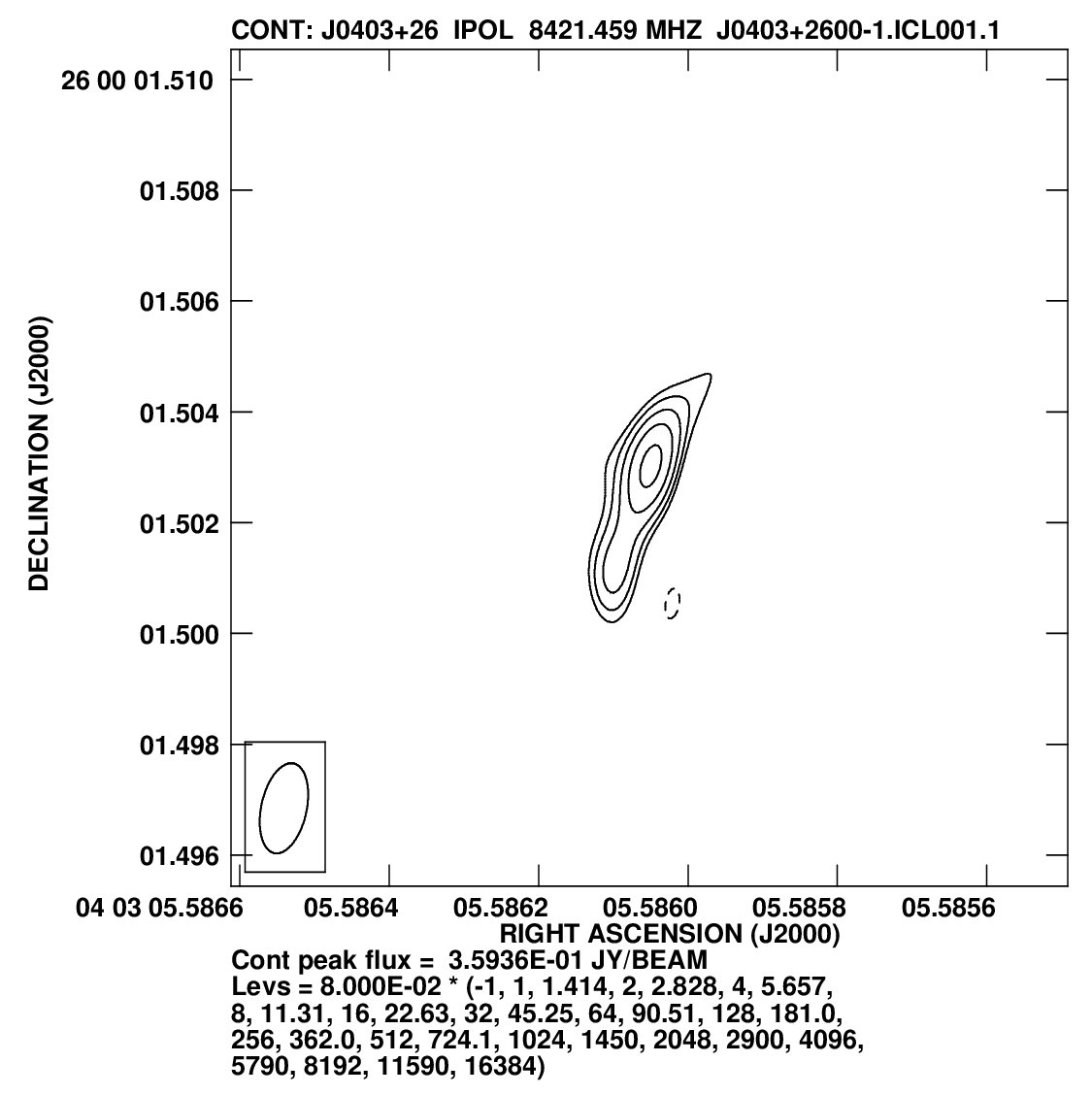}
\end{tabular}
\caption{\footnotesize{
Images of the astronomical target HDE~283572, primary calibrator
J0429+2724 and secondary calibrators J0433+2905, J0408+3032,
J0403+2600 at the first epoch (22/Sep/04; see Tab.\
\ref{tab-observaciones}) produced after steps listed in Sect.\
\ref{c4-basic}.}
\label{fig-reduce}}\end{figure*}

\begin{figure*}[!h]
\centering
\begin{tabular}{cc}
\includegraphics[width=.45\textwidth,angle=0]{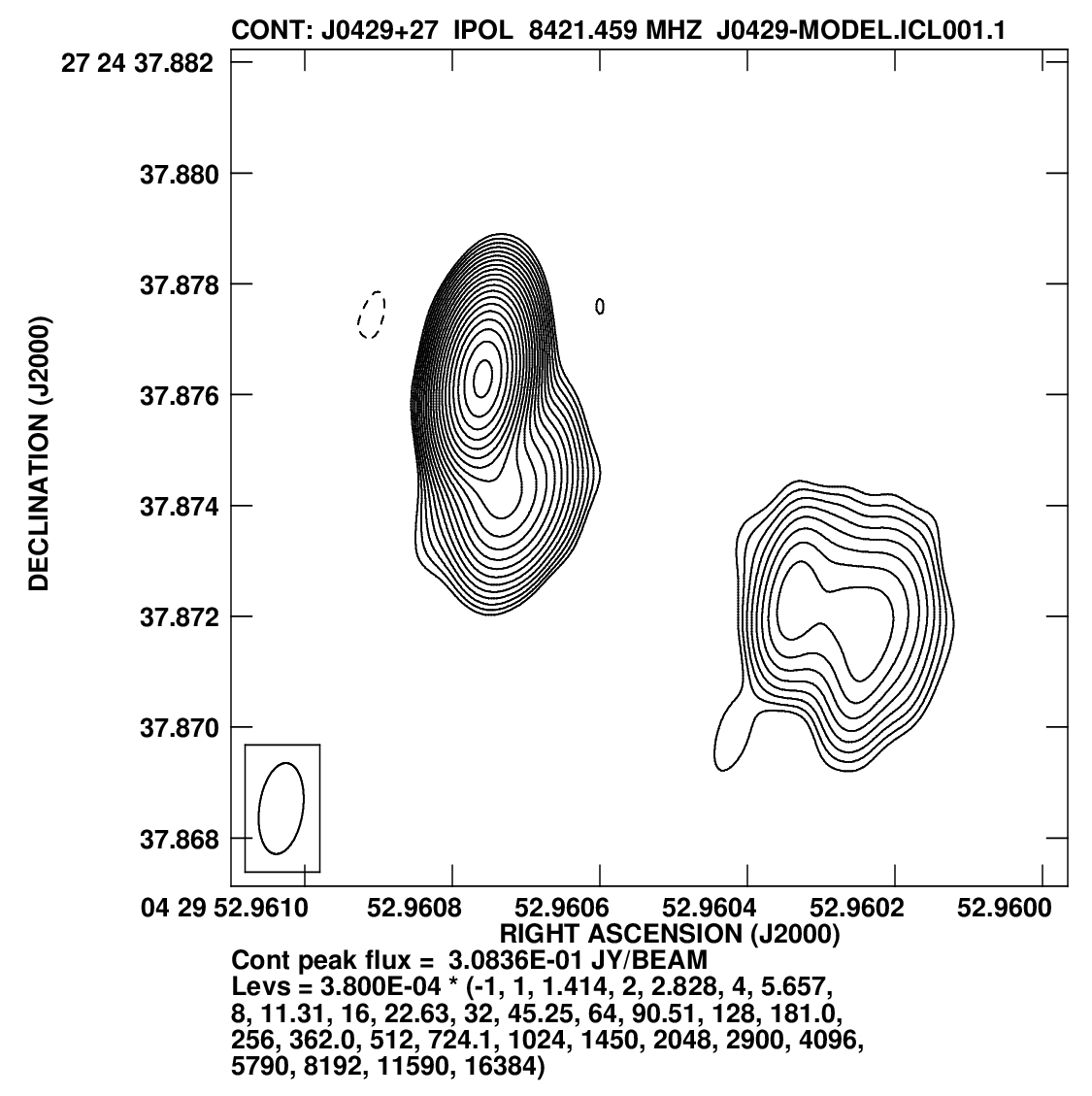}&
\includegraphics[width=.45\textwidth,angle=0]{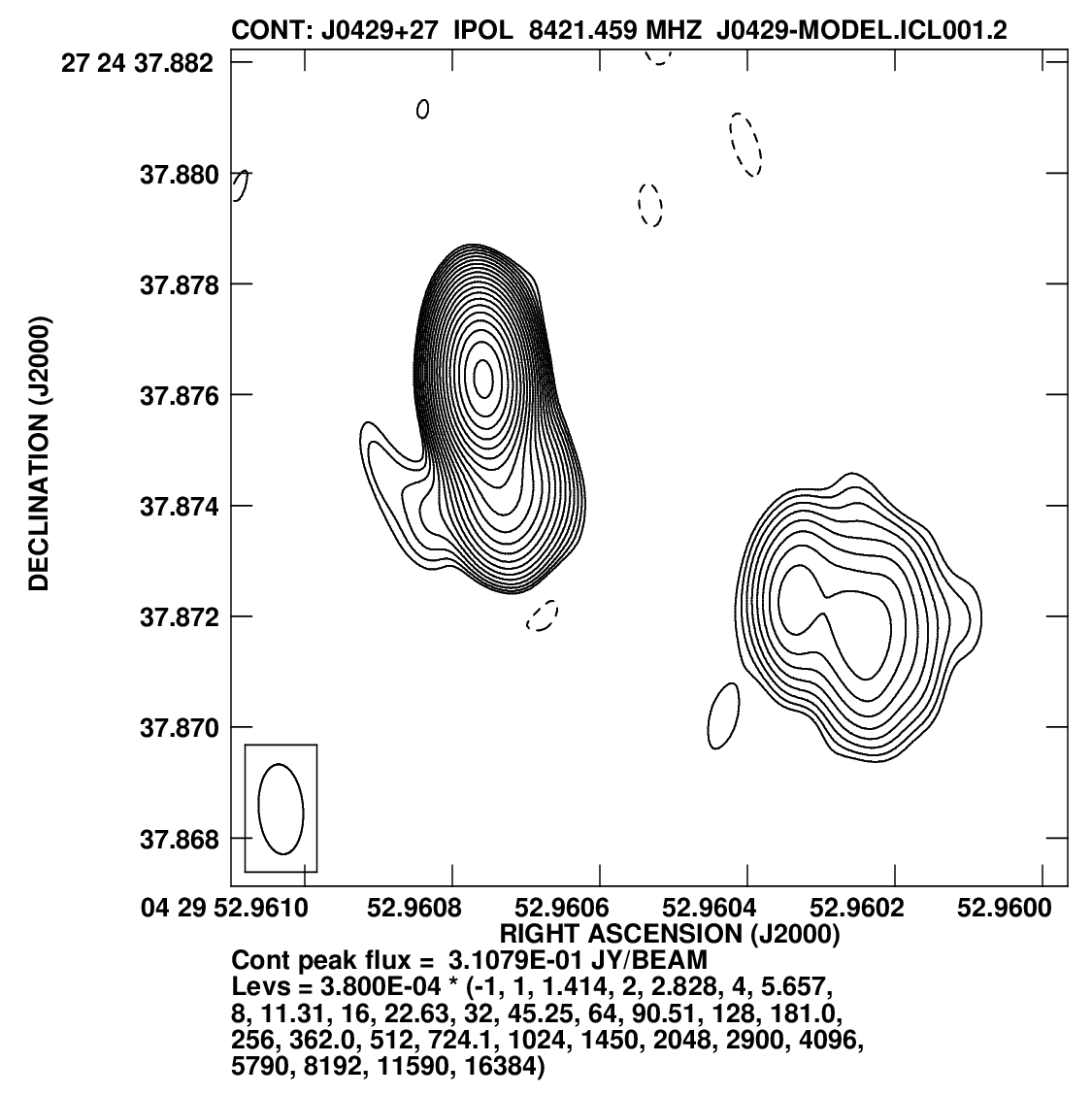}\\
\includegraphics[width=.45\textwidth,angle=0]{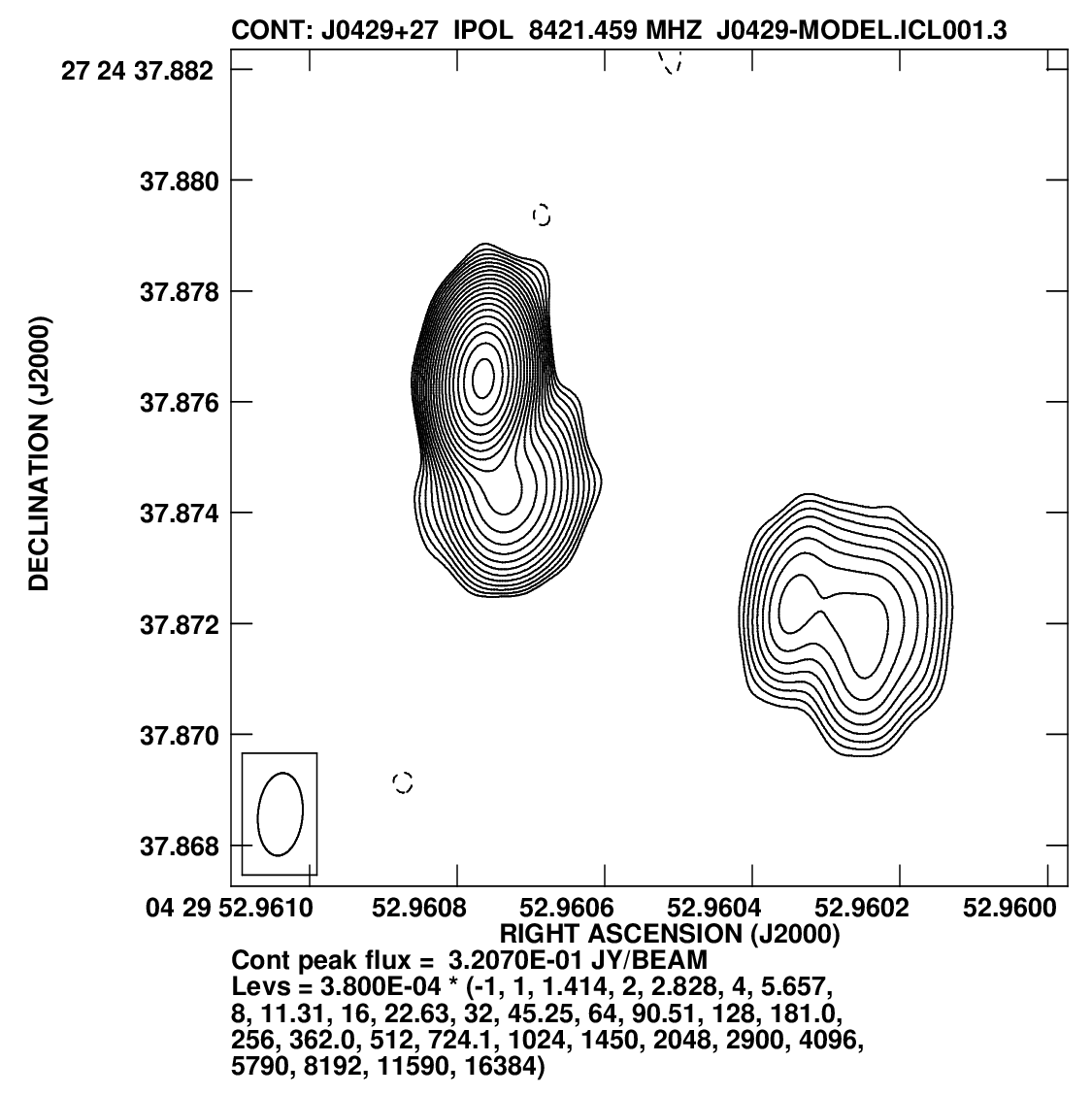}&
\includegraphics[width=.45\textwidth,angle=0]{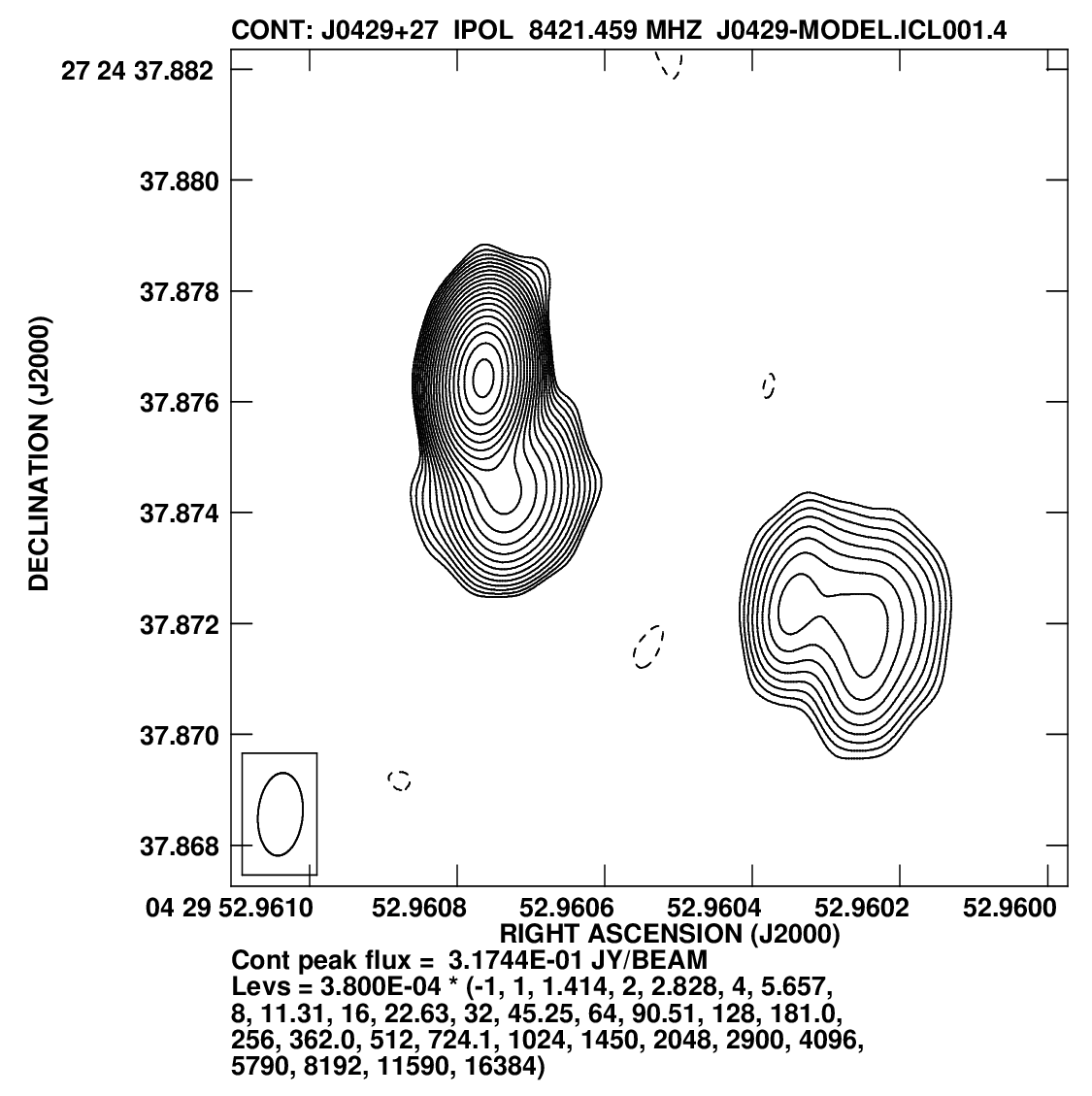}\\
\includegraphics[width=.45\textwidth,angle=0]{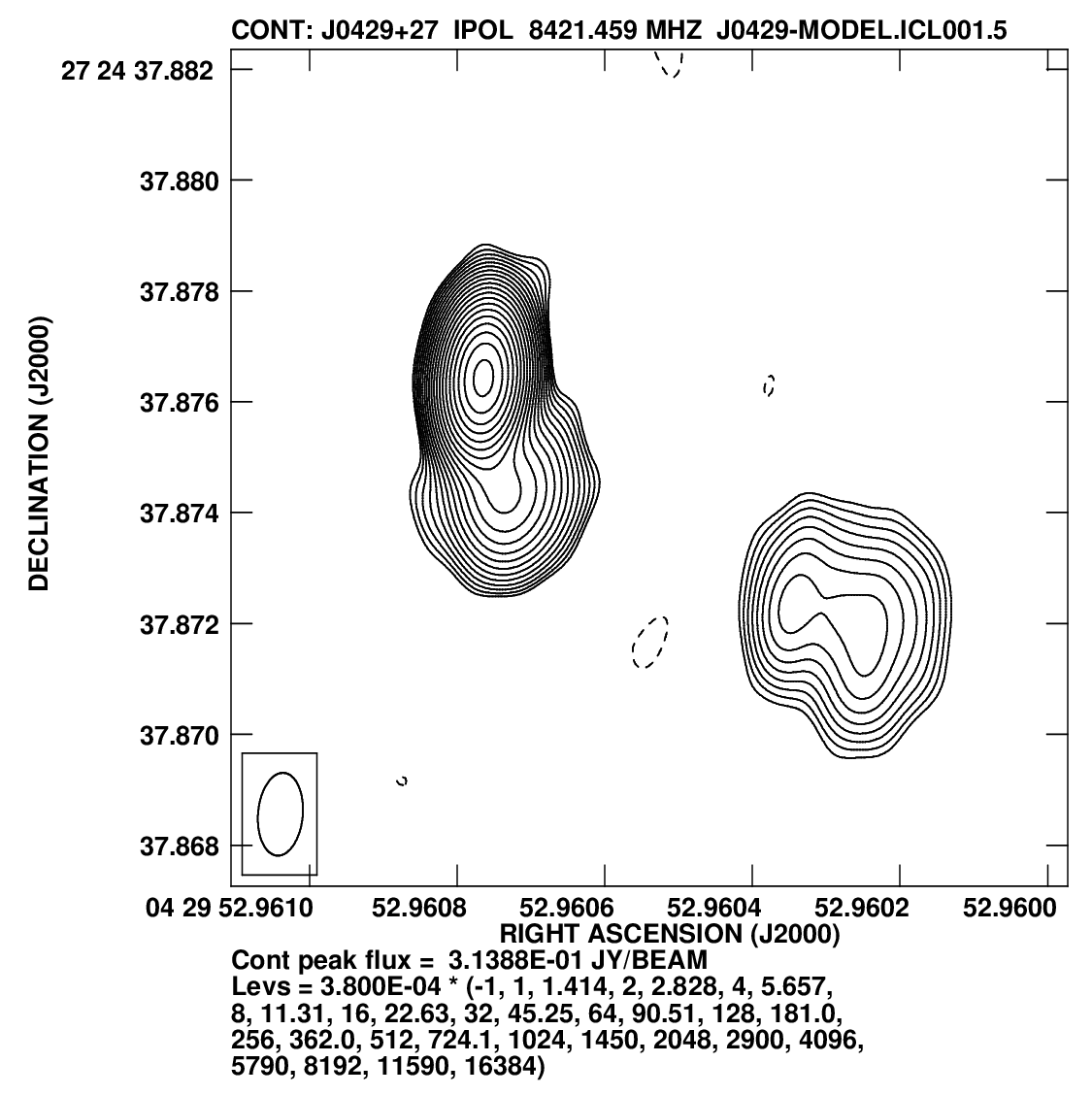}&
\includegraphics[width=.45\textwidth,angle=0]{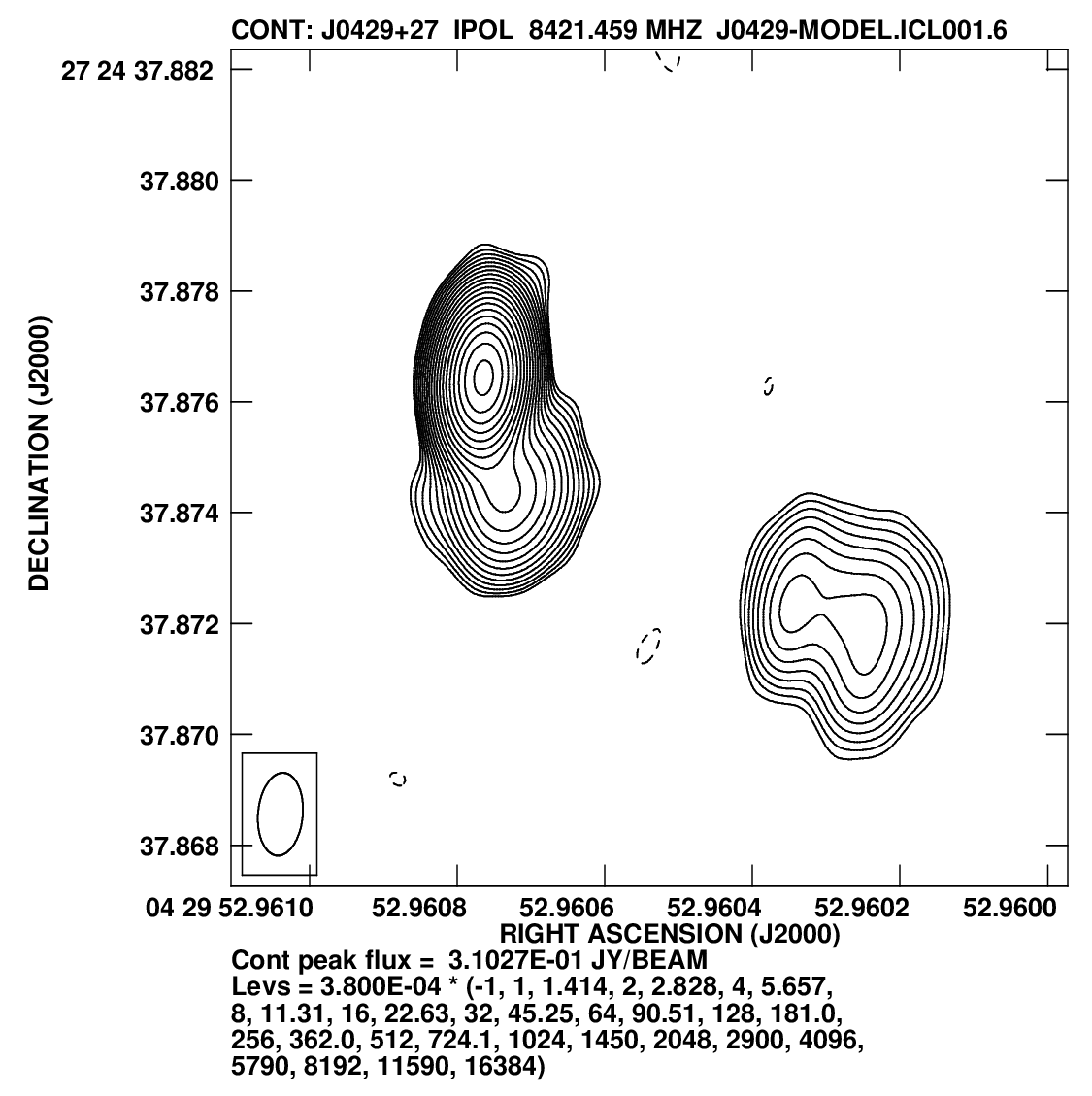}
\end{tabular}
\caption{\footnotesize{
Images of the phase calibrator model J0429+2724 used for HDE~283572 at
all six epochs (22/Sep/04, 06/Jan/05, 30/Mar/05, 23/Jun/05, 23/Sep/05,
24/Dec/05; see Tab.\ \ref{tab-observaciones}). There are produced
after the self-calibration iterations, and are plotted with the same
contour levels.}
\label{fig-model}}\end{figure*}

\begin{figure*}[!h]
\centering
\begin{tabular}{cc}
\includegraphics[width=.45\textwidth,angle=0]{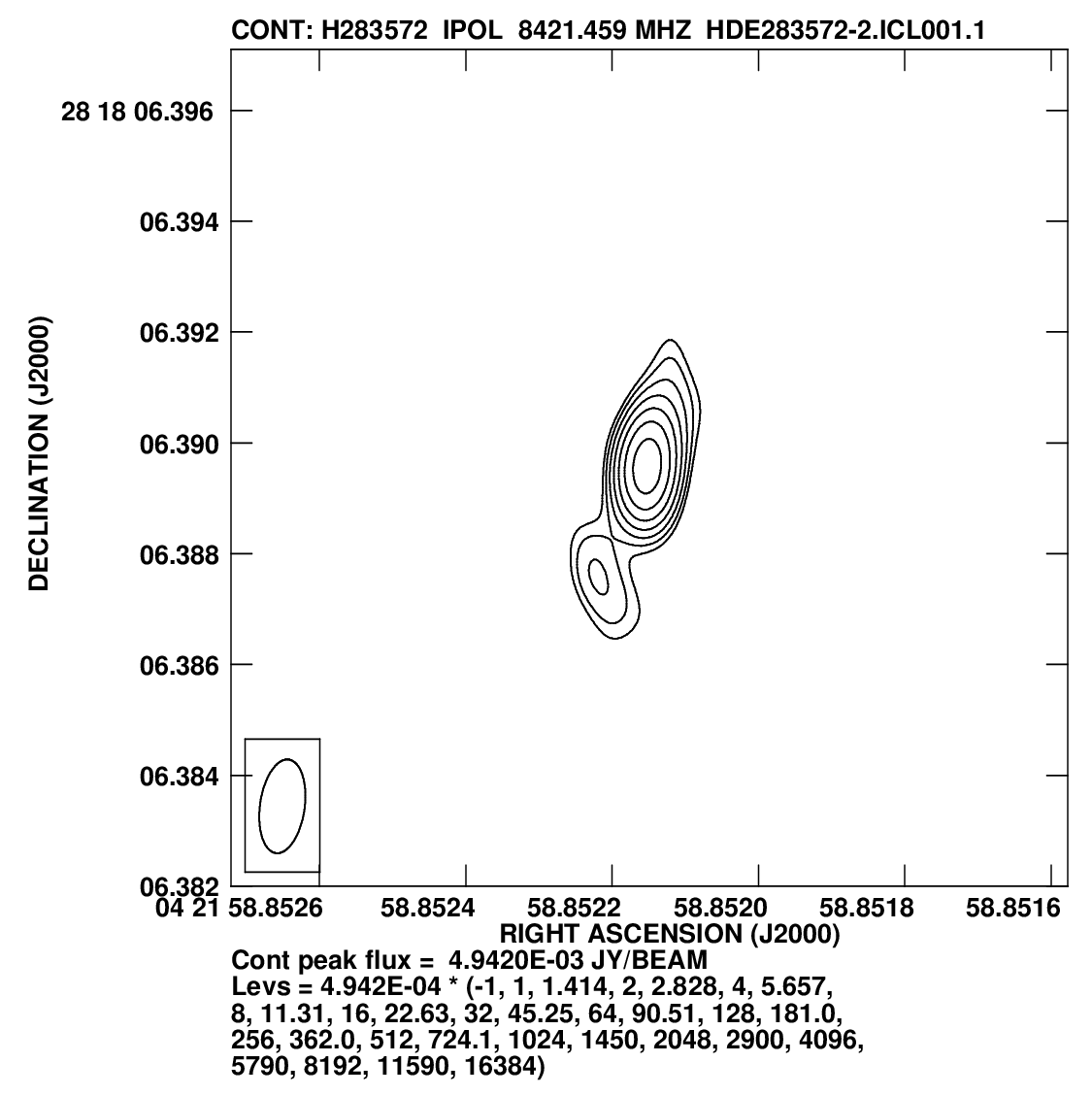}&
\includegraphics[width=.45\textwidth,angle=0]{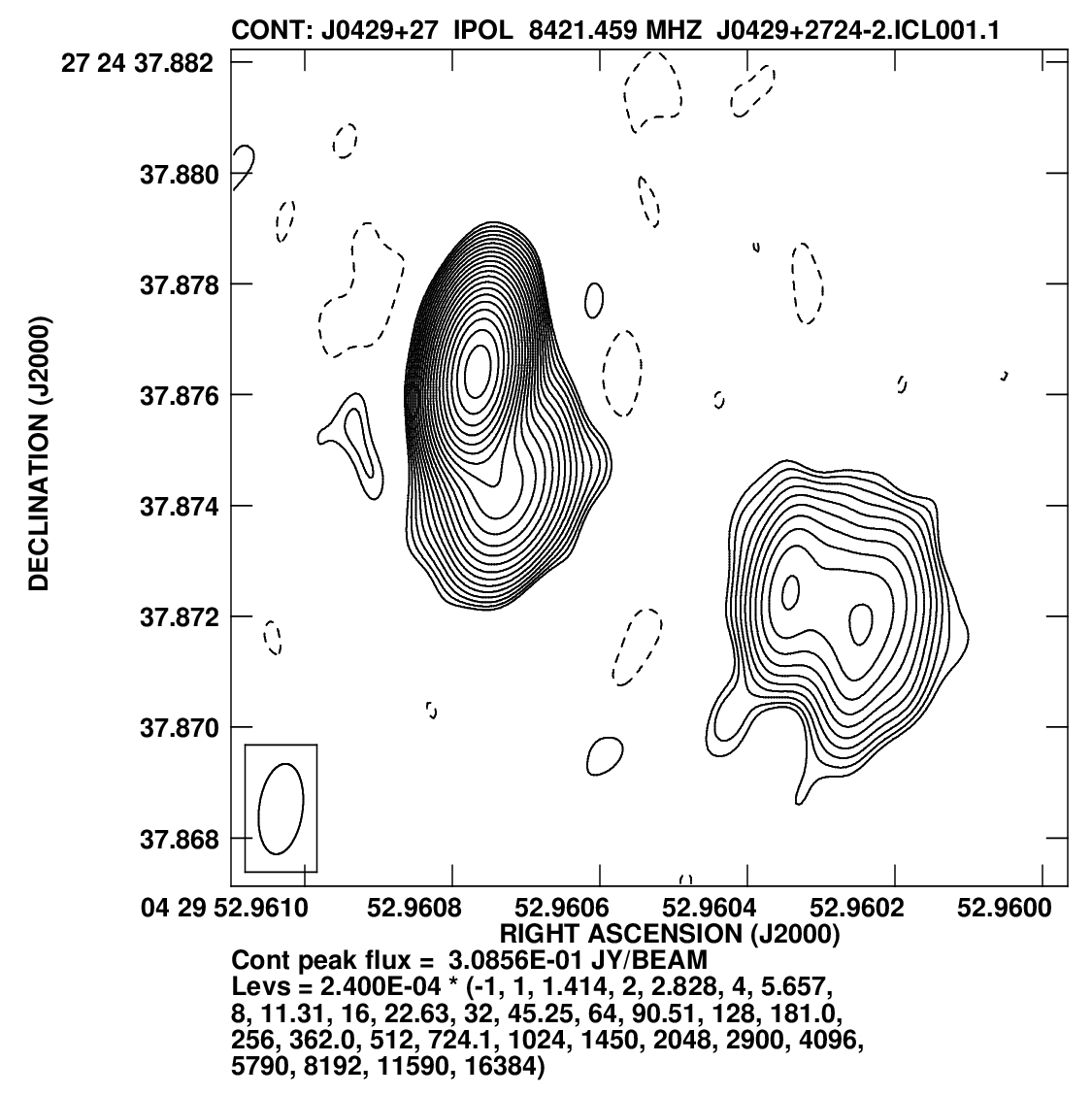}\\
\includegraphics[width=.45\textwidth,angle=0]{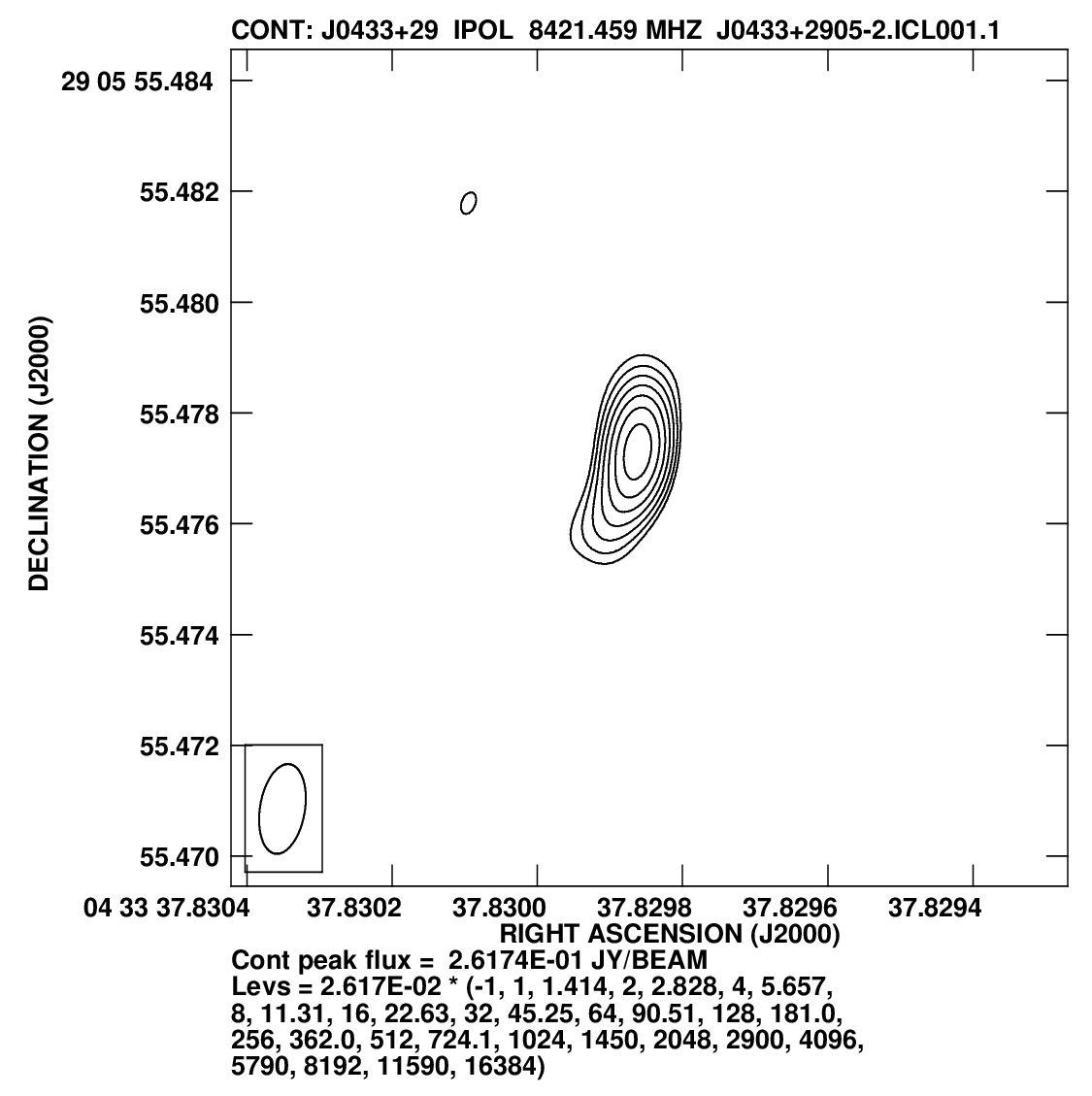}&
\includegraphics[width=.45\textwidth,angle=0]{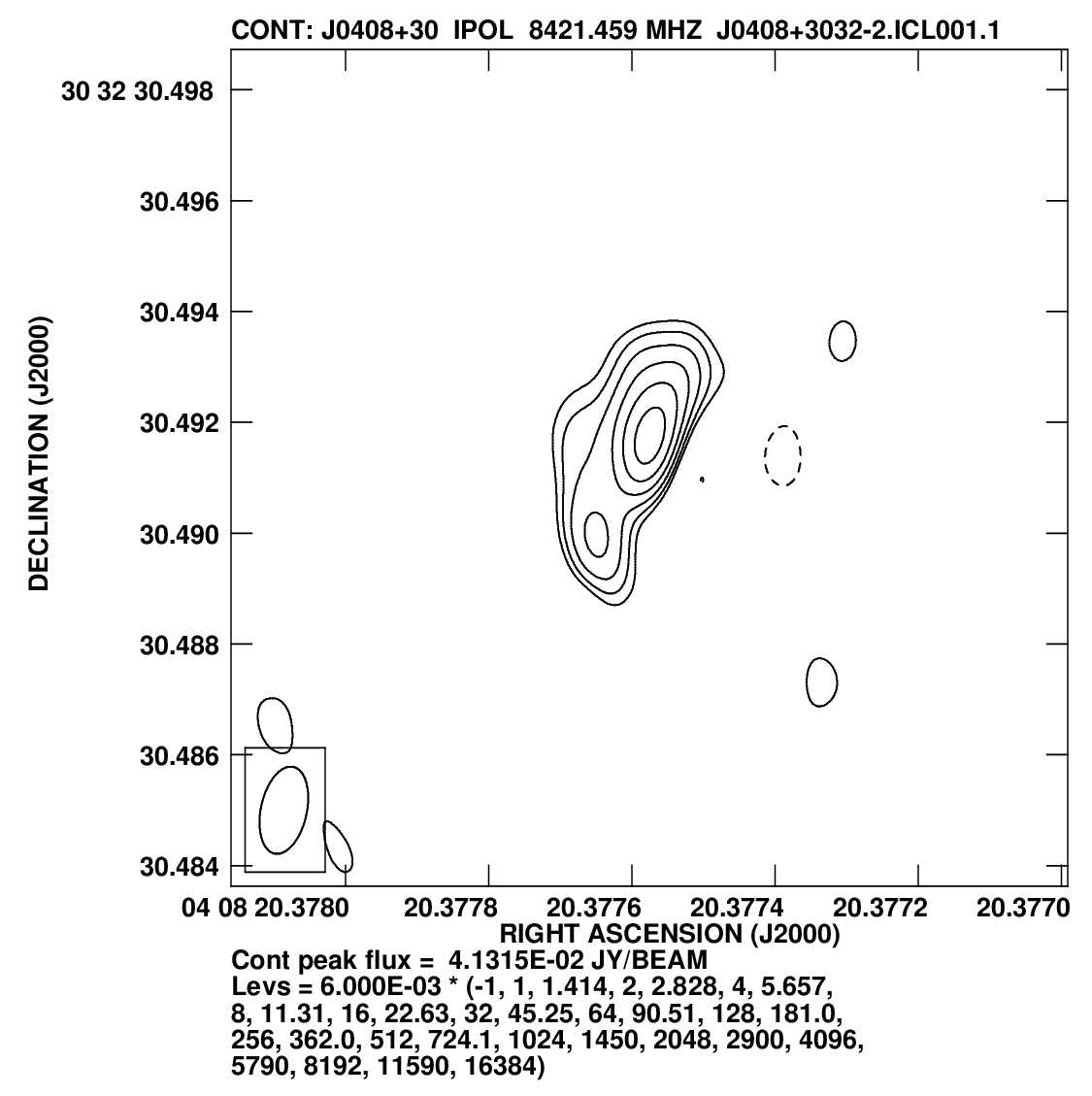}\\
\includegraphics[width=.45\textwidth,angle=0]{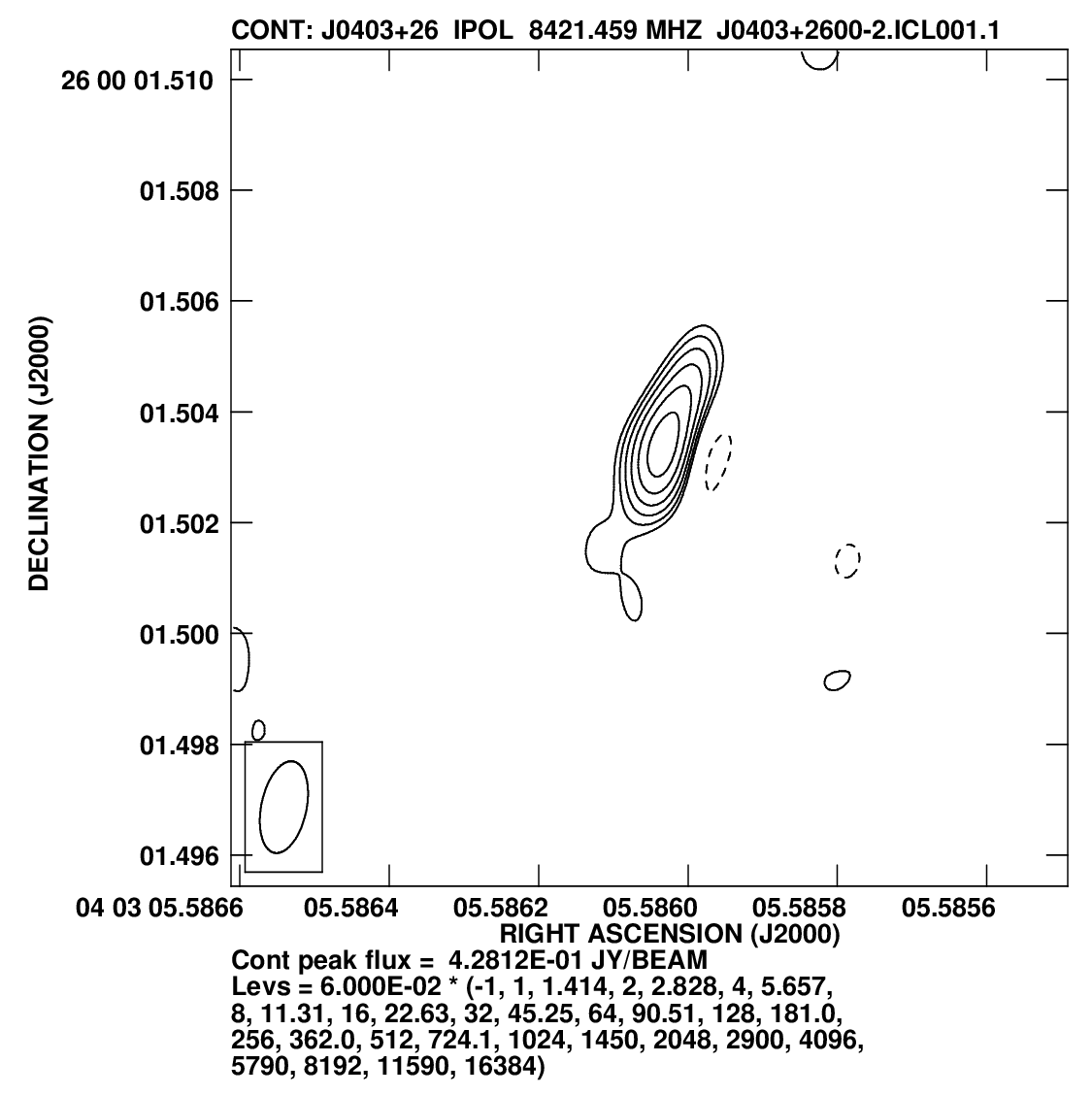}
\end{tabular}
\caption{\footnotesize{
Images of the astronomical target HDE~283572, primary calibrator
J0429+2724 and secondary calibrators J0433+2905, J0408+3032,
J0403+2600 at the first epoch (22/Sep/04; see Tab.\
\ref{tab-observaciones}) produced after steps listed in Sect.\
\ref{c4-re-fringe}.}
\label{fig-redo}}\end{figure*}


\begin{figure*}[!h]
\centering
\begin{tabular}{cc}
\includegraphics[width=.45\textwidth,angle=0]{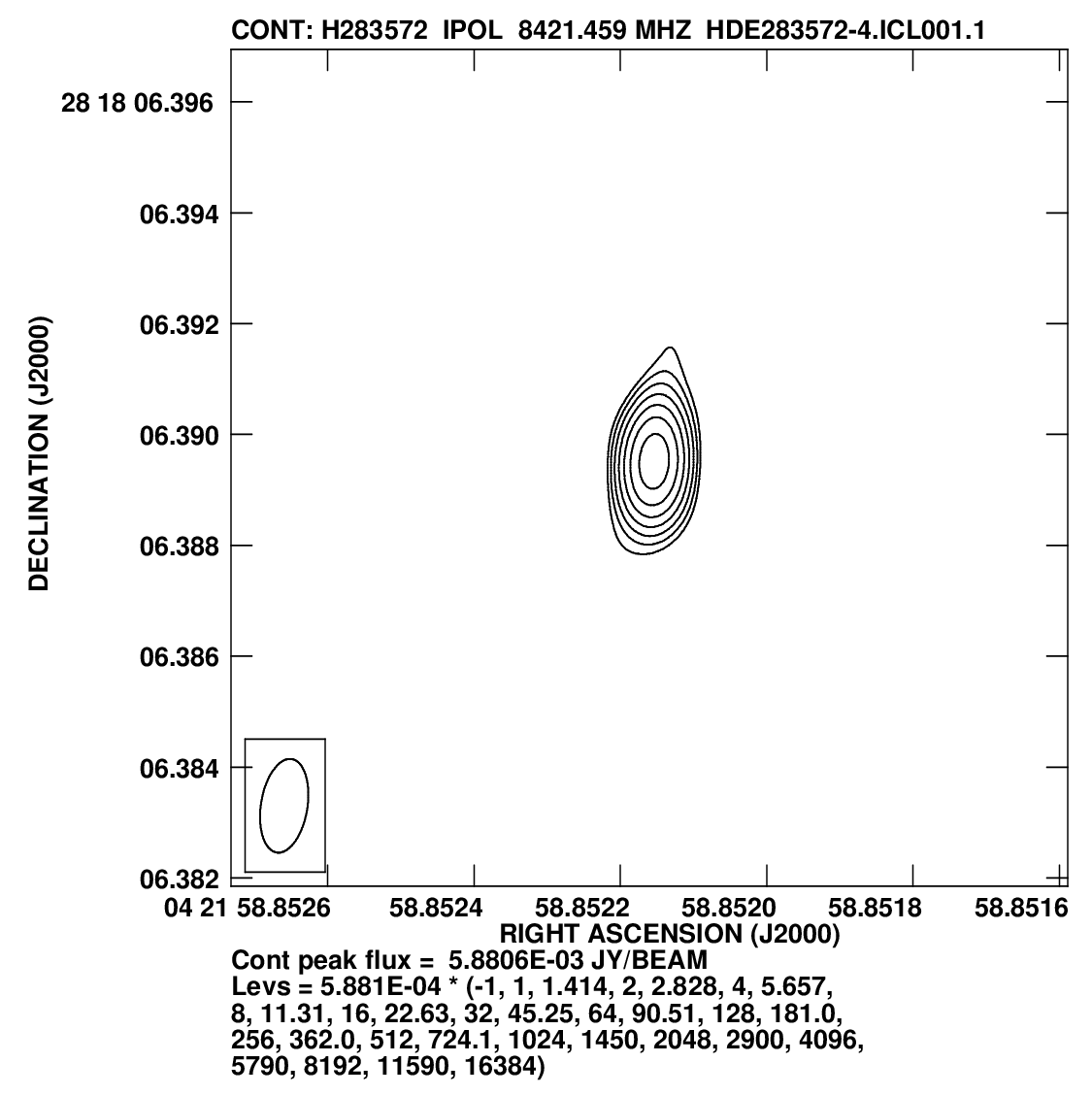}&
\includegraphics[width=.45\textwidth,angle=0]{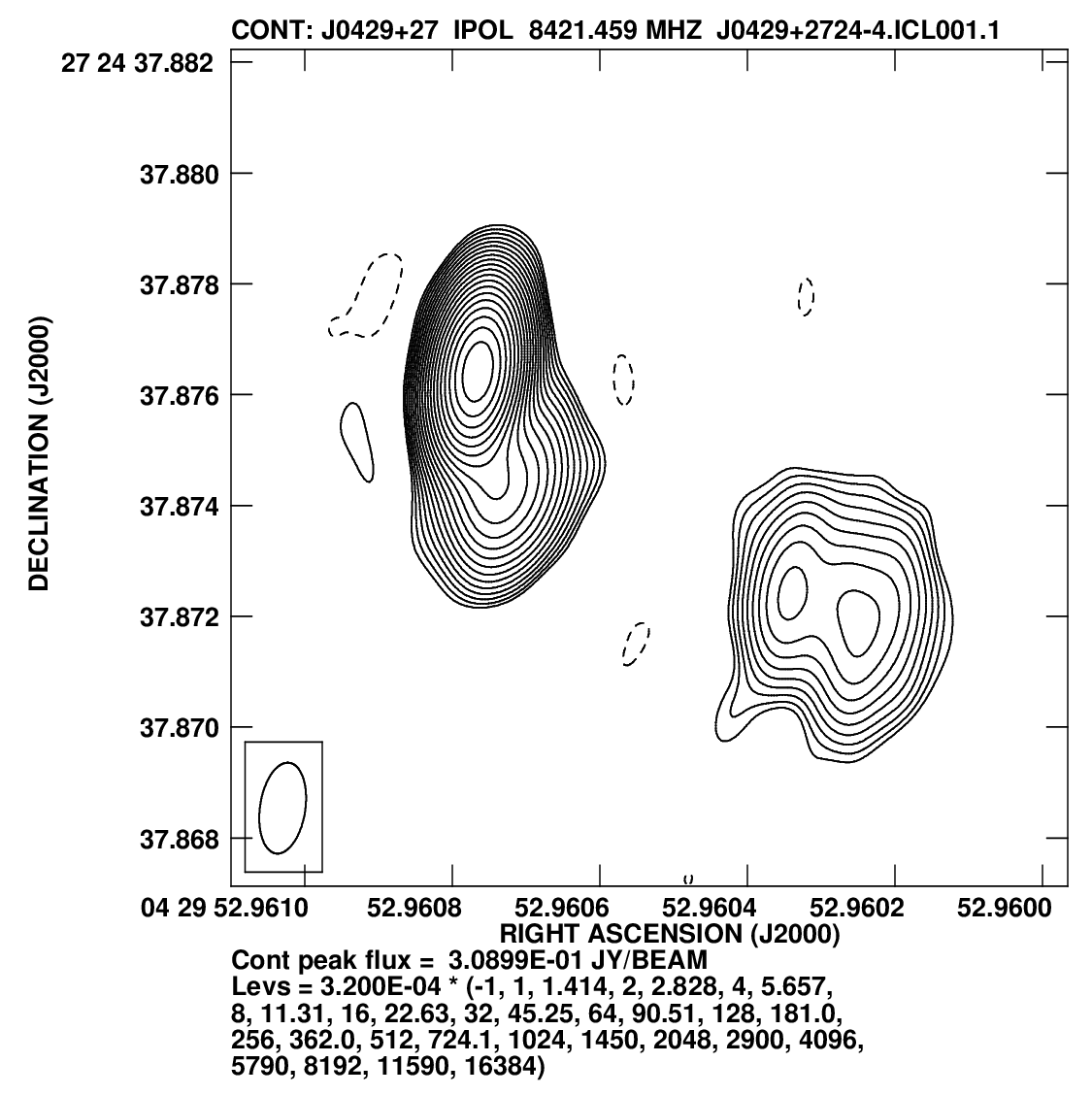}\\
\includegraphics[width=.45\textwidth,angle=0]{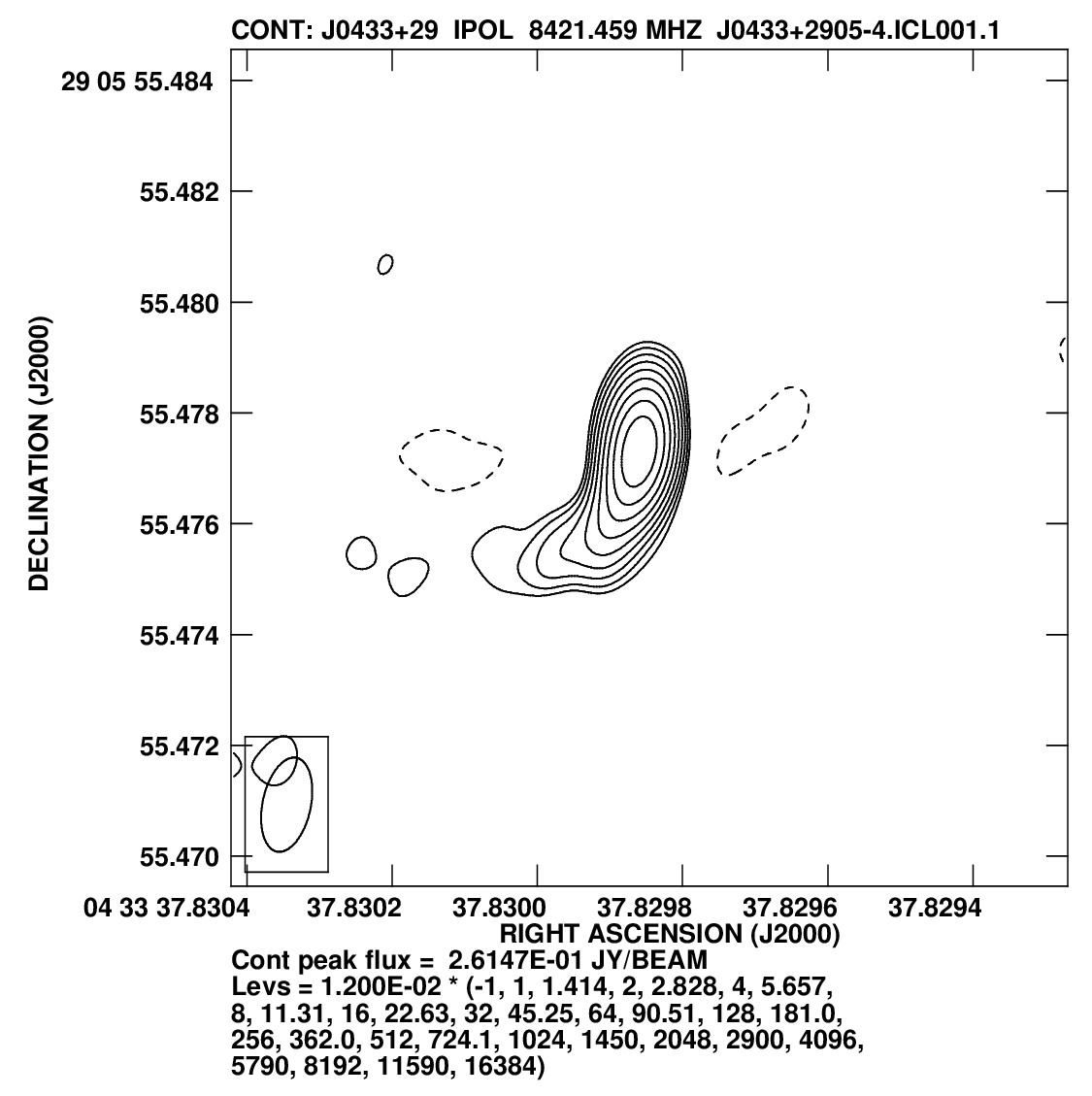}&
\includegraphics[width=.45\textwidth,angle=0]{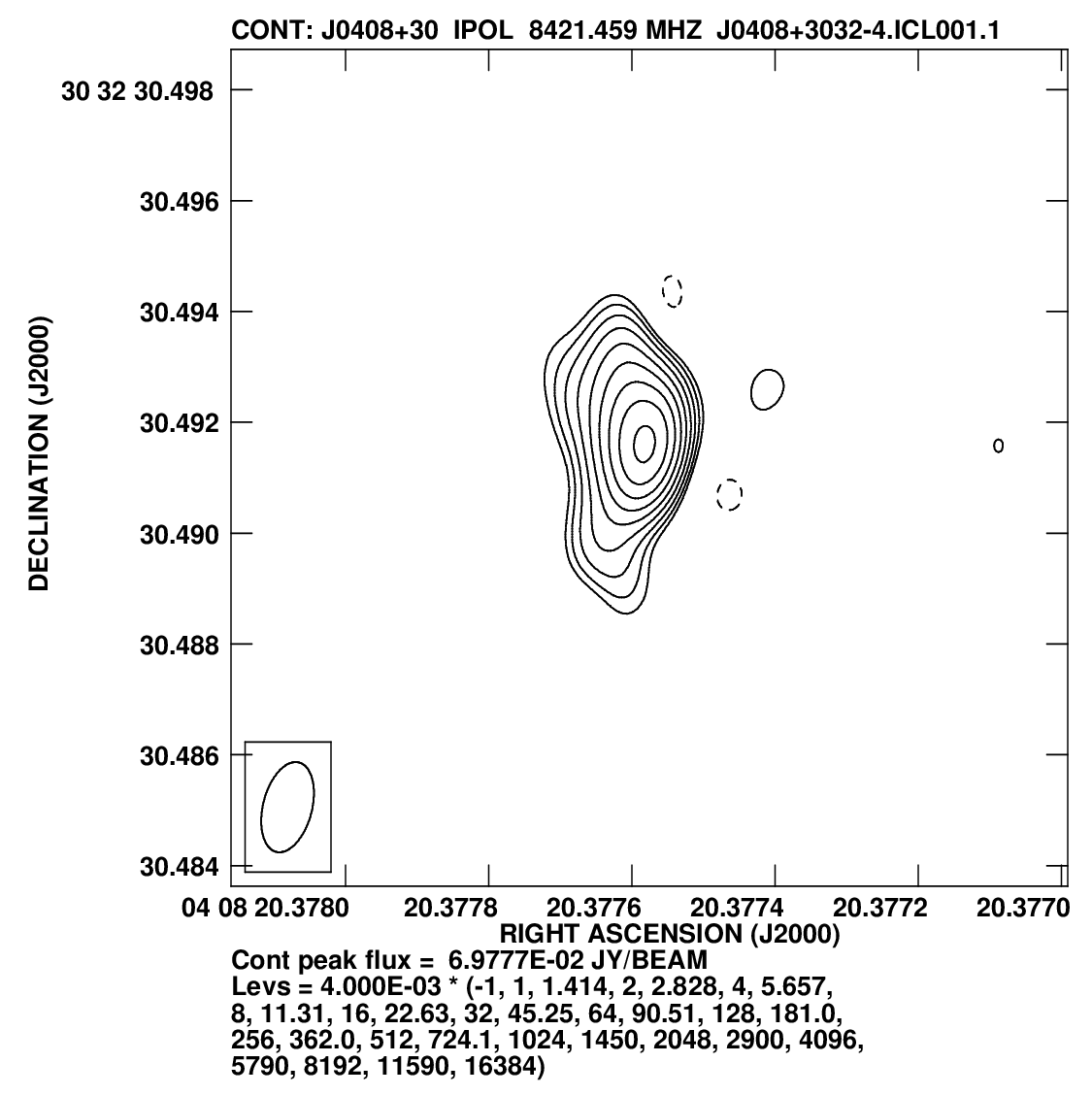}\\
\includegraphics[width=.45\textwidth,angle=0]{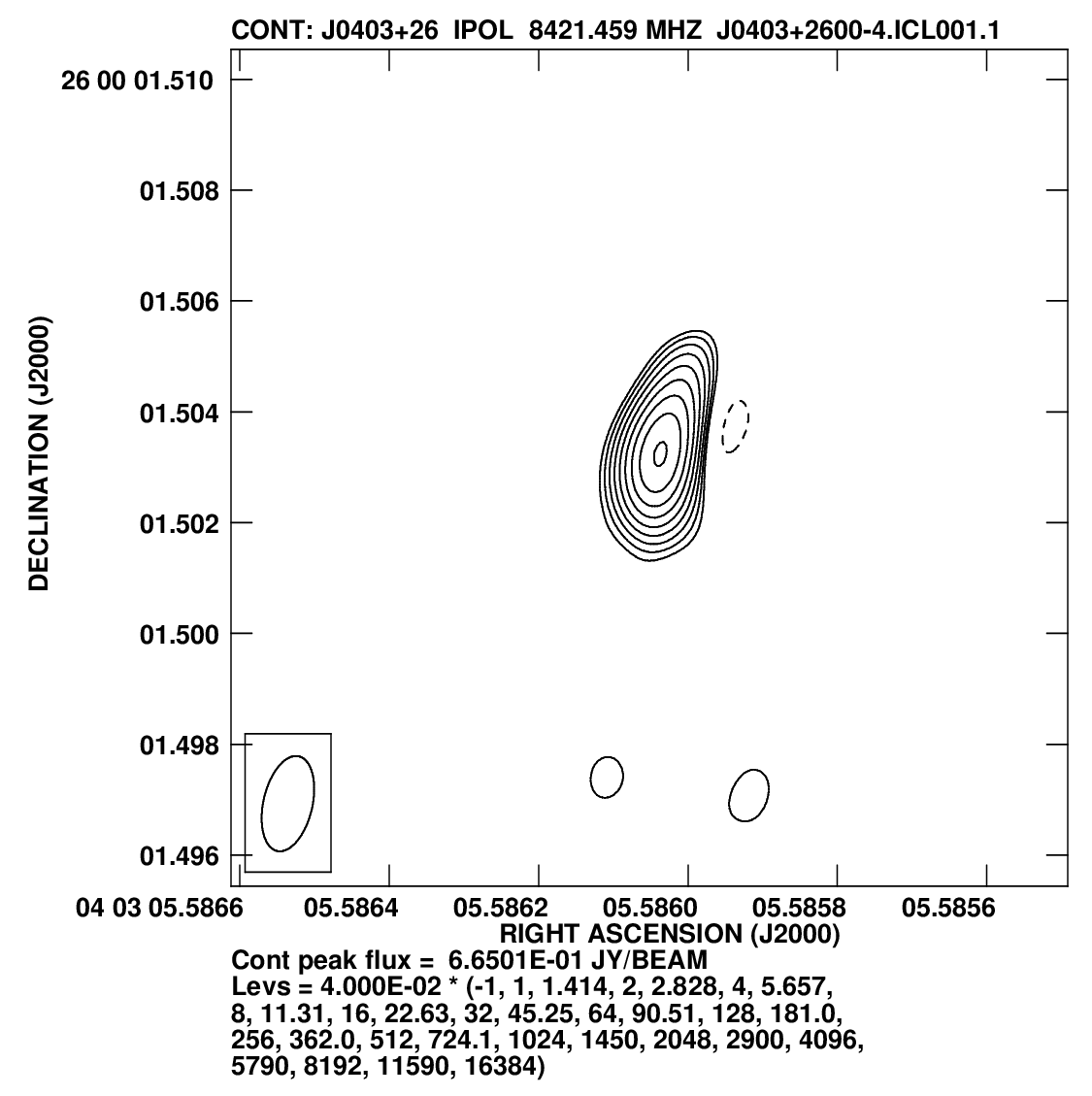}
\end{tabular}
\caption{\footnotesize{
Images of the astronomical target HDE~283572, primary calibrator
J0429+2724 and secondary calibrators J0433+2905, J0408+3032,
J0403+2600 at the first epoch (22/Sep/04; see Tab.\
\ref{tab-observaciones}) produced after \textsf{ATMCA}.}
\label{fig-atmca}}\end{figure*}

\begin{figure*}[!h]
\centering
\begin{tabular}{c}
\includegraphics[width=.45\textwidth,angle=0]{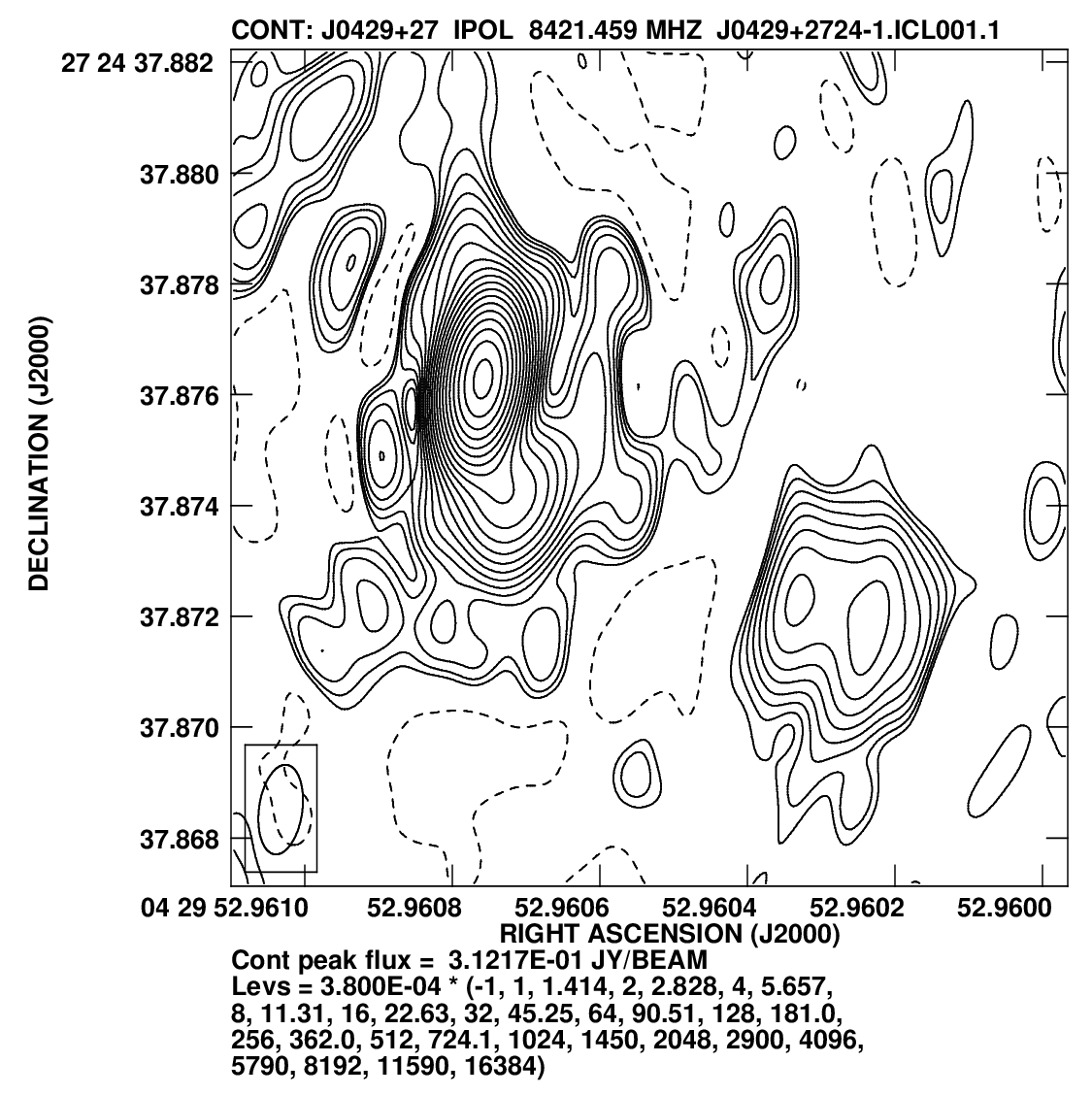}\\
\includegraphics[width=.45\textwidth,angle=0]{cal-model.ps}\\
\includegraphics[width=.45\textwidth,angle=0]{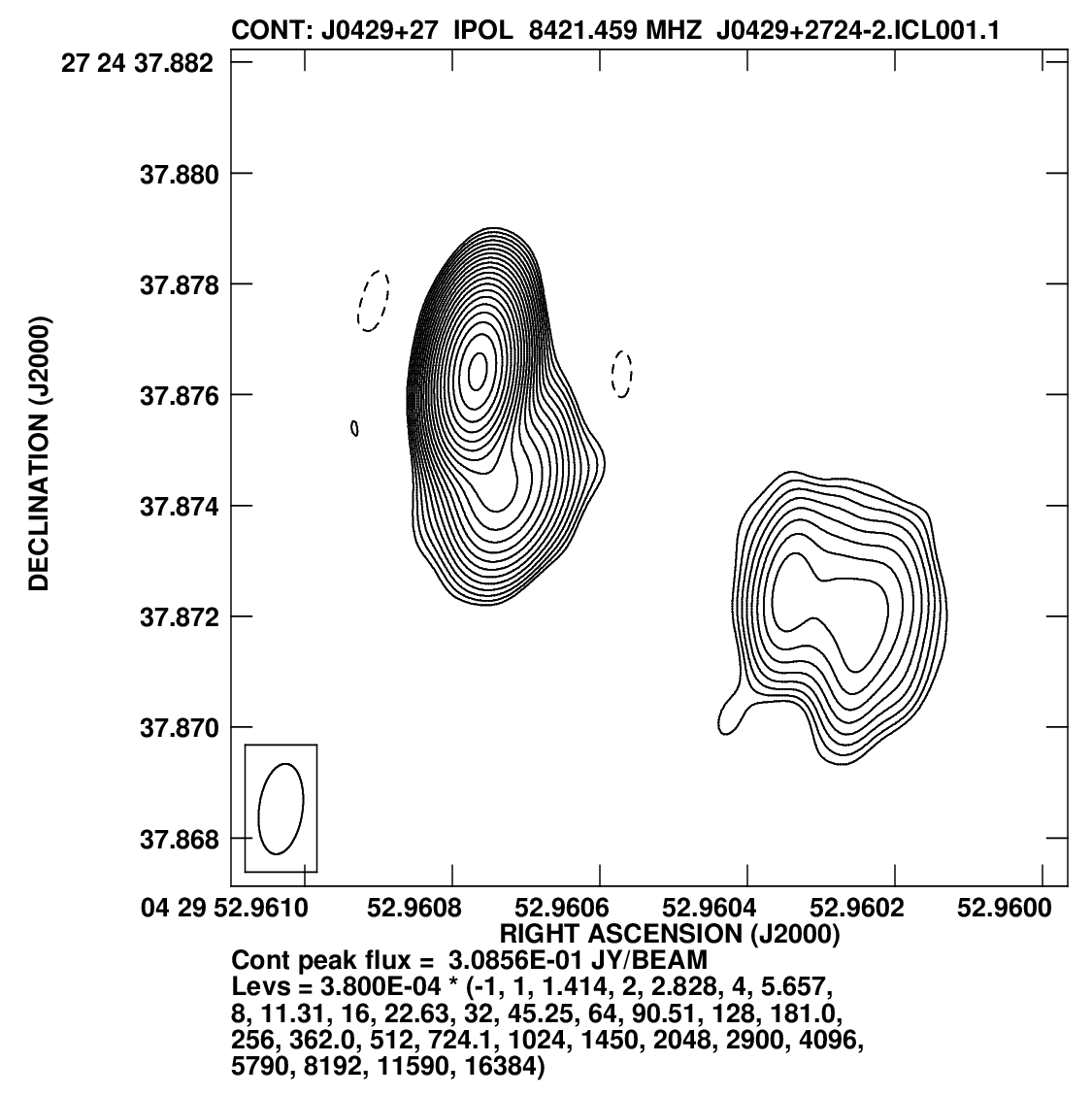}
\end{tabular}
\caption{\footnotesize{
Images of the main calibrator J0429+2724 used to reduce data of
HDE~283572 at the first epoch (22/Sep/04; see Tab.\
\ref{tab-observaciones}). The first map correspond to the image
produced after \textit{basic data reduction} (Sect.\ \ref{c4-basic})
when we assumed the phase calibrator as a punctual source. The second
map is the phase calibrator model produced in
\textit{self-calibration} (Sect.\ \ref{c4-self}; note the structure of
the source). And the third map is produced after \textit{re-fringe fit
main phase calibrator} (Sect.\ \ref{c4-re-fringe}) when we used a real
phase calibrator. Note that all three images are plotted with the same
contour levels.}
\label{fig-cal-mejoras}}\end{figure*}

\begin{figure*}[!h]
\centering
\begin{tabular}{c}
\includegraphics[width=.45\textwidth,angle=0]{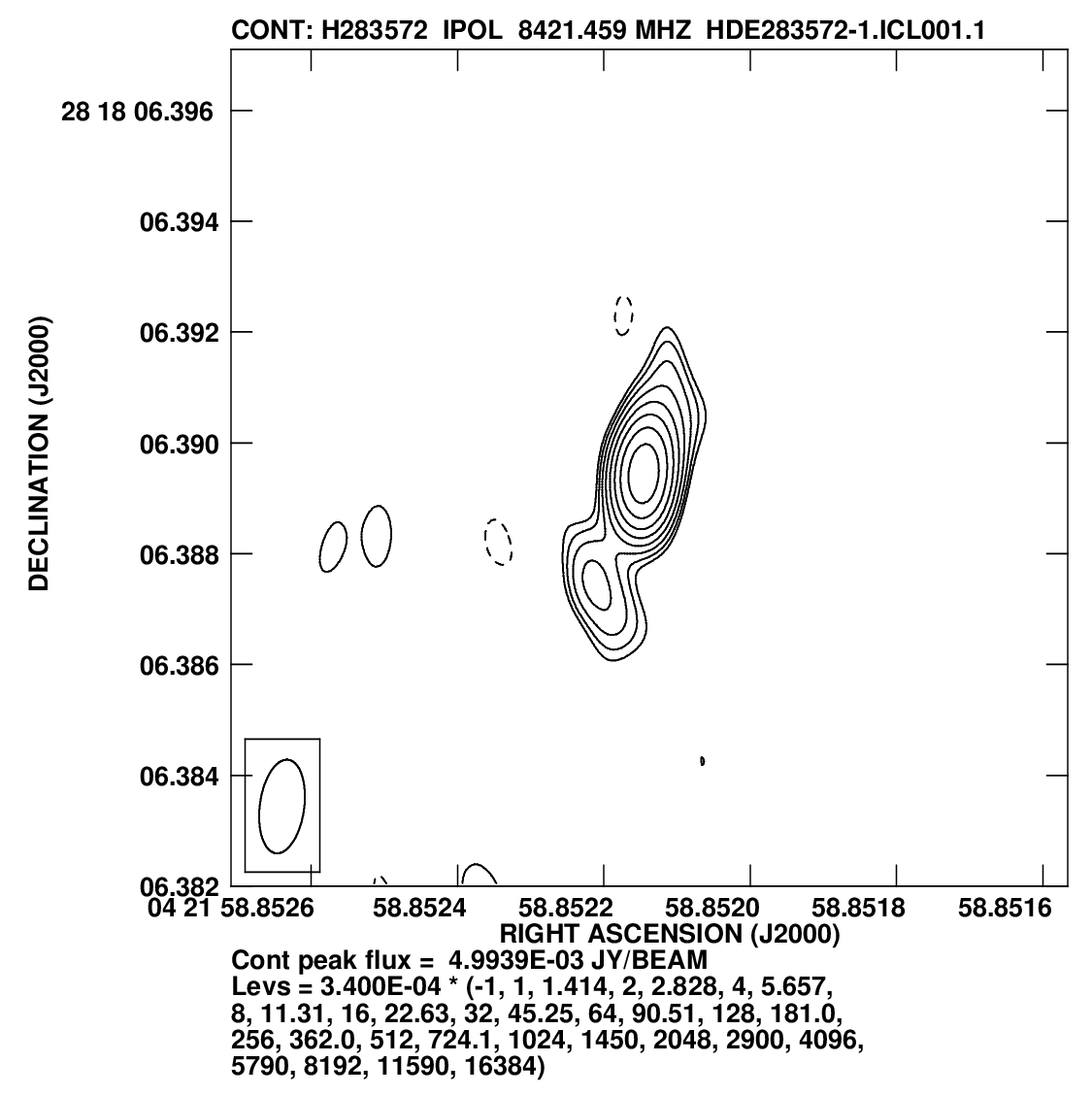}\\
\includegraphics[width=.45\textwidth,angle=0]{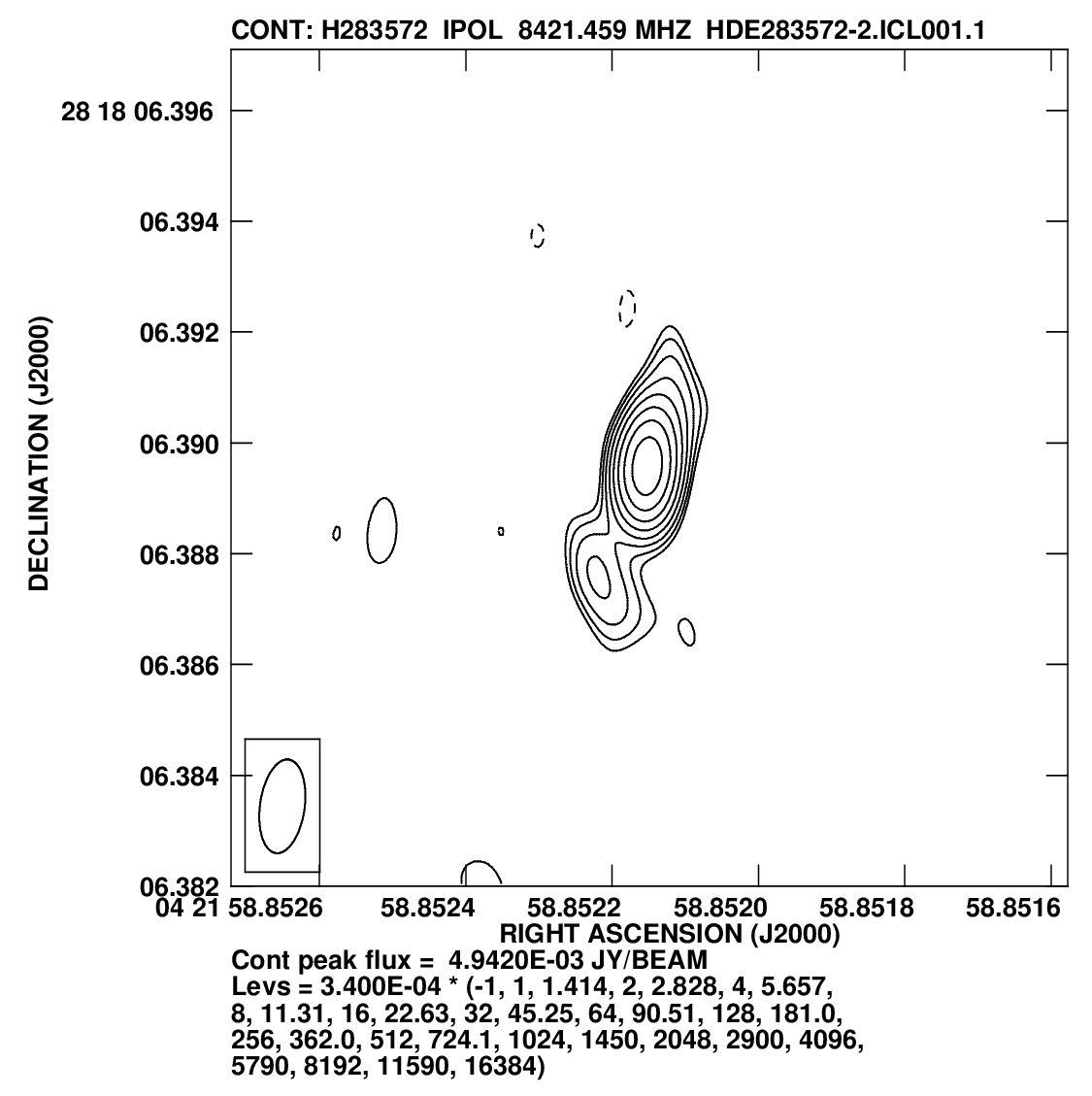}\\
\includegraphics[width=.45\textwidth,angle=0]{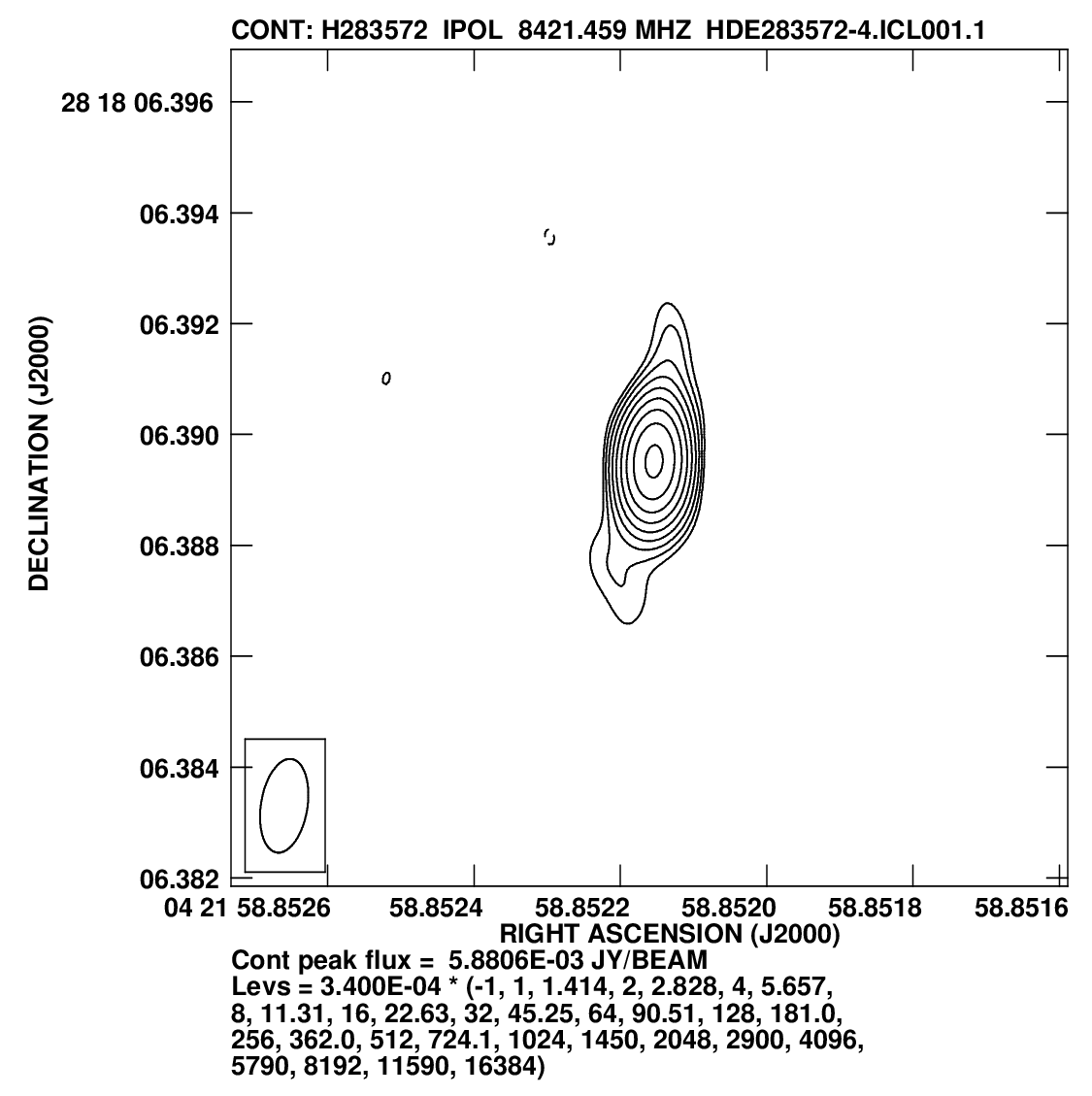}
\end{tabular}
\caption{\footnotesize{
Images of the astronomical target HDE~283572 at the first epoch
(22/Sep/04; see Tab.\ \ref{tab-observaciones}). The first map
correspond to the image produced after \textit{basic data reduction}
(Sect.\ \ref{c4-basic}), the second after \textit{re-fringe fit main
phase calibrator} (Sect.\ \ref{c4-re-fringe}), and the third after
\textit{phase referencing using more than one calibrator}
(Sect.\ \ref{c4-phase}). Note that all three images are plotted with
the same contour levels, and that the peak flux and quality of the
image is always better after each step.}
\label{fig-hde-mejoras}}\end{figure*}

\begin{deluxetable}{cccllr@{$\pm$}lr@{$\pm$}l}
\rotate
\tabletypesize{\scriptsize}
\tablecolumns{9}
\tablewidth{0pc}
\tablecaption{\footnotesize{Improvements in intensities and positions
of the astronomical target HDE~283572 and the main calibrator J0429+2724
at first epoch (22/Sep/04) after each step in the reduction of data.}
\label{tab-mejoras}}
\tablehead{
\colhead{Step\tablenotemark{\dagger}}&
\colhead{Peak intensity}       &
\colhead{Integral intensity}   &
\colhead{$\alpha$ (J2000.0)}   &
\colhead{$\delta$ (J2000.0)}   &
\multicolumn{2}{c}{X-position} &
\multicolumn{2}{c}{Y-position} \\
\multicolumn{1}{c}{}           &
\multicolumn{1}{c}{[mJy/beam]} &
\multicolumn{1}{c}{[mJy]}      &
\multicolumn{1}{c}{}           &
\multicolumn{1}{c}{}           &
\multicolumn{2}{c}{[pixels]}   &
\multicolumn{2}{c}{[pixels]}   }
\startdata
\multicolumn{3}{l}{HDE~283572}&
\multicolumn{1}{c}{$04^{h}21^{m}$}&
\multicolumn{1}{c}{$28^{\circ}18^{'}$}\\\\[-0.2cm]
1 & 4.9091$\pm$0.0668 & 5.9111$\pm$0.131 & \mmsec{58}{85214432}$\pm$\mmsec{0}{00000040333} & \msec{06}{3894682}$\pm$\msec{0}{0000105262} & 1025.256&0.1065 & 1022.366&0.2105\\%
3 & 4.8657$\pm$0.0654 & 5.8169$\pm$0.128 & \mmsec{58}{85215084}$\pm$\mmsec{0}{00000039874} & \msec{06}{3896017}$\pm$\msec{0}{0000103789} & 1023.535&0.1053 & 1025.036&0.2076\\%
4 & 5.8289$\pm$0.0593 & 6.8963$\pm$0.115 & \mmsec{58}{85215374}$\pm$\mmsec{0}{00000030889} & \msec{06}{3895120}$\pm$\msec{0}{0000077977} & 1022.766&0.0816 & 1023.241&0.1559\\%
\\[-0.2cm]\hline\\[-0.2cm]
\multicolumn{3}{l}{J0429+2724}&
\multicolumn{1}{c}{$04^{h}29^{m}$}&
\multicolumn{1}{c}{$27^{\circ}24^{'}$}\\\\[-0.2cm]
1 & 309.33$\pm$0.320 & 329.38$\pm$0.578 & \mmsec{52}{96075807}$\pm$\mmsec{0}{00000002709} & \msec{37}{8762762}$\pm$\msec{0}{0000007421} & 255.981&0.0072 & 256.924&0.0148\\%
2 & 305.66$\pm$0.112 & 322.52$\pm$0.200 & \mmsec{52}{96075821}$\pm$\mmsec{0}{00000000962} & \msec{37}{8762719}$\pm$\msec{0}{0000002637} & 255.942&0.0026 & 256.838&0.0053\\%
3 & 305.89$\pm$0.117 & 323.78$\pm$0.210 & \mmsec{52}{96076520}$\pm$\mmsec{0}{00000001001} & \msec{37}{8764019}$\pm$\msec{0}{0000002746} & 254.082&0.0027 & 259.439&0.0055\\%
4 & 306.36$\pm$0.122 & 323.85$\pm$0.219 & \mmsec{52}{96076520}$\pm$\mmsec{0}{00000001091} & \msec{37}{8764010}$\pm$\msec{0}{0000002878} & 254.080&0.0029 & 259.420&0.0058\\%
\enddata
\tablenotetext{\dagger}{
Step 1=after basic data reduction (Sect.\ \ref{c4-basic}), 2=after self-calibration (Sect.\ \ref{c4-self}), 3=after re-fringe fit main phase calibrator (Sect.\ \ref{c4-re-fringe}), 4=after phase referencing using more than one calibrator (Sect.\ \ref{c4-phase}).}
\end{deluxetable}

\chapter{Results}\label{chap-results}

\begin{quote}
\noindent
In this chapter, we will present the main results of the
thesis. Sect.\ \ref{c5-absolute} is about absolute astrometry, and
includes parallaxes (and therefore, distances) and proper motions for
our seven sources. We use these results to then go on and discuss the
kinematics of the sources in Taurus, and the three-dimensional
structure of the Taurus and Ophiuchus complexes with the distances. In
Sect.\ \ref{c5-variability} we briefly discuss the variability of the
sources in our sample. The implications for the properties of some
stars in the sample are discussed in Sect.\
\ref{c5-implications}. Finally, the relative astrometry is discussed
in Sect.\ \ref{c5-relative} which includes orbital fits for
T~Tau~Sa/Sb and V773~Tau~Aa/Ab systems.
\end{quote}

\medskip

\begin{figure*}[!h]
\centering
\begin{tabular}{cc}
\includegraphics[width=.45\textwidth,angle=0]{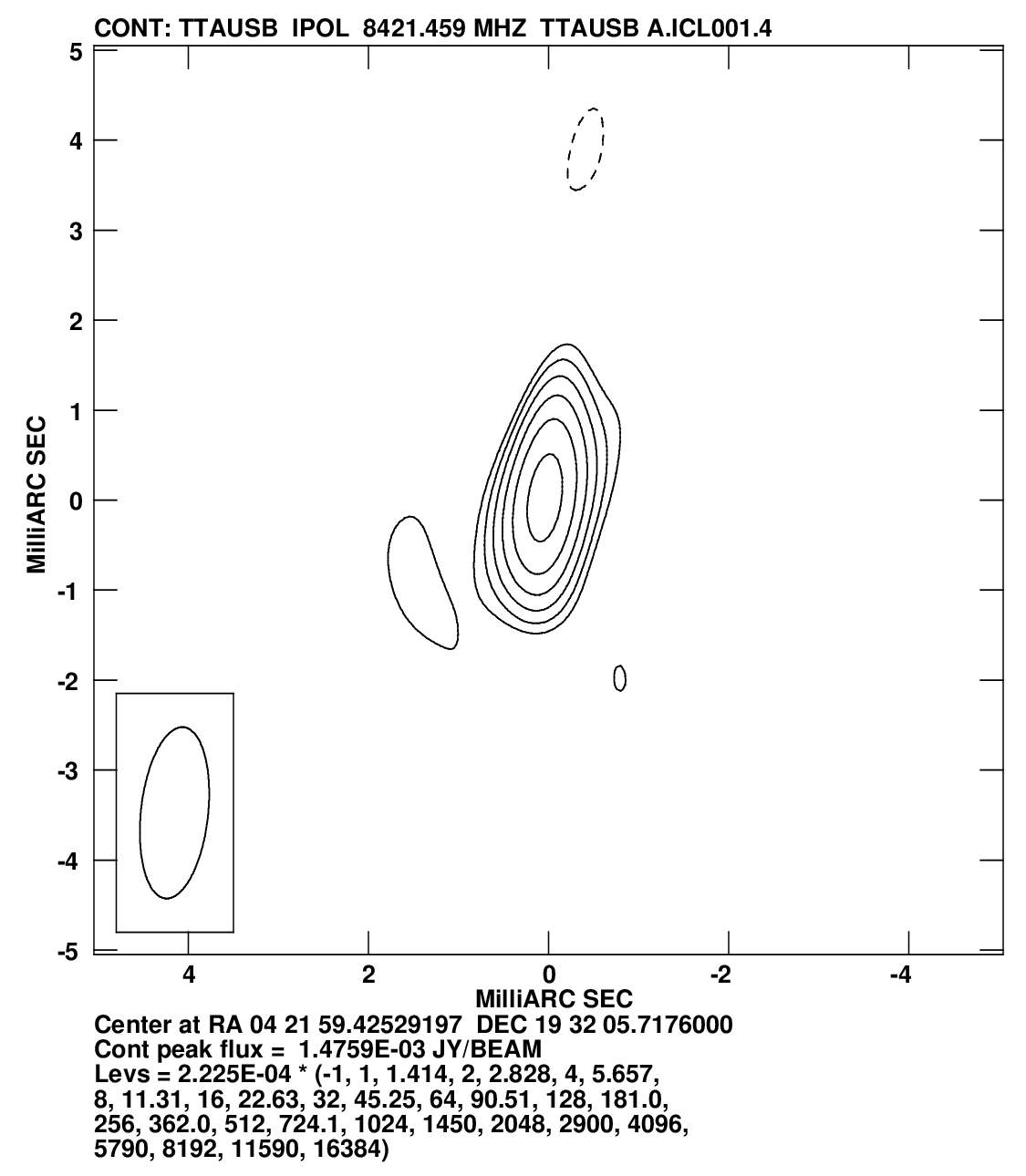} &
\includegraphics[width=.45\textwidth,angle=0]{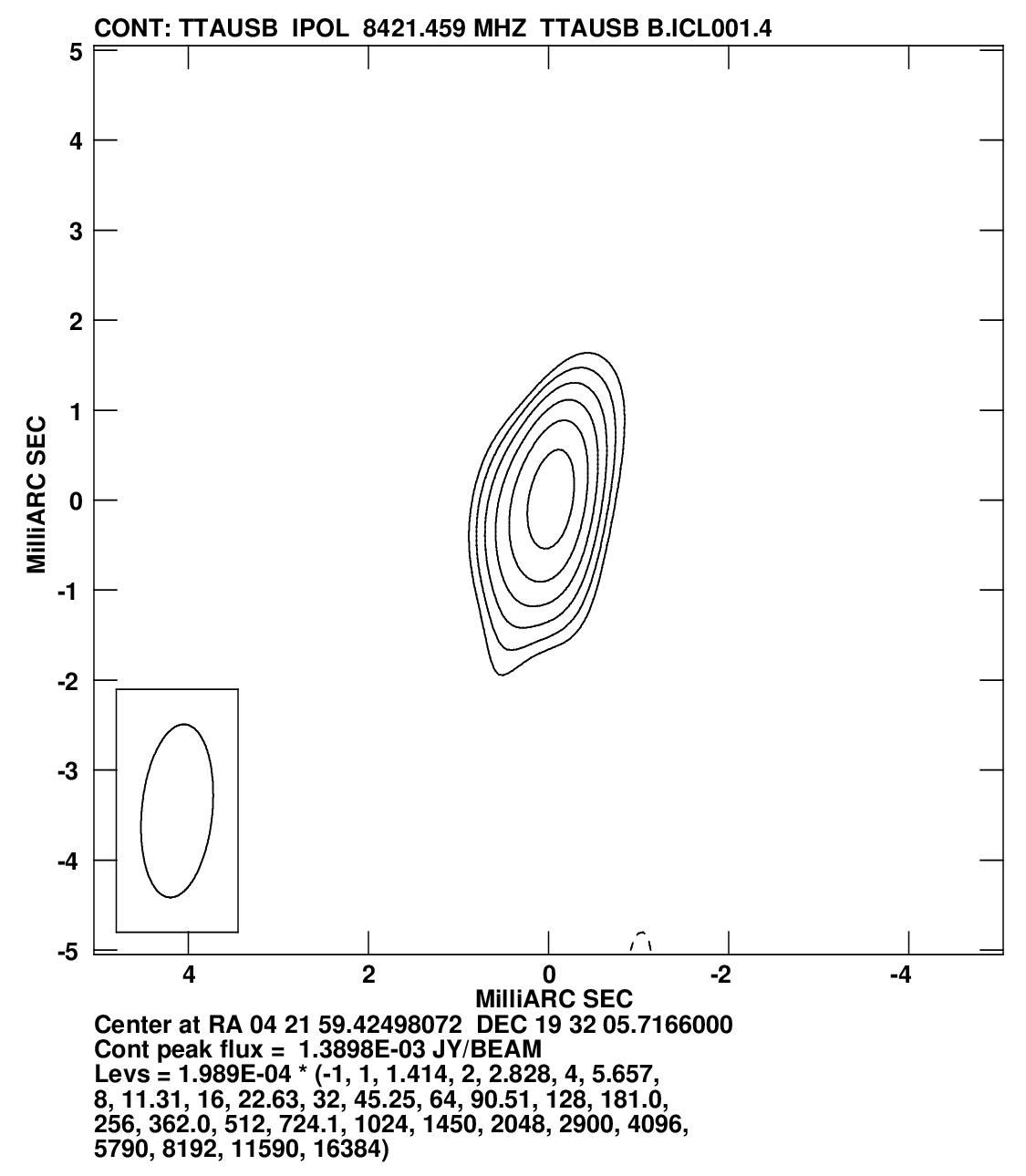} \\
\includegraphics[width=.45\textwidth,angle=0]{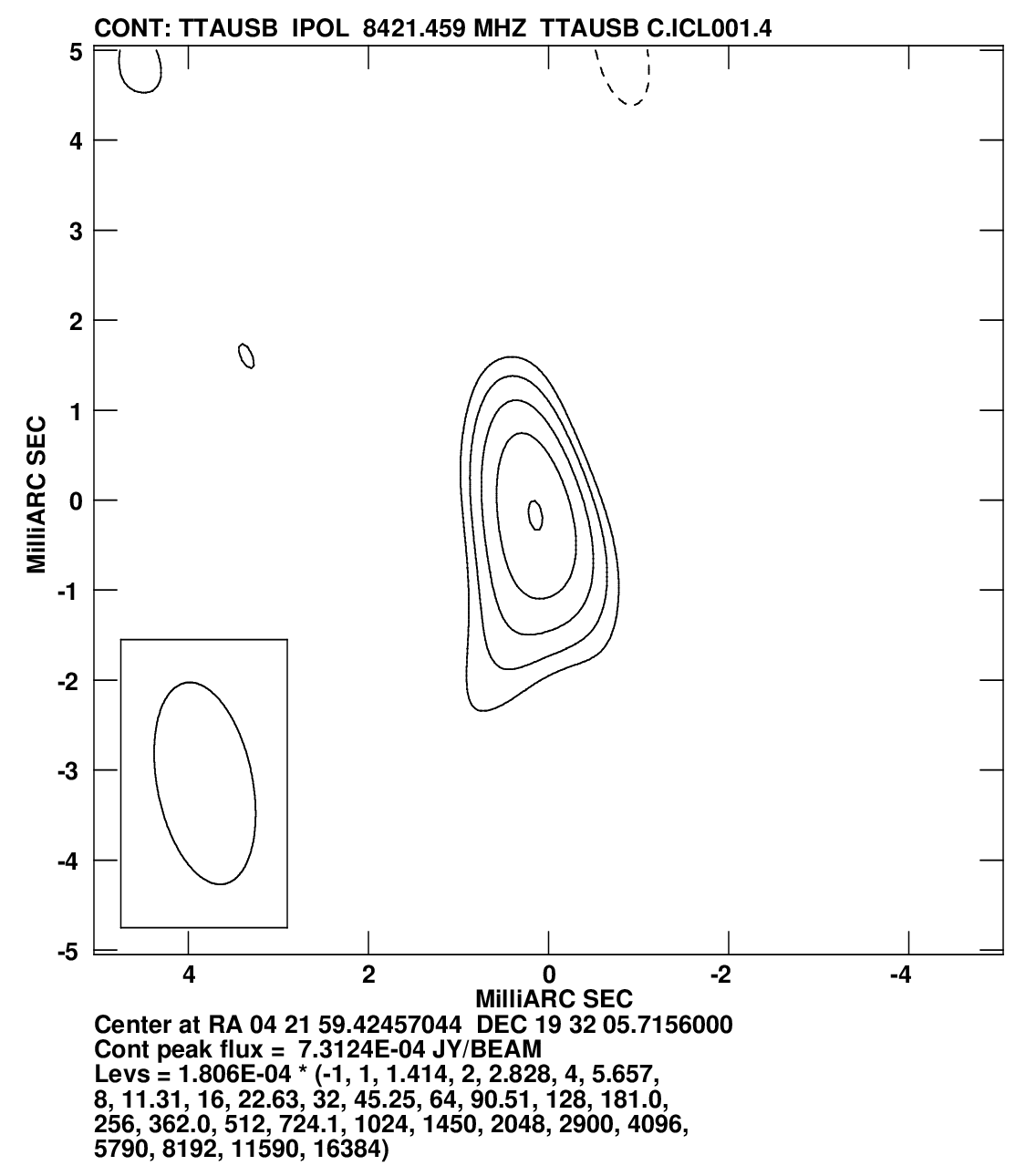} &
\includegraphics[width=.45\textwidth,angle=0]{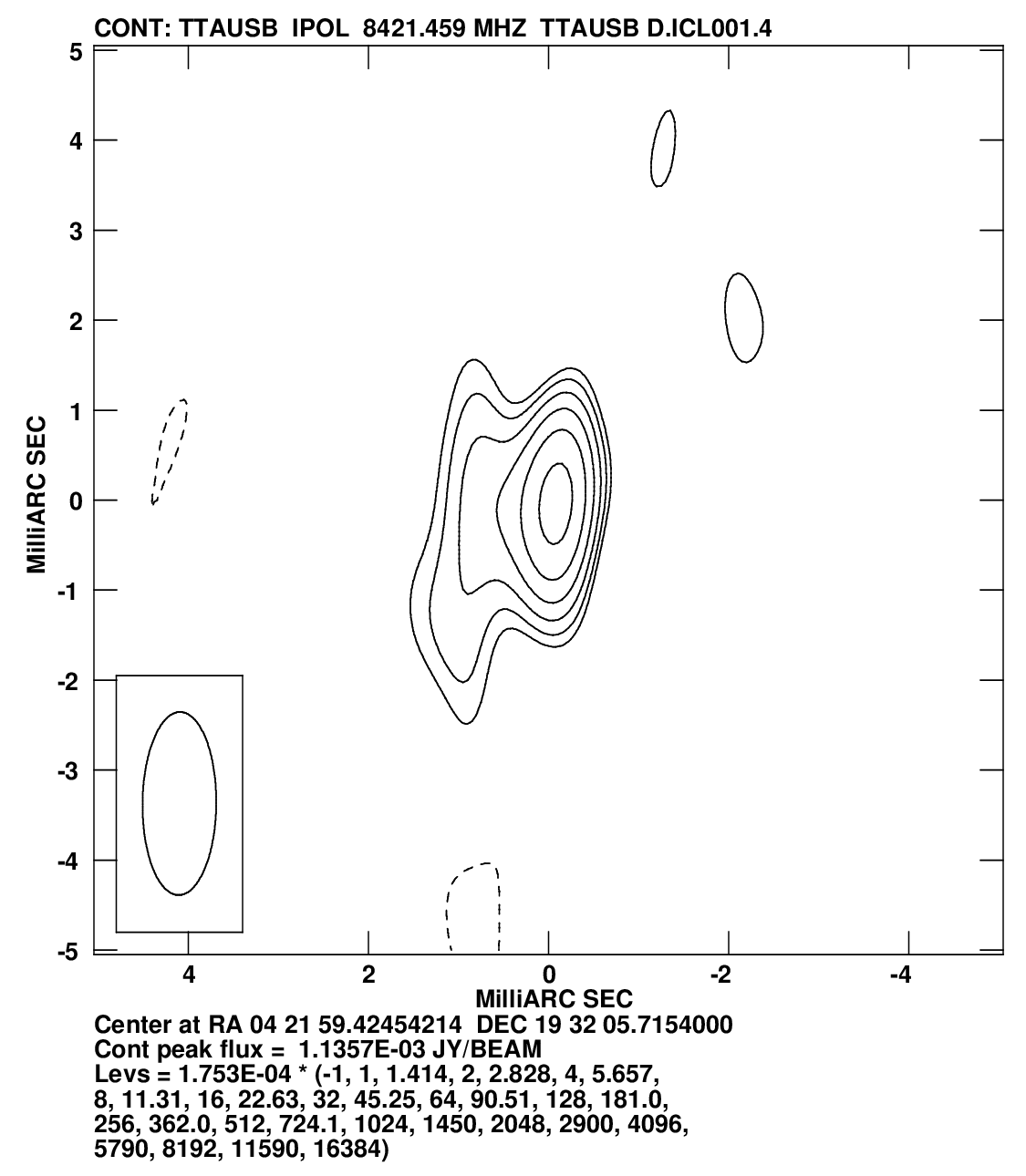} \\
\includegraphics[width=.45\textwidth,angle=0]{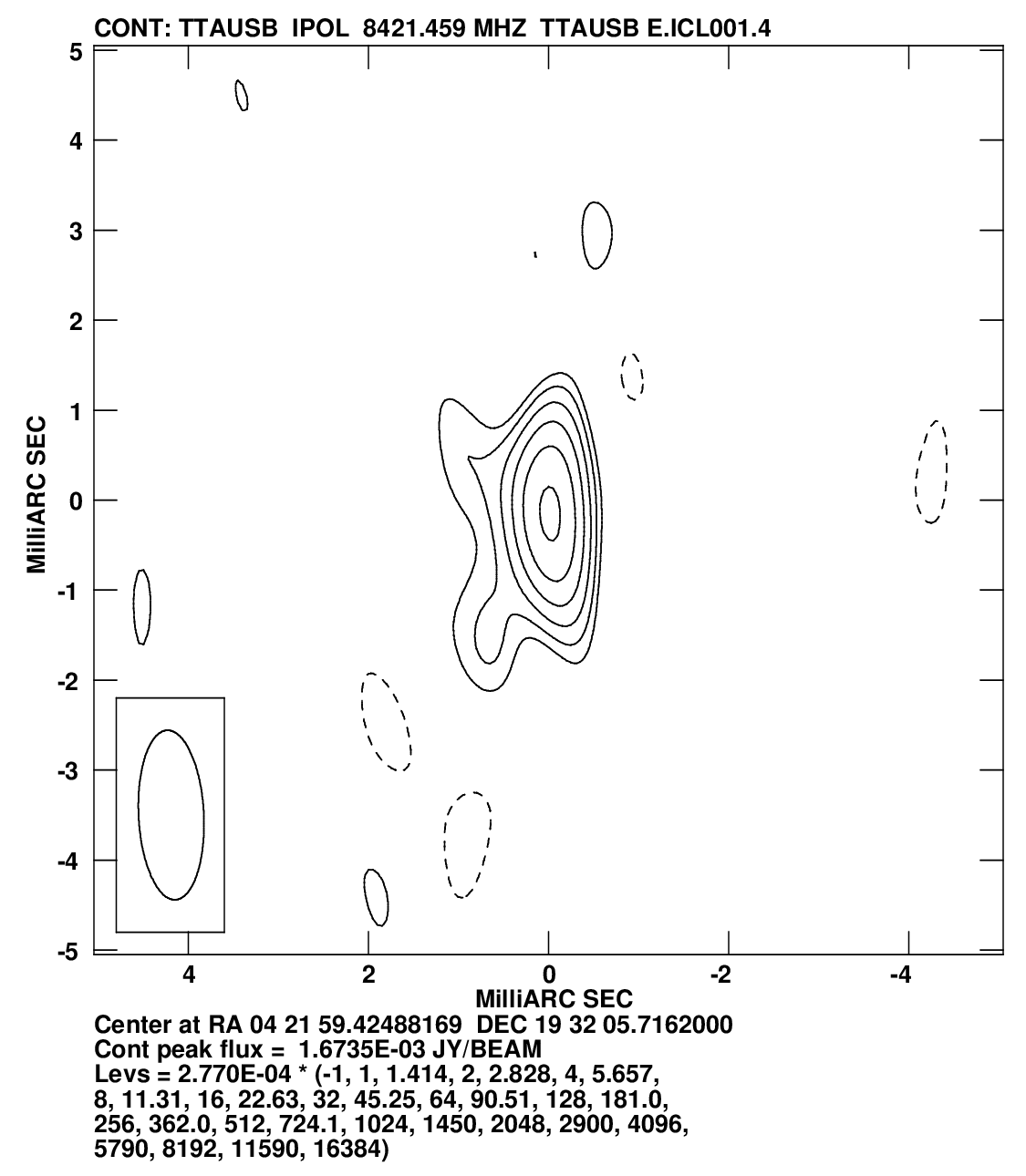} &
\includegraphics[width=.45\textwidth,angle=0]{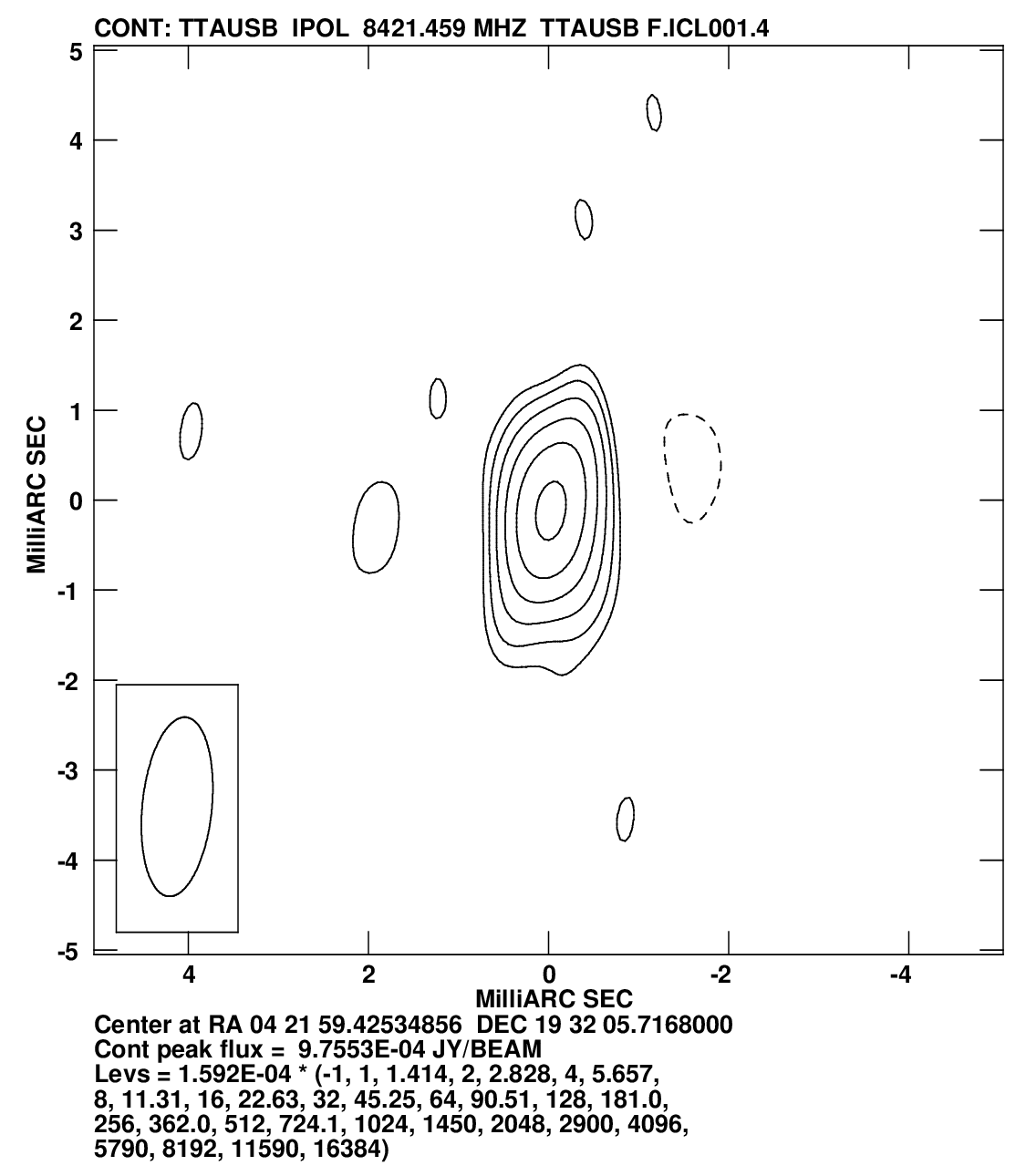}
\end{tabular}
\caption{\footnotesize{
VLBA images of T~Tau~Sb at epochs 1 to 6 listed in Tab.\
\ref{tab-observaciones} and \ref{tab-resultados}.}
\label{map-ttau1}}\end{figure*}

\begin{figure*}[!h]
\centering
\begin{tabular}{cc}
\includegraphics[width=.45\textwidth,angle=0]{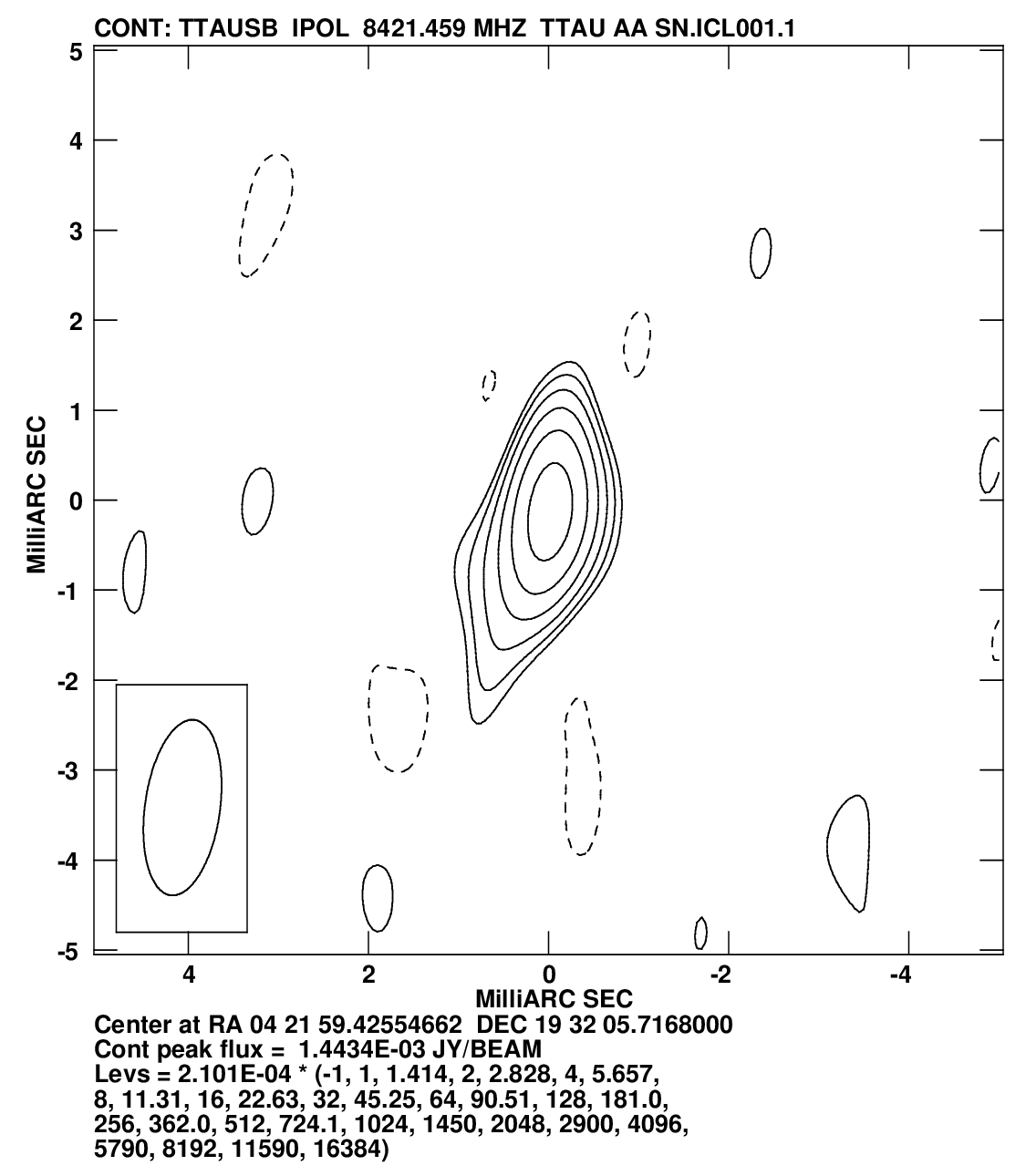} &
\includegraphics[width=.45\textwidth,angle=0]{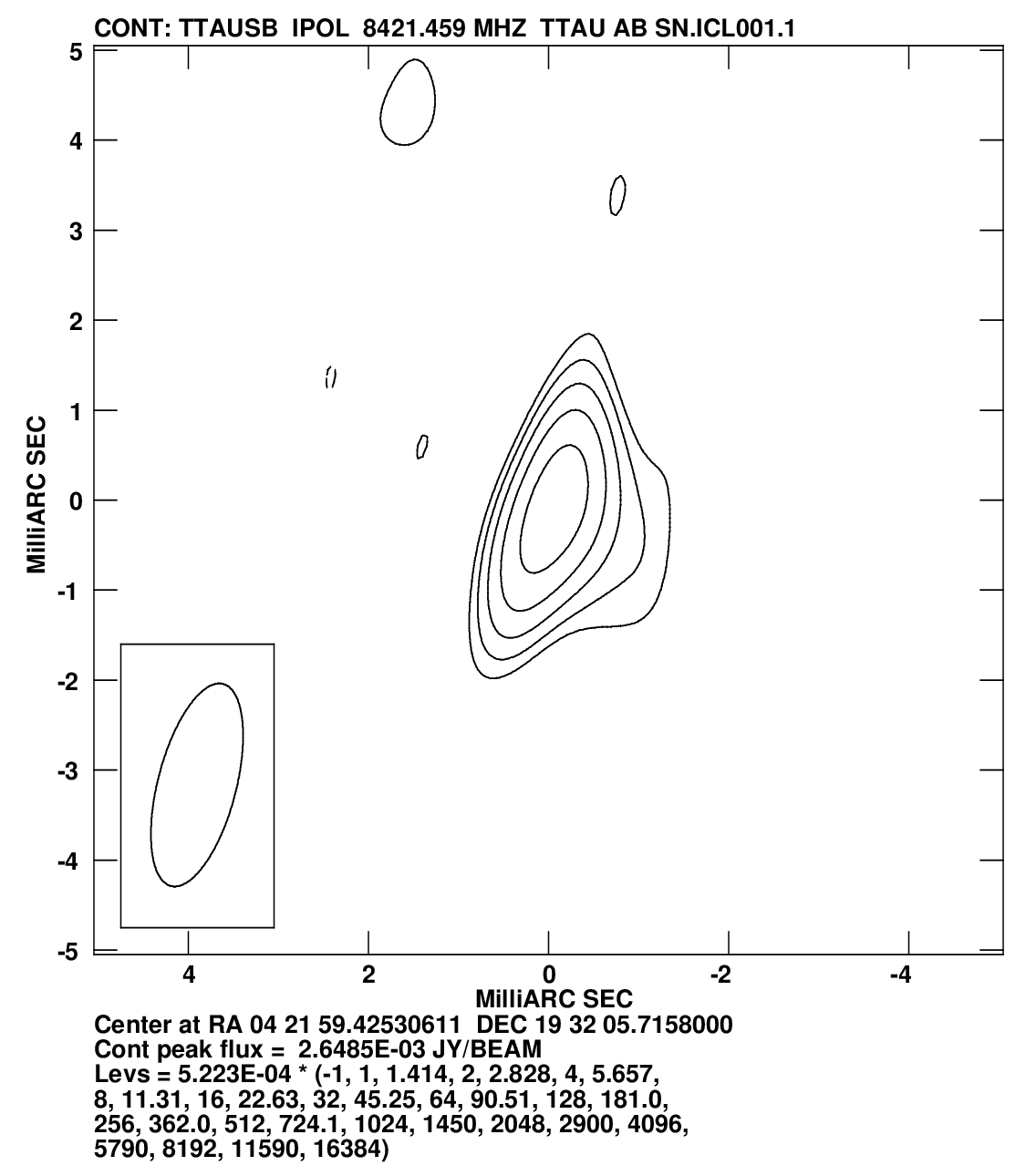} \\
\includegraphics[width=.45\textwidth,angle=0]{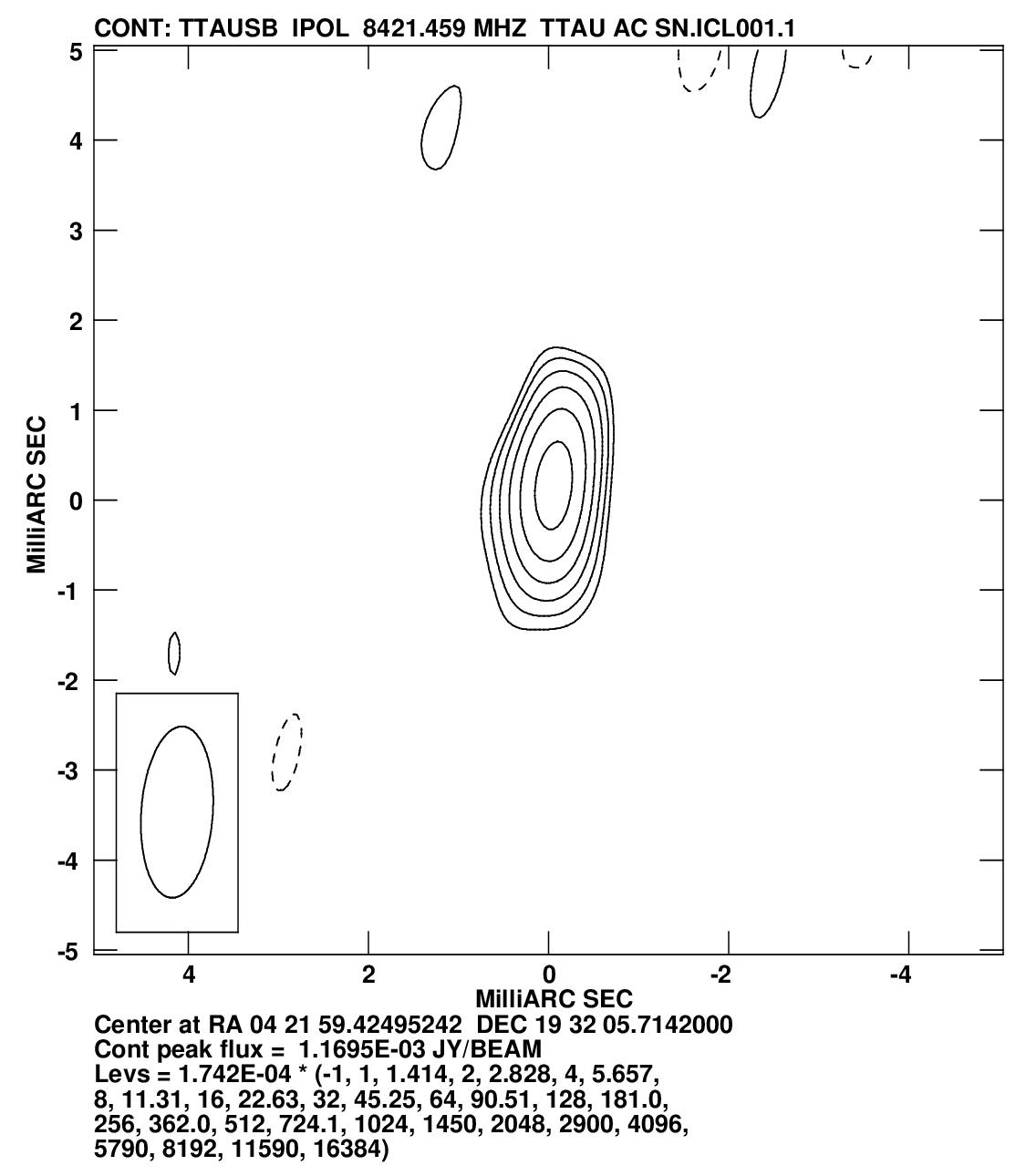} &
\includegraphics[width=.45\textwidth,angle=0]{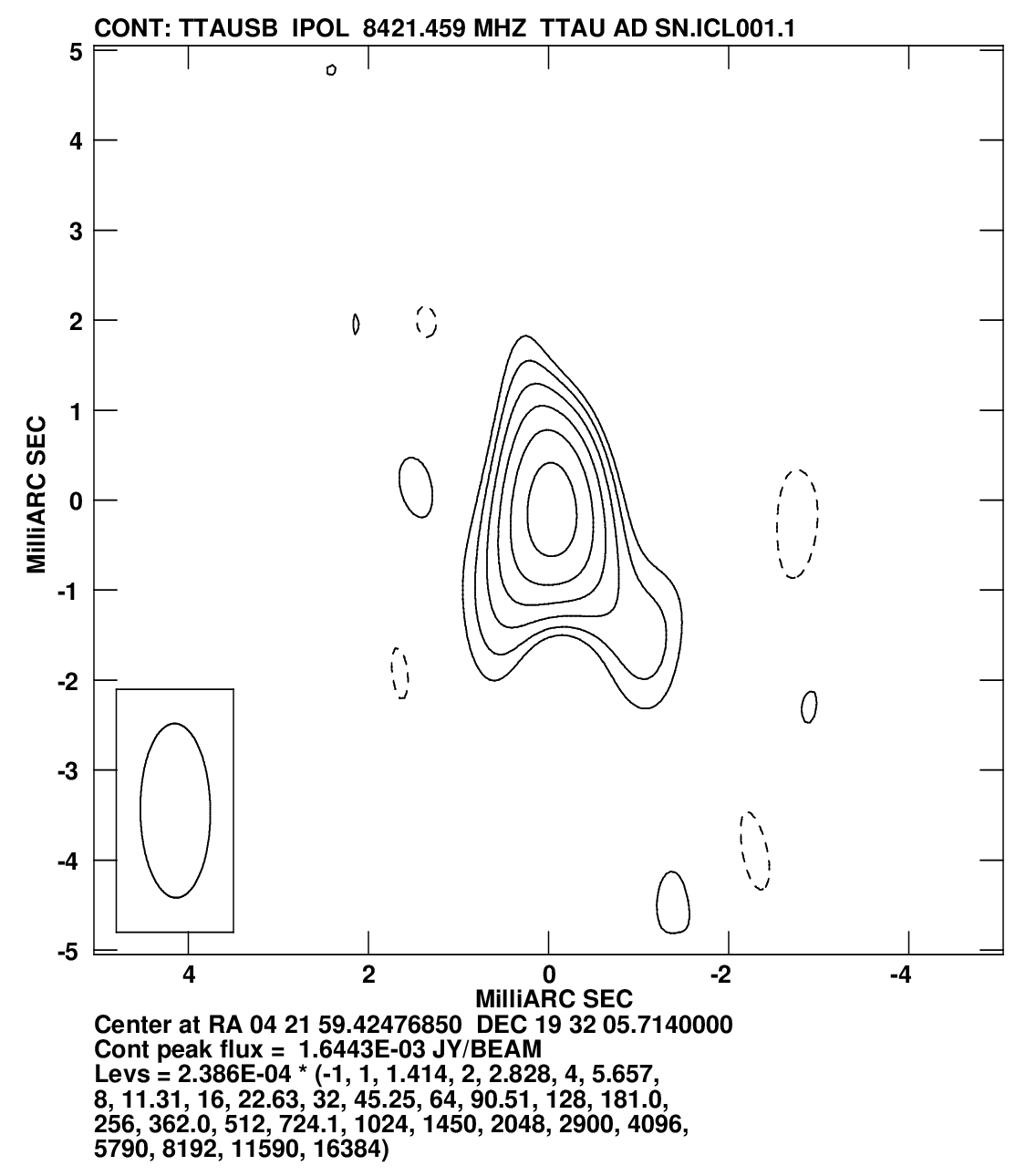} \\
\includegraphics[width=.45\textwidth,angle=0]{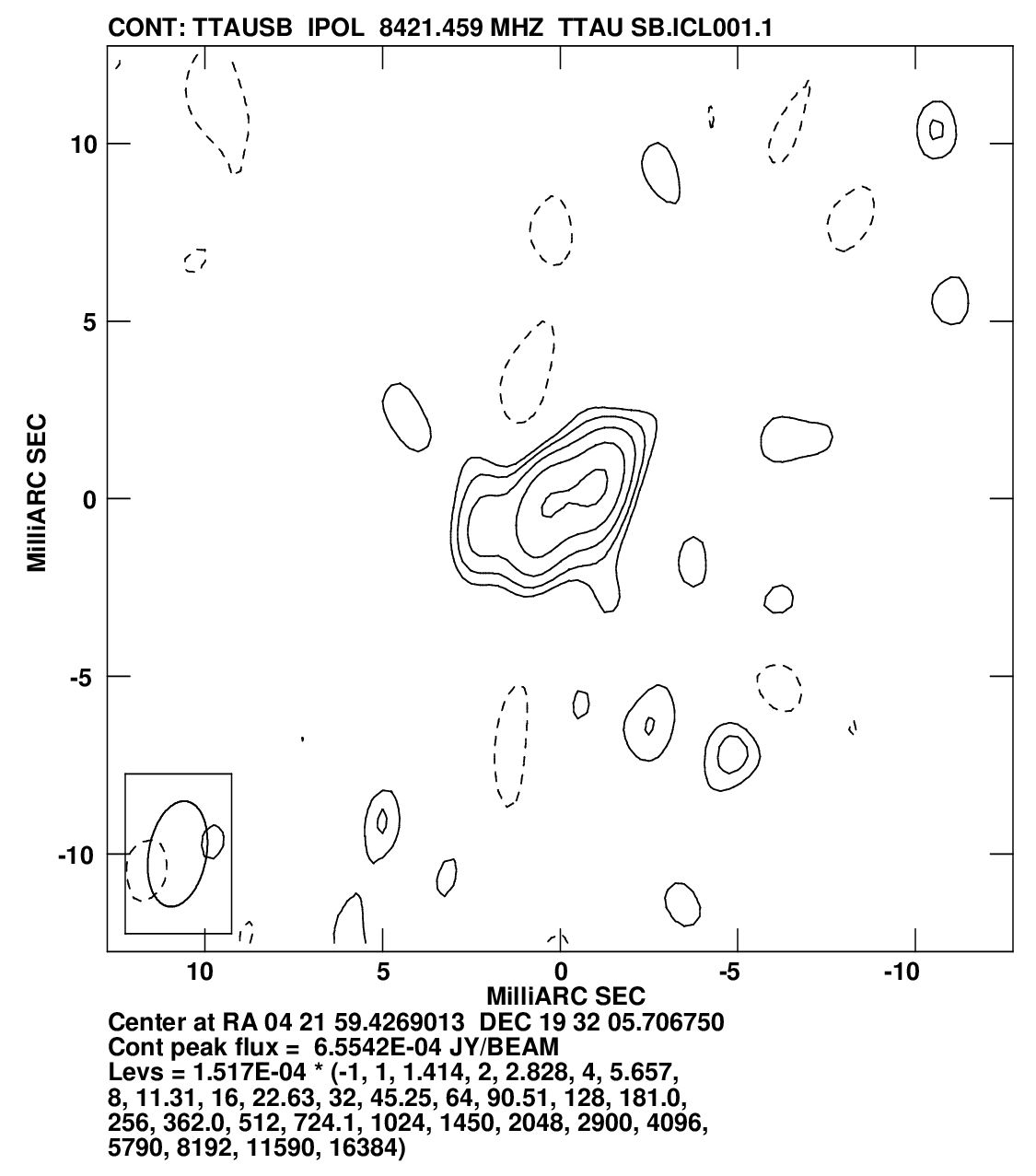} &
\includegraphics[width=.45\textwidth,angle=0]{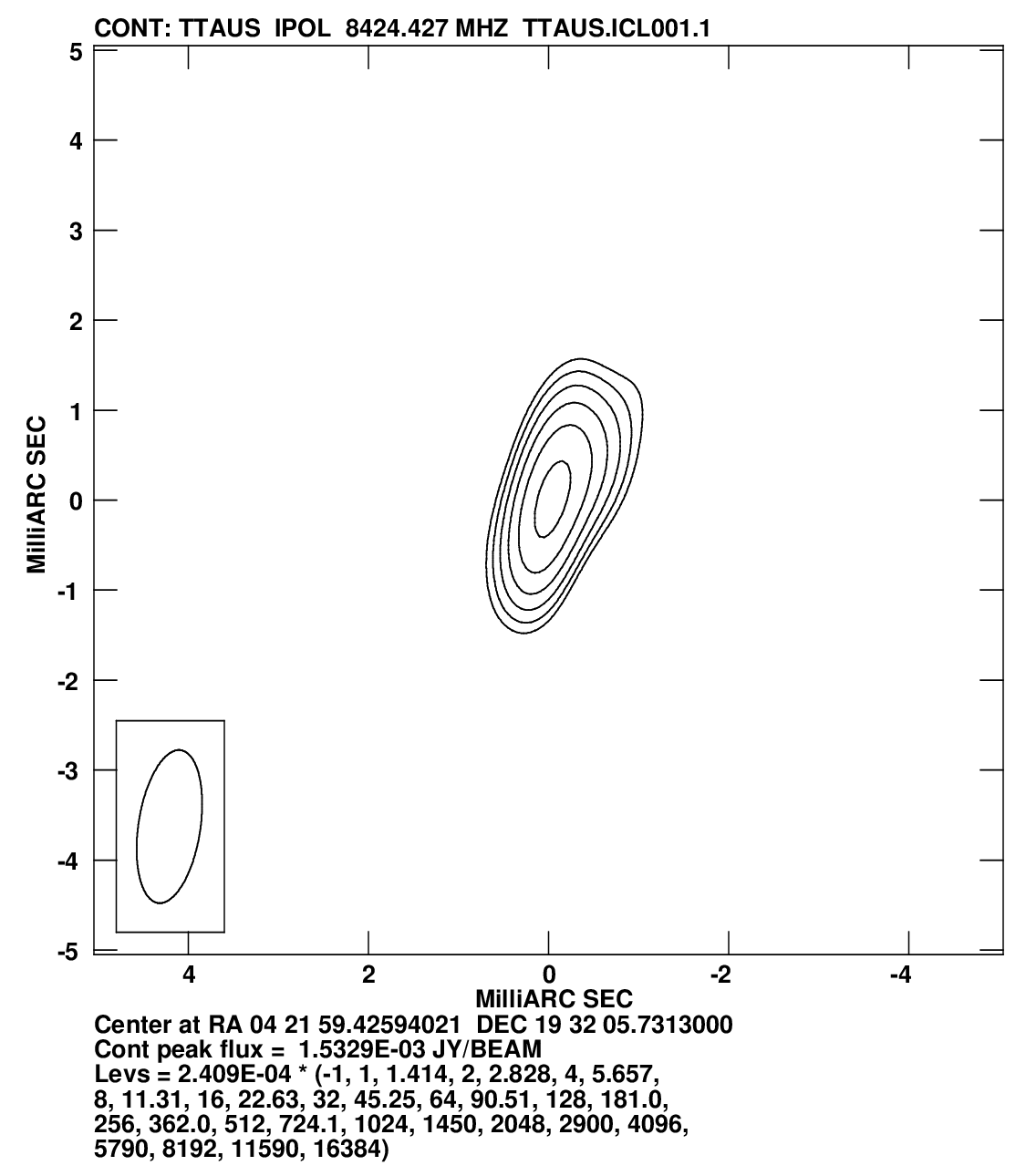}
\end{tabular}
\caption{\footnotesize{
VLBA images of T~Tau~Sb at epochs 7, 8, 9, 10, 13 and 14 listed in
Tab.\ \ref{tab-observaciones} and \ref{tab-resultados}.}
\label{map-ttau2}}\end{figure*}

\begin{figure*}[!h]
\centering
\begin{tabular}{cc}
\includegraphics[width=.45\textwidth,angle=0]{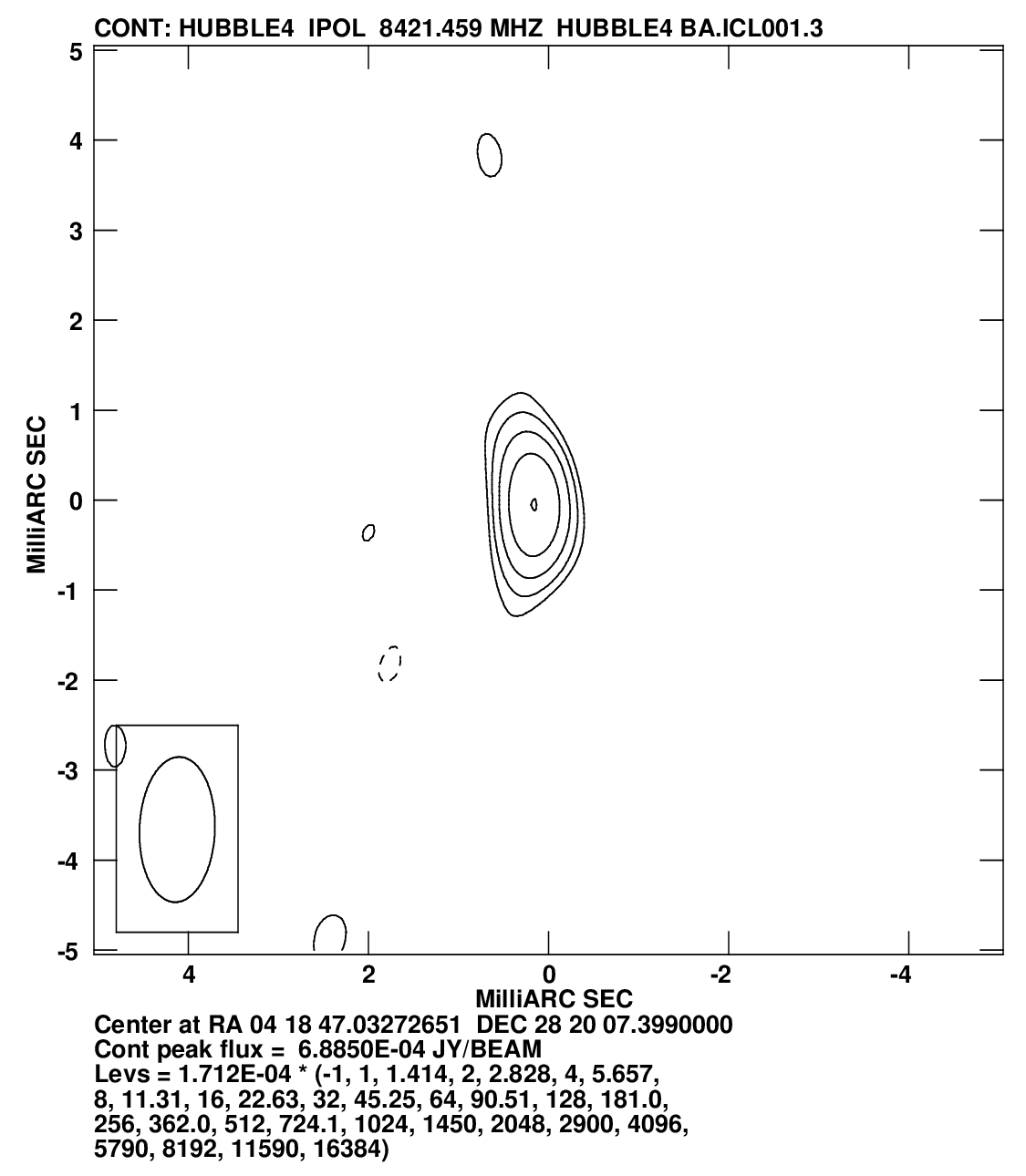} &
\includegraphics[width=.45\textwidth,angle=0]{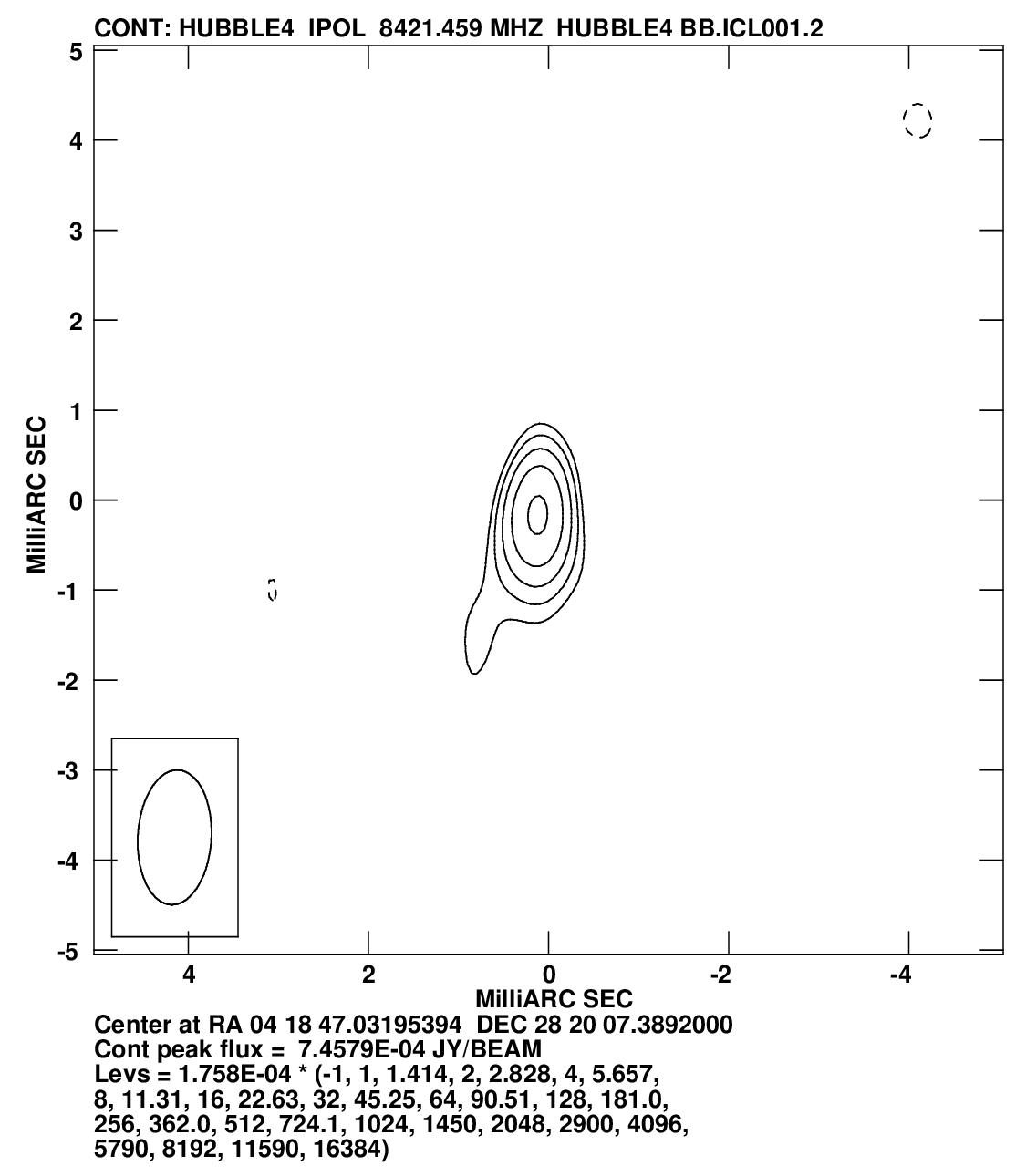} \\
\includegraphics[width=.45\textwidth,angle=0]{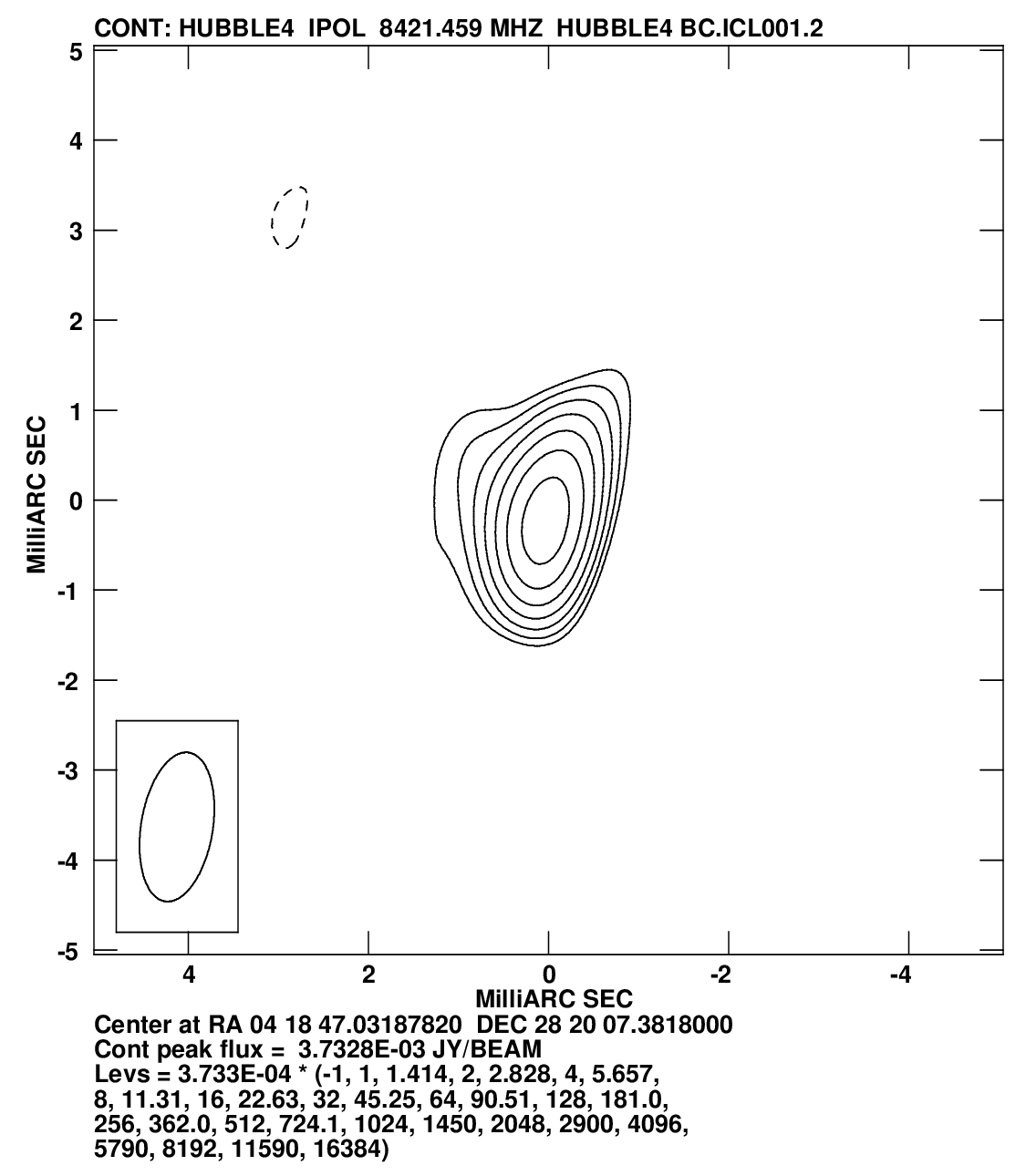} &
\includegraphics[width=.45\textwidth,angle=0]{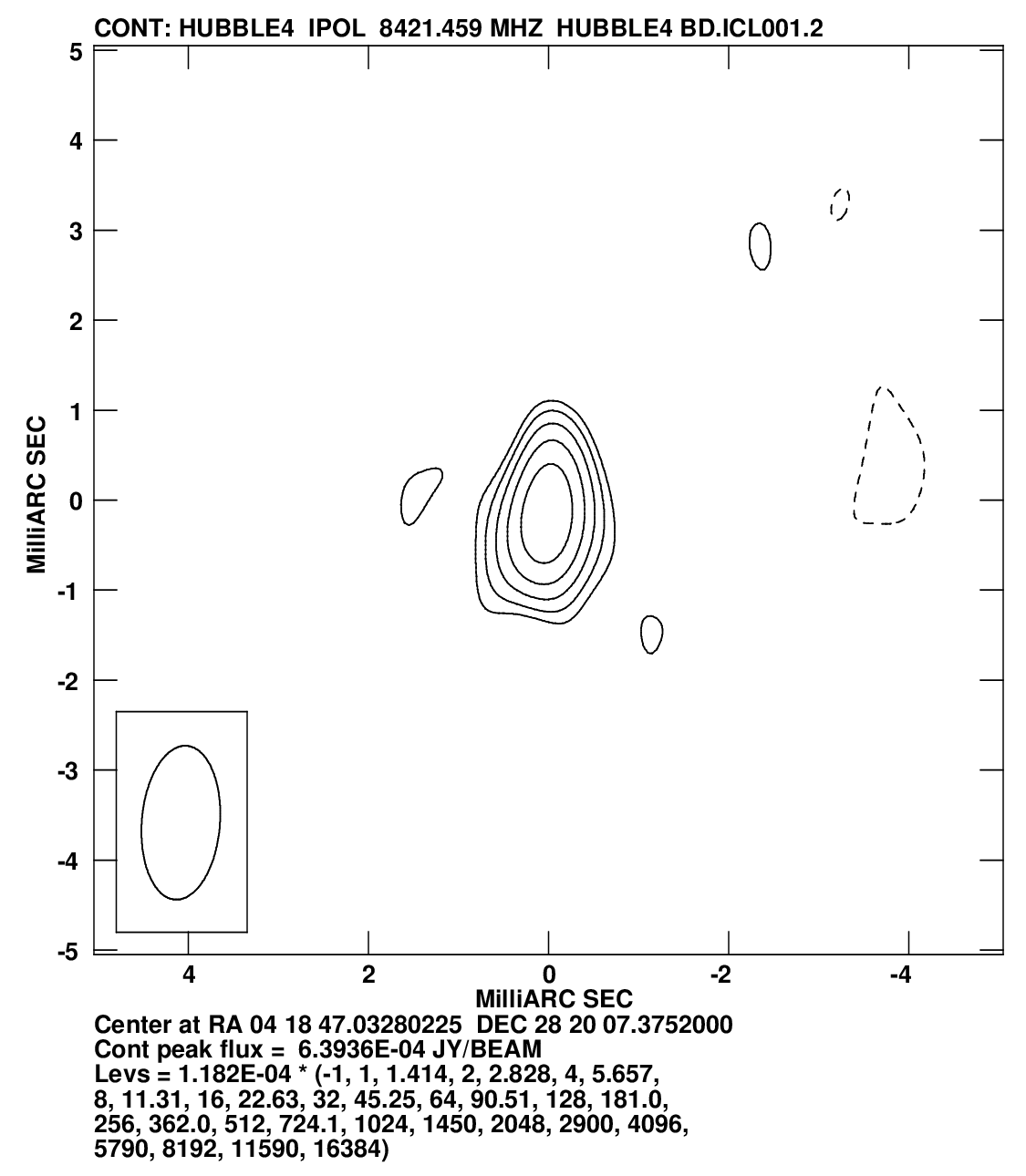} \\
\includegraphics[width=.45\textwidth,angle=0]{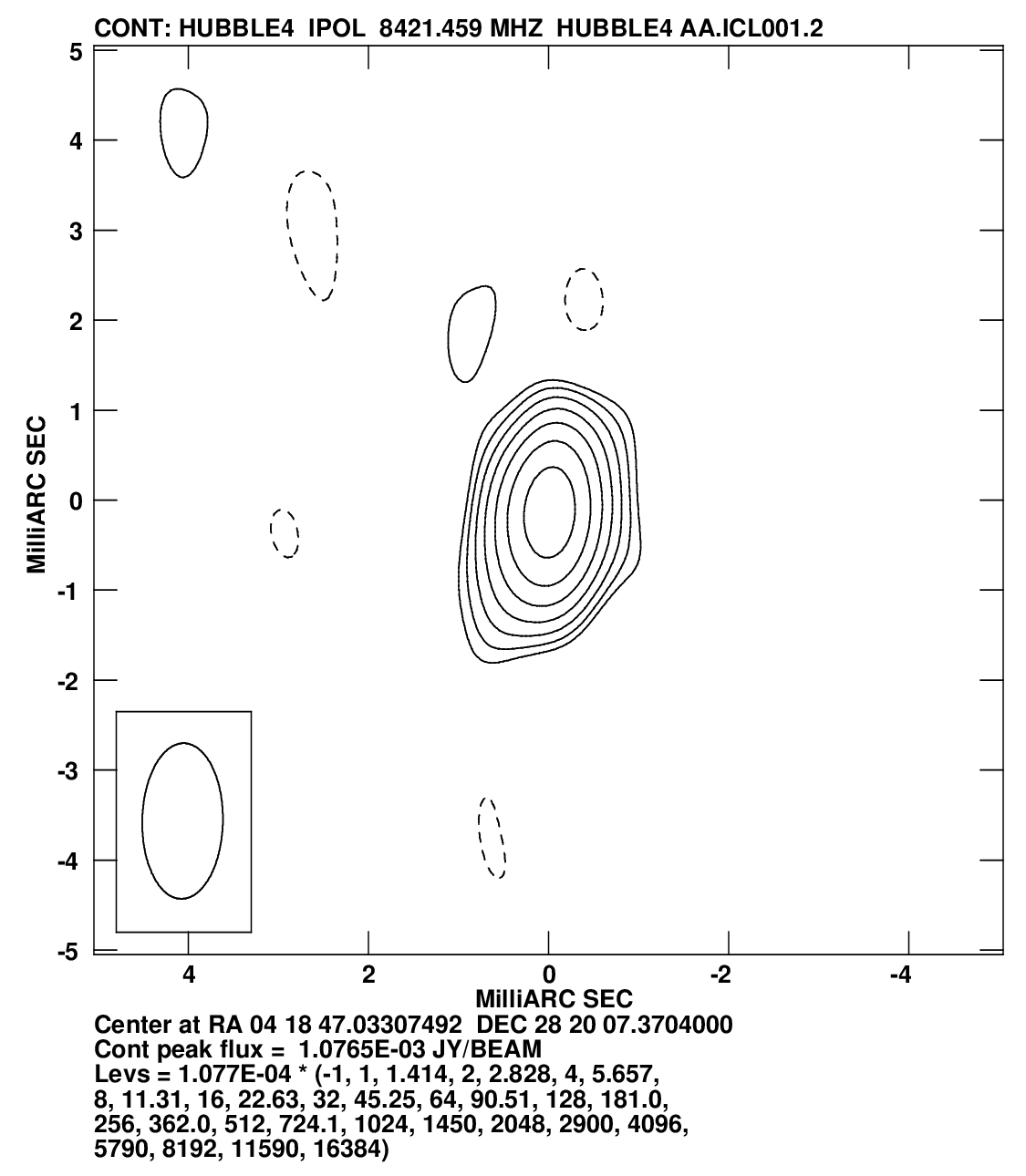} &
\includegraphics[width=.45\textwidth,angle=0]{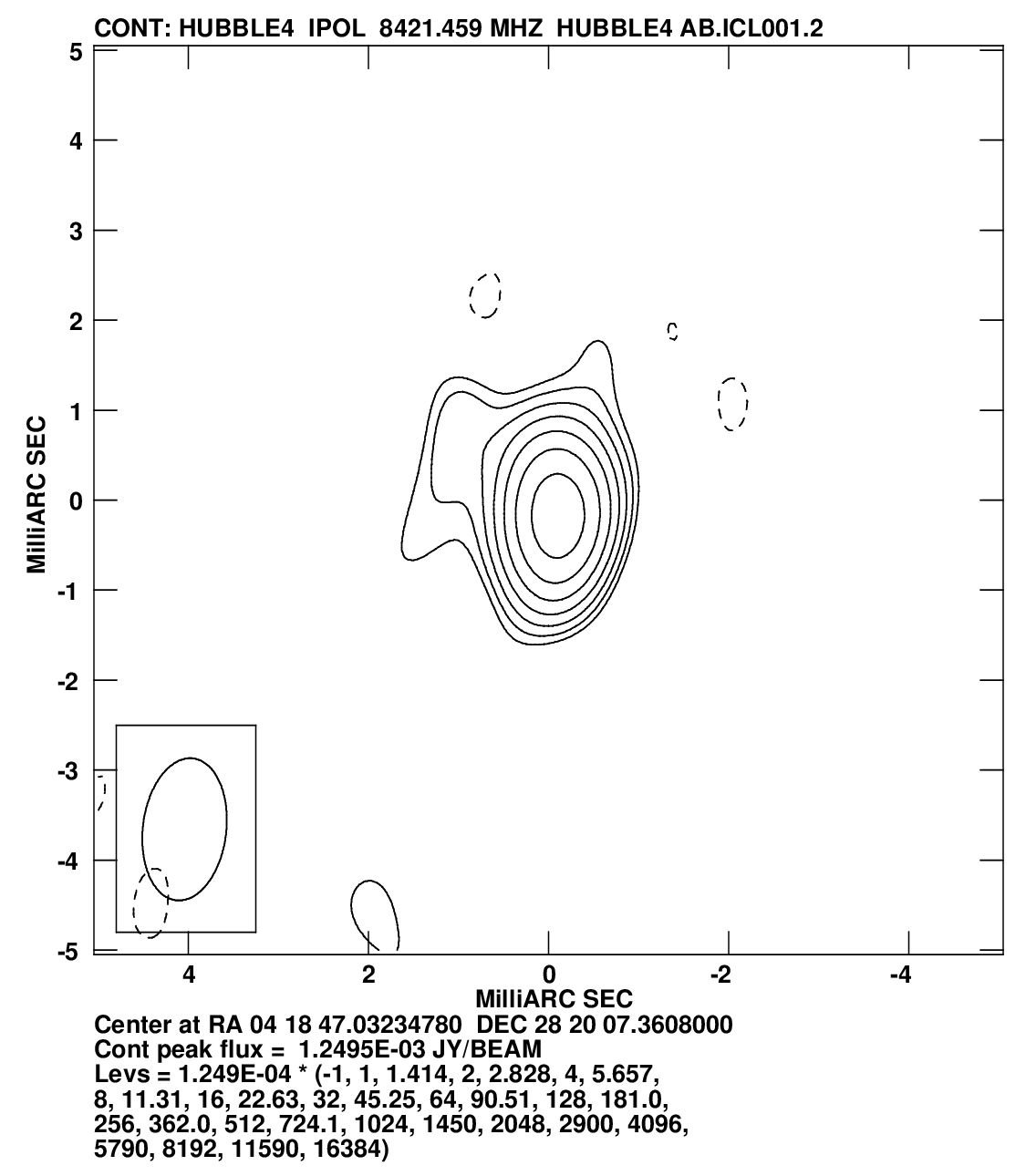}
\end{tabular}
\caption{\footnotesize{VLBA images of Hubble~4 at epochs 1 to 6 listed in Tab.\ \ref{tab-observaciones} and \ref{tab-resultados}.}
\label{map-hubble4}}\end{figure*}

\begin{figure*}[!h]
\centering
\begin{tabular}{cc}
\includegraphics[width=.45\textwidth,angle=0]{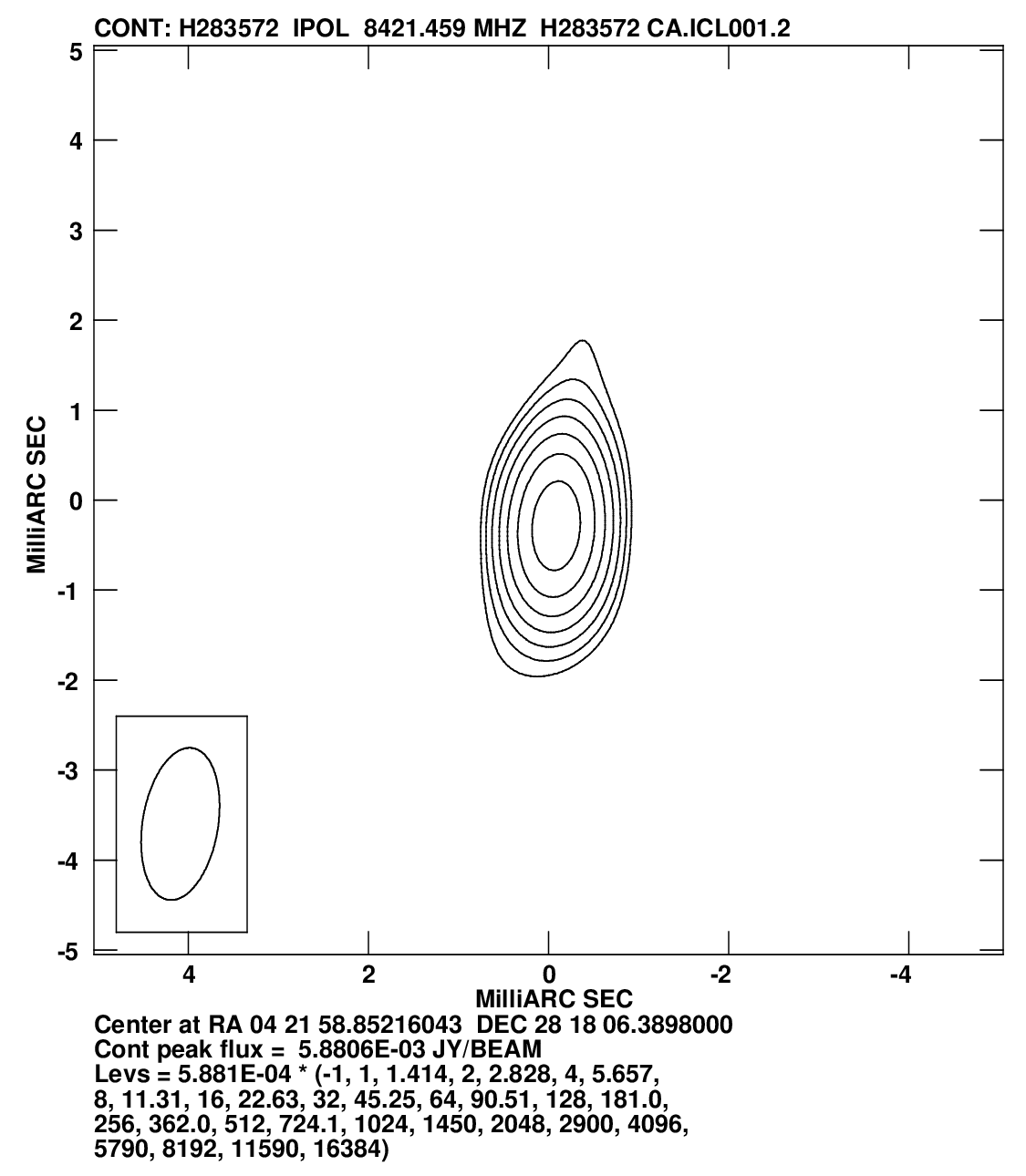} &
\includegraphics[width=.45\textwidth,angle=0]{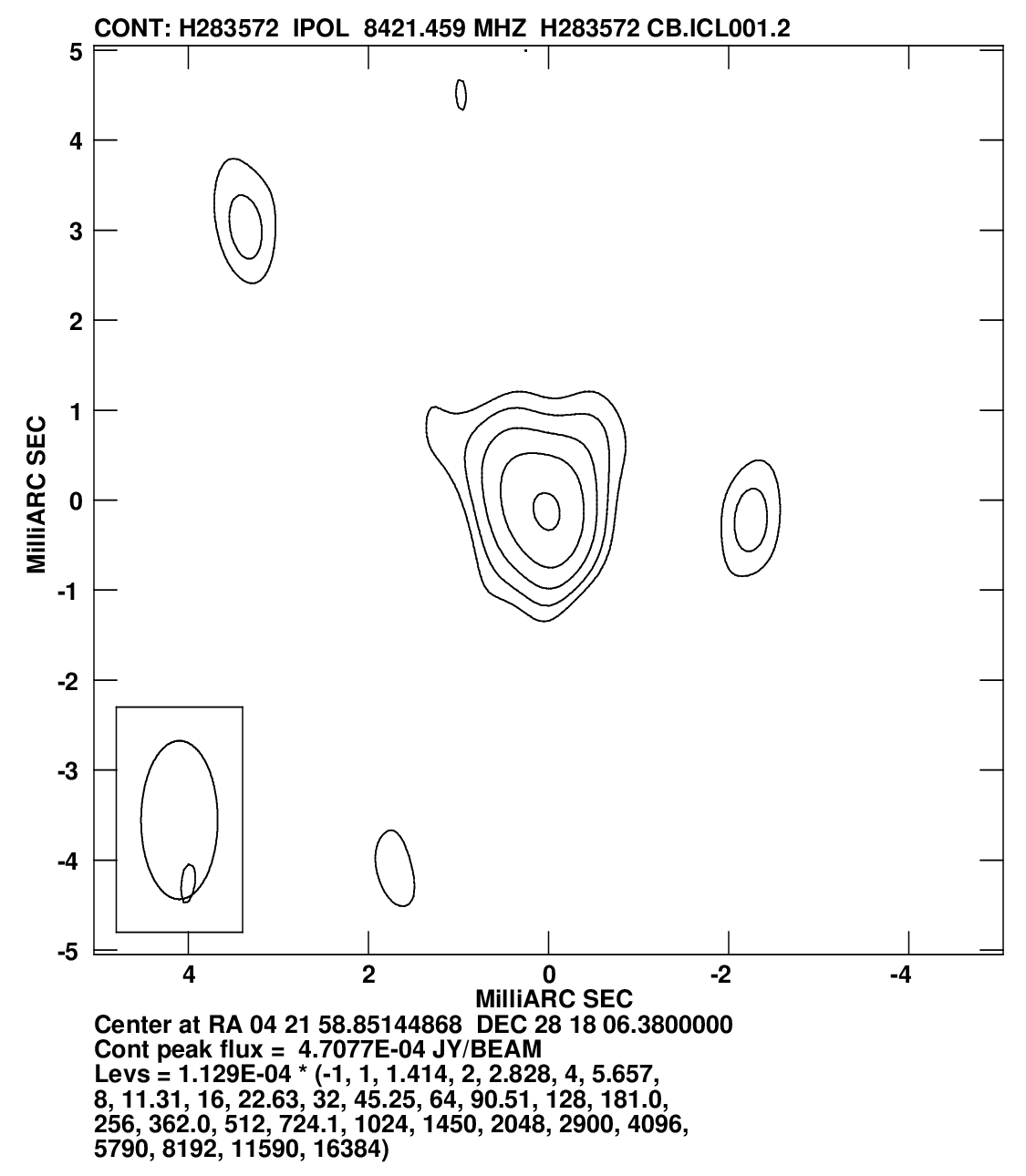} \\
\includegraphics[width=.45\textwidth,angle=0]{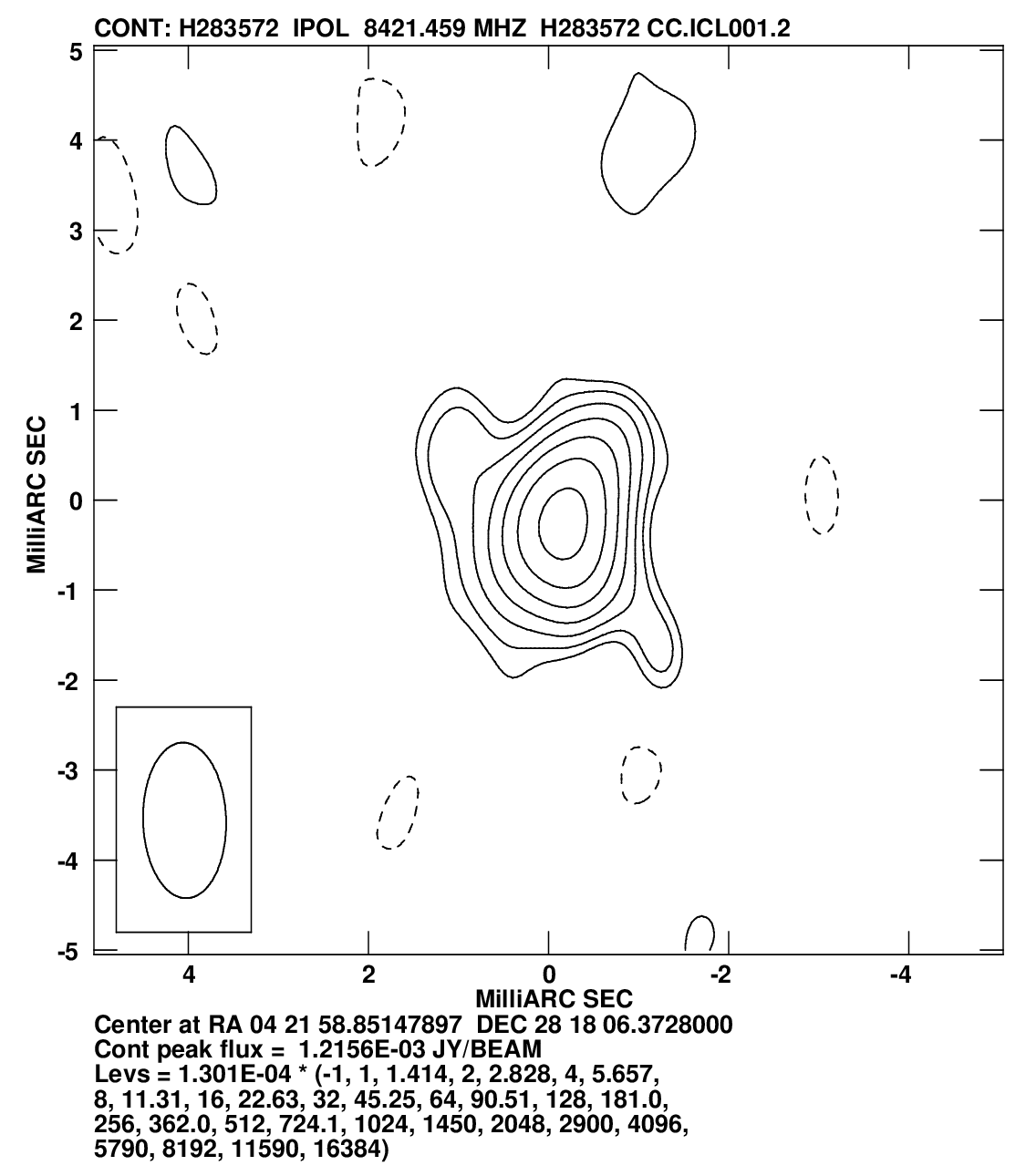} &
\includegraphics[width=.45\textwidth,angle=0]{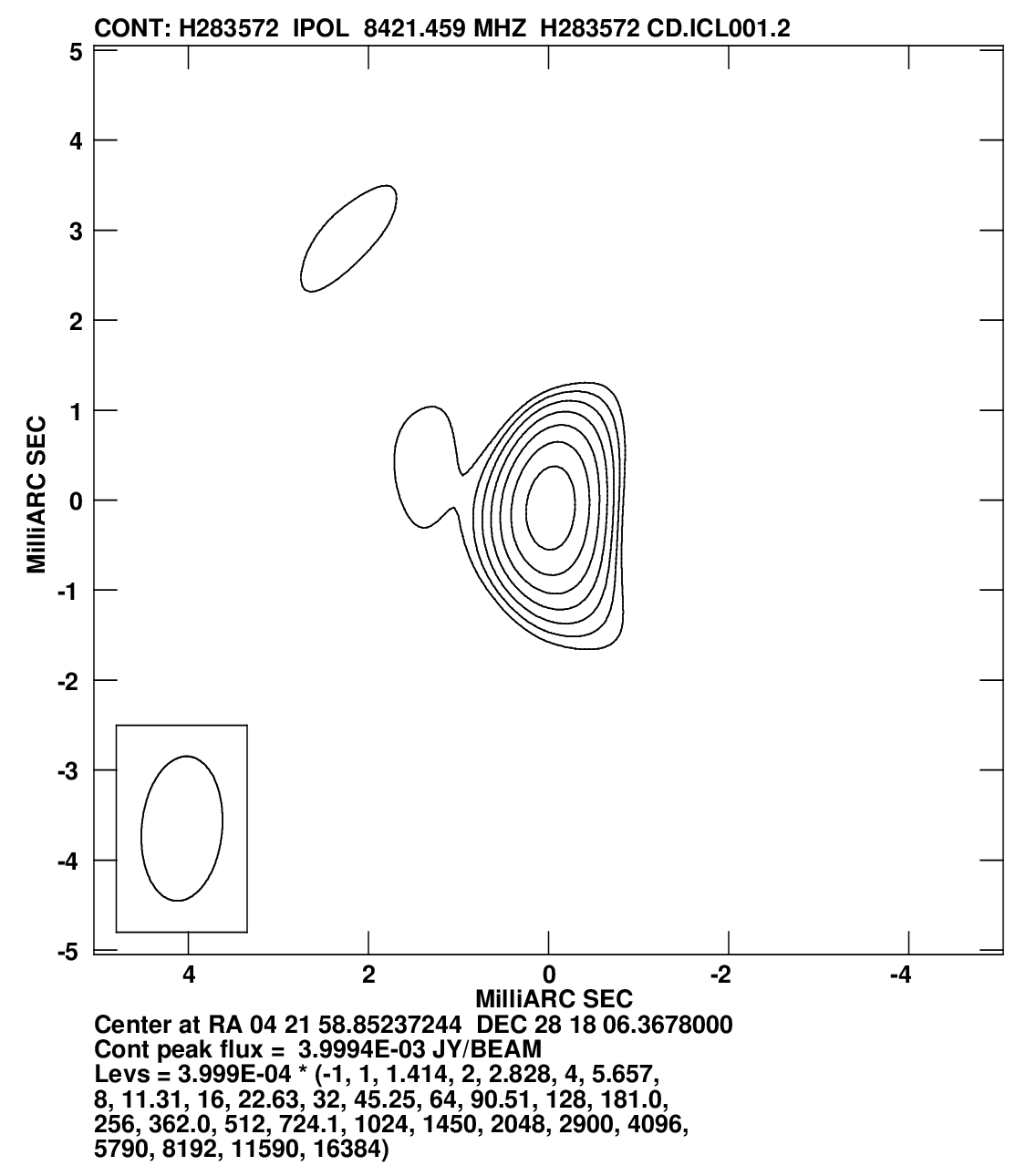} \\
\includegraphics[width=.45\textwidth,angle=0]{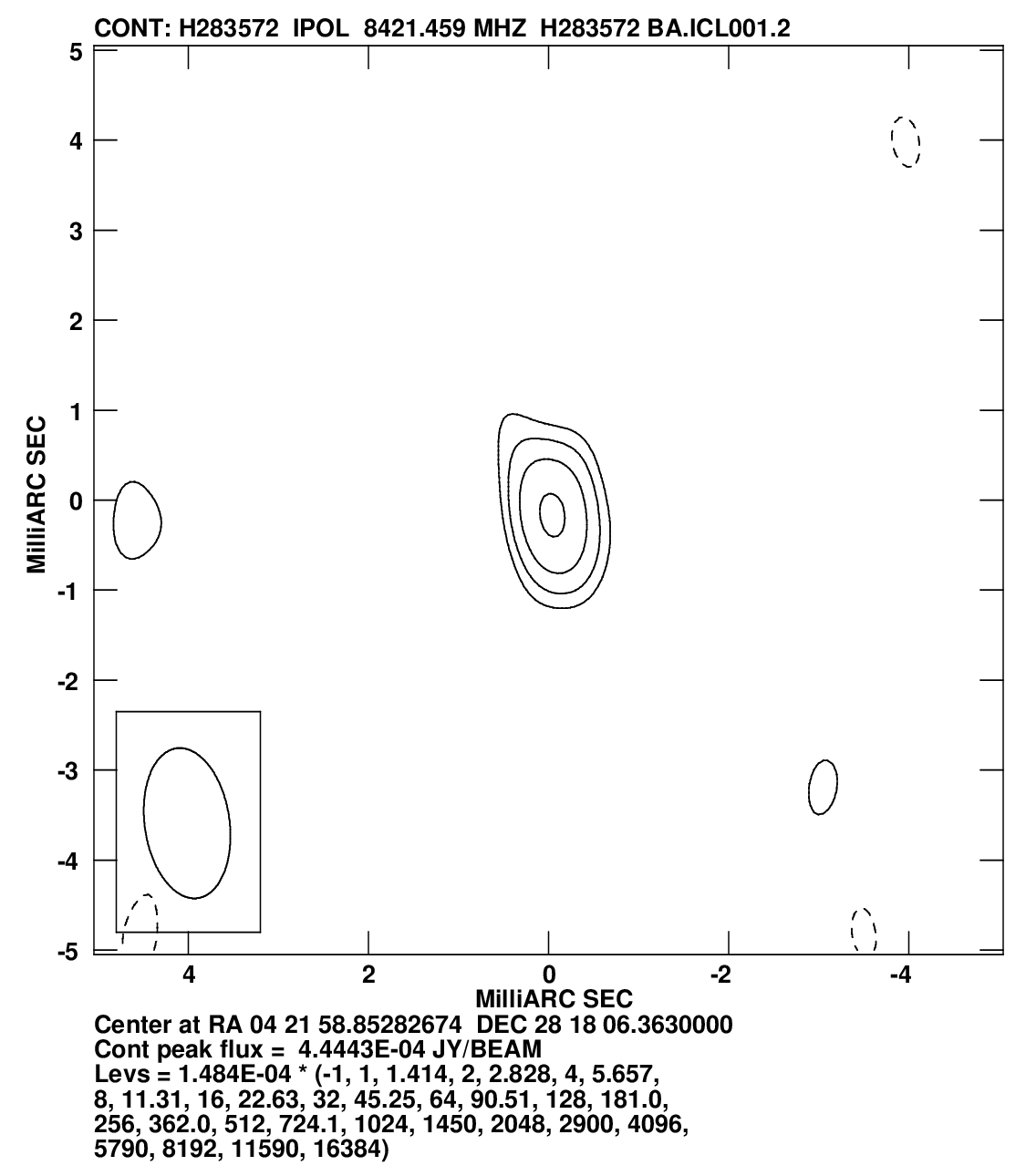} &
\includegraphics[width=.45\textwidth,angle=0]{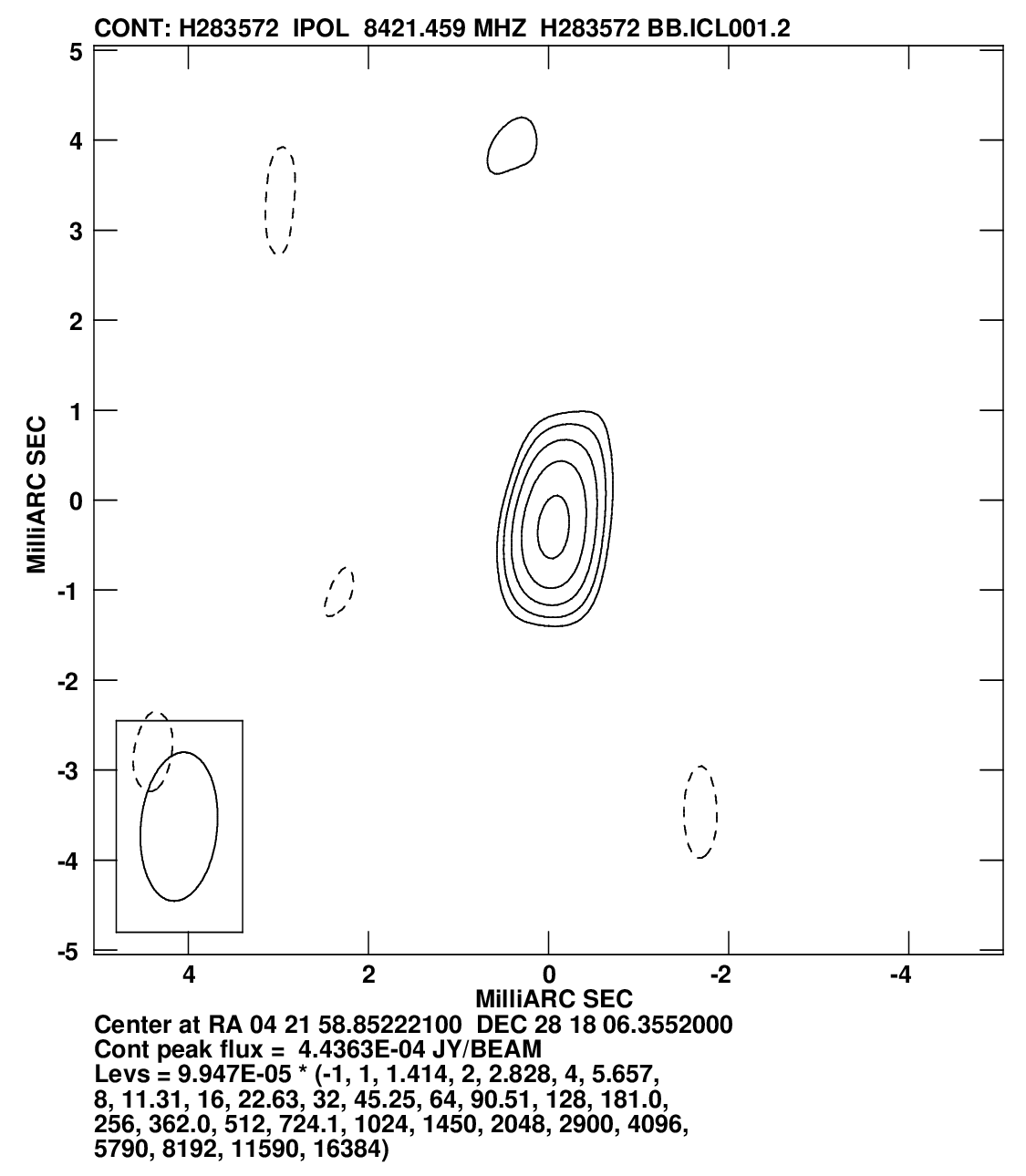}
\end{tabular}
\caption{\footnotesize{
VLBA images of HDE~283572 at epochs 1 to 6 listed in Tab.\
\ref{tab-observaciones} and \ref{tab-resultados}.}
\label{map-hde}}\end{figure*}

\begin{figure*}[!h]
\centering
\begin{tabular}{cc}
\includegraphics[width=.45\textwidth,angle=0]{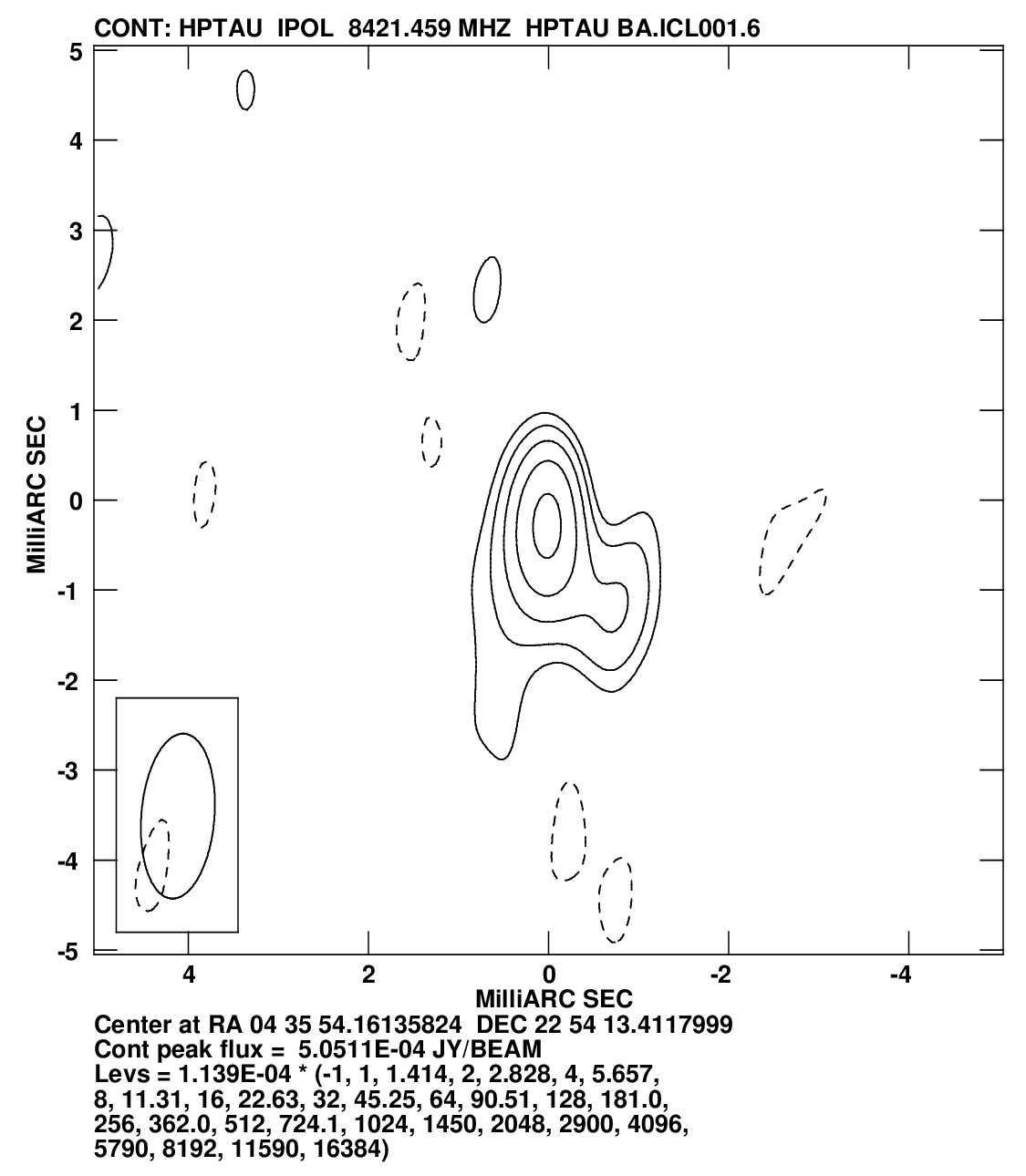} &
\includegraphics[width=.45\textwidth,angle=0]{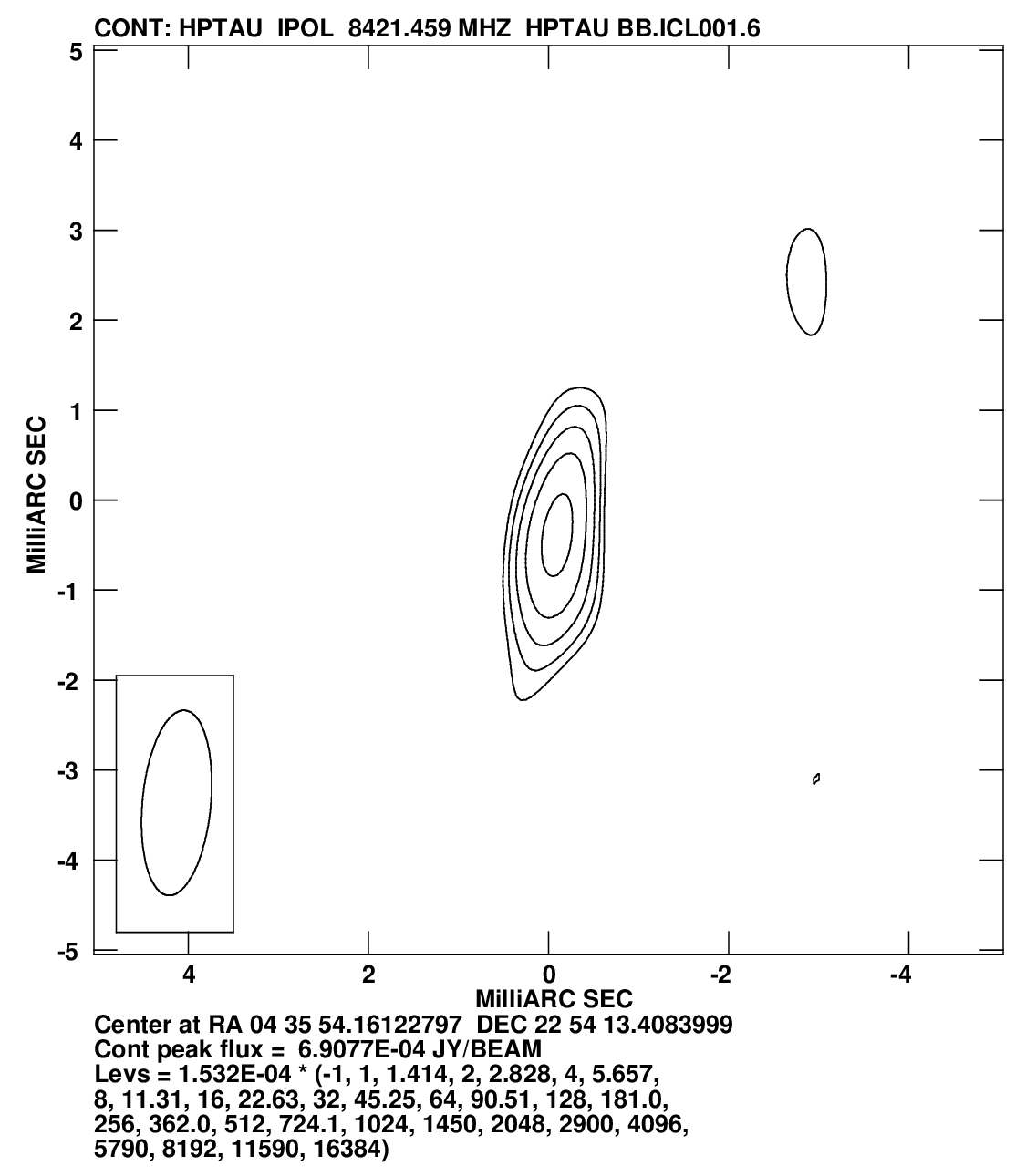} \\
\includegraphics[width=.45\textwidth,angle=0]{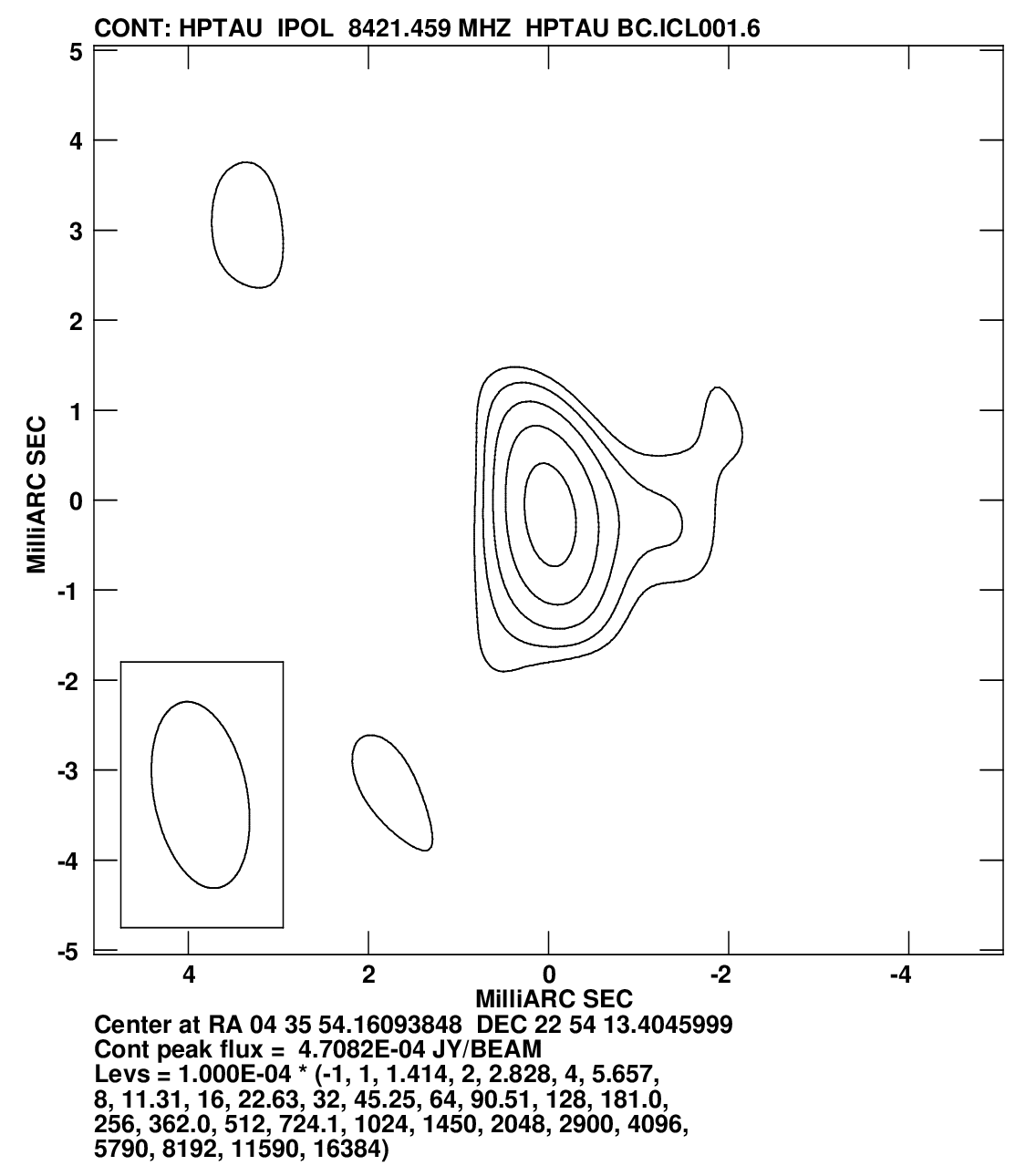} &
\includegraphics[width=.45\textwidth,angle=0]{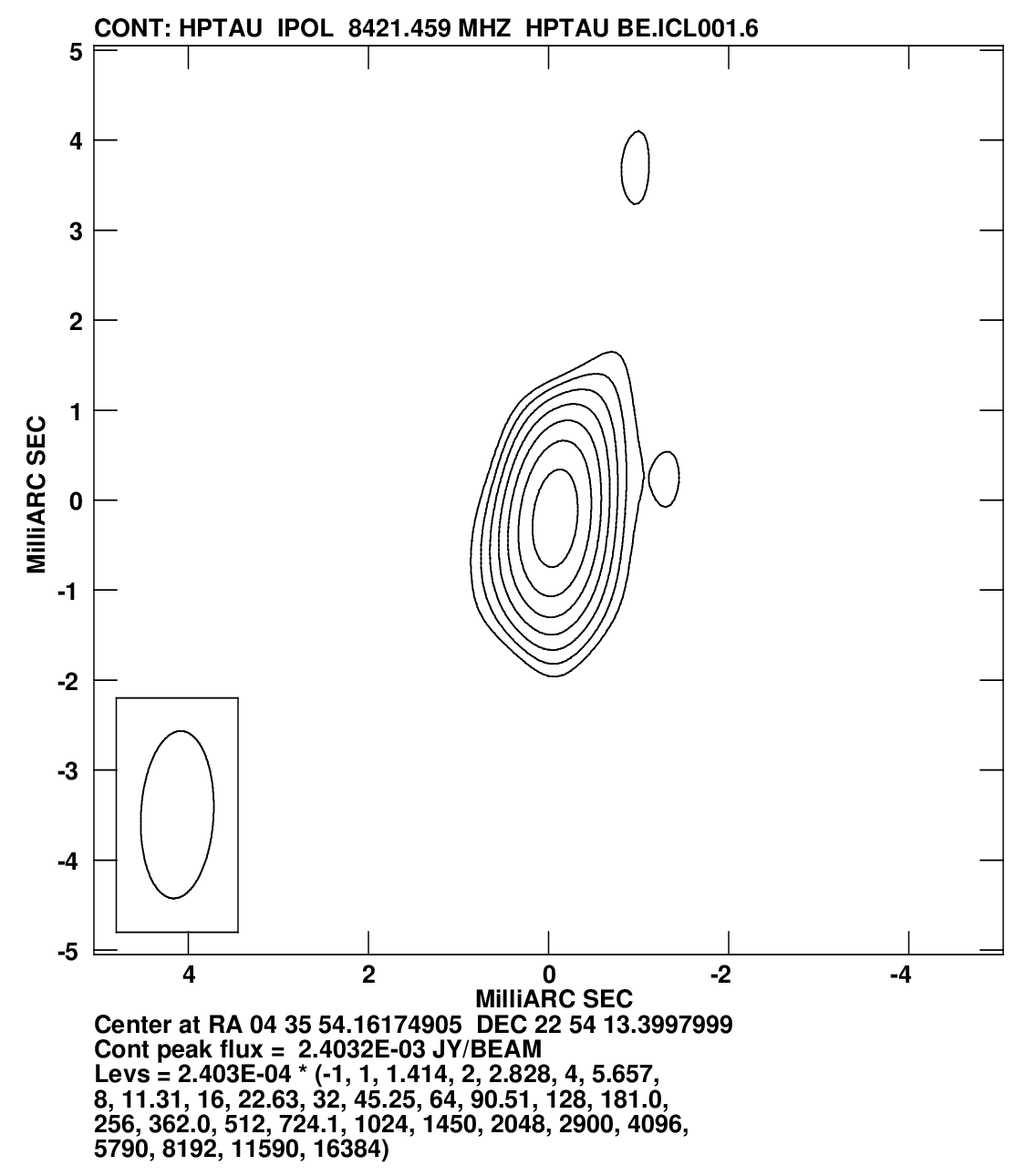} \\
\includegraphics[width=.45\textwidth,angle=0]{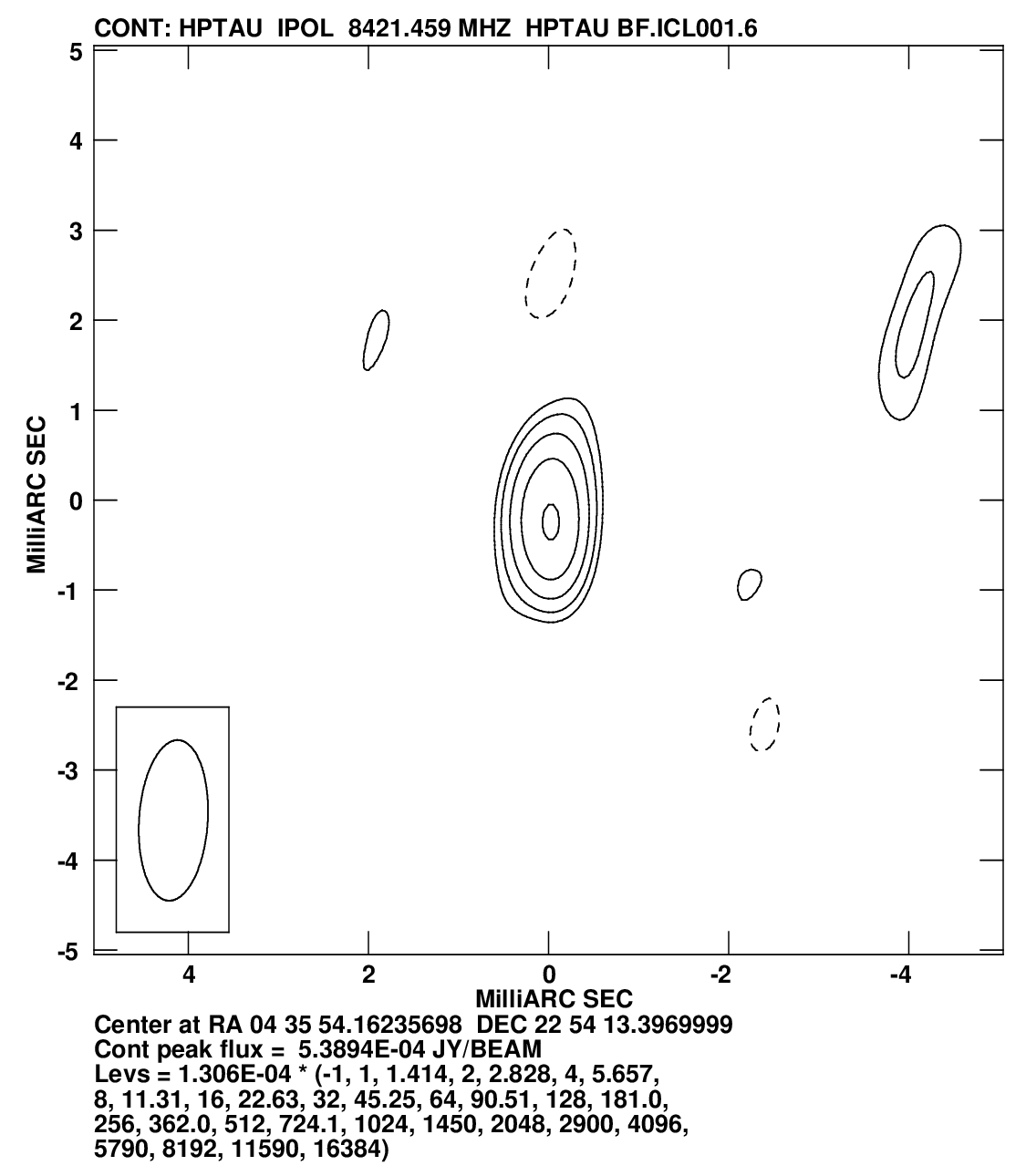} &
\includegraphics[width=.45\textwidth,angle=0]{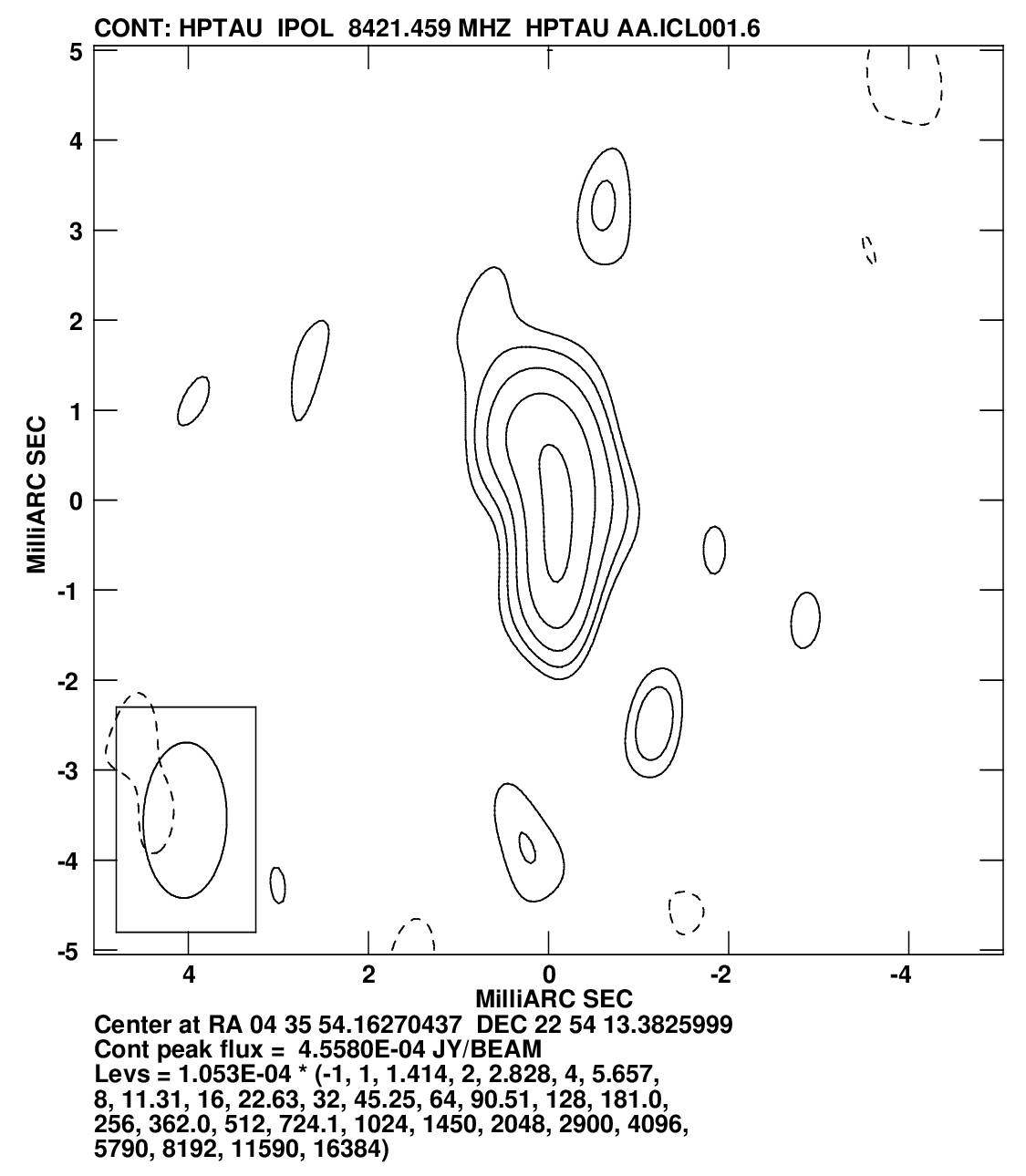}
\end{tabular}
\caption{\footnotesize{
VLBA images of HP~Tau/G2 at epochs 1, 2, 3, 5, 6 and 7 listed in Tab.\
\ref{tab-observaciones} and \ref{tab-resultados}.}
\label{map-hptau}}\end{figure*}

\begin{figure*}[!h]
\centering
\begin{tabular}{cc}
\includegraphics[width=.45\textwidth,angle=0]{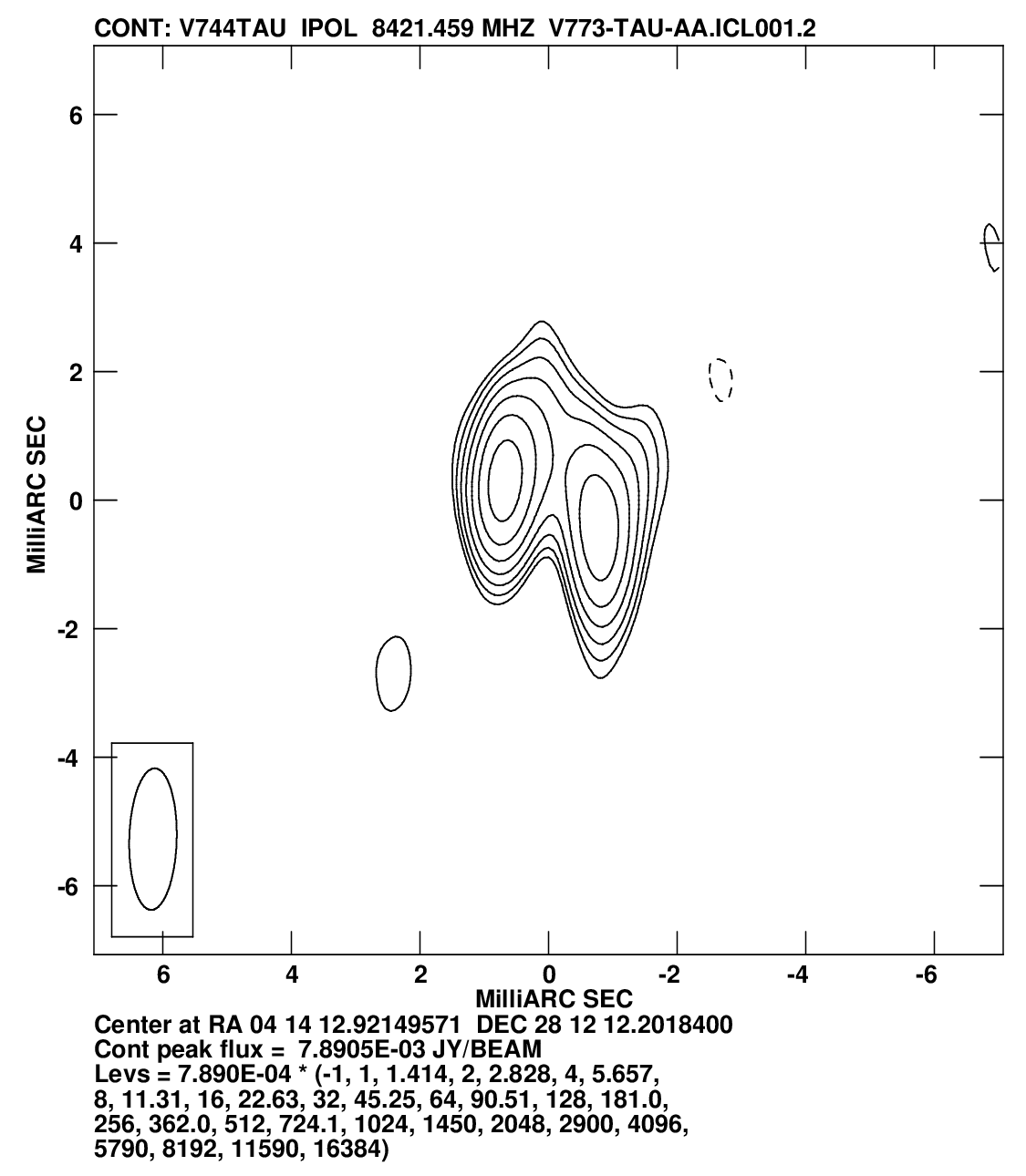} &
\includegraphics[width=.45\textwidth,angle=0]{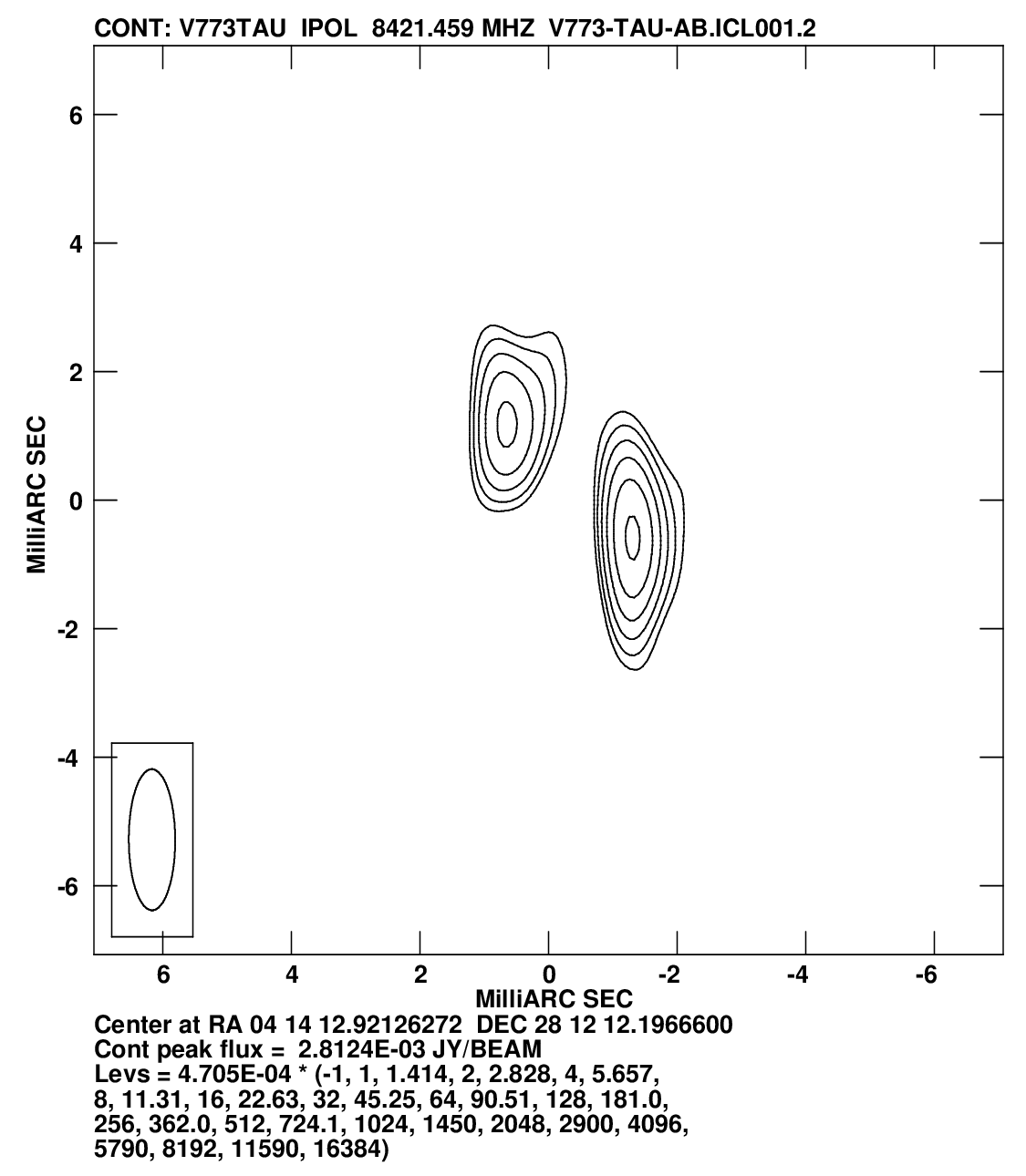} \\
\includegraphics[width=.45\textwidth,angle=0]{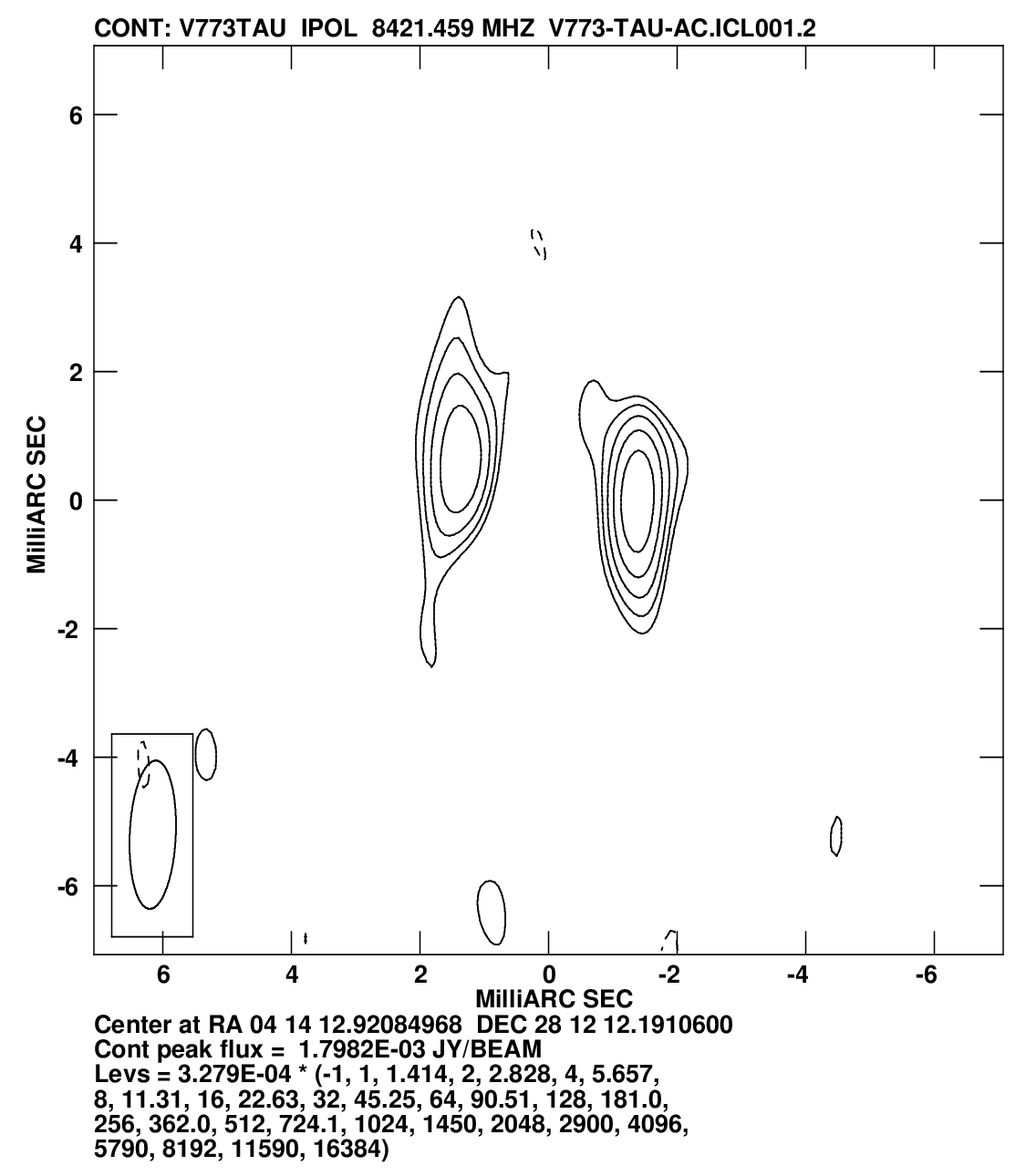} &
\includegraphics[width=.45\textwidth,angle=0]{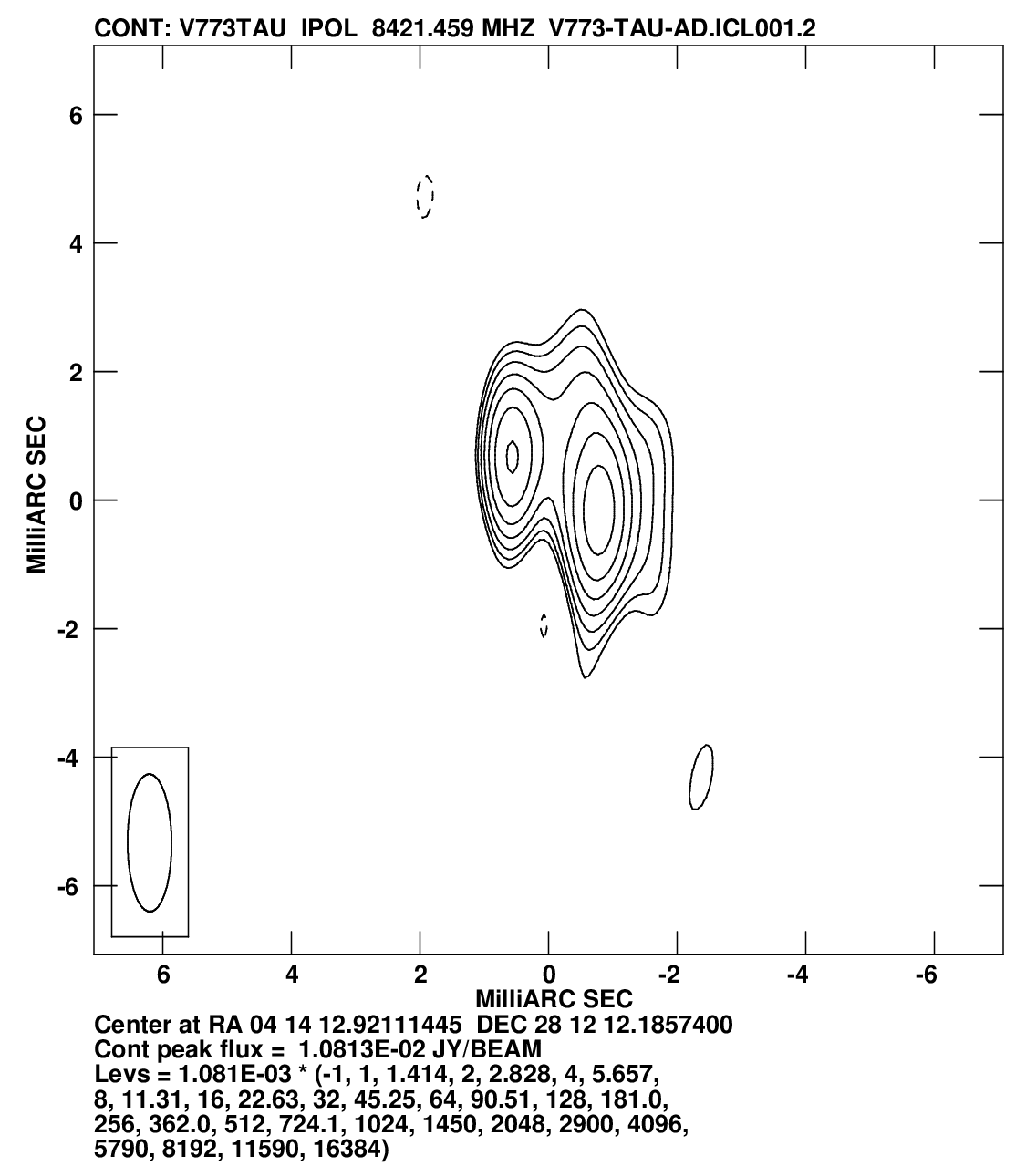} \\
\includegraphics[width=.45\textwidth,angle=0]{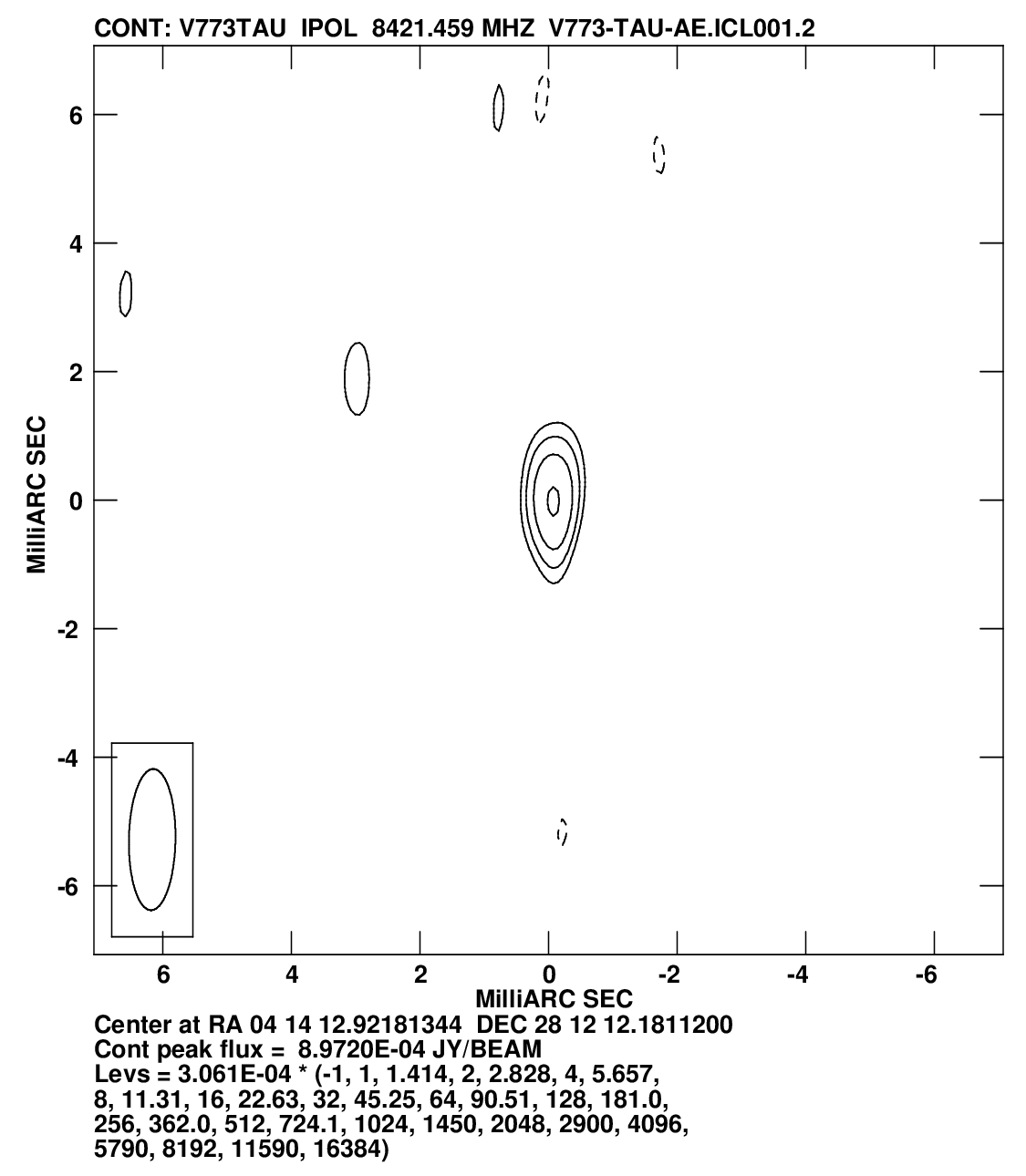} &
\includegraphics[width=.45\textwidth,angle=0]{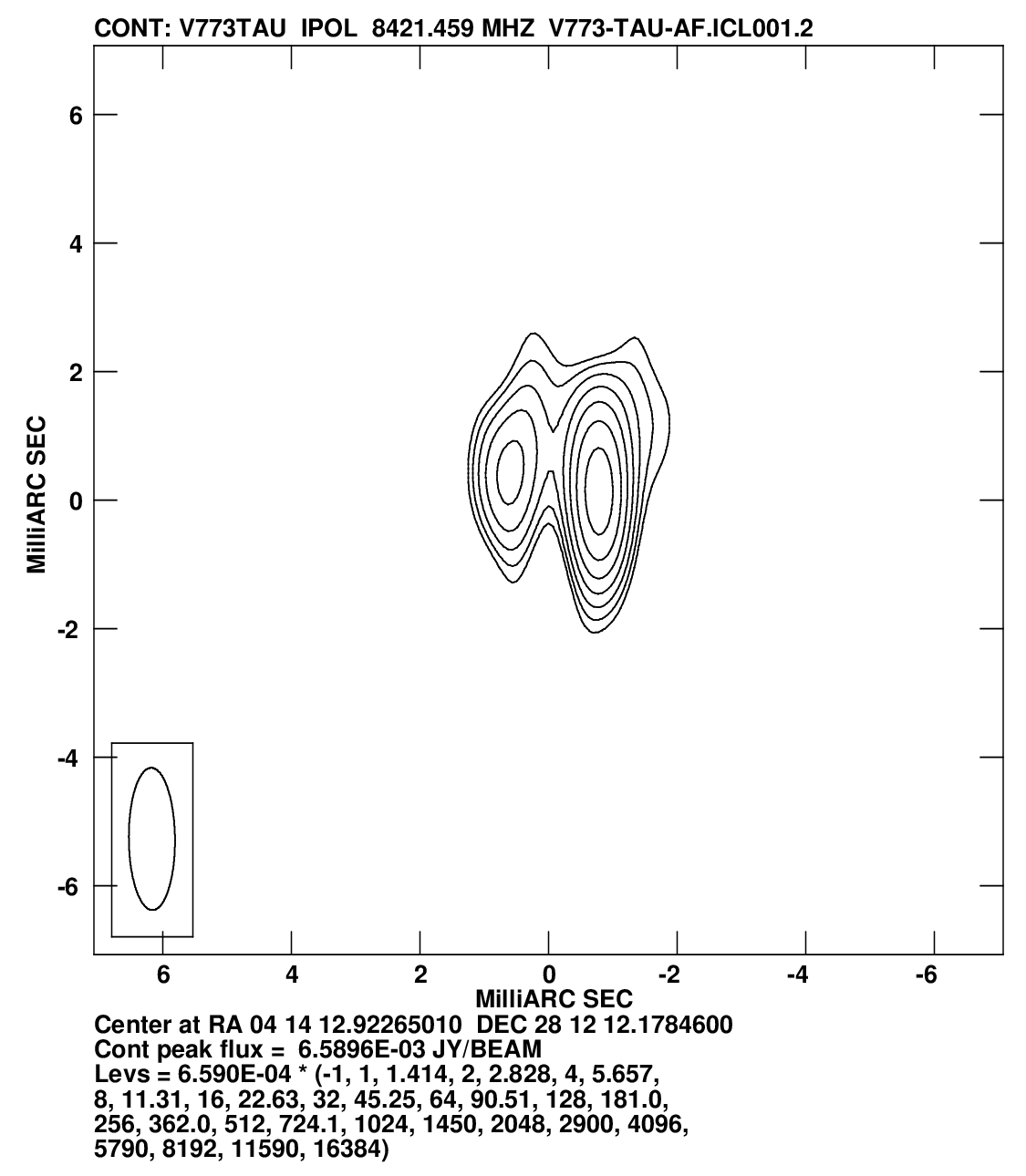}
\end{tabular}
\caption{\footnotesize{
VLBA images of V773~Tau~A at epochs 1 to 6 listed in Tab.\
\ref{tab-observaciones} and \ref{tab-resultados}.}
\label{map-v773tau1}}\end{figure*}

\begin{figure*}[!h]
\centering
\begin{tabular}{cc}
\includegraphics[width=.45\textwidth,angle=0]{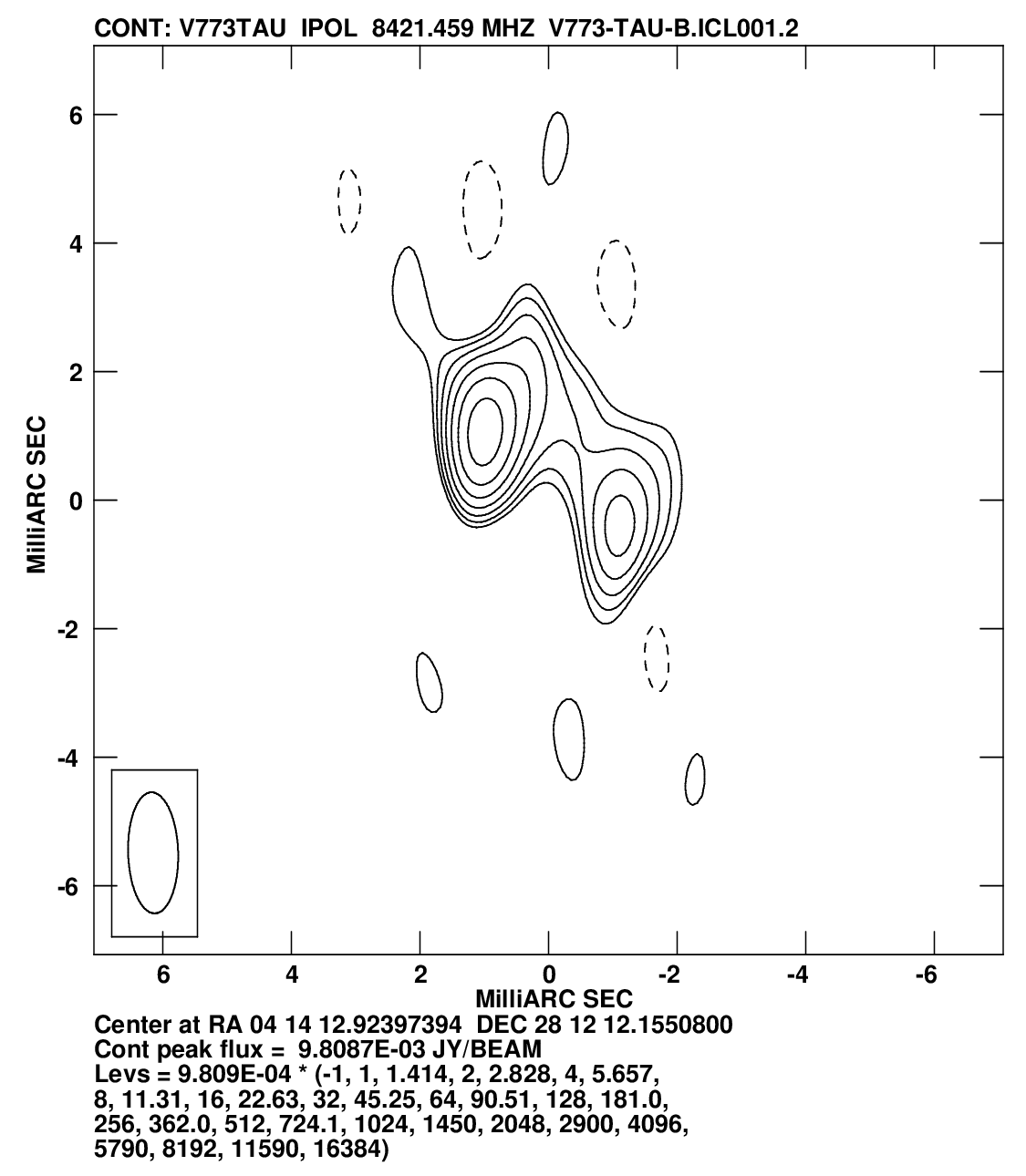} &
\includegraphics[width=.45\textwidth,angle=0]{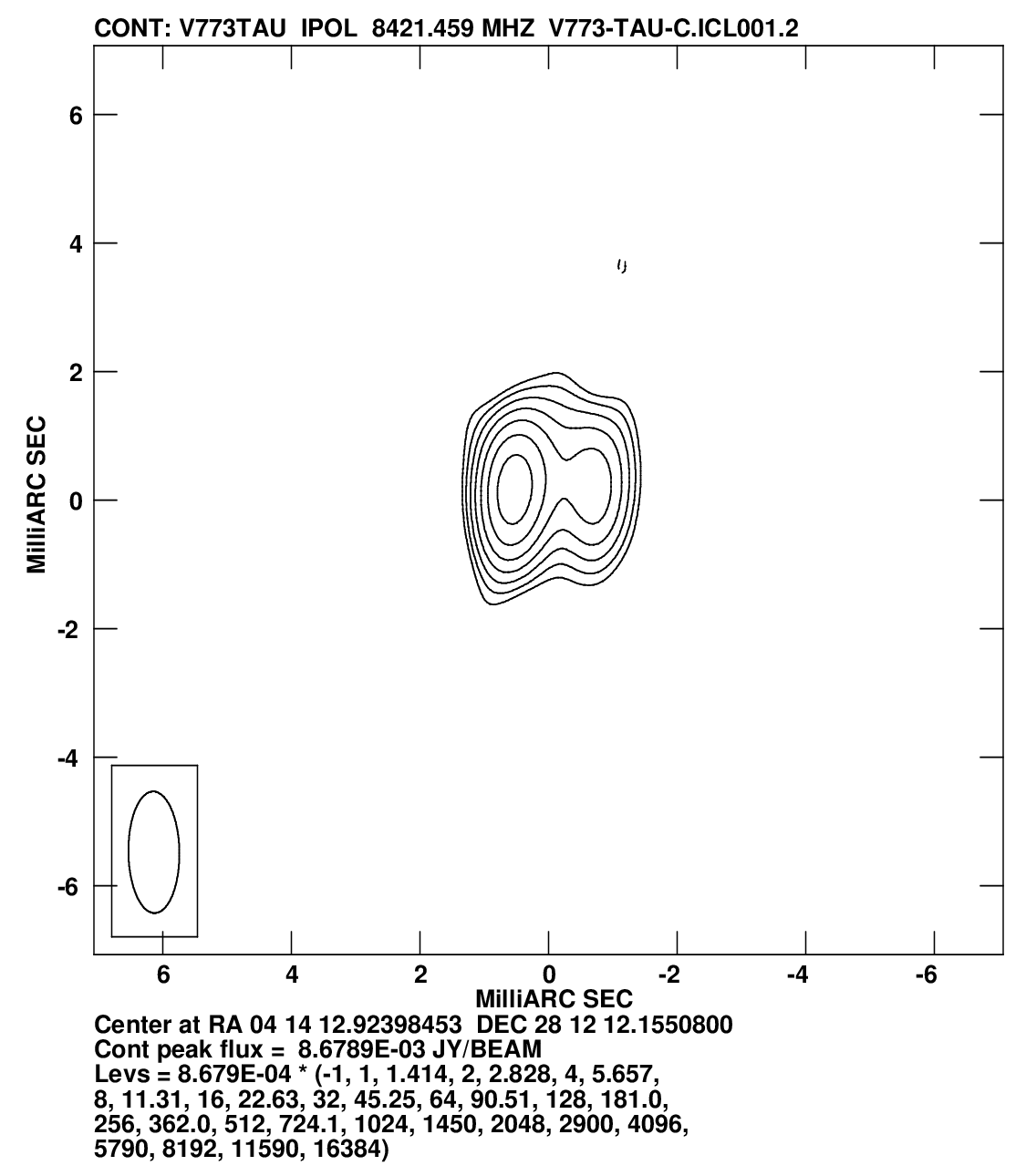} \\
\includegraphics[width=.45\textwidth,angle=0]{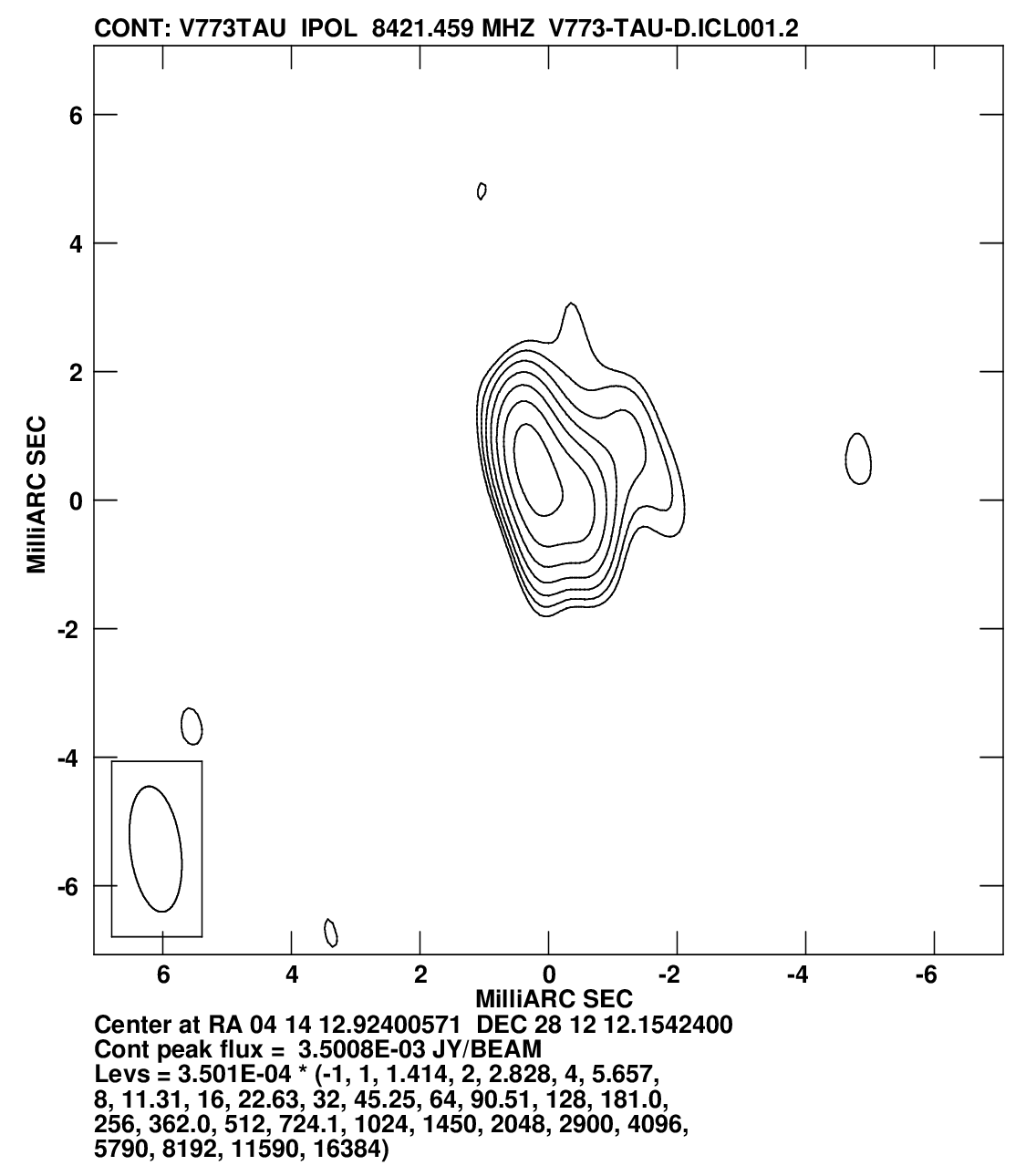} &
\includegraphics[width=.45\textwidth,angle=0]{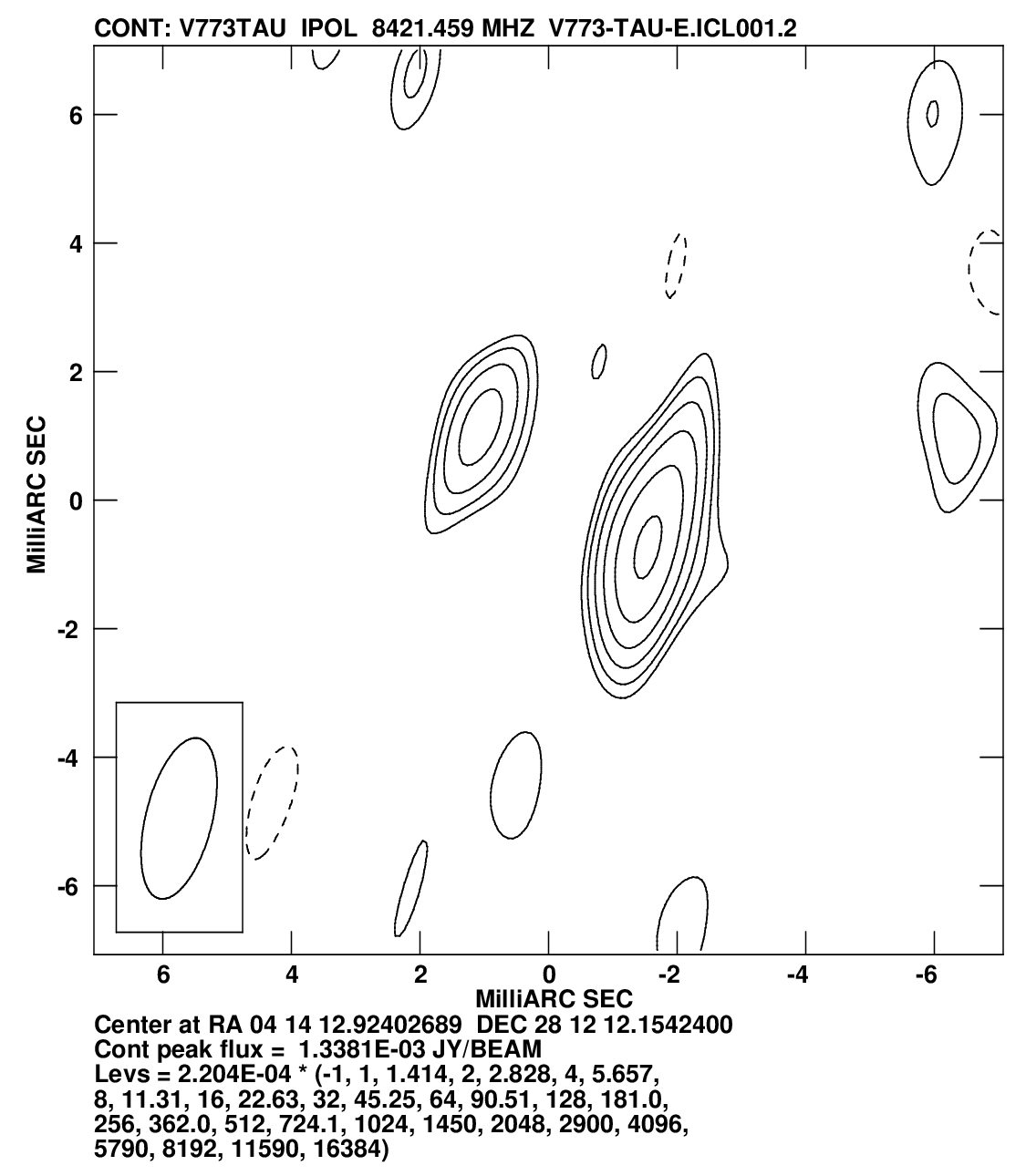} \\
\includegraphics[width=.45\textwidth,angle=0]{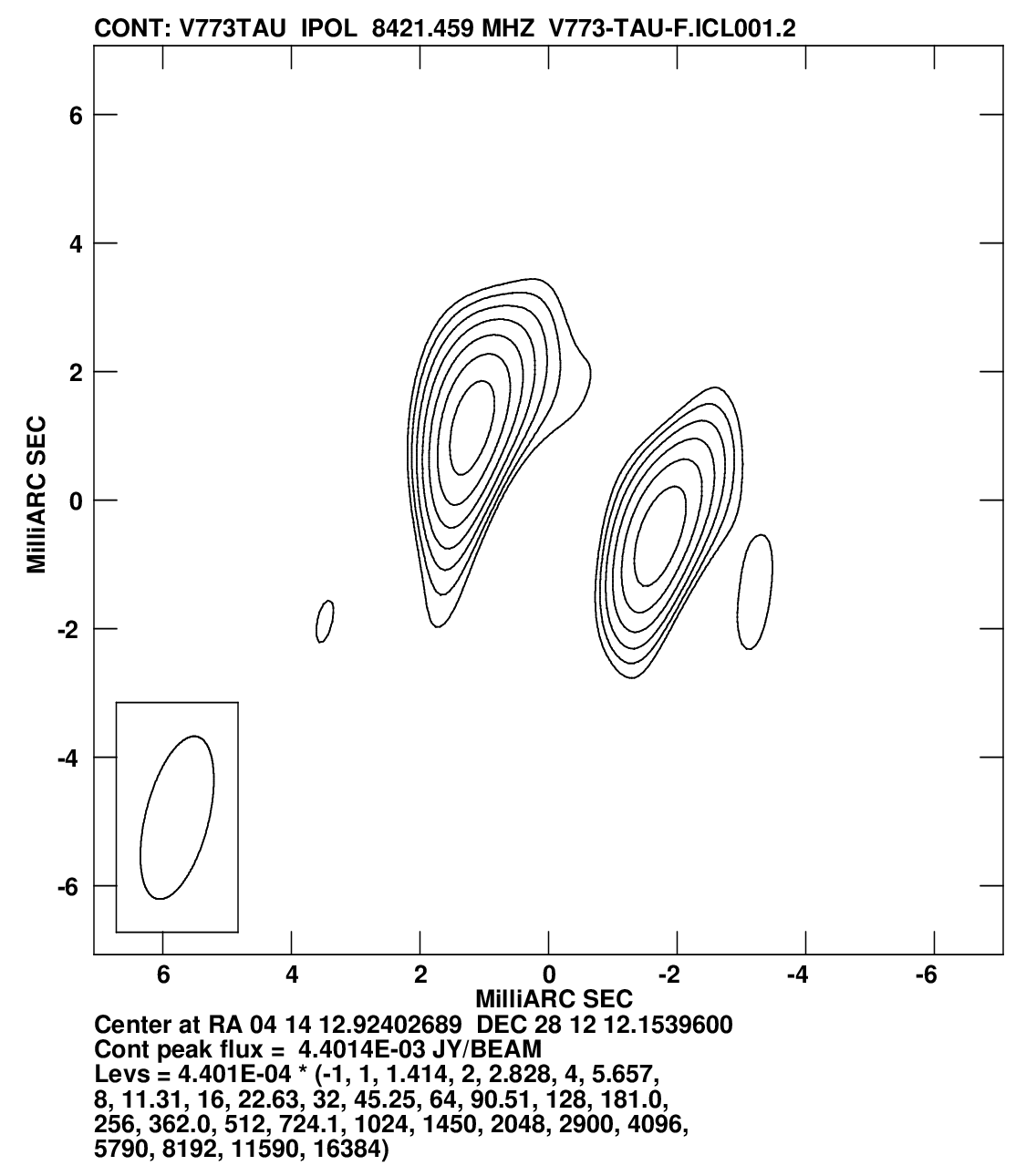} &
\includegraphics[width=.45\textwidth,angle=0]{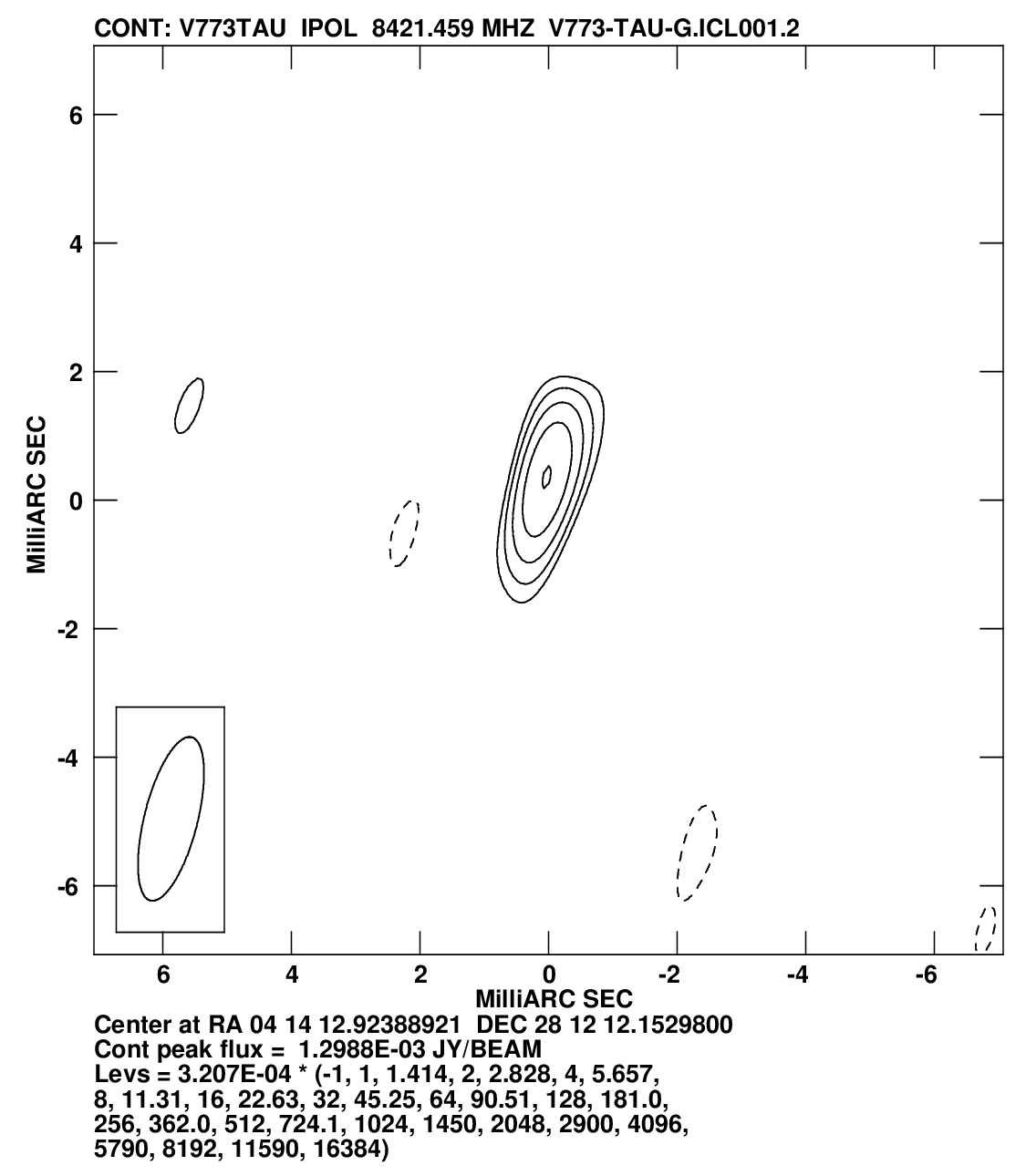}
\end{tabular}
\caption{\footnotesize{
VLBA images of V773~Tau~A at epochs 7 to 12 listed in Tab.\
\ref{tab-observaciones} and \ref{tab-resultados}.}
\label{map-v773tau2}}\end{figure*}

\begin{figure*}[!h]
\centering
\begin{tabular}{cc}
\includegraphics[width=.45\textwidth,angle=0]{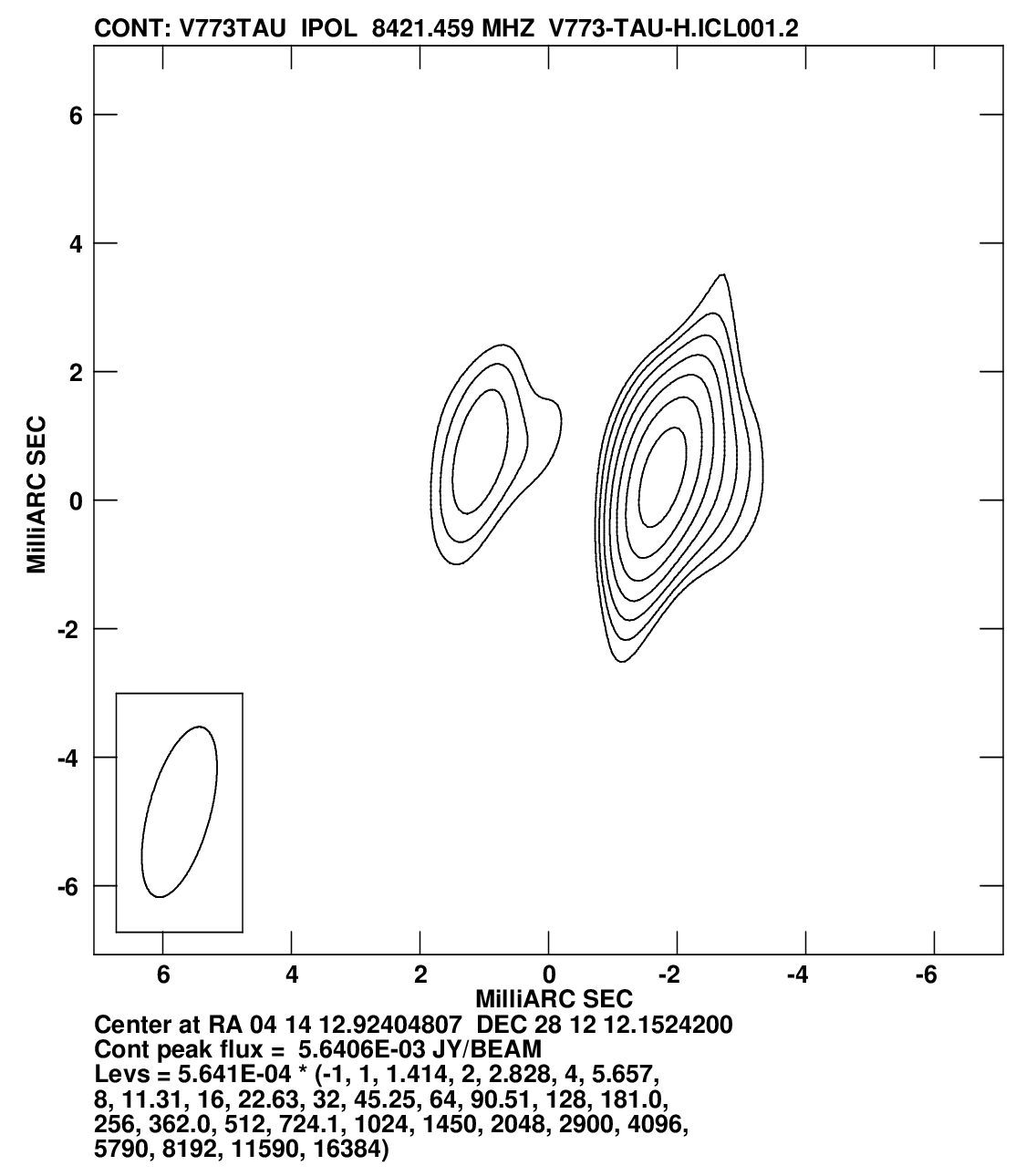} &
\includegraphics[width=.45\textwidth,angle=0]{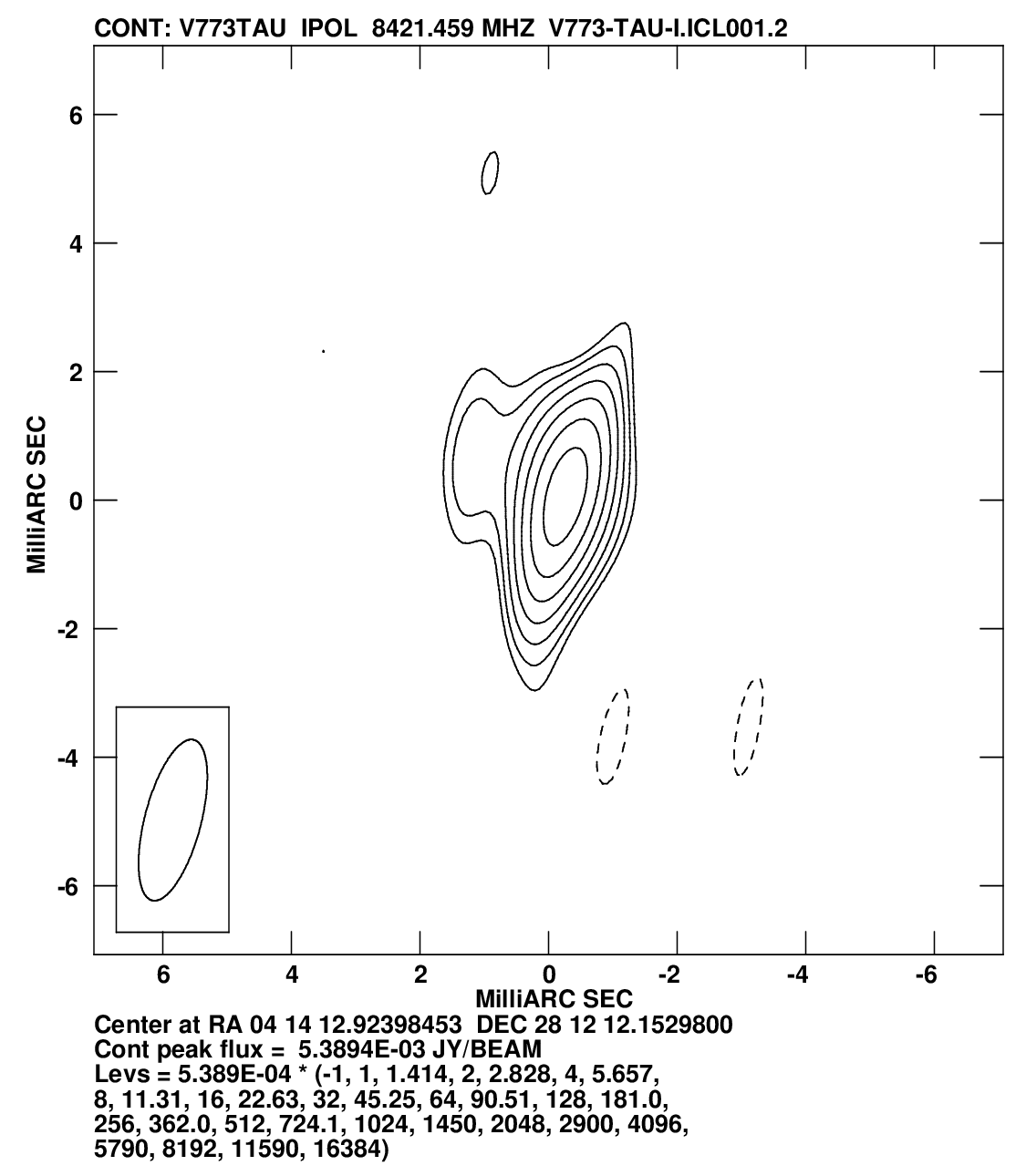} \\
\includegraphics[width=.45\textwidth,angle=0]{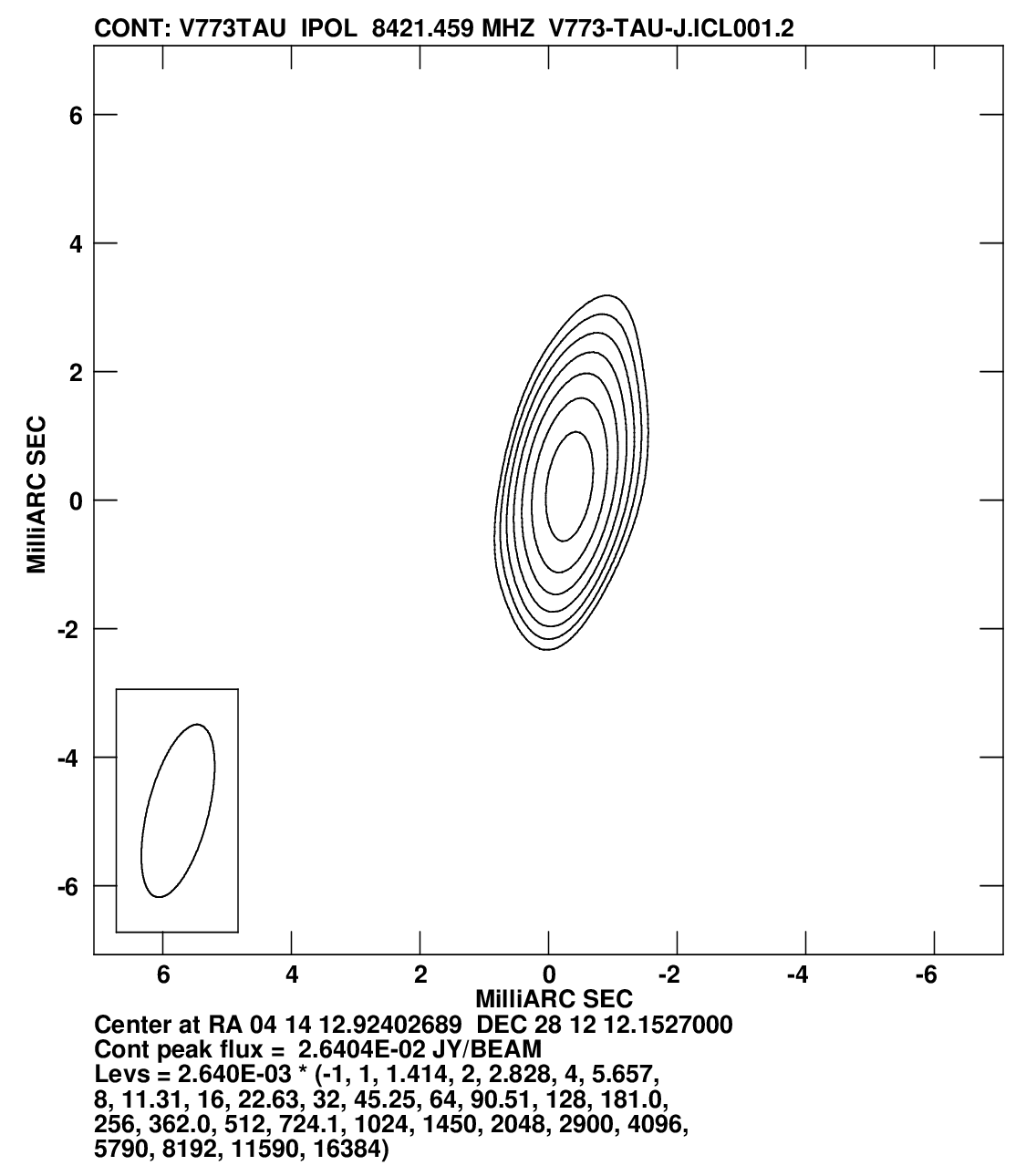} &
\includegraphics[width=.45\textwidth,angle=0]{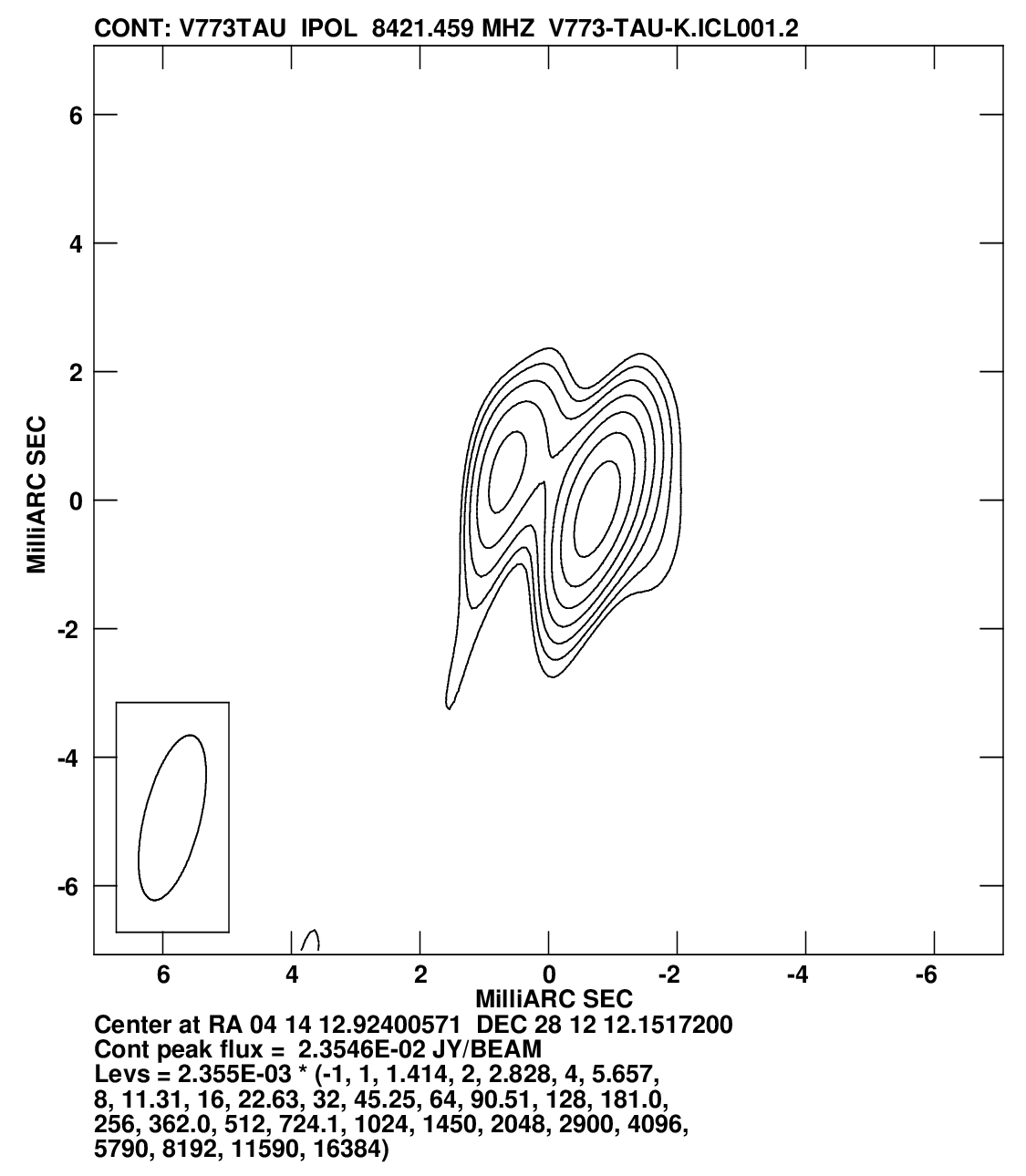} \\
\includegraphics[width=.45\textwidth,angle=0]{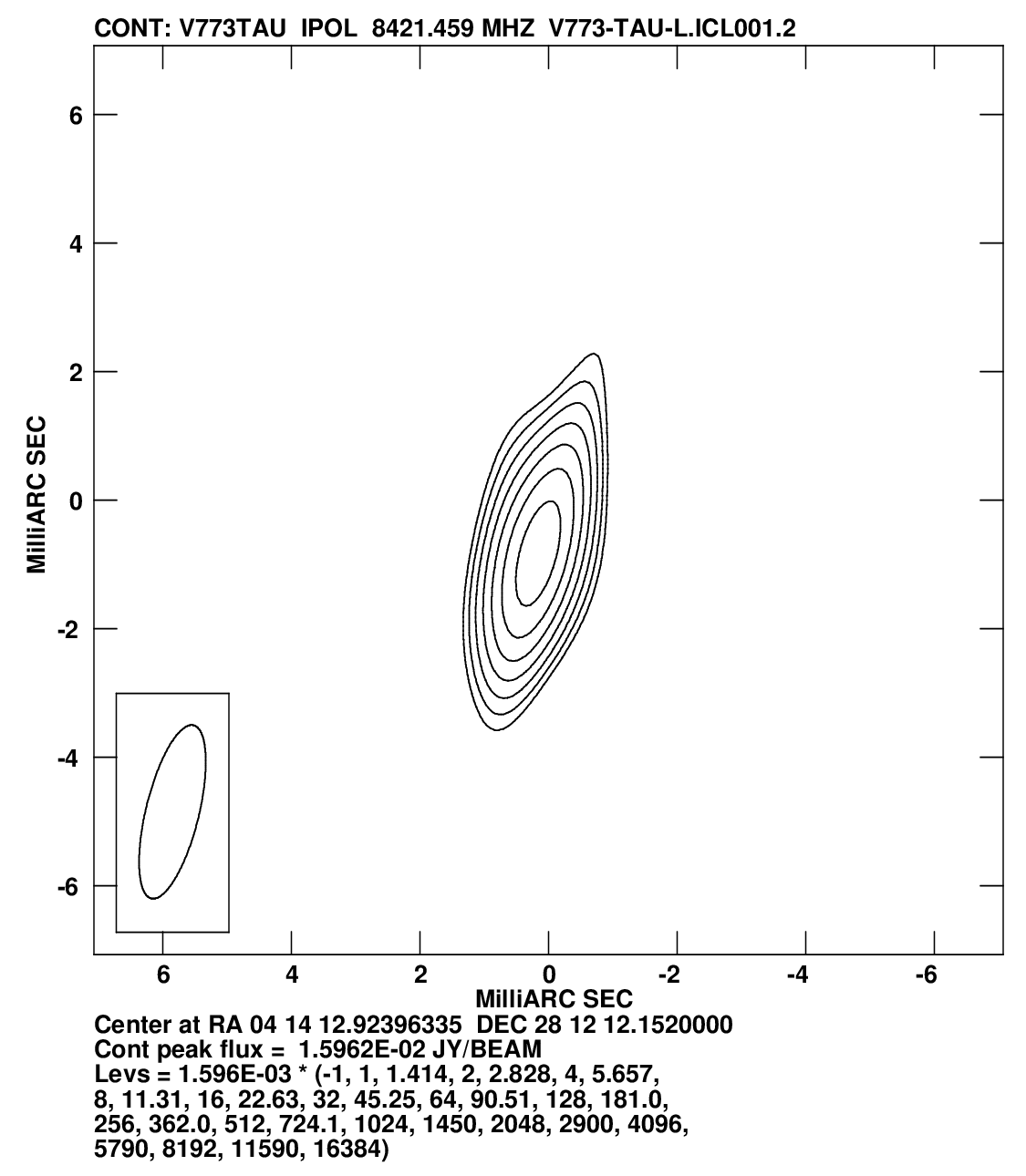} &
\includegraphics[width=.45\textwidth,angle=0]{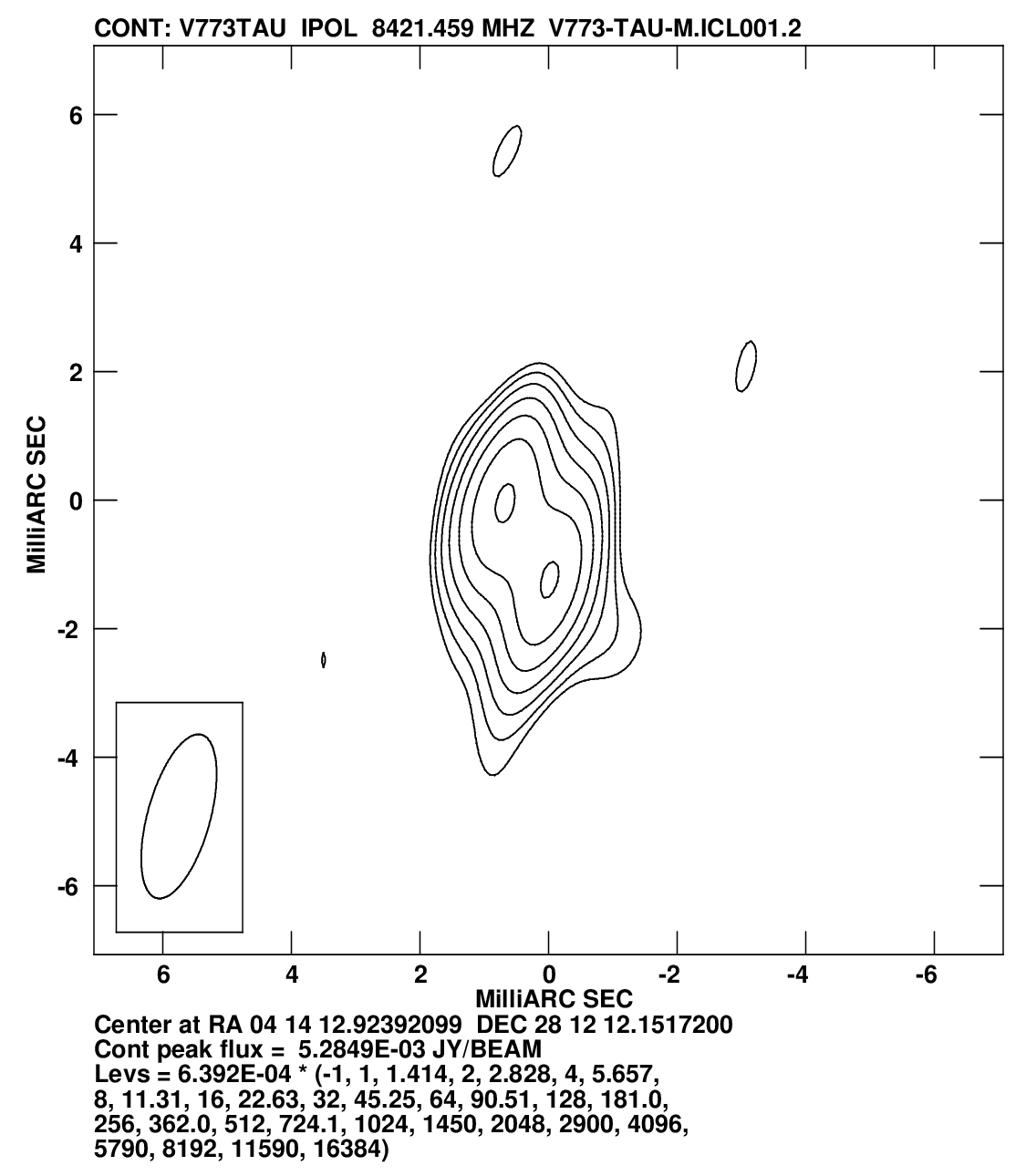}
\end{tabular}
\caption{\footnotesize{
VLBA images of V773~Tau~A at epochs 13 to 18 listed in Tab.\
\ref{tab-observaciones} and \ref{tab-resultados}.}
\label{map-v773tau3}}\end{figure*}

\begin{figure*}[!h]
\centering
\begin{tabular}{cc}
\includegraphics[width=.45\textwidth,angle=0]{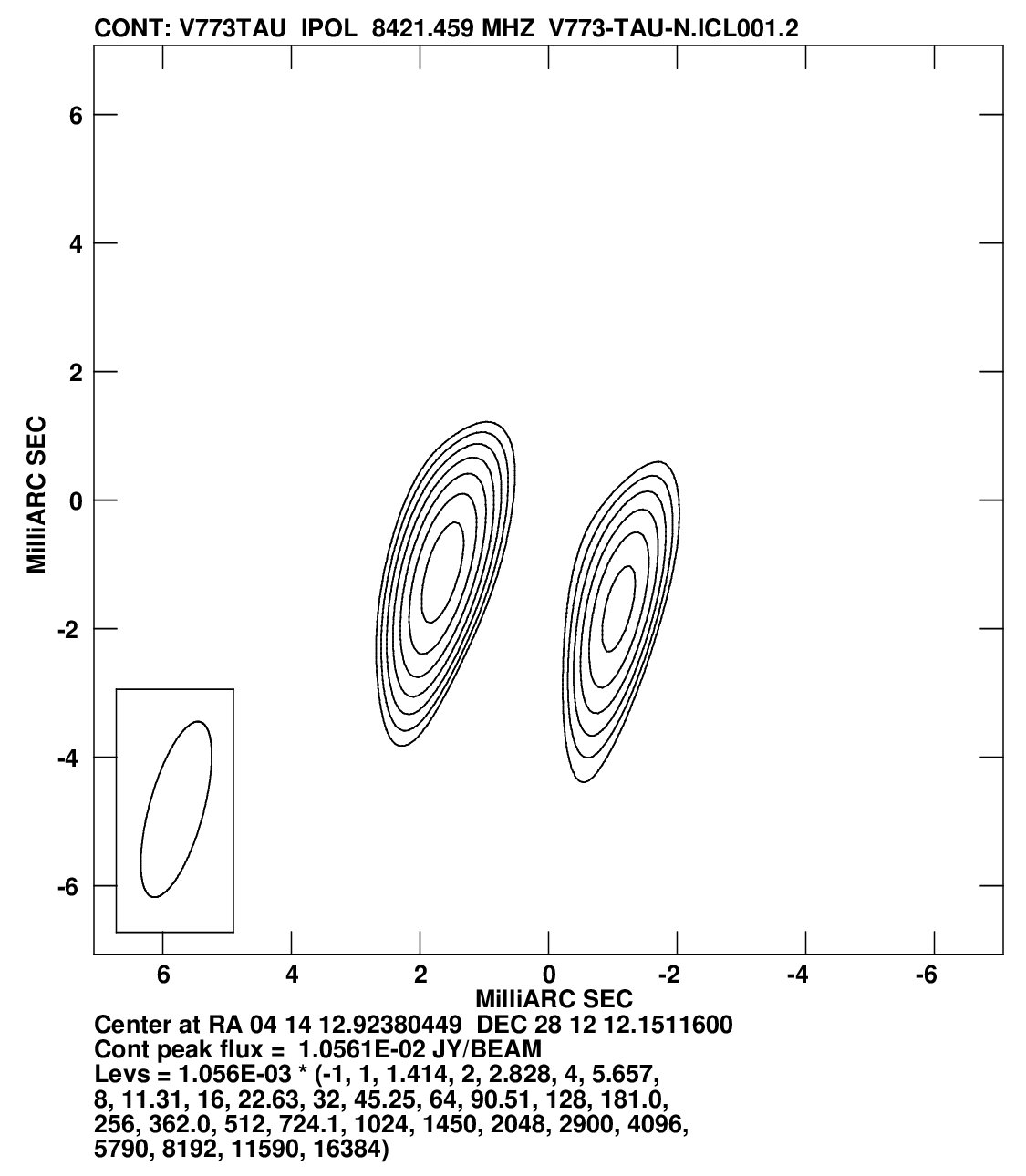} \\
\includegraphics[width=.45\textwidth,angle=0]{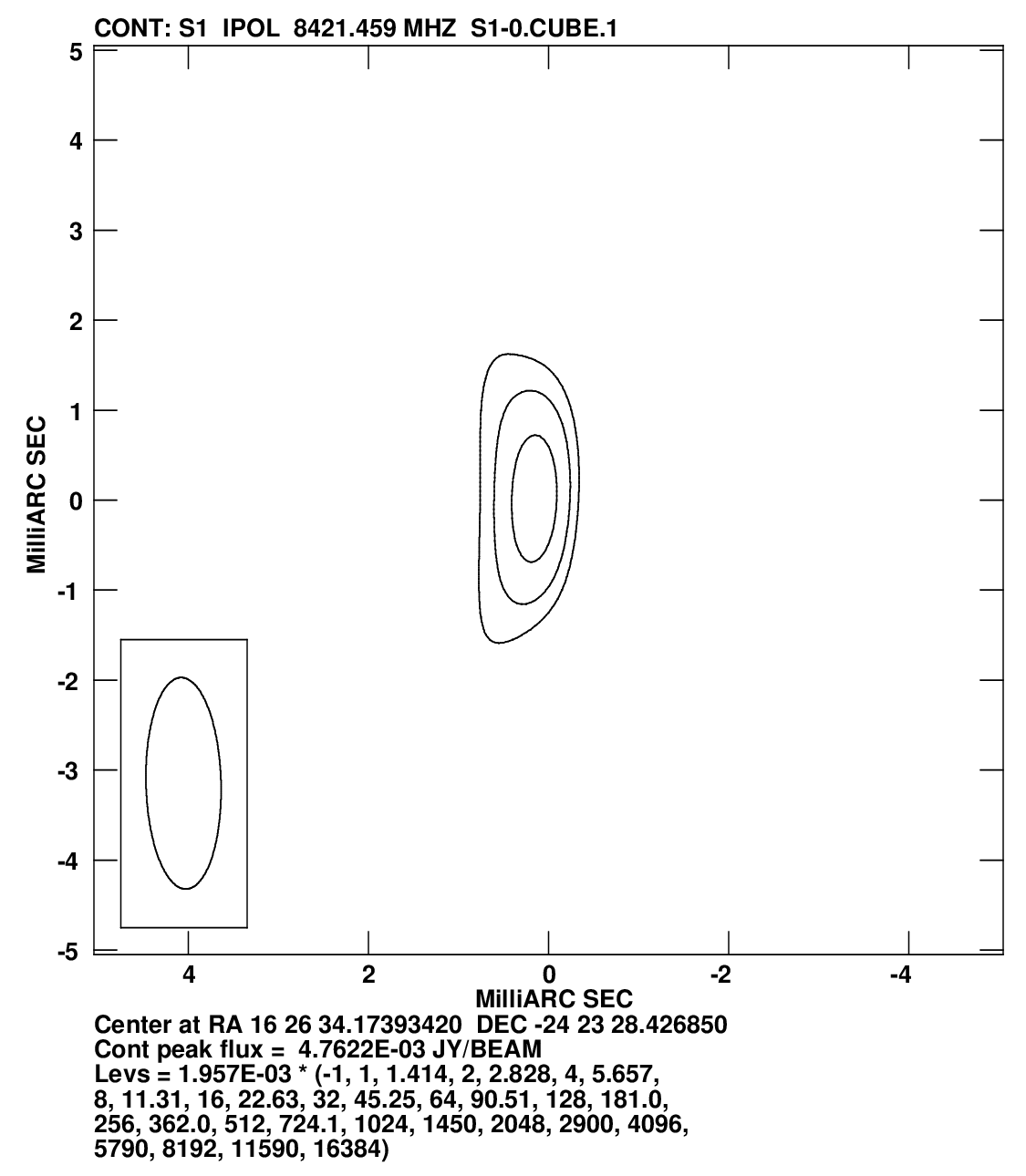} &
\includegraphics[width=.45\textwidth,angle=0]{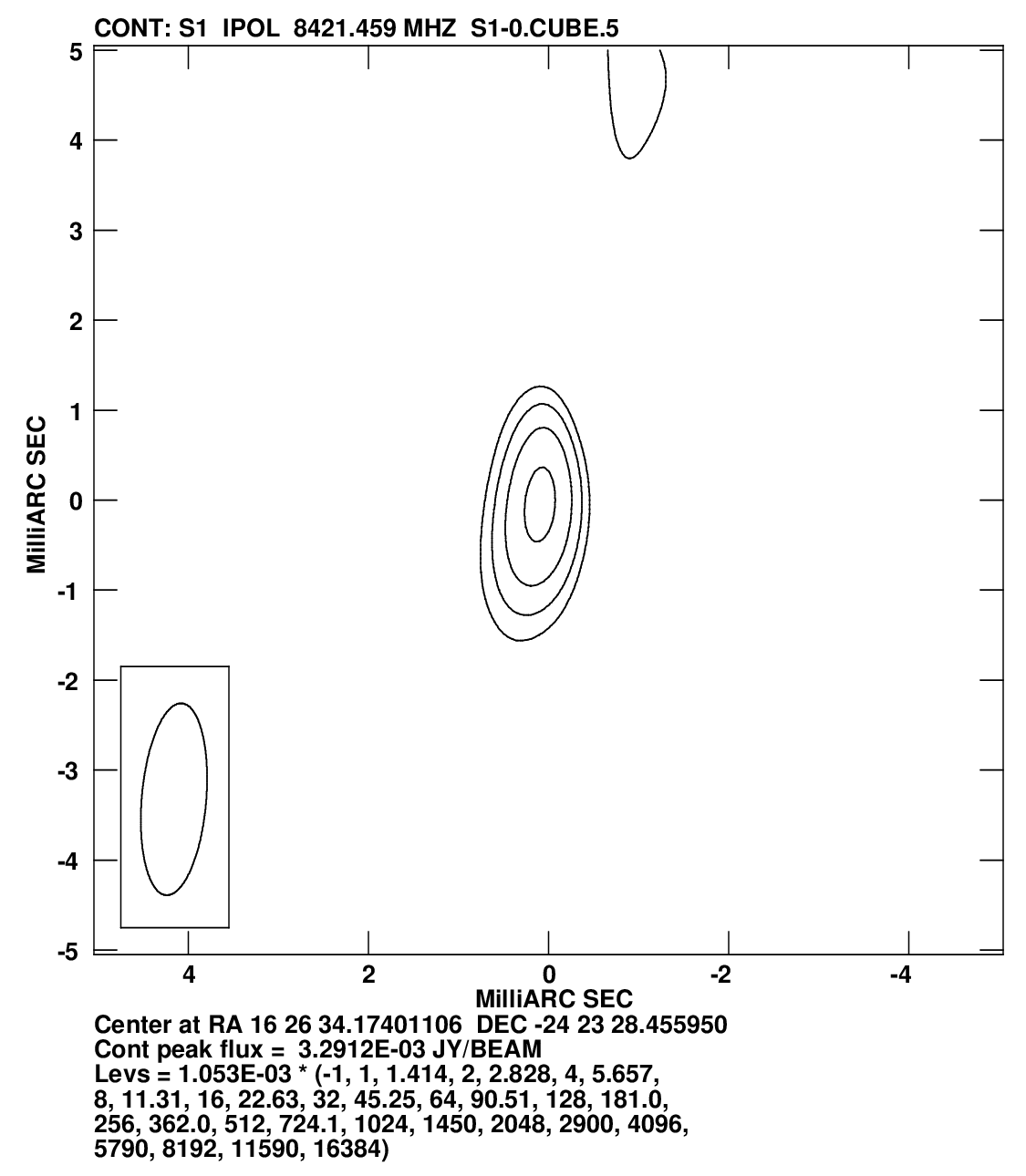} \\
\includegraphics[width=.45\textwidth,angle=0]{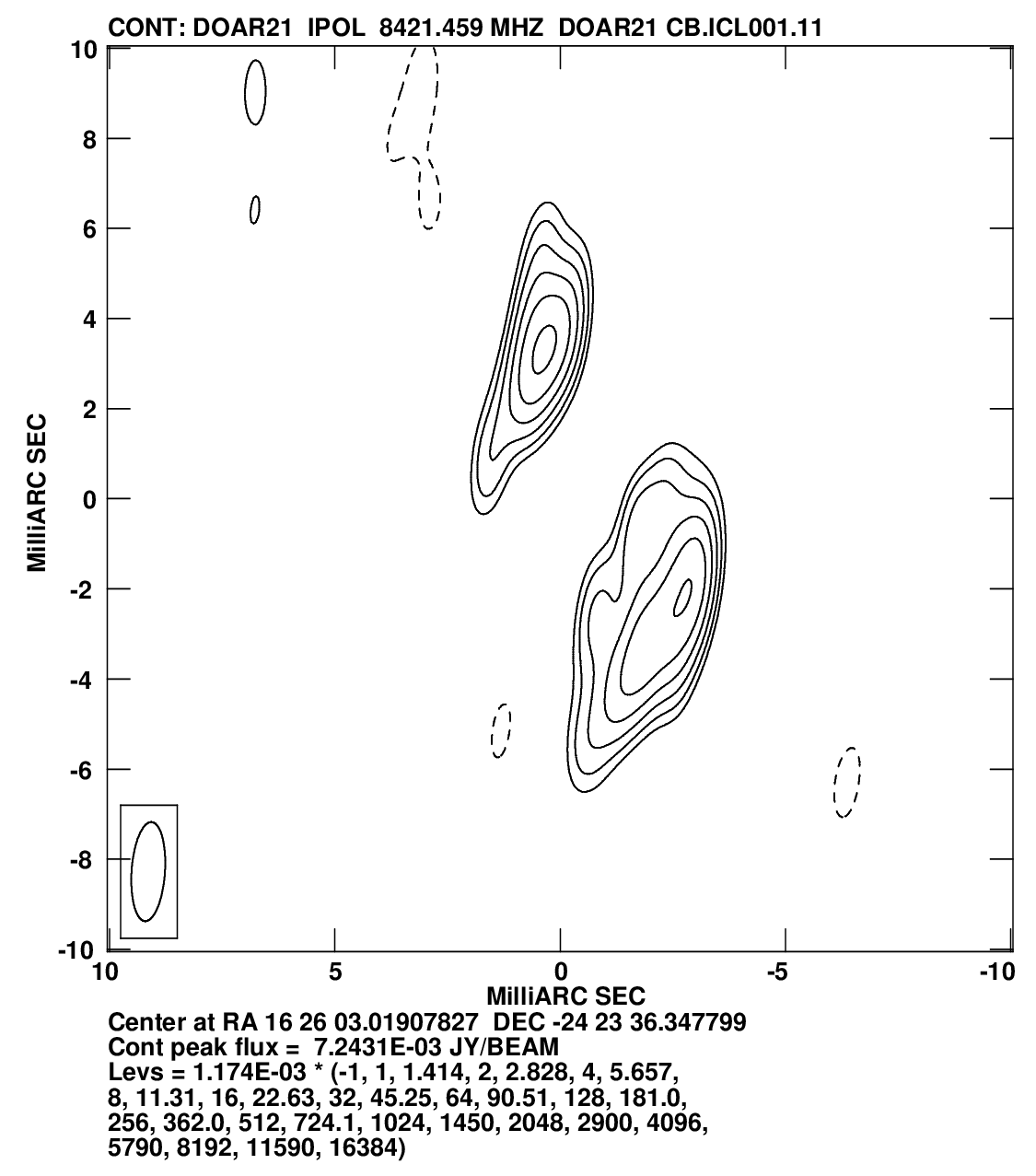} &
\includegraphics[width=.45\textwidth,angle=0]{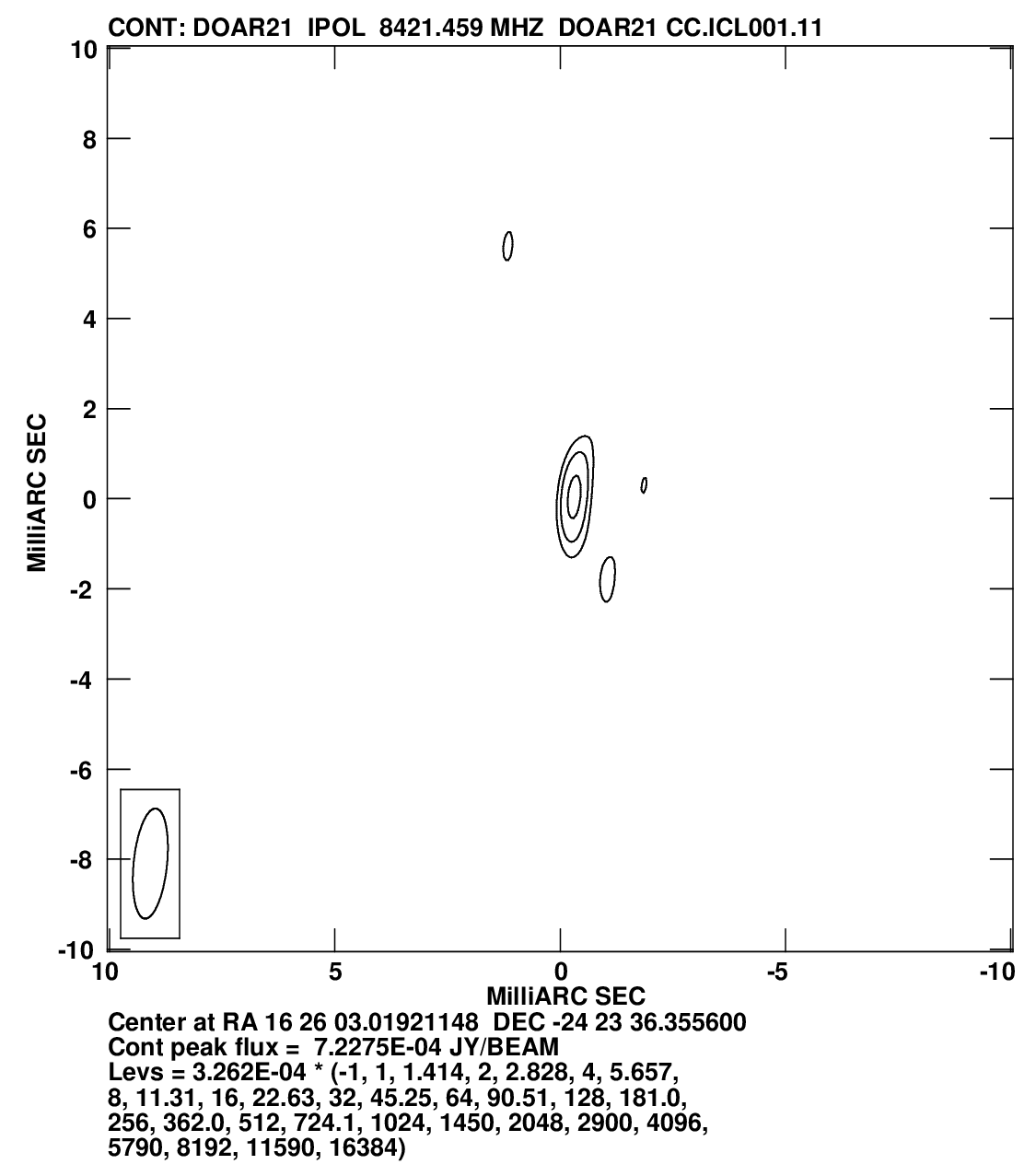}
\end{tabular}
\caption{\footnotesize{
VLBA images of V773~Tau~A (epoch 19), S1 (epochs 1 and 5), and DoAr~21
(epochs 2 and 4) listed in Tab.\ \ref{tab-observaciones} and
\ref{tab-resultados}.}
\label{map-varios}}\end{figure*}

\noindent
Once calibrated, the visibilities corresponding to each source at each
epoch, were imaged with a pixel size of 50 $\mu$as after weights
intermediate between natural and uniform (\textsf{ROBUST} $=$ 0 in
\textsf{IMAGR}) were applied. 67 continuum images were obtained in the
thesis, but here we are showing only 53 contour maps corresponding to:
twelve epochs of T~Tau~Sb (Figs.\ \ref{map-ttau1} and \ref{map-ttau2}),
six of Hubble~4 (Fig.\ \ref{map-hubble4}), six of HDE~283572 (Fig.\
\ref{map-hde}), six of HP~Tau/G2 (Fig.\ \ref{map-hptau}), all 19
epochs of V773~Tau~A system (Figs.\ \ref{map-v773tau1},
\ref{map-v773tau2}, \ref{map-v773tau3} and first panel of Fig.\
\ref{map-varios}), two of S1 (panel two and three of Fig.\
\ref{map-varios}), and two of DoAr~21 (panel four and five of Fig.\
\ref{map-varios}).

\medskip

\noindent
From images, we obtained all parameters needed (see Tab.\
\ref{tab-resultados}) to fit for the parallax and proper motions of
the sources, as well as for relative motions in multiple systems of
our sample. The source absolute positions at each epoch (Cols.\ [3]
and [4] in Tab.\ \ref{tab-resultados}) were determined using a 2D
Gaussian fitting procedure (task \textsf{JMFIT} in \aips). This task
provides an estimate of the position error based on the expected
theoretical astrometric precision of an interferometer (Condon
1997). Tab.\ \ref{tab-resultados} also contains the epoch in Col.\
[1], Julian Day in Col.\ [2], flux densities in Col.\ [5], and the
corresponding brightness temperature in Col.\ [6].

\begin{deluxetable}{clllr@{.}lr@{.}l}
\tabletypesize{\scriptsize}
\tablecolumns{8}
\tablewidth{0pc}
\tablecaption{\footnotesize{Measured source positions and fluxes.}
\label{tab-resultados}}
\tablehead{
\colhead{Epoch}                       &
\colhead{Mean Date}                   &
\colhead{$\alpha$ (J2000.0)}          &
\colhead{$\delta$ (J2000.0)}          &
\multicolumn{2}{c}{$f_\nu$}           &
\multicolumn{2}{c}{$T_{\rm b}$}       \\
\multicolumn{1}{c}{}                  &
\multicolumn{1}{c}{[JD]}              &
\multicolumn{1}{c}{}                  &
\multicolumn{1}{c}{}                  &
\multicolumn{2}{c}{[mJy]}             &
\multicolumn{2}{c}{[10$^6$ K]}        }
\startdata
\multicolumn{2}{l}{T~Tau~Sb}&
\multicolumn{1}{c}{$04^{h}21^{m}$}&\multicolumn{1}{c}{$19^{\circ}32^{'}$}\\\\[-0.2cm]
1  & 2452906.981522 & \mmsec{59}{4252942}$\pm$\mmsec{0}{000001} & \msec{05}{71762}$\pm$\msec{0}{00004} & 1&62$\pm$0.14 & 19&62$\pm$1.70  \\
2  & 2452961.834705 & \mmsec{59}{4249805}$\pm$\mmsec{0}{000002} & \msec{05}{71655}$\pm$\msec{0}{00004} & 1&74$\pm$0.14 & 20&06$\pm$1.61  \\
3  & 2453019.672980 & \mmsec{59}{4245823}$\pm$\mmsec{0}{000004} & \msec{05}{71532}$\pm$\msec{0}{00011} & 0&92$\pm$0.15 &  6&61$\pm$1.08  \\
4  & 2453091.476395 & \mmsec{59}{4245420}$\pm$\mmsec{0}{000002} & \msec{05}{71533}$\pm$\msec{0}{00005} & 1&27$\pm$0.14 & 13&24$\pm$1.46  \\
5  & 2453139.345324 & \mmsec{59}{4248818}$\pm$\mmsec{0}{000002} & \msec{05}{71603}$\pm$\msec{0}{00006} & 1&90$\pm$0.21 & 24&36$\pm$2.69  \\
6  & 2453195.192419 & \mmsec{59}{4253464}$\pm$\mmsec{0}{000002} & \msec{05}{71665}$\pm$\msec{0}{00006} & 1&25$\pm$0.13 & 14&08$\pm$1.46  \\
7  & 2453264.999583 & \mmsec{59}{4255476}$\pm$\mmsec{0}{000002} & \msec{05}{71660}$\pm$\msec{0}{00004} & 1&61$\pm$0.14 & 17&09$\pm$1.49  \\
8  & 2453318.852141 & \mmsec{59}{4252999}$\pm$\mmsec{0}{000002} & \msec{05}{71563}$\pm$\msec{0}{00004} & 3&36$\pm$0.21 & 28&89$\pm$1.81  \\
9  & 2453367.718351 & \mmsec{59}{4249488}$\pm$\mmsec{0}{000002} & \msec{05}{71434}$\pm$\msec{0}{00005} & 1&26$\pm$0.14 & 14&49$\pm$1.61  \\
10 & 2453425.559664 & \mmsec{59}{4247667}$\pm$\mmsec{0}{000002} & \msec{05}{71383}$\pm$\msec{0}{00004} & 2&30$\pm$0.16 & 26&38$\pm$1.83  \\
11 & 2453500.355214 & \mmsec{59}{4251475}$\pm$\mmsec{0}{000006} & \msec{05}{71485}$\pm$\msec{0}{00018} & 1&12$\pm$0.21 & 11&17$\pm$2.09  \\
12 & 2453560.191348 & \mmsec{59}{4256679}$\pm$\mmsec{0}{000002} & \msec{05}{71560}$\pm$\msec{0}{00006} & 1&41$\pm$0.15 & 14&86$\pm$1.58  \\
13 & 2454616.298136 & \mmsec{59}{4268831}$\pm$\mmsec{0}{000011} & \msec{05}{70677}$\pm$\msec{0}{00018} & 1&64$\pm$0.24 &  5&80$\pm$0.85  \\
14 & 2451527.753322 & \mmsec{59}{4259349}$\pm$\mmsec{0}{000001} & \msec{05}{73136}$\pm$\msec{0}{00004} & 2&11$\pm$0.15 & 30&85$\pm$2.19  \\
\\[-0.2cm]\hline\\[-0.2cm]\multicolumn{2}{l}{Hubble~4}&
\multicolumn{1}{c}{$04^{h}18^{m}$}&\multicolumn{1}{c}{$28^{\circ}20^{'}$}\\\\[-0.2cm]
1  & 2453267.991464 & \mmsec{47}{0327419}$\pm$\mmsec{0}{000002} & \msec{07}{39898}$\pm$\msec{0}{00005} & 0&67$\pm$0.11 &  8&60$\pm$1.41 \\
2  & 2453374.699299 & \mmsec{47}{0319609}$\pm$\mmsec{0}{000002} & \msec{07}{38901}$\pm$\msec{0}{00007} & 0&76$\pm$0.15 & 10&72$\pm$2.12 \\
3  & 2453455.478107 & \mmsec{47}{0318775}$\pm$\mmsec{0}{000001} & \msec{07}{38139}$\pm$\msec{0}{00002} & 4&66$\pm$0.23 & 60&53$\pm$2.99 \\
4  & 2453556.202349 & \mmsec{47}{0328115}$\pm$\mmsec{0}{000002} & \msec{07}{37500}$\pm$\msec{0}{00005} & 0&65$\pm$0.12 &  7&58$\pm$1.40 \\
5  & 2453631.994548 & \mmsec{47}{0330740}$\pm$\mmsec{0}{000002} & \msec{07}{37032}$\pm$\msec{0}{00004} & 1&25$\pm$0.11 & 13&93$\pm$1.23 \\
6  & 2453732.719045 & \mmsec{47}{0323418}$\pm$\mmsec{0}{000001} & \msec{07}{36057}$\pm$\msec{0}{00003} & 1&53$\pm$0.10 & 17&95$\pm$1.17 \\
\\[-0.2cm]\hline\\[-0.2cm]\multicolumn{2}{l}{HDE~28357}&
\multicolumn{1}{c}{$04^{h}21^{m}$}&\multicolumn{1}{c}{$28^{\circ}18^{'}$}\\\\[-0.2cm]
1  & 2453270.983223 & \mmsec{58}{8521561}$\pm$\mmsec{0}{000001} & \msec{06}{38942}$\pm$\msec{0}{00001} & 7&13$\pm$0.16 & 86&13$\pm$1.93 \\
2  & 2453376.693842 & \mmsec{58}{8514573}$\pm$\mmsec{0}{000005} & \msec{06}{38002}$\pm$\msec{0}{00009} & 0&92$\pm$0.12 & 10&67$\pm$1.39 \\
3  & 2453460.464438 & \mmsec{58}{8514676}$\pm$\mmsec{0}{000002} & \msec{06}{37253}$\pm$\msec{0}{00004} & 1&71$\pm$0.13 & 18&64$\pm$1.42 \\
4  & 2453545.232390 & \mmsec{58}{8523648}$\pm$\mmsec{0}{000001} & \msec{06}{36785}$\pm$\msec{0}{00001} & 4&23$\pm$0.16 & 51&23$\pm$1.94 \\
5  & 2453636.981174 & \mmsec{58}{8528216}$\pm$\mmsec{0}{000007} & \msec{06}{36318}$\pm$\msec{0}{00014} & 0&52$\pm$0.12 &  5&71$\pm$1.32 \\
6  & 2453728.729982 & \mmsec{58}{8522172}$\pm$\mmsec{0}{000003} & \msec{06}{35481}$\pm$\msec{0}{00007} & 0&51$\pm$0.09 &  6&27$\pm$1.11 \\
\\[-0.2cm]\hline\\[-0.2cm]\multicolumn{2}{l}{HP~Tau/G2}&
\multicolumn{1}{c}{$04^{h}35^{m}$}&\multicolumn{1}{c}{$22^{\circ}54^{'}$}\\\\[-0.2cm]
1  & 2453621.024774 & \mmsec{54}{1613574}$\pm$\mmsec{0}{000003} & \msec{13}{41131}$\pm$\msec{0}{00009} & 0&71$\pm$0.12 &  8&27$\pm$1.40  \\
2  & 2453690.833709 & \mmsec{54}{1612212}$\pm$\mmsec{0}{000002} & \msec{13}{40798}$\pm$\msec{0}{00007} & 0&97$\pm$0.14 & 10&70$\pm$1.54  \\
3  & 2453758.648061 & \mmsec{54}{1609360}$\pm$\mmsec{0}{000004} & \msec{13}{40439}$\pm$\msec{0}{00010} & 0&99$\pm$0.17 &  7&87$\pm$1.35  \\
4  & 2453826.462390 & \mmsec{54}{1610940}$\pm$\mmsec{0}{000003} & \msec{13}{40161}$\pm$\msec{0}{00010} & 0&68$\pm$0.15 &  8&64$\pm$1.91  \\
5  & 2453897.268547 & \mmsec{54}{1617432}$\pm$\mmsec{0}{000001} & \msec{13}{39959}$\pm$\msec{0}{00002} & 3&06$\pm$0.16 & 35&24$\pm$1.84  \\
6  & 2453987.022807 & \mmsec{54}{1623557}$\pm$\mmsec{0}{000002} & \msec{13}{39681}$\pm$\msec{0}{00007} & 1&08$\pm$0.17 & 13&78$\pm$2.17  \\
7  & 2454256.289010 & \mmsec{54}{1627018}$\pm$\mmsec{0}{000004} & \msec{13}{38267}$\pm$\msec{0}{00013} & 0&63$\pm$0.13 &  6&79$\pm$1.40  \\
8  & 2454347.037078 & \mmsec{54}{1633509}$\pm$\mmsec{0}{000003} & \msec{13}{38127}$\pm$\msec{0}{00008} & 0&78$\pm$0.12 &  9&22$\pm$1.42  \\
9  & 2454438.785532 & \mmsec{54}{1631378}$\pm$\mmsec{0}{000004} & \msec{13}{37605}$\pm$\msec{0}{00009} & 0&76$\pm$0.11 &  5&20$\pm$0.75  \\
\\[-0.2cm]\hline\\[-0.2cm]\multicolumn{2}{l}{V773~Tau~A \tablenotemark{\dagger}}&
\multicolumn{1}{c}{$04^{h}14^{m}$}&\multicolumn{1}{c}{$28^{\circ}12^{'}$}\\\\[-0.2cm]
1  & 2453622.001337 & \mmsec{12}{9215438}$\pm$\mmsec{0}{000001} & \msec{12}{20222}$\pm$\msec{0}{00002} & 10&00$\pm$0.32 & 106&59$\pm$3.41  \\
   &		    & \mmsec{12}{9214354}$\pm$\mmsec{0}{000001} & \msec{12}{20152}$\pm$\msec{0}{00003} &  9&22$\pm$0.37 &  98&27$\pm$3.94  \\%
2  & 2453689.813594 & \mmsec{12}{9211625}$\pm$\mmsec{0}{000001} & \msec{12}{19608}$\pm$\msec{0}{00004} &  3&45$\pm$0.23 &  37&97$\pm$2.53  \\
   &     	    & \mmsec{12}{9213067}$\pm$\mmsec{0}{000002} & \msec{12}{19793}$\pm$\msec{0}{00005} &  2&46$\pm$0.23 &  27&07$\pm$2.53  \\%
3  & 2453756.632731 & \mmsec{12}{9207455}$\pm$\mmsec{0}{000002} & \msec{12}{19102}$\pm$\msec{0}{00006} &  2&04$\pm$0.21 &  21&28$\pm$2.19  \\
   &                & \mmsec{12}{9209564}$\pm$\mmsec{0}{000003} & \msec{12}{19180}$\pm$\msec{0}{00010} &  1&90$\pm$0.25 &  19&82$\pm$2.61  \\%
4  & 2453827.438617 & \mmsec{12}{9211553}$\pm$\mmsec{0}{000001} & \msec{12}{18646}$\pm$\msec{0}{00002} &  8&72$\pm$0.39 & 103&98$\pm$4.65  \\
   &		    & \mmsec{12}{9210561}$\pm$\mmsec{0}{000001} & \msec{12}{18575}$\pm$\msec{0}{00003} & 18&38$\pm$0.58 & 219&17$\pm$6.92  \\%
5  & 2453899.242297 & \mmsec{12}{9218088}$\pm$\mmsec{0}{000003} & \msec{12}{18114}$\pm$\msec{0}{00010} &  0&87$\pm$0.18 &   9&53$\pm$1.97  \\
6  & 2453984.010203 & \mmsec{12}{9226916}$\pm$\mmsec{0}{000002} & \msec{12}{17899}$\pm$\msec{0}{00005} &  3&76$\pm$0.32 &  41&00$\pm$3.49  \\
   &		    & \mmsec{12}{9225899}$\pm$\mmsec{0}{000001} & \msec{12}{17865}$\pm$\msec{0}{00003} &  7&74$\pm$0.31 &  84&41$\pm$3.38  \\%
7  & 2454336.046065 & \mmsec{12}{9240420}$\pm$\mmsec{0}{000001} & \msec{12}{15627}$\pm$\msec{0}{00002} & 13&16$\pm$0.50 & 155&73$\pm$5.92  \\
   &		    & \mmsec{12}{9238934}$\pm$\mmsec{0}{000002} & \msec{12}{15493}$\pm$\msec{0}{00006} &  7&89$\pm$0.62 &  93&37$\pm$7.34  \\%
8  & 2454342.029543 & \mmsec{12}{9240252}$\pm$\mmsec{0}{000001} & \msec{12}{15525}$\pm$\msec{0}{00001} &  9&80$\pm$0.29 & 113&89$\pm$3.37  \\
   &		    & \mmsec{12}{9239338}$\pm$\mmsec{0}{000001} & \msec{12}{15533}$\pm$\msec{0}{00003} &  6&05$\pm$0.33 &  70&31$\pm$3.84  \\%
9  & 2454349.010613 & \mmsec{12}{9240072}$\pm$\mmsec{0}{000002} & \msec{12}{15456}$\pm$\msec{0}{00003} &  7&27$\pm$0.34 &  81&47$\pm$3.81  \\
10 & 2454354.994294 & \mmsec{12}{9239067}$\pm$\mmsec{0}{000004} & \msec{12}{15360}$\pm$\msec{0}{00011} &  1&84$\pm$0.23 &  11&95$\pm$1.49  \\
   &		    & \mmsec{12}{9241045}$\pm$\mmsec{0}{000007} & \msec{12}{15535}$\pm$\msec{0}{00015} &  0&76$\pm$0.20 &   4&93$\pm$1.30  \\%
11 & 2454359.980648 & \mmsec{12}{9238938}$\pm$\mmsec{0}{000002} & \msec{12}{15346}$\pm$\msec{0}{00004} &  3&45$\pm$0.22 &  23&50$\pm$1.50  \\
   &		    & \mmsec{12}{9241150}$\pm$\mmsec{0}{000001} & \msec{12}{15516}$\pm$\msec{0}{00003} &  5&02$\pm$0.23 &  34&19$\pm$1.57  \\%
12 & 2454364.966887 & \mmsec{12}{9238903}$\pm$\mmsec{0}{000003} & \msec{12}{15329}$\pm$\msec{0}{00009} &  1&39$\pm$0.20 &  11&09$\pm$1.60  \\
13 & 2454370.950388 & \mmsec{12}{9239105}$\pm$\mmsec{0}{000001} & \msec{12}{15281}$\pm$\msec{0}{00003} &  7&01$\pm$0.28 &  46&41$\pm$1.85  \\
   &		    & \mmsec{12}{9241242}$\pm$\mmsec{0}{000005} & \msec{12}{15324}$\pm$\msec{0}{00010} &  2&16$\pm$0.30 &  14&30$\pm$1.99  \\%
14 & 2454376.934236 & \mmsec{12}{9239656}$\pm$\mmsec{0}{000001} & \msec{12}{15305}$\pm$\msec{0}{00003} &  7&27$\pm$0.31 &  55&56$\pm$2.37  \\
15 & 2454382.917679 & \mmsec{12}{9240014}$\pm$\mmsec{0}{000001} & \msec{12}{15297}$\pm$\msec{0}{00001} & 36&44$\pm$0.51 & 241&16$\pm$3.38  \\
16 & 2454390.896007 & \mmsec{12}{9240531}$\pm$\mmsec{0}{000002} & \msec{12}{15221}$\pm$\msec{0}{00005} & 12&27$\pm$0.81 &  93&05$\pm$6.14  \\
   &		    & \mmsec{12}{9239474}$\pm$\mmsec{0}{000001} & \msec{12}{15161}$\pm$\msec{0}{00002} & 28&89$\pm$0.85 & 219&10$\pm$6.45  \\%
17 & 2454396.879462 & \mmsec{12}{9239765}$\pm$\mmsec{0}{000001} & \msec{12}{15119}$\pm$\msec{0}{00002} & 21&96$\pm$0.57 & 165&74$\pm$4.30  \\
18 & 2454400.868432 & \mmsec{12}{9239199}$\pm$\mmsec{0}{000001} & \msec{12}{15043}$\pm$\msec{0}{00004} &  6&86$\pm$0.34 &  46&48$\pm$2.30  \\
   &		    & \mmsec{12}{9239780}$\pm$\mmsec{0}{000001} & \msec{12}{15179}$\pm$\msec{0}{00004} &  5&31$\pm$0.29 &  35&98$\pm$1.97  \\%
19 & 2454421.811366 & \mmsec{12}{9237229}$\pm$\mmsec{0}{000001} & \msec{12}{14946}$\pm$\msec{0}{00003} &  7&60$\pm$0.27 &  55&67$\pm$1.98  \\
   &		    & \mmsec{12}{9239298}$\pm$\mmsec{0}{000001} & \msec{12}{15003}$\pm$\msec{0}{00002} & 11&49$\pm$0.27 &  84&17$\pm$1.98  \\%
\\[-0.2cm]\hline\\[-0.2cm]\multicolumn{2}{l}{S1}&
\multicolumn{1}{c}{$16^{h}26^{m}$}&\multicolumn{1}{c}{$-24^{\circ}23^{'}$}\\\\[-0.2cm]
1  & 2453545.730989 & \mmsec{34}{1739533}$\pm$\mmsec{0}{000002} & \msec{28}{42695}$\pm$\msec{0}{00006} & 7&03$\pm$0.56 & 62&54$\pm$4.98  \\
2  & 2453628.504016 & \mmsec{34}{1736922}$\pm$\mmsec{0}{000002} & \msec{28}{43209}$\pm$\msec{0}{00006} & 4&56$\pm$0.47 & 46&29$\pm$4.77  \\
3  & 2453722.247609 & \mmsec{34}{1743677}$\pm$\mmsec{0}{000001} & \msec{28}{44149}$\pm$\msec{0}{00004} & 4&35$\pm$0.35 & 43&63$\pm$3.51  \\
4  & 2453810.007447 & \mmsec{34}{1746578}$\pm$\mmsec{0}{000002} & \msec{28}{45127}$\pm$\msec{0}{00005} & 5&33$\pm$0.41 & 54&35$\pm$4.18  \\
5  & 2453889.789004 & \mmsec{34}{1740172}$\pm$\mmsec{0}{000001} & \msec{28}{45594}$\pm$\msec{0}{00002} & 3&29$\pm$0.13 & 37&05$\pm$1.46  \\
6  & 2453969.570572 & \mmsec{34}{1732962}$\pm$\mmsec{0}{000001} & \msec{28}{46260}$\pm$\msec{0}{00005} & 4&35$\pm$0.22 & 36&53$\pm$1.85  \\
\\[-0.2cm]\hline\\[-0.2cm]\multicolumn{2}{l}{DoAr~21}&
\multicolumn{1}{c}{$16^{h}26^{m}$}&\multicolumn{1}{c}{$-24^{\circ}23^{'}$}\\\\[-0.2cm]
1  & 2453621.523501 & \mmsec{03}{0189304}$\pm$\mmsec{0}{000007} & \msec{36}{34339}$\pm$\msec{0}{00013} & 11&78$\pm$1.41 & 116&03$\pm$13.89  \\
2  & 2453691.332297 & \mmsec{03}{0191097}$\pm$\mmsec{0}{000002} & \msec{36}{34450}$\pm$\msec{0}{00005} & 20&34$\pm$1.42 & 216&80$\pm$15.14  \\
3  & 2453744.187627 & \mmsec{03}{0191069}$\pm$\mmsec{0}{000006} & \msec{36}{35580}$\pm$\msec{0}{00023} &  0&39$\pm$0.12 &   2&34$\pm$ 0.72  \\
4  & 2453755.157251 & \mmsec{03}{0191795}$\pm$\mmsec{0}{000003} & \msec{36}{35568}$\pm$\msec{0}{00013} &  0&97$\pm$0.19 &   9&25$\pm$ 1.81  \\
5  & 2453822.971950 & \mmsec{03}{0189625}$\pm$\mmsec{0}{000007} & \msec{36}{36192}$\pm$\msec{0}{00020} &  1&49$\pm$0.28 &   6&18$\pm$ 1.16  \\
6  & 2453890.786284 & \mmsec{03}{0182041}$\pm$\mmsec{0}{000002} & \msec{36}{36376}$\pm$\msec{0}{00010} &  1&92$\pm$0.23 &  16&86$\pm$ 2.02  \\
7  & 2453971.565109 & \mmsec{03}{0169857}$\pm$\mmsec{0}{000004} & \msec{36}{36996}$\pm$\msec{0}{00016} &  1&45$\pm$0.32 &  13&14$\pm$ 2.90  \\
\enddata
\tablenotetext{\dagger}{For 13 epochs we detect a double source,
thus we present here the position for both the primary and secondary
components, and the flux densities detected for each one.}
\end{deluxetable}

\section{Absolute astrometry}\label{c5-absolute}

The displacement of a source on the celestial sphere is the
combination of its trigonometric parallax ($\pi$) and its proper
motion ($\mu$). In what follows, we will have to consider three
different situations in terms of proper motions.

\begin{enumerate}

\item
Single Stars

Two of our target stars are apparently single (HDE~283572 and
Hubble~4), and one (HP~Tau/G2) is a member of a multiple system with
an orbital period so much longer than the timespan covered by the
observations that the effect of the companions can safely be
ignored. Thus, in these three cases, the proper motion can be assumed
to be linear and uniform, and the right ascension ($\alpha$) and the
declination ($\delta$) vary as a function of time $t$ as:
\begin{equation}\begin{split}\label{fit1}
\alpha(t) &= \alpha_{0}+(\mu_{\alpha}\cos\delta)t+\pi f_{\alpha}(t)\\%
\delta(t) &= \delta_{0}+\mu_{\delta}t+\pi f_{\delta}(t),
\end{split}\end{equation}
where $\alpha_0$ and $\delta_0$ are the coordinates of the source at a
given reference epoch, $\mu_\alpha$ and $\mu_\delta$ are the
components of the proper motion, and $f_\alpha$ and $f_\delta$ are the
projections over $\alpha$ and $\delta$, respectively, of the
parallactic ellipse.

\item
Binary Systems

One source (T~Tau~Sb) is a member of a binary system with an orbital
period longer than the timespan of our observations but not by a huge
factor. While a full Keplerian fit would again, in principle, be
needed, we found that including a constant acceleration term provides
an adequate description of the trajectory. The fitting functions in
this case are of the form:
\begin{equation}\begin{split}\label{fit2}
\alpha(t) &= \alpha_{0}+(\mu_{\alpha0}\cos\delta)t+{1\over2}
(a_{\alpha}\cos\delta)t^{2}+\pi f_{\alpha}(t)\\%
\delta(t) &= \delta_{0}+\mu_{\delta0}t+{1\over2}
a_{\delta}t^{2}+\pi f_{\delta}(t),
\end{split}\end{equation}
where $\mu_{\alpha0}$ and $\mu_{\delta0}$ are the proper motions at a
reference epoch, and $a_{\alpha}$ and $a_{\delta}$ are the projections
of the uniform acceleration.

\item
Tight Binaries

Three of our sources (V773~Tau~A, S1 and DoAr~21) are compact binary
systems with an orbital period of the order of the timespan covered by
the observations. In such a situation, one should fit simultaneously
for the uniform proper motion of the center of mass and for the
Keplerian orbit of the system. This requires more observations than
are needed to fit only for a uniform proper motion. Additional data
were collected to adequately constrain the required fits only for V773
Tau, and in this case, the orbital path was fully characterized. For
S1 and DoAr~21, we will present preliminary results based on Eqs.\
\ref{fit1} where the Keplerian motion is not included. We will see
momentarily that the main effect of not fitting for the Keplerian
orbit is an increase in the final uncertainty on the distance.

\end{enumerate}

\medskip

\noindent
The astrometric parameters were determined by least-squares fitting
the data points with either Eqs.\ \ref{fit1} or Eqs.\ \ref{fit2} using
a Singular Value Decomposition (SVD) scheme. To check our results, we
also performed two other fits to the data, a linear one based on the
associated normal equations, and a non-linear one based on the
Levenberg-Marquardt algorithm. They gave results identical to those
obtained using the SVD method.

\begin{deluxetable}{llr@{$\pm$}lr@{$\pm$}lc}
\tabletypesize{\scriptsize}
\tablecolumns{7}
\tablewidth{0pc}
\tablecaption{\footnotesize{Astrometric parameters. Reference epochs, source positions at the reference epochs, and number of observations used for each fit.}
\label{tab-astrometry1}}
\tablehead{
\multicolumn{1}{c}{~~~~~~~~~Source~~~~~~~~~}      &
\multicolumn{1}{c}{Reference}                     &
\multicolumn{2}{c}{$\alpha_{0}$}                  &
\multicolumn{2}{c}{$\delta_{0}$}                  &
\multicolumn{1}{c}{No.}                           \\
\multicolumn{1}{c}{}                              &
\multicolumn{1}{c}{Epoch \tablenotemark{\ddagger}}&
\multicolumn{2}{c}{[at the reference epoch]}      &
\multicolumn{2}{c}{[at the reference epoch]}      &
\multicolumn{1}{c}{Obs.}                          }
\startdata
T~Tau~Sb$^{\rm (a)}$ \dotfill                                 &
J2004.627                                                     &
\mbox{\dechms{04}{21}{59}{425081}} & \mbox{\mmsec{0}{000005}} &
\mbox{\decdms{19}{32}{05}{71566}}  & \mbox{\msec{0}{000030}}  &
12/14                                                         \\%
\\[-0.3cm]
T~Tau~Sb$^{\rm (b)}$ \tablenotemark{\dagger} \dotfill         &
J2004.627                                                     &
\mbox{\dechms{04}{21}{59}{425065}} & \mbox{\mmsec{0}{000002}} &
\mbox{\decdms{19}{32}{05}{71566}}  & \mbox{\msec{0}{000400}}  &
12/14                                                         \\%
\\[-0.3cm]
Hubble~4 \dotfill                                             &
J2005.355                                                     &
\mbox{\dechms{04}{18}{47}{032414}} & \mbox{\mmsec{0}{000001}} &
\mbox{\decdms{28}{20}{07}{37920}}  & \mbox{\msec{0}{000200}}  &
6/6                                                           \\%
\\[-0.3cm]
HDE~2835724 \dotfill                                          &
J2005.355                                                     &
\mbox{\dechms{04}{21}{58}{852030}} & \mbox{\mmsec{0}{000020}} &
\mbox{\decdms{28}{18}{06}{37128}}  & \mbox{\msec{0}{000050}}  &
6/6                                                           \\%
\\[-0.3cm]
HP~Tau/G2$^{\rm (a)}$ \dotfill                                &
J2006.810                                                     &
\mbox{\dechms{04}{35}{54}{162033}} & \mbox{\mmsec{0}{000003}} &
\mbox{\decdms{22}{54}{13}{49345}}  & \mbox{\msec{0}{000020}}  &
9/9                                                           \\%
\\[-0.3cm]
HP~Tau/G2$^{\rm (b)}$ \dotfill                                &
J2006.810                                                     &
\mbox{\dechms{04}{35}{54}{162030}} & \mbox{\mmsec{0}{000002}} &
\mbox{\decdms{22}{54}{13}{49362}} & \mbox{\msec{0}{000014}}   &
8/9                                                           \\%
\\[-0.3cm]
V773~Tau~A$^{\rm (a)}$ \dotfill                               &
J2006.780                                                     &
\mbox{\dechms{04}{14}{12}{922271}} & \mbox{\mmsec{0}{000010}} &
\mbox{\decdms{28}{12}{12}{17442}}  & \mbox{\msec{0}{000140}}  &
7/19                                                          \\%
\\[-0.3cm]
V773~Tau~A$^{\rm (b)}$ \dotfill                               &
J2006.780                                                     &
\mbox{\dechms{04}{14}{12}{922269}} & \mbox{\mmsec{0}{000009}} &
\mbox{\decdms{28}{12}{12}{17454}}  & \mbox{\msec{0}{000080}}  &
19/19                                                         \\%
\\[-0.3cm]
S1 \dotfill                                                   &
J2006.061                                                     &
\mbox{\dechms{16}{26}{34}{174127}} & \mbox{\mmsec{0}{000026}} &
\mbox{\decdms{-24}{23}{28}{44498}} & \mbox{\msec{0}{000280}}  &
6/6                                                           \\%
\\[-0.3cm]
DoAr~21 \dotfill                                              &
J2006.167                                                     &
\mbox{\dechms{16}{26}{03}{018535}} & \mbox{\mmsec{0}{000020}} &
\mbox{\decdms{-24}{23}{36}{35830}} & \mbox{\msec{0}{000220}}  &
7/7                                                           \\%
\enddata
\tablenotetext{\ddagger}{The reference epoch was taken at the mean of each
set of observations.}
\end{deluxetable}

\begin{deluxetable}{lr@{$\pm$}lr@{$\pm$}lr@{$\pm$}lr@{$\pm$}l}
\tabletypesize{\scriptsize}
\tablecolumns{9}
\tablewidth{0pc}
\tablecaption{\footnotesize{Astrometric parameters. Proper motions, parallaxes, and distances.}
\label{tab-astrometry2}}
\tablehead{
\multicolumn{1}{c}{~~~~~~~~~Source~~~~~~~~~} &
\multicolumn{2}{c}{$\mu_{\alpha}\cos\delta$} &
\multicolumn{2}{c}{$\mu_{\delta}$}           &
\multicolumn{2}{c}{$\pi$}                    &
\multicolumn{2}{c}{$d$}                      \\
\multicolumn{1}{c}{}                         &
\multicolumn{2}{c}{[mas yr$^{-1}$]}          &
\multicolumn{2}{c}{[mas yr$^{-1}$]}          &
\multicolumn{2}{c}{[mas]}                    &
\multicolumn{2}{c}{[pc]}                     }
\startdata
T~Tau~Sb$^{\rm (a)}$ \dotfill                                 &
4.00     & 0.12 & $-$1.18 & 0.05                              &
6.90     & 0.09 & 145.0   & 2.0                               \\%
\\[-0.3cm]
T~Tau~Sb$^{\rm (b)}$ \tablenotemark{\dagger} \dotfill         &
4.02     & 0.03 & $-$1.18 & 0.05                              &
6.82     & 0.03 & 146.7   & 0.6                               \\%
\\[-0.3cm]
Hubble~4 \dotfill                                             &
4.30     & 0.05 & $-$28.90 & 0.30                             &
7.53     & 0.03 & 132.8    & 0.5                              \\%
\\[-0.3cm]
HDE~2835724 \dotfill                                          &
8.88     & 0.06 & $-$26.60 & 0.10                             &
7.78     & 0.04 & 128.5    & 0.6                              \\%
\\[-0.3cm]
HP~Tau/G2$^{\rm (a)}$ \dotfill                                &
13.90    & 0.06 & $-$15.60 & 0.30                             &
6.19     & 0.07 & 161.6    & 1.7                              \\%
\\[-0.3cm]
HP~Tau/G2$^{\rm (b)}$ \dotfill                                &
13.85    & 0.03 & $-$15.40 & 0.20                             &
6.20     & 0.03 & 161.2    & 0.9                              \\%
\\[-0.3cm]
V773~Tau~A$^{\rm (a)}$ \dotfill                               &
17.28    & 0.19 & $-$23.74 & 0.14                             &
7.53     & 0.26 & 132.8    & 4.5                              \\%
\\[-0.3cm]
V773~Tau~A$^{\rm (b)}$ \dotfill                               &
17.27    & 0.14 & $-$23.79 & 0.09                             &
7.57     & 0.20 & 132.0    & 3.5                              \\%
\\[-0.3cm]
S1 \dotfill                                                   &
$-$3.88  & 0.87 & $-$31.55 & 0.69                             &
8.55     & 0.50 & 116.9    & 7.1                              \\%
\\[-0.3cm]
DoAr~21 \dotfill                                              &
$-$26.47 & 0.92 & $-$28.23 & 0.73                             &
8.20     & 0.37 & 121.9    & 5.5                              \\%
\enddata
\tablenotetext{\dagger}{The acelerations terms obtained for this fit were
$a_{\alpha}=1.53\pm0.13$ mas yr$^{-2}$\\ and $a_{\delta}=0.00\pm0.19$
mas yr$^{-2}$.}
\end{deluxetable}

\subsection{Parallax and proper motions}\label{c5-absolute-parallax}

\textbf{T~Tau~Sb}---
In order to measure the trigonometric parallax of T~Tau~Sb, we
collected observations roughly every 2 months between 2003 September
and 2005 July (first 12 observations listed in Tab.\
\ref{tab-observaciones}). Ten of this twelve observations are shown in
Fig.\ \ref{map-ttau1}, and in panels 1 to 4 of Fig.\
\ref{map-ttau2}. As time went by, we decide to obtain new
observations to fit an orbit to T~Tau~Sb around T~Tau~Sa. One
additional observation was collected in 2008 May (epoch 13 in Tab.\
\ref{tab-observaciones}, panel 5 in Fig.\ \ref{map-ttau2}), and we
made use of data from VLBA archive (epoch 14 in Tab.\
\ref{tab-observaciones}, panel 6 of Fig.\ \ref{map-ttau2}). The source
was detected in all 14 observations. In this section we will explain
how we used the first 12 observations to fit for the parallax and
proper motions, and the orbit of T~Tau~S system, where we used all 14
epochs, is explained in Sect.\
\ref{c5-relative}.

\medskip

\noindent
All the measured source positions and fluxes are listed in Tab.\
\ref{tab-resultados}. Since T~Tau~Sb is a member of a double stellar
system (T~Tau~Sa/Sb) with an orbital period ($\lesssim$ 20 yr) longer
than the timespan of our observations ($\sim$ 2 yr), we performed two
astrometric fits: one where we assumed that the proper motion is
linear and uniform, and other that includes a constant acceleration.

\medskip

\begin{figure*}[!h]
\centerline{\includegraphics[height=1\textwidth,angle=-90]{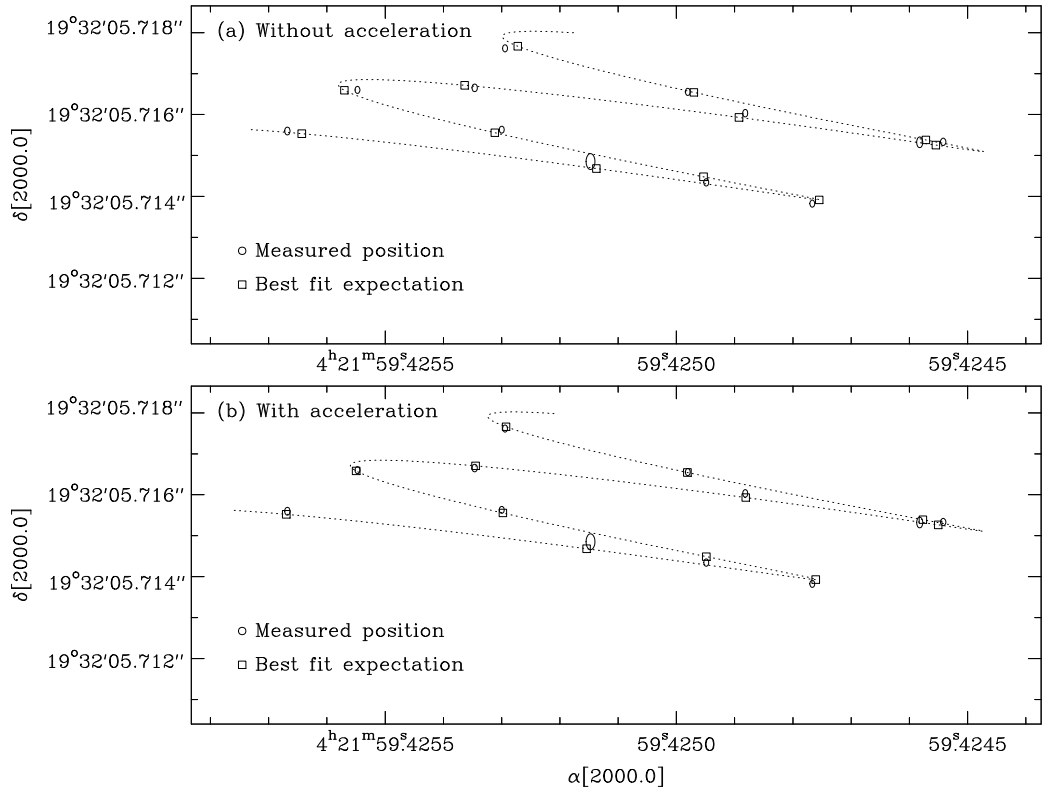}}
\caption
{\footnotesize{Measured positions of T~Tau~Sb and best fit without (a)
and with (b) acceleration terms. The observed positions are shown as
ellipses, the size of which represents the magnitude of the
errors. Note the very significant improvement when acceleration terms
are included.}\label{fig-ttau-1}}\end{figure*}

\noindent
\textit{First astrometric fit}---
The fit to the data points by Eqs.\ \ref{fit1} (Fig.\
\ref{fig-ttau-1}a) yields the astrometric parameters listed in Tabs.\
\ref{tab-astrometry1} and \ref{tab-astrometry2} labelled as
T~Tau~Sb$^{\rm (a)}$. This corresponds to a distance of $145\pm2$
pc. The post-fit r.m.s.\, however, is not very good (particularly in right
ascension: $\sim$ 0.2 mas), as the fit does not pass through many of
the observed positions (see Fig.\ \ref{fig-ttau-1}a). As a matter of
fact, 75 $\mu$as and 16.5 $\mu$s had to be added quadratically to the
formal errors listed in Cols.\ [3] and [4] of Tab.\
\ref{tab-resultados} to obtain a reduced $\chi^{2}$ of 1 in both right
ascension and declination; the errors on the fitted parameters listed
in Tabs.\ \ref{tab-astrometry1} and \ref{tab-astrometry2} include this
systematic contribution. These large systematic errors most certainly
reflects the fact that the proper motion of T~Tau~Sb is not uniform
because it belongs to a multiple system.

\medskip

\noindent
\textit{Second astrometric fit}---
The fit where acceleration terms are included is significantly better
(Fig.\ \ref{fig-ttau-1}b) with a post-fit r.m.s.\ of 60 $\mu$as in right
ascension and 90 $\mu$as in declination. It yields the parameters
listed in Tabs.\ \ref{tab-astrometry1} and \ref{tab-astrometry2}
labelled as T~Tau~Sb$^{\rm (b)}$. To obtain a reduced $\chi^{2}$ of 1
in both right ascension and declination, one must add quadratically
3.8 $\mu$s and 75 $\mu$as to the statistical errors listed in Cols.\
[3] and [4] of Tab.\ \ref{tab-resultados}. The uncertainties on the
astrometric parameters (Tabs.\ \ref{tab-astrometry1} and
\ref{tab-astrometry2}), include this systematic contribution. Note
also that the reduced $\chi^{2}$ for the fit without acceleration
terms is almost 8 if the latter systematic errors (rather than those
mentioned earlier) are used. The trigonometric parallax obtained when
acceleration terms are included, corresponds to a distance of
$146.7\pm0.6$ pc, somewhat smaller than, but within 1$\sigma$ of the
distance obtained by \textit{Hipparcos} ($d=177^{+68}_{-39}$
pc). Note that the relative error of our distance is about 0.4\%,
against nearly 30\% for the \textit{Hipparcos} result, an improvement
of almost two orders of magnitude.

\medskip

\begin{figure*}[!h]
\centerline{\includegraphics[width=1\textwidth,angle=270]{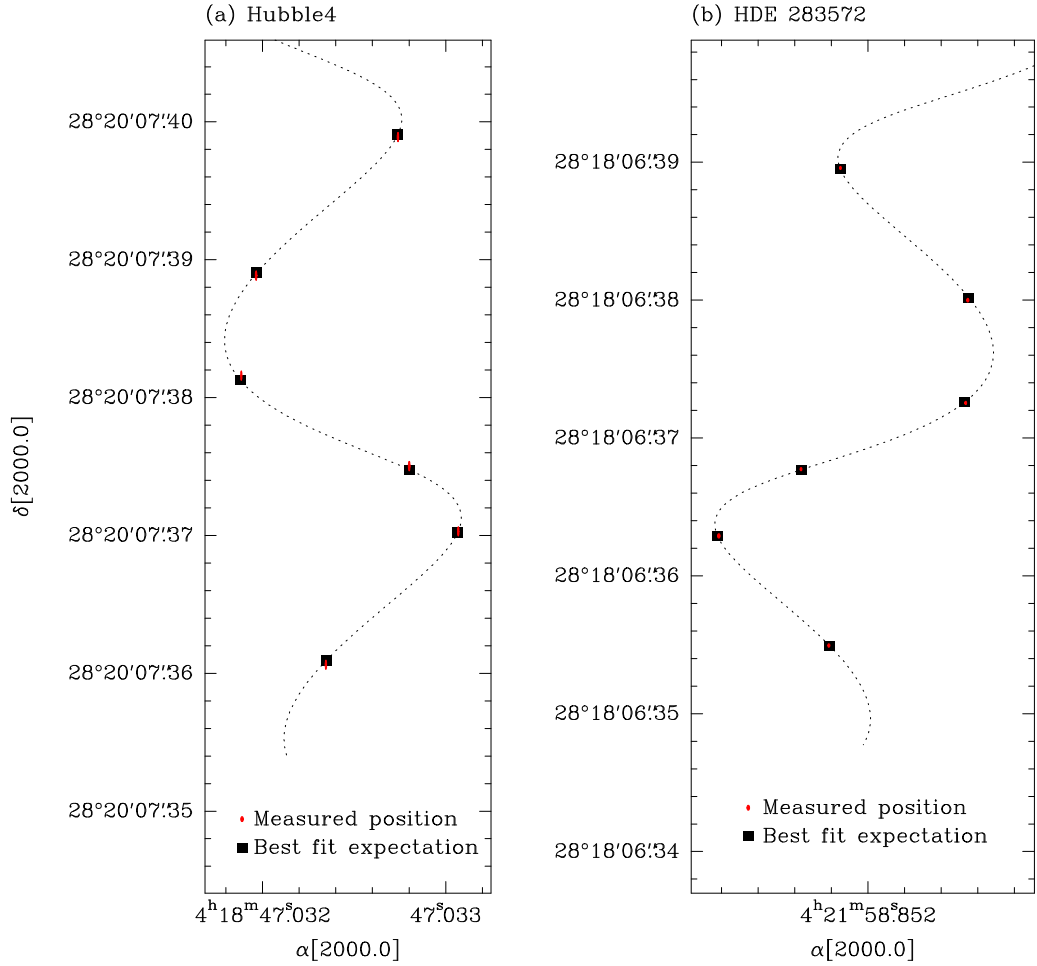}}
\caption
{\footnotesize{Measured positions and best fit for (a) Hubble~4, and
(b) HDE~283572. The observed positions are shown as ellipses, the size
of which represents the error bars.}\label{fig-hubble-1}}\end{figure*}

\noindent
\textbf{Hubble~4 and HDE~283572}---
We made use of all 6 observations listed in Tab.\
\ref{tab-observaciones} for each source (see contour maps in Fig.\
\ref{map-hubble4} and \ref{map-hde}). In Tab.\ \ref{tab-resultados} we
provide the measured source positions and flux densities. Since both
sources appear to be isolated, we considered linear and uniform proper
motions (Eqs.\ \ref{fit1}). The best fits (Fig.\
\ref{fig-hubble-1}) give the parameters listed in Tabs.\
\ref{tab-astrometry1} and \ref{tab-astrometry2}. The measured parallaxes
correspond to distances of 132.8 $\pm$ 0.5 pc for Hubble~4, and 128.5
$\pm$ 0.6 pc for HDE~283572. The post-fit r.m.s.\ (dominated by the
remaining systematic errors) is quite good for HDE~283572: 60 $\mu$as
and 90 $\mu$as in right ascension and declination, respectively. For
Hubble~4, on the other hand, the residual is good in right ascension
(40 $\mu$as), but large in declination (240 $\mu$as). To obtain a
reduced $\chi^2$ of one both in right ascension and declination, one
must add quadratically 3.1 $\mu$s and 340 $\mu$as to the formal errors
in Cols.\ [3] and [4] of in Tab.\ \ref{tab-resultados} for Hubble~4,
and 4.3 $\mu$s and 115 $\mu$as for HDE~283572. All the errors quoted
include these systematic contributions.

\medskip

\begin{figure*}[!h]
\centerline{\includegraphics[width=.5\textwidth,angle=270]{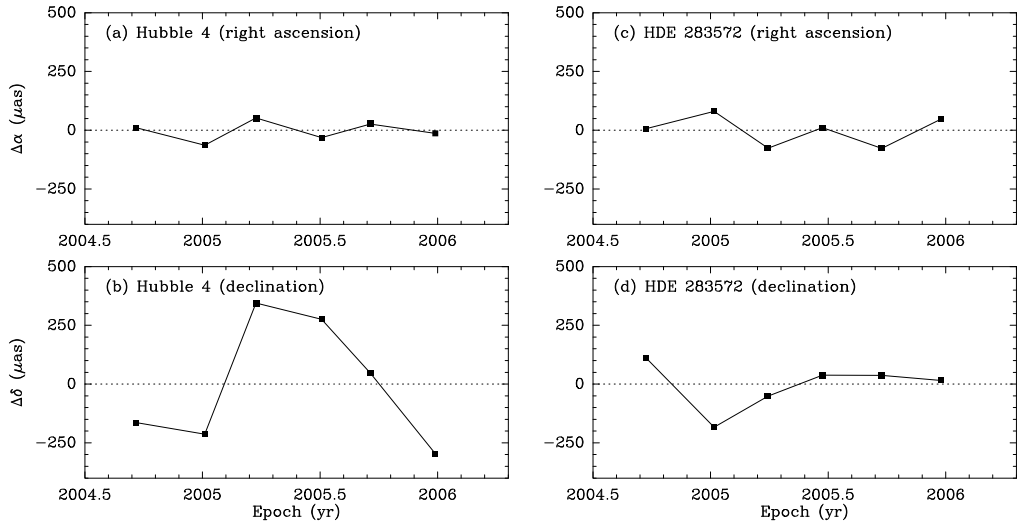}}
\caption
{\footnotesize{Post-fit residuals for Hubble~4 (left) and HDE~283572
(right) in right ascension (top) and declination
(bottom).}\label{fig-hubble-2}}\end{figure*}

\noindent
The origin of the large declination residual for Hubble~4 (which does
not affect strongly the parallax determination, because the latter is
dominated by the right ascension measurements) is not entirely
clear. The fact that the residual is only (or, at least, mostly)
detected in declination (Fig.\ \ref{fig-hubble-2}) would suggest a
calibration issue.  Indeed, astrometric fitting of phase-referenced
VLBI observations is usually worse in declination than in right
ascension (e.g.\ Fig.\ 1 in Chatterjee et al.\ 2004) as a result of
residual zenith phase delay errors (Reid et al.\ 1999). We consider
this possibility fairly unlikely here, however, because such a problem
would have been detected during the multi-source calibration, and
because the observations and reduction of Hubble~4 and HDE~283572
(which does not appear to be affected by any calibration issue) were
performed following identical protocols and over the same period of
time. Another element that argues against a calibration problem is
that the large residual is not the result of one particularly
discrepant observation: in addition to the fit mentioned above where
all 6 observations of Hubble~4 are taken into account, we made 5 fits
where we sequentially discarded one of the epochs. All 5 fits gave
similar astrometric parameters, and a similarly large declination
residual. Thus, we argue that this large residual might be real,
rather than related to a calibration problem. At the distance of
Hubble~4, 240 $\mu$as correspond to 0.032 AU, or about 7
\Rsun. Hubble~4 is estimated to have a radius of about 3.4 \Rsun\
(Johns-Krull et al.\ 2004), so the amplitude of the residual is just
about 2 $R_*$. Keeping this figure in mind, at least two mechanisms
could potentially explain the large declination residual: (i) the
magnetosphere of Hubble~4 could be somewhat more extended than its
photosphere, and the residuals could reflect variations in the
structure of the magnetosphere; (ii) Hubble~4 could have a companion,
and the residuals could reflect the corresponding reflex motion. Let
us examine the pros and cons of these two possibilities.

\medskip

\noindent
If the residuals were related to a variable extended magnetosphere,
one would expect the emission to be occasionally somewhat extended.
Interestingly, Phillips et al.\ (1991) reported that Hubble~4 was
slightly resolved in their VLBI data, and we find it to be resolved
also in at least two of our own observations. On the other hand, if
the emission were related to variations in the magnetosphere, one
would expect to see variations with the periodicity of the rotational
period of the star (about 12/sin{\it i} days --Johns-Krull et al.\
2004). Given that the separation between our successive observations
is typically three months, we would expect the residuals to be
essentially random. Instead, those residuals seem to show a
periodicity of about 1.2 years (see Fig.\ \ref{fig-hubble-2}). This
would be more consistent with our alternative proposal that the
residuals be related to the reflex motion of Hubble~4 due to the
presence of an unseen companion. The semi-major axis corresponding to
a period of 1.2 yr and a mass of 0.7 \Msun\ (see below) is just about
1 AU. Since the ratio between the amplitude of the reflex motion and
that of the orbital path is the inverse of the ratio between the mass
of the primary and that of the companion, the mass of the companion
would have to be 0.7(0.032/1) = 0.02 \Msun. The companion would then
have to be a very low-mass star, or a brown dwarf. Note, however, that
the residuals are relatively poorly constrained with the existing
data, and that additional observations aimed --in particular-- at
confirming the periodicity in the residuals will be needed to resolve
this issue.

\medskip

\noindent
\textbf{HP~Tau/G2}---
We detected the source at all 9 observations listed in Tab.\
\ref{tab-observaciones} (see also contour maps in Fig.\
\ref{map-hptau}). Since HP~Tau/G2 is a member of a triple system, we
should in principle describe its proper motion as the combination of
the uniform motion of the center of mass and a Keplerian orbit. This
is not necessary, however, because the orbital period of the system
must be very much longer than the timespan covered by our
observations. If we assume that the total mass of the HP~Tau/G2 --
HP~Tau/G3 system is 2--3
\Msun\ and that the current observed separation is a good estimate of
the system's semi-major axis, then the orbital period is expected to
be 35,000 to 45,000 yr. This is indeed very much longer that the 2 yr
covered by our observations, and the acceleration terms can be safely
ignored. Thus we made use of Eqs.\ \ref{fit1} for the astrometric
fits. Since the source was elongated in the north-south direction
during the seventh observation (see panel 6 in Fig.\ \ref{map-hptau}),
two different fits were made: one where the seventh epoch was
included, and one where it was ignored.

\medskip

\begin{figure*}[!h]
\centerline{\includegraphics[width=0.8\textwidth,angle=270]{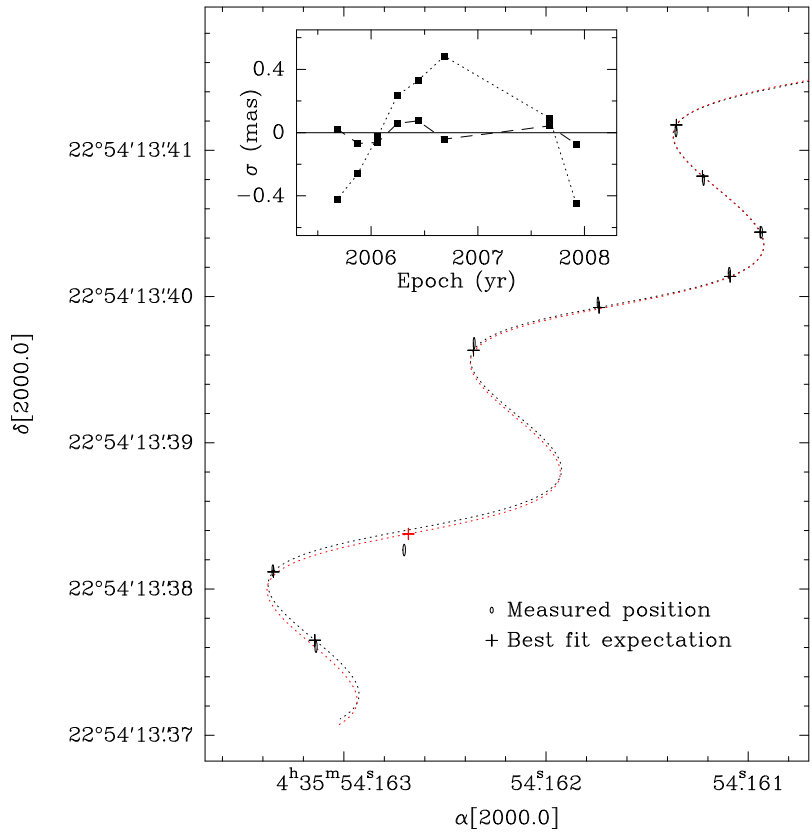}}
\caption
{\footnotesize{Measured positions and best fits for HP~Tau/G2. The
observed positions are shown as ellipses, the size of which represents
the error bars. Two fits are shown: the dotted black line corresponds
to the fit where the 7th epoch is ignored, whereas the dotted red line
is the fit where it is included. Note that the 7th observation falls
significantly to the south of either fit. The inset shows the fit
residuals (of the fit without the 7th epoch) in right ascension
(dashed line) and declination (dotted line). Note the large residuals
in declination.}\label{fig-hptau-1}}\end{figure*}

\noindent
\textit{First astrometric fit}---
When the seventh epoch is included, we obtain the astrometric
parameters listed in Tabs.\ \ref{tab-astrometry1} and
\ref{tab-astrometry2} labelled as HP~Tau/G2$^{(a)}$. This corresponds
to a distance of 161.6 $\pm$ 1.7 pc. The post-fit r.m.s.\ in this case
is 0.12 mas in right ascension and 0.51 mas in declination. To obtain
a reduced $\chi^2$ of one in both right ascension and declination, one
must add quadratically 8.8 $\mu$s and 0.59 mas in right ascension and
declination, respectively, to the errors in Cols.\ [3] and [4] of
Tab.\ \ref{tab-resultados}. The uncertainties on the parameters
include these systematic contributions. Note that the seventh epoch
contributes significantly to the total post-fit r.m.s.\ since the
position corresponding to that observation is farther from the fit
(both in right ascension and declination) than that at any other epoch
(see Fig.\ \ref{fig-hptau-1}).

\medskip

\noindent
\textit{Second astrometric fit}---
If the seventh observation is ignored, the best fit yields the
parameters listed in Tabs.\ \ref{tab-astrometry1} and
\ref{tab-astrometry2} labelled as HP~Tau/G2$^{(b)}$. All these
parameters are consistent within 1$\sigma$ with those obtained when
the seventh observation is included. The corresponding distance in
this case is 161.2 $\pm$ 0.9 pc, and the post-fit r.m.s.\ is 0.058 mas
in right ascension and 0.33 mas in declination, significantly better
than in the previous fit. Indeed, to obtain a reduced $\chi^2$ of one
in both right ascension and declination, one must only add
quadratically 3.65 $\mu$s and 0.38 mas to the formal errors in Cols.\
[3] and [4] of Tab.\ \ref{tab-resultados}. Again, the uncertainties on
the parameters include these systematic contributions.

\medskip

\noindent
As mentioned earlier, the source during the seventh epoch was
extended, and the astrometry consequently less reliable. Since the fit
when this epoch is ignored is clearly much better than that when it is
included, we consider the second fit above our best result. It is
noteworthy that, whether or not the seventh epoch is included, the
post-fit r.m.s.\ and the systematic error contribution that must be
added to the uncertainties quoted in Tab.\ \ref{tab-resultados}, are
much larger in declination than in right ascension. Fortunately, this
large declination contribution does not strongly affect the distance
determination, because the strongest constraints on the parallax come
from the right ascension measurements. Interestingly, this is, with
Hubble~4, the second source for which we find large systematic
declination residuals. In the case of Hubble~4, we argued that the
large post-fit declination r.m.s.\ might trace the reflex motion
caused by an unseen companion, because a periodicity of about 1.2 yr
could be discerned in the residuals. In the present source, the case
for a periodicity is less clear (Fig.\ \ref{fig-hptau-1}, inset), but
the residuals are clearly not random. Interestingly, the large
residuals are in the same north-south direction as the extension of
the source seen during our 7th observation. This orientation might,
therefore, correspond to a preferred direction of the system along
which it tends to vary more strongly. Additional observations will
clearly be necessary to settle this issue.

\medskip

\noindent
\textbf{V773~Tau~A}---
The source was detected at all 19 epochs listed in Tab.\
\ref{tab-observaciones}, but it was found to be double at 13 epochs,
and single for the remaining 6. In Figs.\ \ref{map-v773tau1},
\ref{map-v773tau2}, \ref{map-v773tau3}, and in panel 1 of Fig.\
\ref{map-varios} are shown all 19 contour maps of the system. The
source positions (including both primary and secondary when the source
is double) are listed in Tab.\ \ref{tab-resultados}. It is possible to
use Eqs.\ \ref{fit1} for a binary system if we know the position of
the barycenter of the system. To deduce the coordinates of the
barycenter from the positions listed in Tab.\ \ref{tab-resultados}, we
need to take into account the orbital motion of the system. The orbital
fit for V773~Tau~A is discussed in Sect.\ \ref{c5-relative}, and we
will make reference to this section in the next few paragraphs. We,
again, performed two different astrometric fits.

\begin{figure*}[!h]
\centerline{\includegraphics[angle=0,width=1\textwidth]{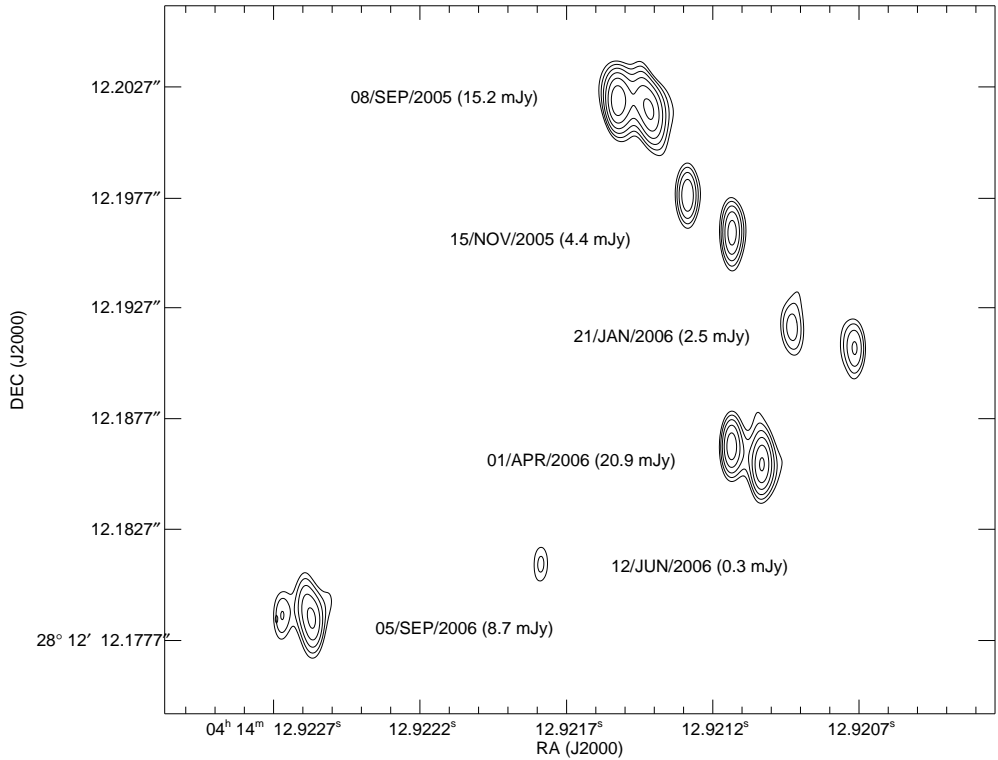}}
\caption
{\footnotesize{3.6 cm VLBA images of V773~Tau~A system at six
different epochs. The separation between V773~Tau~Aa and V773~Tau~Ab
changes from epoch to epoch, and the flux density changes
significantly with separation. The motion seen in the image is real,
it is a combination of parallactic, orbital, and proper motion of the
center of mass.}\label{fig-v773-3}}\end{figure*}

\medskip

\noindent
\textit{First astrometric fit}---
The only observations for which obtaining the coordinates of the
barycenter of the system can unambiguously be done are those when the
source was double. For the corresponding 13 observations (epochs 1, 2,
3, 4, 6, 7, 8, 10, 11, 13, 16, 18, 19; see Tab.\ \ref{tab-fases} and
also Figs.\ \ref{map-v773tau1}, \ref{map-v773tau2},
\ref{map-v773tau3}, and \ref{map-varios}), the preferred strategy
would be to combine the {\em measured} separation between the primary
and the secondary (let us call it $\mathbf r$), and the masses deduced
from the orbital fit. The separation $\mathbf{r}_1$ between the
primary to the barycenter is then given by
\begin{equation}\begin{split}\label{ec-separacion}
\mathbf{r}_1 &= {m_s \over m_s + m_p} \mathbf{r}.
\end{split}\end{equation}
Here, $m_p$ and $m_s$ are the masses of the primary and secondary,
respectively. This strategy works well as long as the separation
between the radio sources properly traces the separation between the
stars. This is a valid assumption except near periastron (we will come
back to this point in Sect.\ \ref{c5-variability} when we discuss the
variability of V773~Tau~A). As a consequence, we initially considered
only the seven observations when the source was double and not near
periastron (epochs 2, 3, 10, 11, 13, 18, 19; see Tab.\
\ref{tab-fases}). These observations are sufficiently well spread in
time to provide a reasonable characterization of the trigonometric
parallax and proper motion of the barycenter of V773~Tau~A. The best
fit to these ``best seven observations'' yields the parameters shown
in the Tabs.\ \ref{tab-astrometry1} and \ref{tab-astrometry2} labeled
as V773~Tau~A$^{(a)}$. The post-fit r.m.s.\ is 0.26 mas and 0.24 mas in
right ascension and declination, respectively. The parallax estimated
above corresponds to a distance $d=132.8^{+4.8}_{-4.5}$.

\medskip

\noindent
\textit{Second astrometry fit}---
The astrometric fit obtained above only includes about a third of the
VLBA observations available. Although the position of the barycenter
of the system is deduced in a somewhat less direct fashion for the
other two thirds of the data, it would still be desirable to
incorporate them in a global astrometric fit. Moreover, the
astrometric behavior of the source is important to help constrain the
origin of the radio emission and the structure of the system at some
epochs.

\medskip

\noindent
The six observations where the source was found to be double, and
which were not included in the last fit, were taken when the system
was near periastron (epochs 1, 4, 6, 7, 8, 16; see Tab.\
\ref{tab-fases}). In such cases, the separation between the radio
sources tended to be systematically smaller than the separation
between the stars. This likely occurs because of enhanced flaring
activity on either of the two stars, or on both of them. If both stars
experience this enhanced activity, then the radio sources associated
with them are likely to be similarly displaced from their respective
star. The best strategy to obtain the position of the barycenter would
still be to use Eq.\ \ref{ec-separacion} using the {\em measured}
separation between the stars for the separation $\mathbf{r}$, because
both $\mathbf{r}$ and $\mathbf{r_1}$ are expected to be shrunk by a
similar factor. If the enhanced activity only occurs on one of the
stars, however, then only for that object will the radio source be
offset from the star. A better strategy to derive the position of the
barycenter of the system in this case would be to identify the star
not affected by enhanced activity, and calculate the separation
between that star and the barycenter using the {\em prediction} of the
orbit model presented in Sect.\ \ref{c5-relative} (which provides the
true separation between the stars) rather than the measured separation
between the radio sources. To test for these different possibilities,
we ran three tests (all of which include the ``best seven
observations'', in addition to the six resolved observations near
periastron: epochs 2, 3, 10, 11, 13, 18, 19, 1, 4, 6, 7, 8, 16). In
the first one, we assumed that both stars experienced enhanced
activity and that the position of the barycenter should be obtained
using the {\em measured} separation between the radio sources. In the
second, we assumed that only the secondary suffered from enhanced
activity, and used the measured position of the primary and the
separation between the primary and the barycenter predicted by the
orbit model to estimate the position of the barycenter. Finally, in
the third test, we assumed that only the primary was affected by
enhanced activity, and used the position of the secondary and a
predicted secondary-barycenter separation. The astrometric parameters
obtained from these three tests are very similar to one another, and
well within one sigma of the results of the fit presented above
(\textit{First astrometric fit} with the ``best seven
observations''). In terms of post-fit r.m.s., the best result is
obtained when the position of the barycenter is deduced from the {\em
measured} separation between the sources. This indicates that both
stars experience enhanced activity near periastron, a result that was
to be expected since {\em both} components of the system become
brighter near periastron.

\medskip

\noindent
When the radio source associated with V773~Tau~A is single, one must
again distinguish between two different situations. For two of these
observations, the source is weak ($\sim$ 1 mJy), and the system was
near apoastron (epochs 5, 12; see Tab.\ \ref{tab-fases}). Both sources
are known to become dimmer near apoastron, and one of them has clearly
faded below our detection limit in these two cases. It is {\it a
priori} unclear whether the detected source corresponds to the primary
or the secondary. To find out, we estimated the position of the
barycenter of the system using our orbit model (Sect.\
\ref{c5-relative}) assuming first that the detected source was the
primary and then that it was the secondary. The positions derived
assuming that the detected source is the primary fall almost exactly
on the astrometry fits, whereas the positions deduced assuming that
the detected source is the secondary fall several mas away from the
fit. We conclude that in these two cases, the detected source is the
primary. For the remaining 4 observations, the source is single, but
bright (at least several mJy; epochs 9, 14, 15, 17). All four of these
observations were obtained at an orbit phase of about 0.2 before or
after periastron (see Tab.\ \ref{tab-fases}). In this situation, the
two stars are located in projection almost exactly north-south of each
other, with a projected separation of about 1 mas. Since our
resolution in the north-south direction is about 2 mas, we do not
expect to resolve the two stars in this situation, but instead to
detect a single source slightly elongated in the north-south
direction. This is in fact what happens. Indeed, the mean deconvolved
FWHM size of the emission for these four epochs is 1.33 mas, 50\%
larger than the corresponding figure (0.89 mas) when the source is
double or single but near apoastron. In this situation, the measured
position of the source is likely to provide a good approximation (to
within a few tenths of a mas) of the position of the barycenter of the
system.

\medskip

\noindent
To refine the determination of the parallax and proper motion of
V773~Tau~A, we performed a final astrometric fit which includes:

\begin{itemize}

\item[-]
 The ``best seven observations'' used to obtain the \textit{First
 astrometric fit}.

\item[-]
 The six observations corresponding to double sources near
 periastron. The position of the barycenter in this case was
 calculated using Eq.\ \ref{ec-separacion} and the true separation for
 $\mathbf{r}$.

\item[-]
 The two observations where the source was single and near apoastron.
 In this case, we obtained the position of the barycenter using a
 separation between the primary and the barycenter deduced from our
 orbit model.

\item[-]
 The four epochs at $\Delta \phi$ $\sim$ 0.2, for which the position
 of the barycenter was assumed to be that of the (single) source
 itself.

\end{itemize}

\begin{figure*}[!b]
\centerline{\includegraphics[width=0.8\textwidth,angle=270]{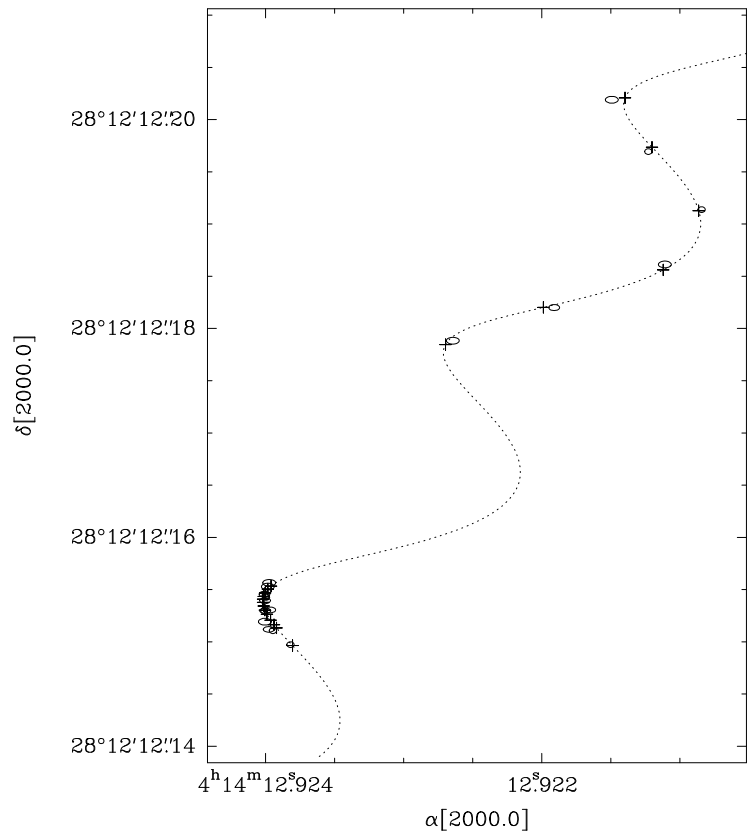}}
\caption
{\footnotesize{Measured positions and best fit for V773~Tau A. The
observed positions are shown as ellipses, the size of which represents
the error bars.}\label{fig-v773-1}}\end{figure*}

\medskip

\noindent
The results of this fit are shown in Tabs.\ \ref{tab-astrometry1} and
\ref{tab-astrometry2} labeled as V773~Tau~A$^{(b)}$, and are very
similar to those of our initial fit. All the parameters are well
within one sigma of one another. The distance corresponding to this
global fit is 132.0$^{+3.5}_{-3.4}$, in excellent agreement with the
fit to the best seven VLBA observations. We note, finally, that the
distance obtained here is only marginally consistent with the the
result obtained by Lestrade et al.\ (1999; $d=148.4\pm5.3$ pc). We
attribute the discrepancy to the fact that Lestrade did not take into
account the binarity of the source in their astrometric fits.

\medskip

\noindent
\textbf{S1 and DoAr~21}---
We detected the sources in all observations of S1 and DoAr~21 (listed
in Tab.\ \ref{tab-observaciones}; see contour maps in Fig.\
\ref{map-varios}). DoAr~21 was found to be double at several epochs,
whereas S1 was always single. The source positions are listed in Tab.\
\ref{tab-resultados} (when DoAr~21 was a double source, we used the
average position of both primary and secondary). We have seen that the
extended magnetosphere of S1, and the reflex motions of both S1 and
DoAr~21 are likely to produce significant shifts in the positions of
the source photocenters. While the effect of an extended magnetosphere
might be to produce a random jitter, the reflex orbital motions ought
to generate oscillations with a periodicity equal to that of the
orbital motions. Six or seven observations, however, are currently
insufficient to properly fit full Keplerian orbits. Instead, we
represent the possible systematic calibration errors as well as the
jitter due to extended magnetospheres and the oscillations due to
reflex motions, by a constant error term (the value of which will be
determined below) that we add quadratically to the errors given in
Cols.\ [3] and [4] of Tab.\ \ref{tab-resultados}.

\begin{figure*}[!h]
\centerline{\includegraphics[height=1\textwidth,angle=-90]{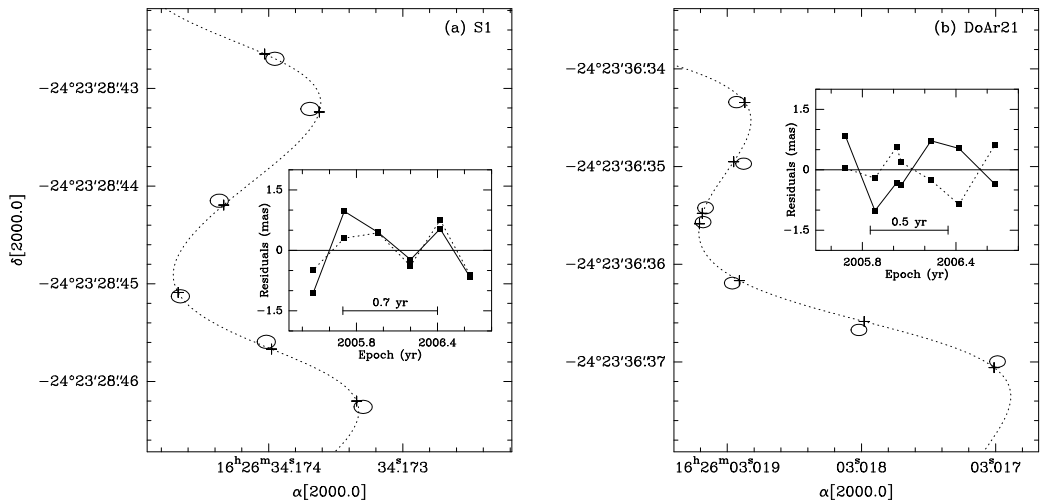}}
\caption
{\footnotesize{Measured positions and best fit for (a) S1, and (b)
DoAr~21. The observed positions are shown as ellipses, the size of
which represents the magnitude of the errors. The positions at each
epoch expected from the best fits are shown as $+$ signs. The insets
show the residuals (fit-observation) in right ascension (full line)
and declination (dotted line).}\label{fig-oph-1}}\end{figure*}

\medskip

\noindent
The displacements of both S1 and DoAr~21 were assumed to be uniform
and linear (Eqs.\ \ref{fit1}), and the best fit for S1 (Fig.\
\ref{fig-oph-1}a) and DoAr~21 (Fig.\ \ref{fig-oph-1}b) yields the
astrometric parameters listed in Tabs.\ \ref{tab-astrometry1} and
\ref{tab-astrometry2}. To obtain a reduced $\chi^2$ of 1 in both right
ascension and declination, one must add quadratically 0.062 ms, 0.67
mas for S1, and 0.053 ms, 0.57 mas for DoAr~21 to the statistical
errors listed in Cols.\ [3] and [4] of Tab.\
\ref{tab-resultados}. These figures include all the unmodeled sources
of positional shifts mentioned earlier. Interestingly, the residuals
of the fit to the S1 data (inset in Fig.\ \ref{fig-oph-1}a) are not
random, but seem to show a $\sim$ 0.7 yr periodicity, as expected from
the reflex motions. Similarly, the residuals from the fit to DoAr~21
seem to show a periodicity of $\sim$ 1.2 yr (Fig.\
\ref{fig-oph-1}b, inset), within the range of expected orbital periods
of that system. This suggests that the errors are largely dominated by
the unmodeled binarity of both sources, and that additional
observations designed to provide a better characterization of the
orbits ought to improve significantly the precision on the
trigonometric parallax determinations. The distance to S1 deduced from
the parallax calculated is 116.9$^{+7.2}_{-6.4}$ pc, while the
distance deduced for DoAr~21 is 121.9$^{+5.8}_{-5.3}$ pc.

\subsection{Kinematics of the sources in Taurus}\label{c5-absolute-kinematics}

\begin{figure*}[!h]
\centerline{\includegraphics[width=.8\textwidth,angle=270]{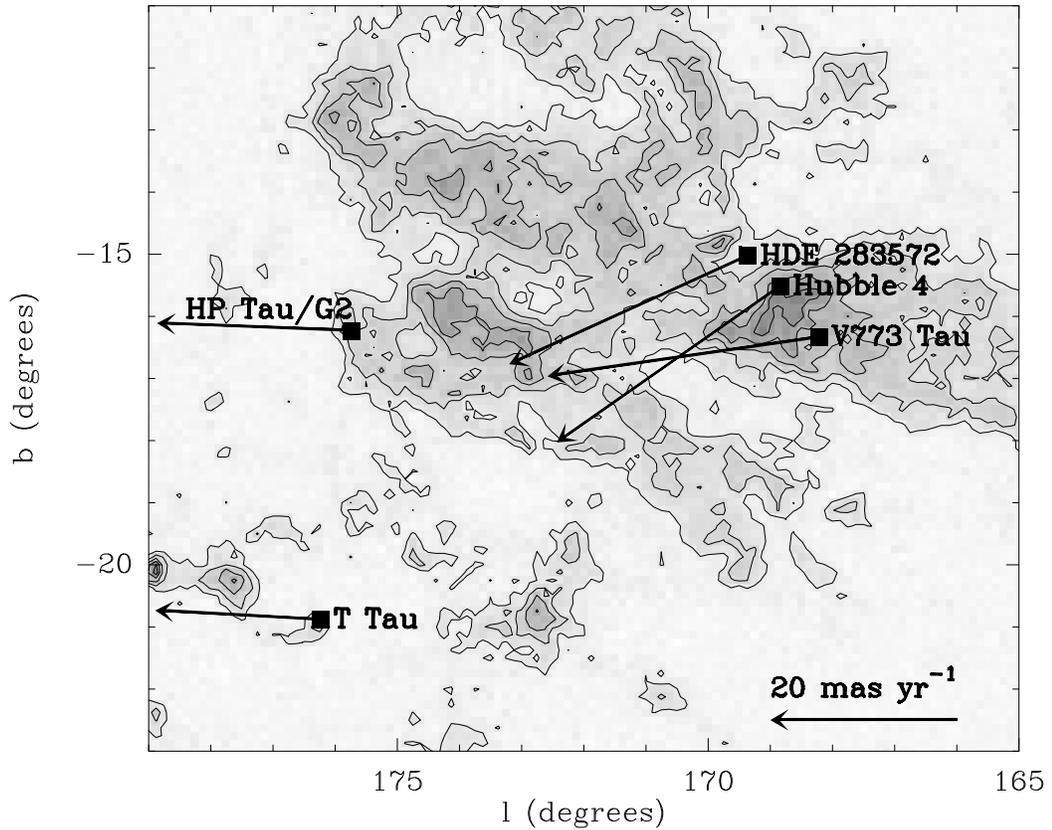}}
\caption
{\footnotesize{Positions and proper motions of T~Tau, Hubble~4,
HDE~283572, and HP~Tau/G2 superposed on the CO(1-0) map of Taurus from
Dame et al.\ (2001).}\label{fig-sfrtauro-2}}\end{figure*}

For Galactic sources, it is interesting to express the proper motions
in Galactic coordinates rather than in the equatorial system naturally
delivered by the VLBA. The results for four sources observed in Taurus
with the VLBA are given in columns 3 and 4 of Tab.\
\ref{tab-kinematics}. Interestingly, the proper motion of HP~Tau/G2 is
very similar to that of T~Tau, but significantly different from those
of Hubble~4 and HDE~283572, (which are themselves very similar to each
other). HP~Tau/G2 and T~Tau also happen to both be located on the
eastern side of the Taurus complex, whereas Hubble~4 and HDE~283572
are both around Lynds 1495 near the center of the complex (Fig.\
\ref{fig-sfrtauro-2}).

\begin{deluxetable}{lcrrccccccc}
\rotate
\tabletypesize{\scriptsize}
\tablecolumns{11}
\tablewidth{0pc}
\tablecaption{\footnotesize{Radial velocities, proper motions, heliocentric,
and peculiar velocities in Galactic coordinates for 4 sources in Taurus.}
\label{tab-kinematics}}
\tablehead{
\multicolumn{1}{c}{Source} &
\multicolumn{1}{c}{$V_r$} &  
\multicolumn{1}{c}{$\mu_\ell$cos$(b)$} &  
\multicolumn{1}{c}{$\mu_b$} & 
\multicolumn{1}{c}{$U$} & \multicolumn{1}{c}{$V$} & \multicolumn{1}{c}{$W$} & 
\multicolumn{1}{c}{$u$} & \multicolumn{1}{c}{$v$} & \multicolumn{1}{c}{$w$} &
References\tablenotemark{a} \\
\multicolumn{1}{c}{} &
\multicolumn{1}{c}{[km s$^{-1}$]} &
\multicolumn{2}{c}{[mas yr$^{-1}$]} &
\multicolumn{3}{c}{[km s$^{-1}$]} & 
\multicolumn{3}{c}{[km s$^{-1}$]} &
\multicolumn{1}{c}{} }
\startdata
T~Tau\tablenotemark{b}    & 19.1$\pm$1.2 & $+$17.76$\pm$0.03 &  $+$0.99$\pm$0.04 & $-$19.09 & $-$11.27 & $-$6.30  & $-$9.09 & $-$6.02 & $+$0.87 & 1,3 \\%
\\[-0.3cm]
Hubble~4                  & 15.0$\pm$1.7 & $+$23.94$\pm$0.12 & $-$16.74$\pm$0.15 & $-$14.96 & $-$12.66 & $-$14.30 & $-$4.96 & $-$7.41 & $-$7.13 & 1,3 \\%
\\[-0.3cm]
HDE~283572                & 15.0$\pm$1.5 & $+$25.53$\pm$0.05 & $-$11.61$\pm$0.06 & $-$15.89 & $-$13.07 & $-$10.84 & $-$5.89 & $-$7.82 & $-$3.67 & 1,2 \\%
\\[-0.3cm]
HP~Tau/G2                 & 17.7$\pm$1.8 & $+$20.90$\pm$0.07 &  $+$0.82$\pm$0.10 & $-$18.59 & $-$14.65 & $-$4.50  & $-$8.59 & $-$9.40 & $+$2.67 & 1,2 \\%
\enddata
\tablenotetext{a}{1=This work; 2=Walter et al.\ 1988; 3=Hartmann et al.\ 1986}
\tablenotetext{b}{The radial velocity and proper motions used here are those of T~Tau~N.
The radial velocities for T~Tau~Sa and T~Tau~Sb are available in Duchêne et al.\ (2002) and are very similar.}
\end{deluxetable}

\medskip

\noindent
The fifth star in Taurus, V773~Tau~A, is located about a degree
south-west of Hubble~4, and the distance are similar to those of
Hubble~4 and HDE~283572 (see Tab.\ \ref{tab-astrometry2}). Thus, we
expected the proper motions of V773~Tau~A to be like those for
Hubble~4 and HDE~283572. However the observed proper motion of
V773~Tau~A is from the northwest to southeast direction (see Fig.\
\ref{fig-v773-1}). We suspect that the peculiar proper motion of
V773~Tau~A is due to its companions, particularly V773~Tau~B. We are
currently analyzing this possibility, and here we will take only the
other four sources to describe the kinematics of Taurus.

\medskip

\begin{figure*}[!t]
\centerline{\includegraphics[width=.3\textwidth,angle=270]{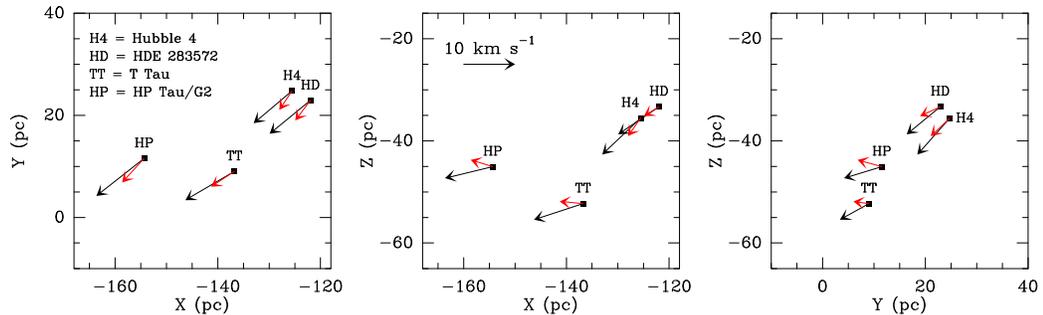}}
\caption
{\footnotesize{Heliocentric velocities (black arrows) and peculiar
velocities (red arrows) for the four stars in Taurus with VLBA-based
distance determinations.}\label{fig-sfrtauro-1}}\end{figure*}

\noindent
Knowing the distance to the sources with high accuracy, it is possible
to transform the observed proper motions into transverse velocities.
Combining this information with radial (Heliocentric) velocities taken
from the literature (second column of Tab.\ \ref{tab-kinematics}), it
becomes possible to construct the three-dimensional velocity
vectors. It is common to express these vectors on a rectangular
$(X,Y,Z)$ coordinate system centered in the Sun, with $X$ pointing
towards the Galactic center, $Y$ in the direction of Galactic
rotation, and $Z$ towards the Galactic North Pole. In this system, the
coordinates of the Heliocentric velocities will be written
$(U,V,W)$. As a final step, it is also possible to calculate the
peculiar velocity of the stars. This involves two stages: first, the
peculiar motion of the Sun must be removed to transform the
Heliocentric velocities into velocities relative to the LSR. Following
Dehnen \& Binney (1998), we will use $u_0$ = +10.00 km s$^{-1}$, $v_0$
= +5.25 km s$^{-1}$, and $w_0$ = +7.17 km s$^{-1}$ for the peculiar
velocity of the Sun expressed in the coordinate system defined
above. The second stage consists in estimating the difference in
circular velocity between Taurus and the Sun, so the peculiar
velocities are expressed relative to the LSR appropriate for Taurus,
rather than relative to the LSR of the Sun. This was done assuming the
rotation curve of Brand \& Blitz (1993), and represents a small
correction of only about 0.3 km s$^{-1}$. We will write $(u,v,w)$ the
coordinates of the peculiar velocity of the sources. Both $(U,V,W)$
and $(u,v,w)$ are given in Tab.\ \ref{tab-kinematics} for the four
sources considered here. Their projections onto the $(X,Y)$, $(X,Z)$,
and $(Y,Z)$ planes are shown in Fig.\ \ref{fig-sfrtauro-1}.

\medskip

\noindent
The mean heliocentric velocity and the velocity dispersion of the four
sources are:
\begin{equation}\begin{split}\label{ec-UVW}
U &= -17.1 \pm 1.7 \mbox{~km s$^{-1}$}\nonumber \\%
V &= -12.9 \pm 1.2 \mbox{~km s$^{-1}$}\nonumber \\%
W &= -9.0 \pm 3.8 \mbox{~km s$^{-1}$}.\nonumber
\end{split}\end{equation}
These values are similar to those reported by Bertout \& Genova (2006)
for a larger sample of young stars in Taurus with optically measured
proper motions. Note that the velocity dispersion in the $W$ direction
is somewhat artificially high because Hubble~4 and HDE~283572 on the
one hand, and T~Tau and HP~Tau/G2 on the other, clearly have different
vertical velocities. They likely belong to two different kinematic
sub-groups.

\medskip

\noindent
The mean peculiar velocity of the four sources considered here is:
\begin{equation}\begin{split}\label{ec-uvw}
u &= -7.1 \pm 1.7 \mbox{~km s$^{-1}$}\nonumber \\%
v &= -7.7 \pm 1.2 \mbox{~km s$^{-1}$}\nonumber \\%
w &= -1.8 \pm 3.8 \mbox{~km s$^{-1}$}.\nonumber
\end{split}\end{equation}

\medskip

\noindent
We argue that this is a good estimate of the mean peculiar velocity of
the Taurus complex. This velocity is almost entirely in the $(X,Y)$
plane. Thus, although Taurus is located significantly out of the
mid-plane of the Galaxy (about 40 pc to its south), it appears to be
moving very little in the vertical direction. The motion in the
$(X,Y)$ plane, on the other hand, is fairly large, leading to a total
peculiar velocity $(u^2+v^2+w^2)^{0.5}$ $=$ 10.6 km
s$^{-1}$. According to Stark \& Brand (1989), the one-dimensional
velocity dispersion of giant molecular clouds within 3 kpc of the Sun
is about 8 km s$^{-1}$. As a consequence, each component of the
peculiar velocity of a given molecular cloud is expected to be of that
order, and our determination of the mean peculiar velocity of Taurus
is in reasonable agreement with that prediction. Another useful
comparison is with the velocity dispersion of young main sequence
stars. For the bluest stars in their sample (corresponding to early A
stars), Dehnen \& Binney (1998) found velocity dispersions of about 6
km s$^{-1}$ in the vertical direction, and of 10-14 km s$^{-1}$ in the
$X$ and $Y$ directions. The young stars in Taurus are significantly
younger than typical main sequence early A stars, so one would expect
young stars in Taurus to have peculiar velocities somewhat smaller
than 6 km s$^{-1}$ in the vertical direction, and than 10--14 km
s$^{-1}$ in the $X$ and $Y$ directions. This is indeed what is
observed. Note, however, that Taurus is not among the star-forming
regions with the smallest peculiar velocities. In Orion, Gómez et al.\
(2005) found a difference between expected and observed proper motions
smaller than 0.5 km s$^{-1}$.

\subsection{Three-dimensional structure of Taurus}\label{c5-absolute-3dtaurus}

Two of the five objects observed in Taurus have measured Hipparcos
parallaxes (Bertout et al.\ 1999): T~Tau~Sb with $5.66\pm1.58$ mas,
and HDE~283572 with $7.81\pm1.30$ mas. Our results are consistent with
these values, but one to two orders of magnitude more accurate. Also,
the parallax of both Hubble~4 and HDE~283572 were estimated by Bertout
\& Genova (2006) using a modified convergent point method. Their
results ($8.12\pm1.50$ mas for Hubble~4 and $7.64\pm1.05$ mas for
HDE~283572) are also consistent with our results, but again more than
one order of magnitude less accurate. The distance to V773~Tau~A had
been obtained using radio VLBI observations before ($d=148.4\pm5.3$
pc; Lestrade et al.\ 1999). Our result is only marginally consistent
with this earlier figure, and we argue that the discrepancy is due to
the fact that Lestrade et al.\ (1999) did not model the binarity of
the source in their analysis. We obtained a mostly independent
estimate of the distance to V773~Tau~A by modeling the physical orbit
of the binary using a combination of optical radial velocity
measurements, Keck Interferometer observations, and our own VLBA data
(see Sect.\ \ref{c5-relative}). The distance obtained by this
alternative method was found $134.5\pm3.2$ pc, in excellent agreement with
the value obtained from our parallax measurement, but again only
marginally consistent with the older VLBI value.

\medskip

\noindent
Taking the mean of the five VLBA-based parallax measurements
available, we can estimate the mean parallax to the Taurus complex to
be $\bar{\pi}$ = 7.18 mas. This corresponds to a mean distance of
$\bar{d}=139$ pc, in good agreement with previous estimates (Kenyon et
al.\ 1994). The angular size of Taurus is about 10$^\circ$,
corresponding to a physical size of roughly 25 pc. It would be natural
to expect that the depth of Taurus might be similar, since HP~Tau/G2
is about 30 pc farther than Hubble~4, HDE~283572 or V773~Tau~A. This
has a trivial but important consequence: even if the mean distance of
the Taurus association were known to infinite accuracy, one would
still make errors as large as 10--20\% by using the mean distance
indiscriminately for all sources in Taurus. To reduce this systematic
source of error, one needs to establish the three-dimensional
structure of the Taurus association, and observations similar to those
presented here currently represent the most promising avenue toward
that goal. Indeed, the observations of the five stars presented here
already provide some hints of what the three-dimensional structure of
Taurus might be. Hubble~4, HDE~283572, and V773~Tau~A which were found
to be at about 130 pc, are also located in the same portion of Taurus,
near Lynds 1495. T~Tau~Sb is located in the southern part of Taurus
near Lynds 1551, its tangential velocity is clearly different from
that of Hubble~4, HDE~283572, and V773~Tau~A, and it appears to be
somewhat farther from us. Finally, HP~Tau/G2 is located near the
(Galactic) eastern edge of Taurus, and it is the farthest of the four
sources considered here. Although additional observations are needed
to draw definite conclusions, our data, therefore, suggest that the
region around Lynds 1495 corresponds to the near side of the Taurus
complex at about 130 pc, while the eastern side of Taurus corresponds
to the far side at 160 pc. The region around Lynds 1551 and T~Tau~Sb
appears to be at an intermediate distance of about 147 pc.

\medskip

\noindent
Taurus has long been known to present a filamentary structure. The two
main filaments are roughly parallel to one another, and have an axis
ratio of about 7:1. Our observations suggest that these filaments are
oriented nearly along the line of sight, i.e.\ roughly along the
Galactic center--anticenter axis. This peculiar orientation might
indeed explain the low star-forming efficiency of Taurus compared with
other nearby star-forming regions (Ballesteros-Paredes et al., 2009).

\subsection{Three-dimensional structure of Ophiuchus}\label{c5-absolute-3doph}

Ophiuchus is composed of a compact core, only about 2 pc across, and
filamentary structures (called ``streamers'') extending (in
projection) to about 10 pc. The Ophiuchus core is sufficiently compact
that we do not expect to resolve any structure along the line of
sight. Thus, the weighted mean of the two parallaxes for S1 and
DoAr~21 is 8.33 $\pm$ 0.30, corresponding to a distance of
120.0$^{+4.5}_{-4.3}$ pc. Since both S1 and DoAr~21 are {\it bone
fide} members of the Ophiuchus core, this figure must represent a good
estimate of the distance to Ophiuchus. It is in good agreement with
several recent determinations (e.g.\ de Geus et al.\ 1989, Knude \&
Hog 1998, Lombardi et al.\ ibid), but with a significantly improved
relative error of 4\%.

\begin{figure*}[!t]
\centerline{\includegraphics[height=0.5\textwidth,angle=0]{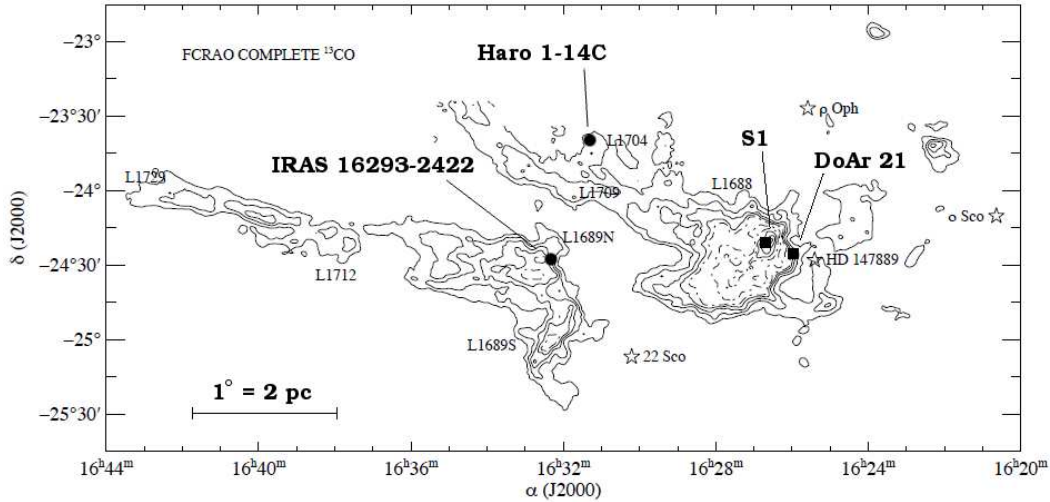}}
\caption
{\footnotesize{Positions of four stars superposed on the $^{13}$CO
map of Ophiuchus from Ridge et al.\ (2006). The two stars studied here
(S1 and DoAr~21) are shown as a squares; Haro 1-14C and IRAS
16293--2422 are shown as a circles.}\label{fig-sfroph}}\end{figure*}

\medskip

\noindent
There could potentially be distance gradients of several parsecs
across the streamers. We note, however, that Schaefer et al.\ (2008)
determined the physical orbit of the binary system Haro 1-14C, and
deduced a distance of $111\pm19$ pc, in good agreement with our
determination. Haro 1-14C is located in the northern streamer
associated with the darks clouds L1709/L1704 (see Fig.\
\ref{fig-sfroph}), so the result of Schaefer et al.\ (2008) suggests
that that streamer is, if anything, somewhat closer that the core.
This is, indeed, in agreement with recent results of Lombardi et al.\
(2008). On the other hand, Imai et al.\ (2007) used the Japanese VLBI
system (VERA) to determine the parallax to the very young protostar
IRAS 16293--2422 deeply embedded in the southern Ophiuchus streamer
(in L1689N, see Fig.\ \ref{fig-sfroph}). They obtain a distance of
$178^{+18}_{-37}$ pc, which would be more consistent with the older
value of 165 pc. Even including the streamers, Ophiuchus is only 10 pc
across in projection, so it is unlikely to be 60 pc deep. Thus, if the
results of Imai et al.\ (2007) are confirmed, they would indicate the
existence of several unrelated star-forming regions along the line of
sight. More observations --some of which have already been collected
and partially analyzed-- will be necessary to settle this issue.

\section{Variability}\label{c5-variability}

In 1953 Haro \& Morgan discovered that flares are common in T~Tauri
stars and stellar associations, implying that the flare process is
stronger in the younger stars. Since T~Tauri stars present high levels
of magnetic activity (Feigelson \& Montmerle, 1999), and that radio
emission generally depends on the characteristics of the magnetic
field (see Chapter \ref{chap-emission}), the variability due to flares
is expected for non-thermal sources.

\begin{deluxetable}{lrrrrrrrrr}
\tabletypesize{\scriptsize}
\tablecolumns{10}
\tablewidth{0pc}
\tablecaption{\footnotesize{Mean brightness temperature and mean fluxes for all sources in the thesis.}\label{tab-flux}}
\tablehead{
\multicolumn{1}{c}{~~~~~~~~~Source~~~~~~~~~}          &
\multicolumn{1}{c}{$\bar{T_{\rm b}}$}                 &
\multicolumn{1}{c}{$\bar{f_\nu}$}                     &
\multicolumn{1}{c}{$rms$}                             &
\multicolumn{1}{c}{$f_\nu^{\rm ~max}$}                &
\multicolumn{1}{c}{$f_\nu^{\rm ~min}$}                &
\multicolumn{1}{c}{}                                  &
\multicolumn{1}{c}{~~~$F_{1}$ \tablenotemark{*}~~~}   &
\multicolumn{1}{c}{~~~$F_{2}$ \tablenotemark{**}~~~}  &
\multicolumn{1}{c}{~~~$F_{3}$ \tablenotemark{***}~~~} \\[0.1cm]
\cline{3-6}                                         \\[-0.335cm]
\cline{8-10}                                        \\[-0.2cm]
\multicolumn{1}{c}{}           &
\multicolumn{1}{c}{[$10^7$ K]} &
\multicolumn{4}{c}{[~mJy~]}    &
\multicolumn{1}{c}{}           &
\multicolumn{3}{c}{[~\%~]}     }
\startdata
T~Tau~Sb\dotfill    &  17.68 &  1.68 &  0.59 &  2.30 & 0.92 &&   82  &   37  &  45  \\%
Hubble~4\dotfill    &  19.89 &  1.59 &  1.41 &  4.66 & 0.65 &&  252  &  193  &  59  \\%
HDE~283572\dotfill  &  29.78 &  2.50 &  2.43 &  7.13 & 0.51 &&  265  &  185  &  80  \\%
HP~Tau/G2\dotfill   &  11.75 &  1.07 &  0.72 &  3.06 & 0.63 &&  227  &  186  &  41  \\%
V773~Tau~A\dotfill  & 127.75 & 14.34 & 11.15 & 41.16 & 0.87 &&  281  &  187  &  94  \\%
V773~Tau~Aa\dotfill &  52.42 &  6.15 &  3.92 & 36.44 & 0.87 &&  578  &  493  &  86  \\%
V773~Tau~Ab\dotfill &  97.36 &  8.25 &  7.50 & 28.89 & 0.76 &&  341  &  250  &  91  \\%
S1\dotfill          &  46.73 &  4.81 &  1.15 &  7.03 & 3.29 &&   72  &   46  &  32  \\%
DoAr~21\dotfill     &  54.37 &  5.48 &  7.09 & 20.34 & 0.39 &&  364  &  271  &  93  \\%
\enddata
\tablenotetext{*~~}{$F_1=\bar{f_{\nu}}^{-1}[f_{\nu}^{\rm max}-f_{\nu}^{\rm min}]\times100$}
\tablenotetext{**~}{$F_2=\bar{f_{\nu}}^{-1}[f_{\nu}^{\rm max}-\bar{f_{\nu}}]\times100$}
\tablenotetext{***}{$F_3=\bar{f_{\nu}}^{-1}[\bar{f_{\nu}}-f_{\nu}^{\rm min}]\times100$}
\end{deluxetable}

\medskip

\begin{figure*}[!h]
\centering
\begin{tabular}{cc}
\includegraphics[width=.4\textwidth,angle=0]{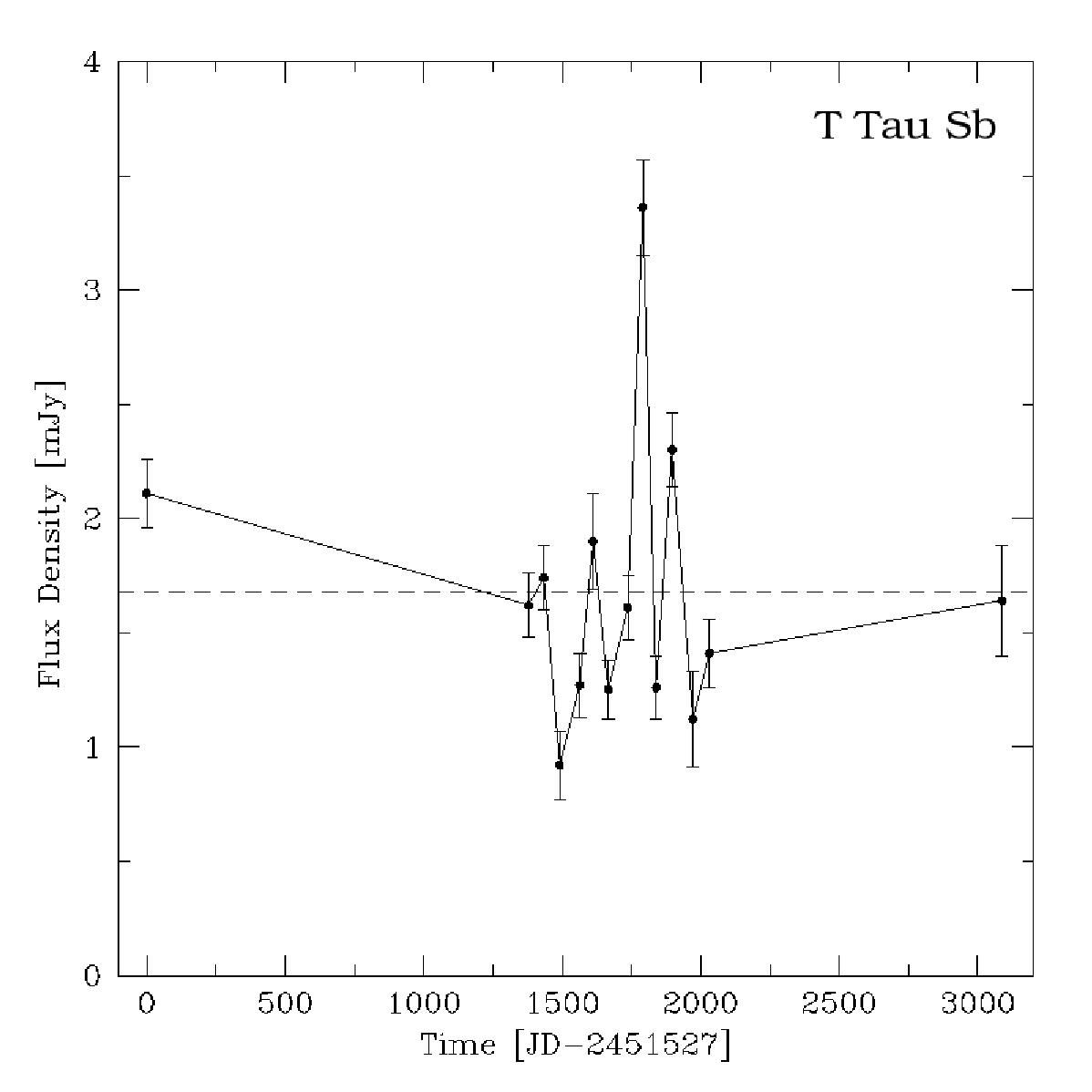}     &
\includegraphics[width=.4\textwidth,angle=0]{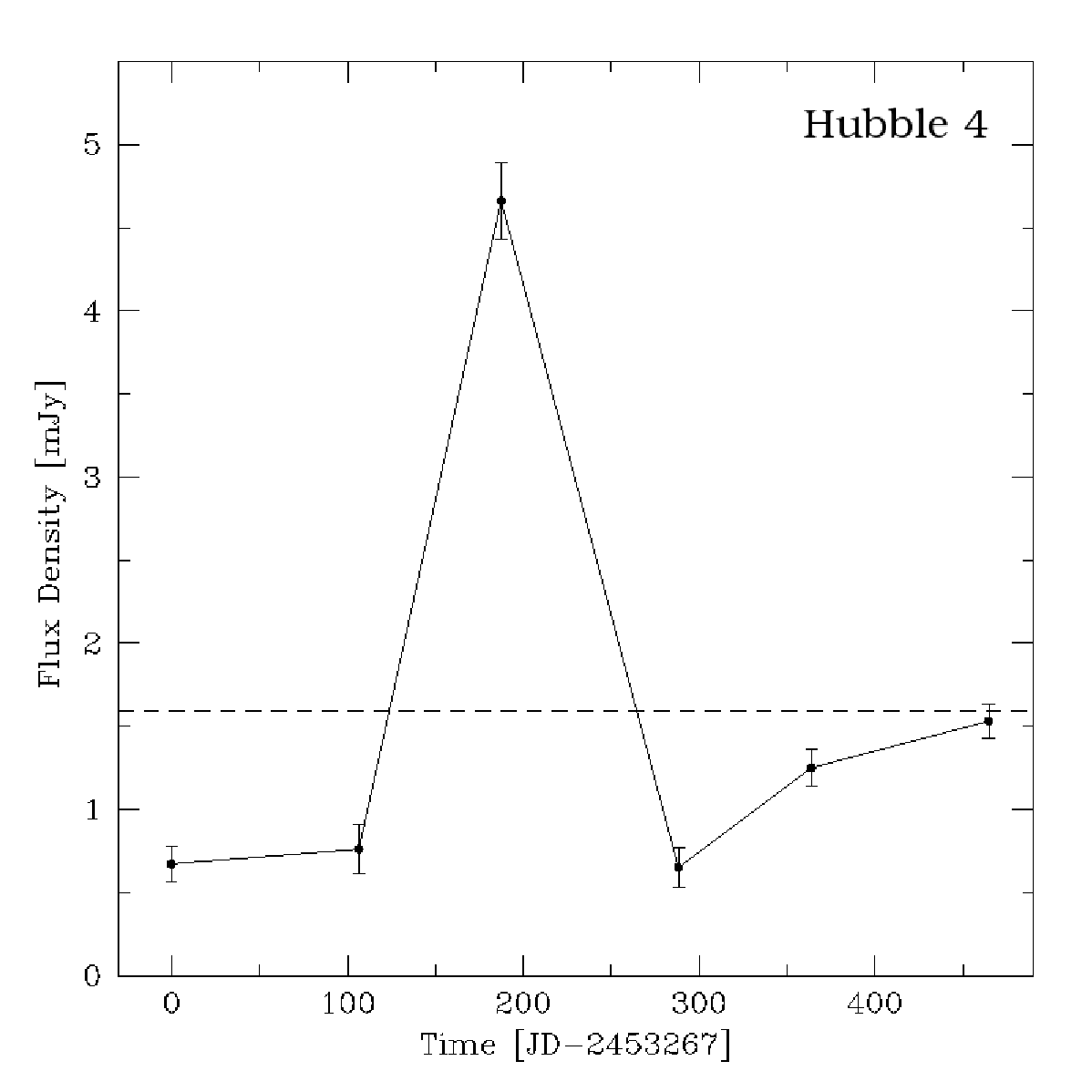}  \\
\includegraphics[width=.4\textwidth,angle=0]{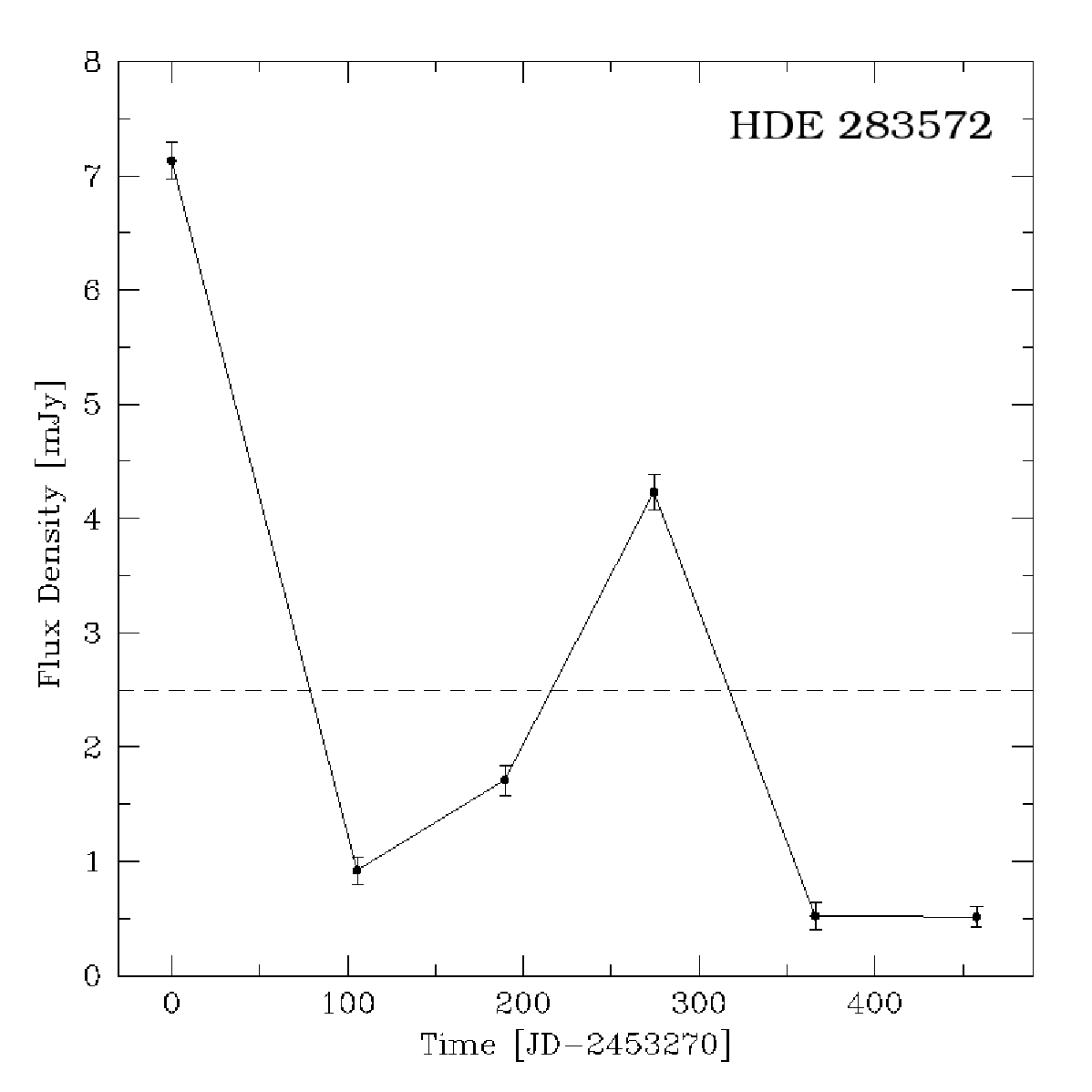}&
\includegraphics[width=.4\textwidth,angle=0]{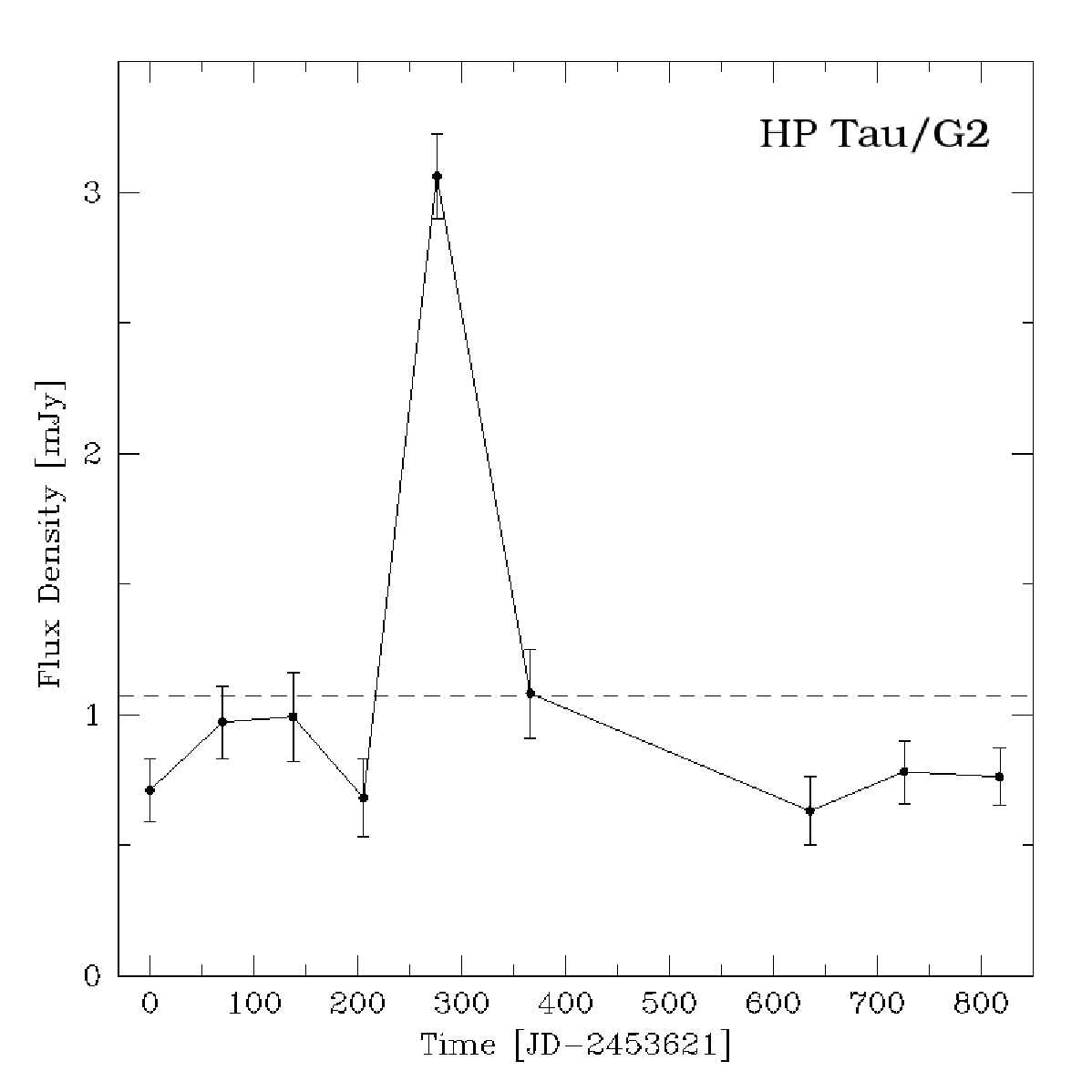}    \\
\includegraphics[width=.4\textwidth,angle=0]{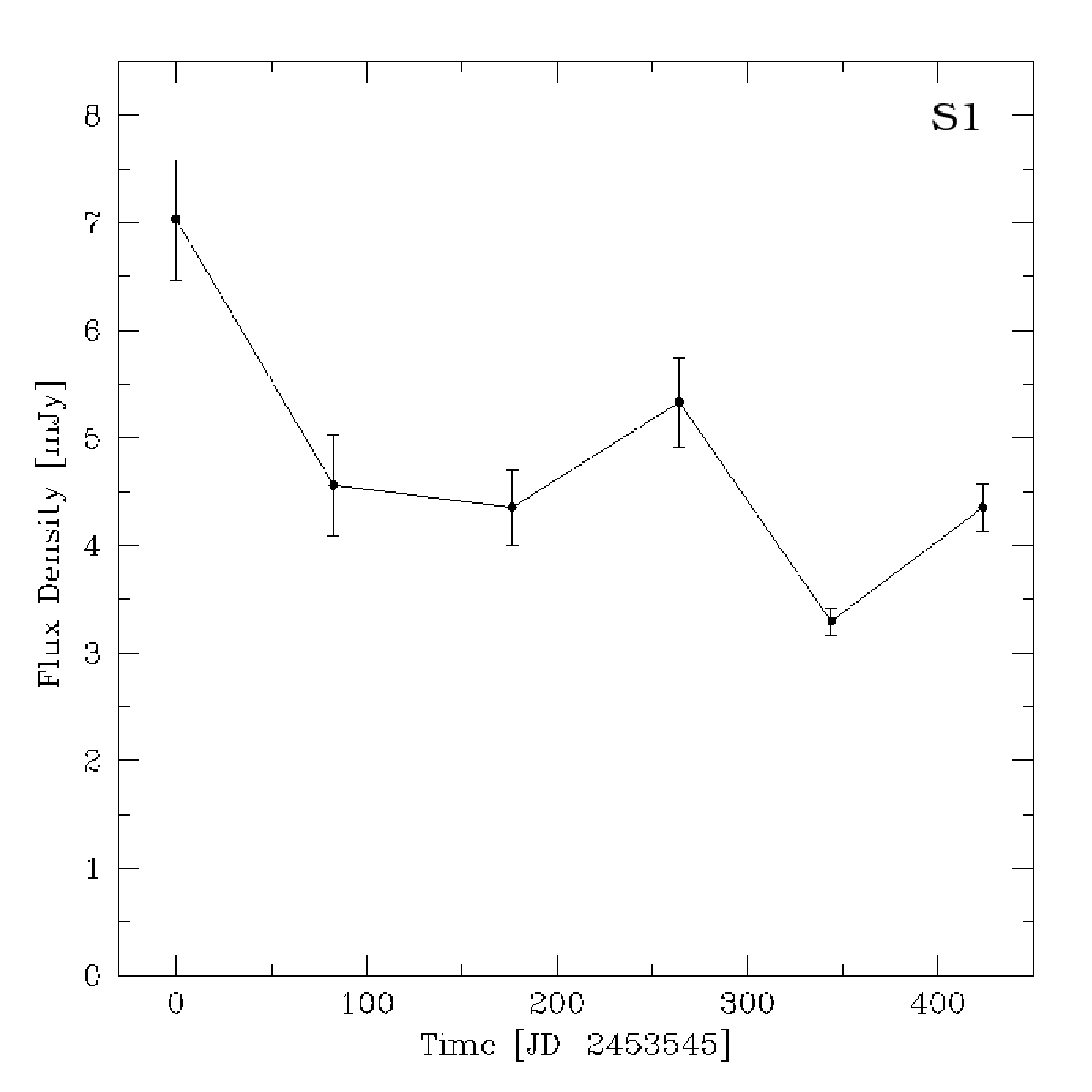}       &
\includegraphics[width=.4\textwidth,angle=0]{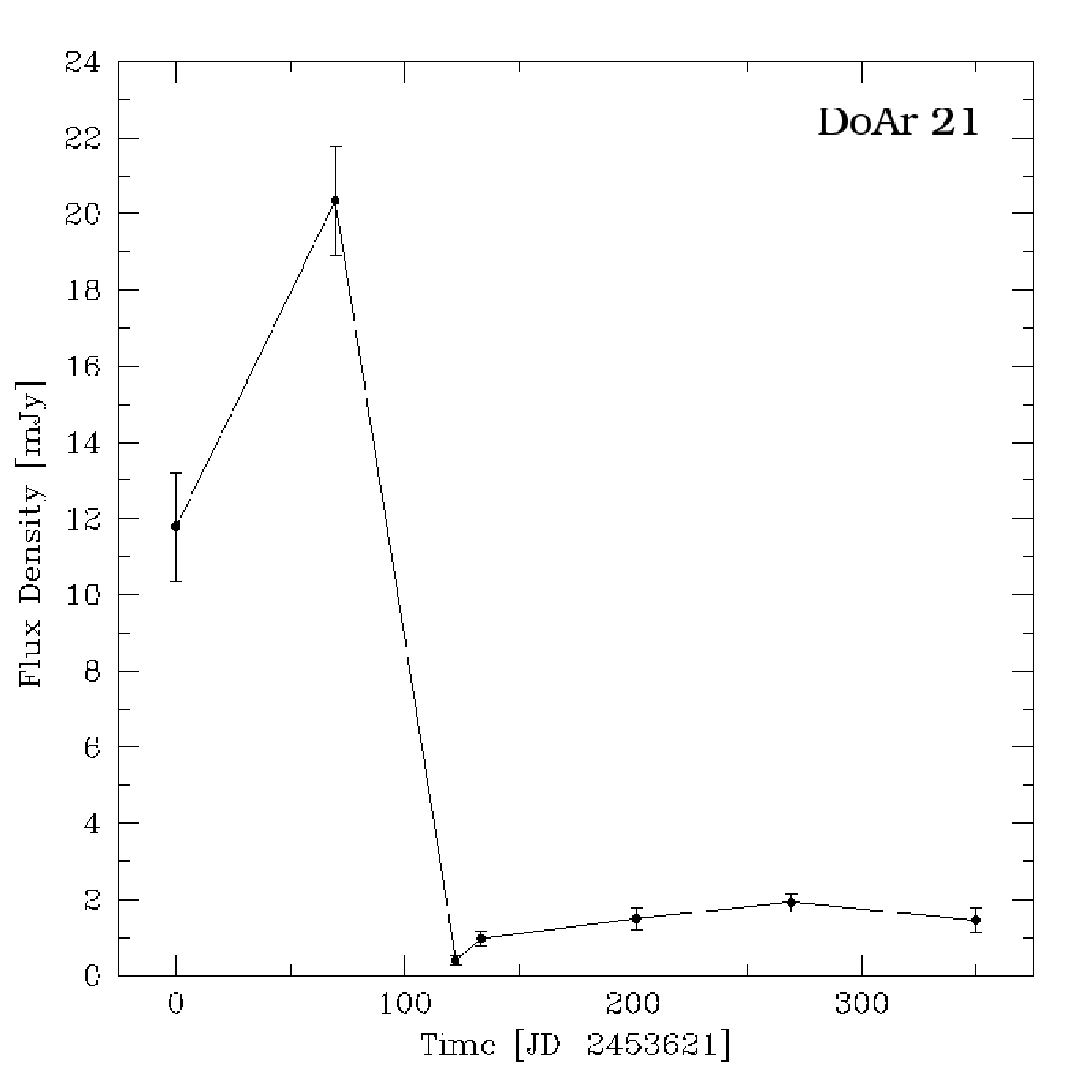}
\end{tabular}
\caption
{\footnotesize{Flux evolution of T~Tau~Sb, Hubble~4, HDE~283572,
HP~Tau/G2, S1, and DoAr~21. The horizontal dotted lines in correspond
to the mean flux density of the observations. Note that all six
sources have variability on timescales from months to
years.}\label{fig-6flujos}}\end{figure*}

\begin{figure*}[!b]
\centering
\begin{tabular}{cc}
\includegraphics[width=.4\textwidth,angle=0]{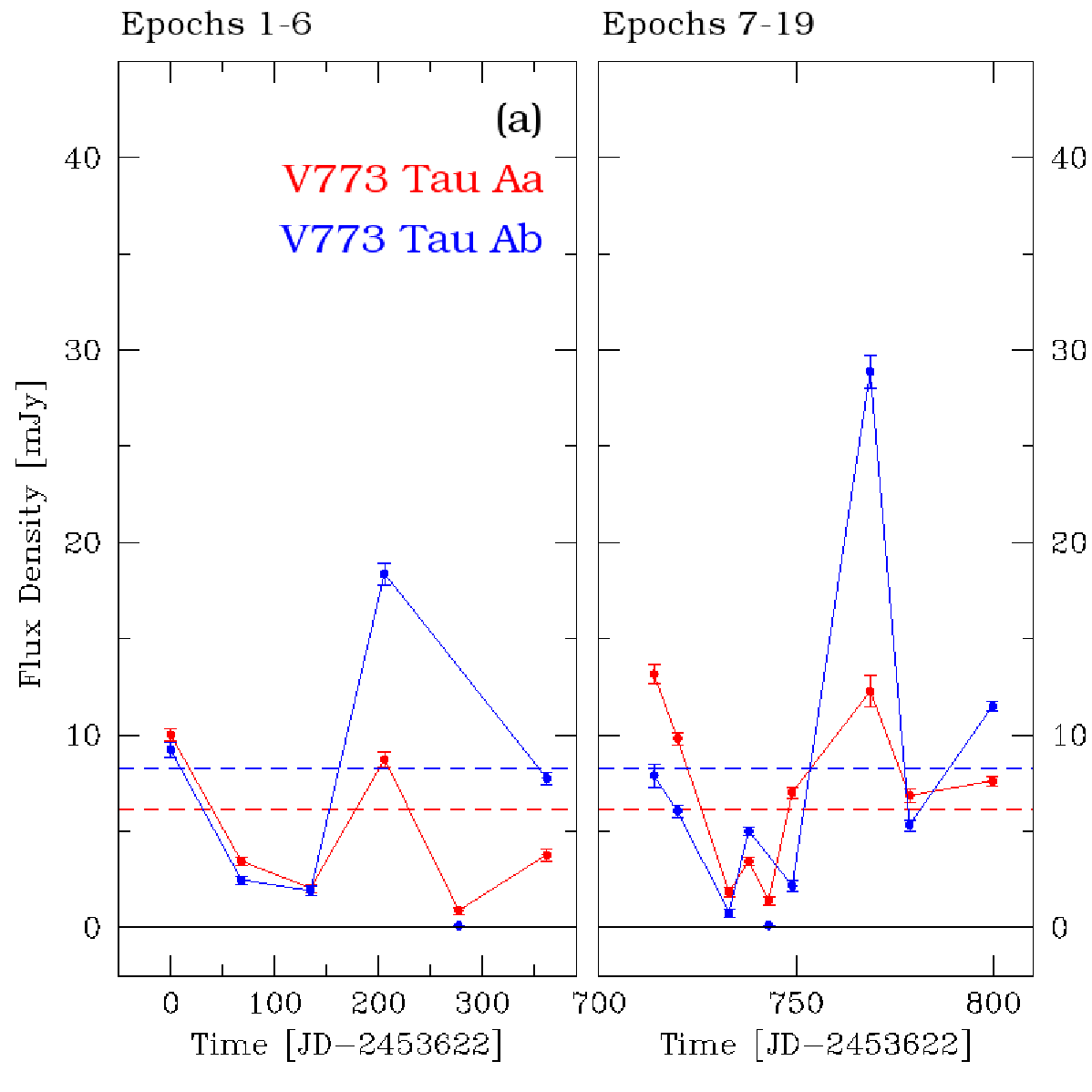}       &
\includegraphics[width=.4\textwidth,angle=0]{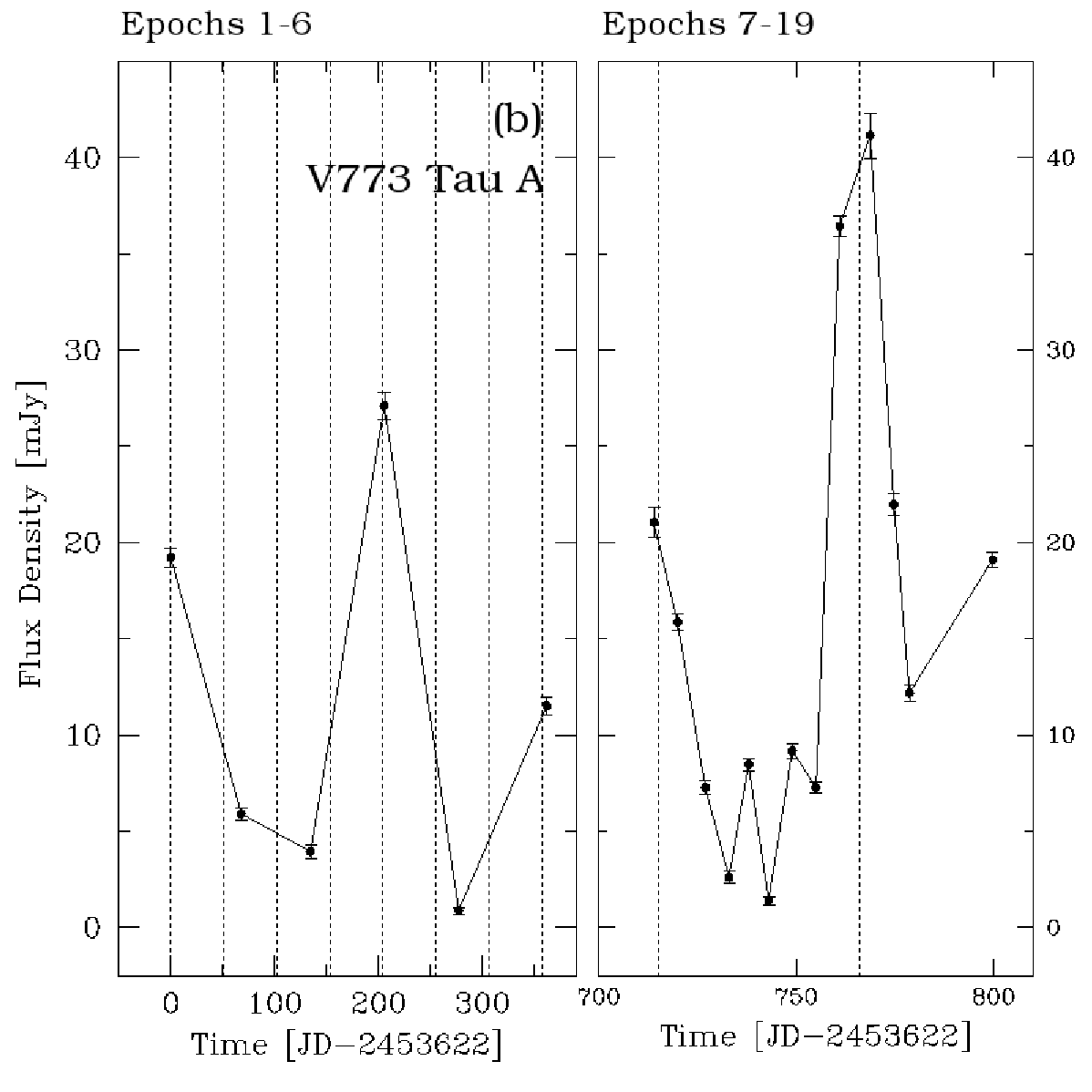}
\end{tabular}
\includegraphics[width=.5\textwidth,angle=0]{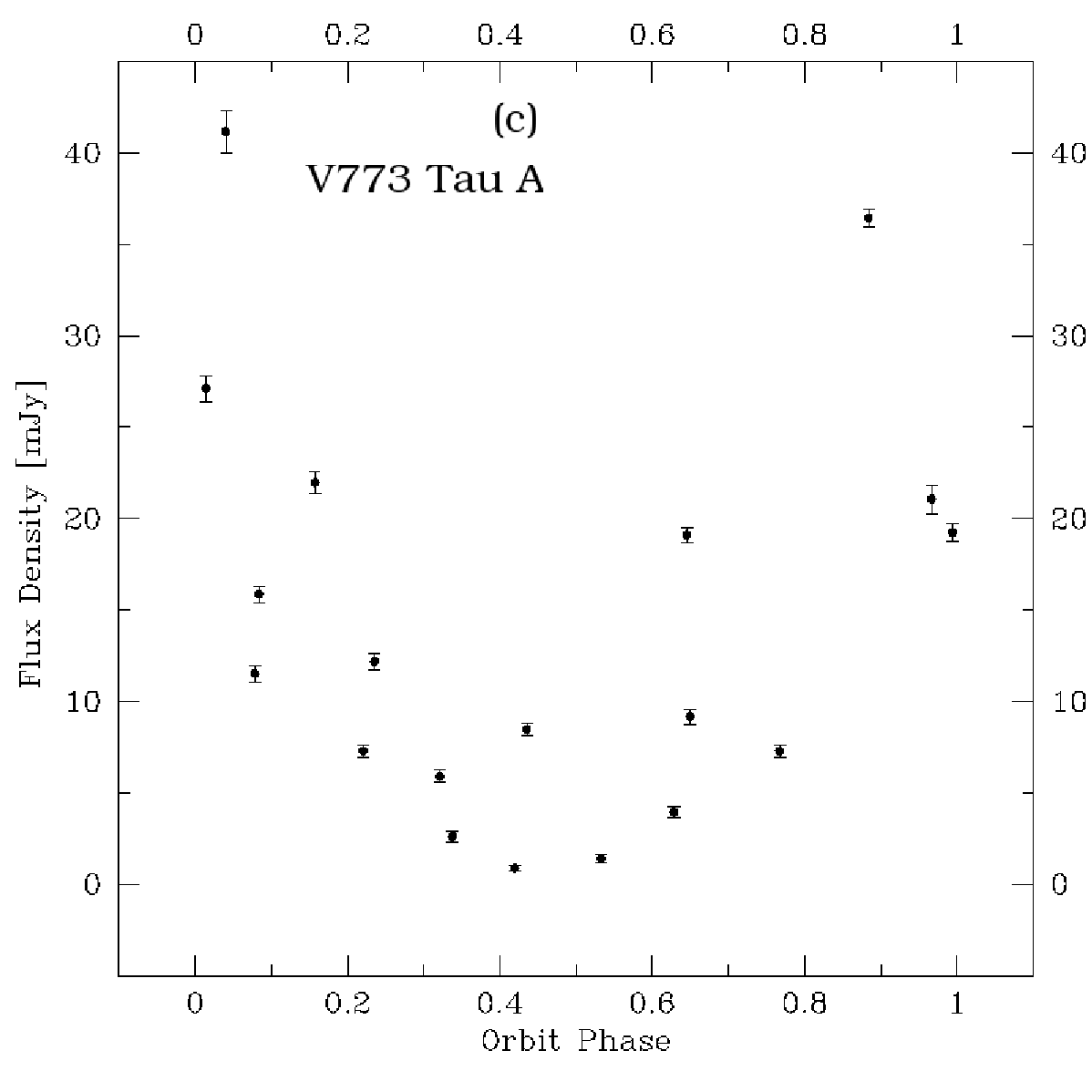}
\caption
{\footnotesize{In left side of (a) and (b): the first 6 epochs of
V773~Tau~A that were obtained every 3 months. In the right side of (a)
and (b): the last 13 epochs that were obtained over one orbital period
of the system (observations every $\sim4$ days). (a) Flux evolution
for each member of the system. It was not taken in to account the flux
for the epochs where we did not resolve the sources (epochs 9, 14, 15,
17). For the primary (in \textit{red}) we used the flux for the
remaining 15 epochs, and for the secondary (in \textit{blue}) we used
the flux for the 13 epochs where we detected the source (see Sect.\
\ref{c5-absolute}). The \textit{horizontal dotted lines} correspond to
the mean flux density of the observations. Note that the flux for the
primary is lower than that for the secondary in almost all epochs. (b)
Total flux evolution in the system at all 19 epochs. The periastron
passages are shown in \textit{vertical dotted lines}. Note that the
total flux at periastron is highest than that at apoastron. (c) Total
flux density of the system vs.\ orbit
phase.}\label{fig-flujo-v773-1}}\end{figure*}

\noindent
In Fig.\ \ref{fig-6flujos} we present the flux evolution for T~Tau~Sb,
Hubble~4, HP~Tau/G2, S1, and DoAr~21. A similar figure for V773~Tau~A
system is shown in Fig.\ \ref{fig-flujo-v773-1}. The dotted horizontal
lines in each plot correspond to the mean flux density of all
observations. All seven sources have variability on timescales from
months to years. In some cases the sources underwent flaring events
that move the average flux up. To give a classification scheme for the
young stars in the thesis, we constructed Table \ref{tab-flux} that
contains the mean brightness temperature and the mean flux density in
Cols.\ [2] and [3]. Additional columns were added in order to
understand the degree of variability for each source: maximum and
minimum flux densities in Cols.\ [5] and [6]; the difference between
maximum and minimum flux density in Col.\ [7]; the difference between
maximum and mean flux density in Col.\ [8]; finally, the difference
between minimum and mean flux density in Col.\ [9]. Note that the last
three columns are weighted by the mean flux density. With these values
in mind, we can give the next classification scheme in accordance with
the flux evolution shown in light curves as well.

\medskip

\noindent
\textbf{Small variability}---
The mean fluxes of T~Tau~Sb and S1 in our data are 1.68 and 4.81 mJy
with a dispersion of 0.59 and 1.15 mJy, respectively. The values for
$r.m.s.$, $F_1$, $F_2$ and $F_3$ in Tab.\ \ref{tab-flux} are smaller for
both T~Tau~Sb and S1 respect to the other sources, this shows that
T~Tau~Sb and S1 are variable at low levels.

\medskip

\noindent
\textbf{Single flares}---
HP~Tau/G2, Hubble~4 and HDE~283572 have small $r.m.s.$ but high $F_1$,
these values mean that the flux of the sources were fairly constant
around a few mJy at all of our observations, except for one epoch (or
two in the case of HDE~283572) where the sources underwent flaring
events.

\medskip

\noindent
\textbf{Multiple flares}---
V773~Tau~A and DoAr~21 were detected with highly variable flux
densities. The values for $r.m.s.$, $F_1$, $F_2$, and $F_3$ are high for
both V773~Tau~A and DoAr~21 sources. In the case of V773~Tau~A the
flux was found to be highest near periastron, and lowest around
apoastron (see Fig.\ \ref{fig-flujo-v773-1}d), in agreement with the
results by Massi et al.\ (2002, 2006), who showed that the variability
had the same periodicity as the orbital motion, with the radio flux
being highest at periastron. Interestingly, DoAr~21 was found to be
double during our second observation. This suggests that the same
mechanism that enhances the radio emission when the two binary
components are nearest, might be at work in both objects. The
separation between the two components of DoAr~21 in our second
observation is about 5 mas. This value, of course, corresponds to the
projected separation; the actual distance between them must be
somewhat larger. Moreover, if the mechanisms at work in DoAr~21 and
V773~Tau~A are similar, then DoAr~21 must have been near periastron
during our second epoch, and the orbit must be somewhat eccentric. As
a consequence of these two effects, the semi-major axis of the orbit
is likely to be a few times larger than the measured separation
between the components at our second epoch, perhaps 10 to 15 mas. At
the distance of DoAr~21, this corresponds to 1.2 to 1.8 AU. For a mass
of 2.2 \Msun, the corresponding orbital period is 0.4 to 1.3 yr, and
one would expect the source to oscillate with this kind of
periodicity.

\begin{figure*}[!t]
\centering
\begin{tabular}{cc}
\includegraphics[width=.5\textwidth,angle=0]{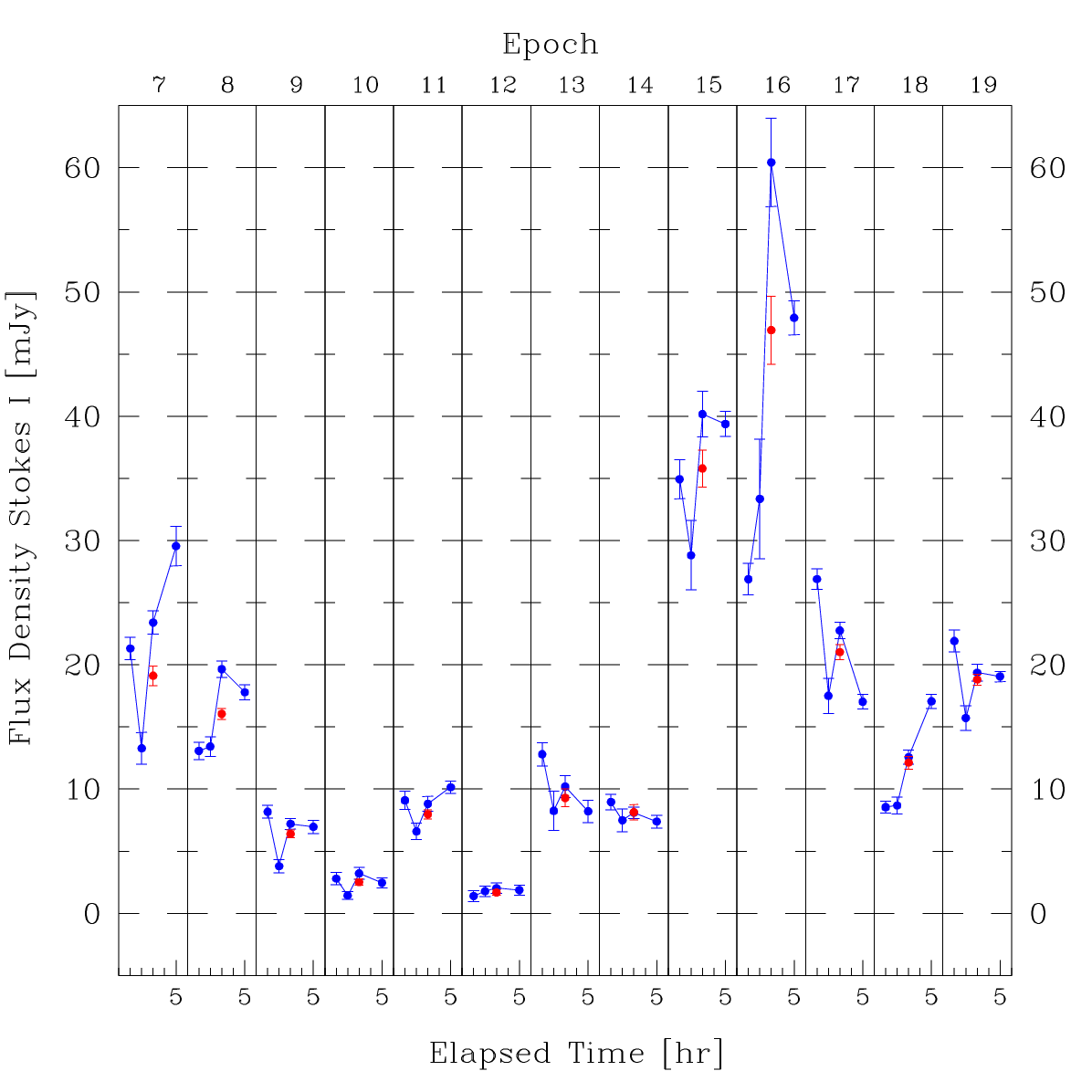} &
\includegraphics[width=.5\textwidth,angle=0]{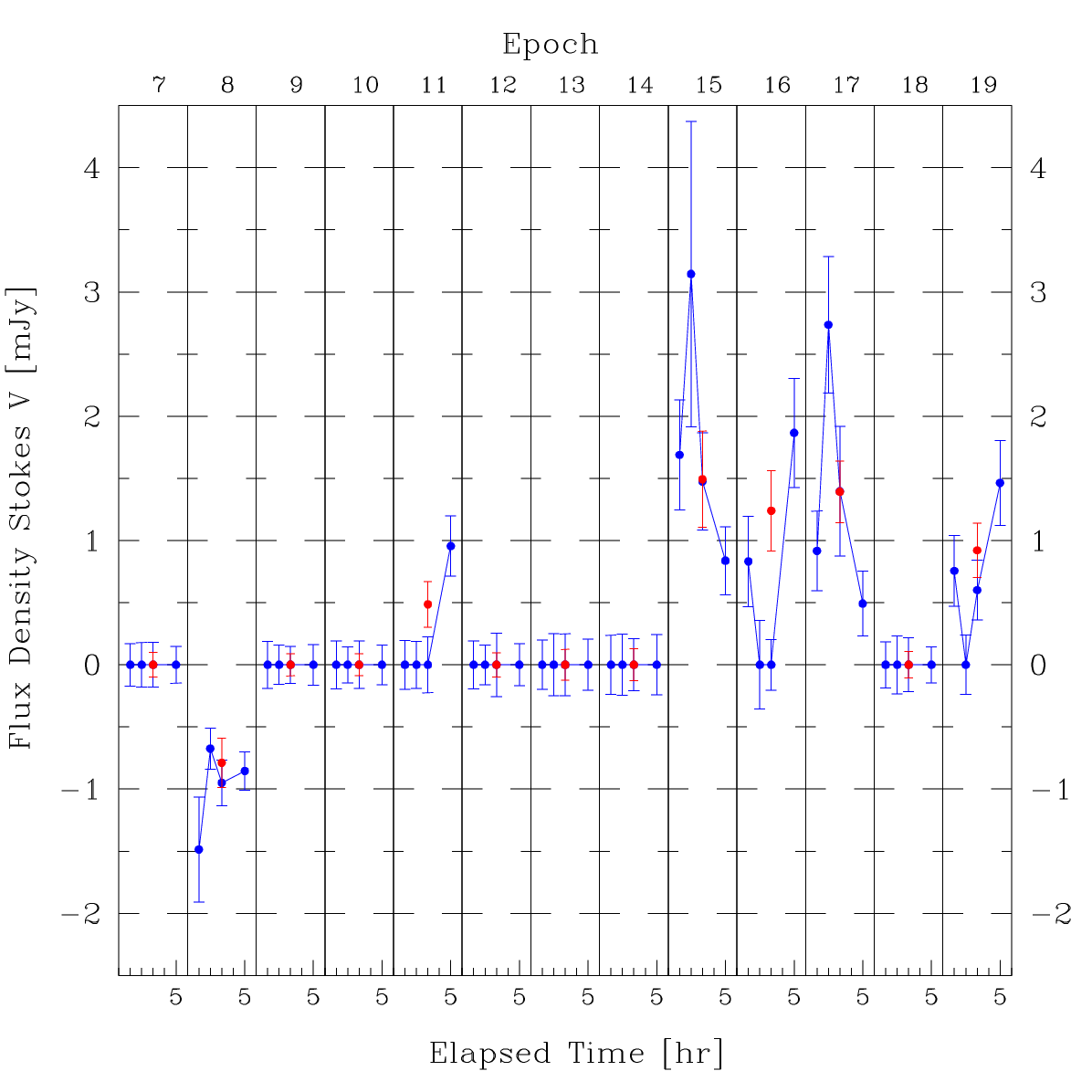}
\end{tabular}
\caption
{\footnotesize{The last 13 epochs of V773~Tau~A system were observed
for 5 hours each one. Each panel in figures correspond to one
epoch. \textit{Blue points} show the flux evolution in timescales of 5
hours, while \textit{red points} correspond to the average total flux
for the whole epoch. Total flux density (stokes I) on left side, and
circular polarization (stokes V) on right side.
}\label{fig-flujo-v773-23}}\end{figure*}

\medskip

\noindent
\textbf{Short timescale variability}---
As we mentioned before, the flux density for V773~Tau~A system tends
to increase when the sources are near periastron. This is clear in
Fig.\ \ref{fig-flujo-v773-1}. In order to investigate how is the flux
evolution at short timescales, we used epochs from 7 to 19 for imaging
the system each hour. Unfortunately it was not possible for the first
six observations because our observation time was only 2 hours (see
Tab.\ \ref{tab-observaciones}). In Fig.\ \ref{fig-flujo-v773-23} is
shown the flux density (left side) and circular polarization (right
side) of V773~Tau~A system. Each plot have 13 panels, where each one
corresponds to 1 epoch. Blue points shown the flux evolution in
timescales of 5 hours. Red points corresponds to the average total
flux (exactly the same flux showed in Fig.\
\ref{fig-flujo-v773-1}b). It is clear that more flux variations
(stokes I, and stokes V) were detected when V773~Tau~Aa and
V773~Tau~Ab are closer (epochs 7, 8, 15, 16, 17), suggesting a
magnetospheric interaction between those stars.

\section{Implications for the properties of the stars}\label{c5-implications}

\subsection{T~Tau system}\label{c5-implications-ttau}

Having obtained an improved distance estimate to the T~Tauri system,
we are now in a position to refine the determination of the intrinsic
properties of each of the components of that system. Since the orbital
motion between T~Tau~N and T~Tau~S is not yet known to very good
precision, we will use synthetic spectra fitting to obtain the
properties of T~Tau~N. For the very obscured T~Tau~S companion, on the
other hand, we will refine the mass determinations based on the
orbital fit obtained by Duchêne et al.\ (2006).

\medskip

\begin{figure*}[!t]
\centerline{\includegraphics[height=.5\textwidth,angle=0]{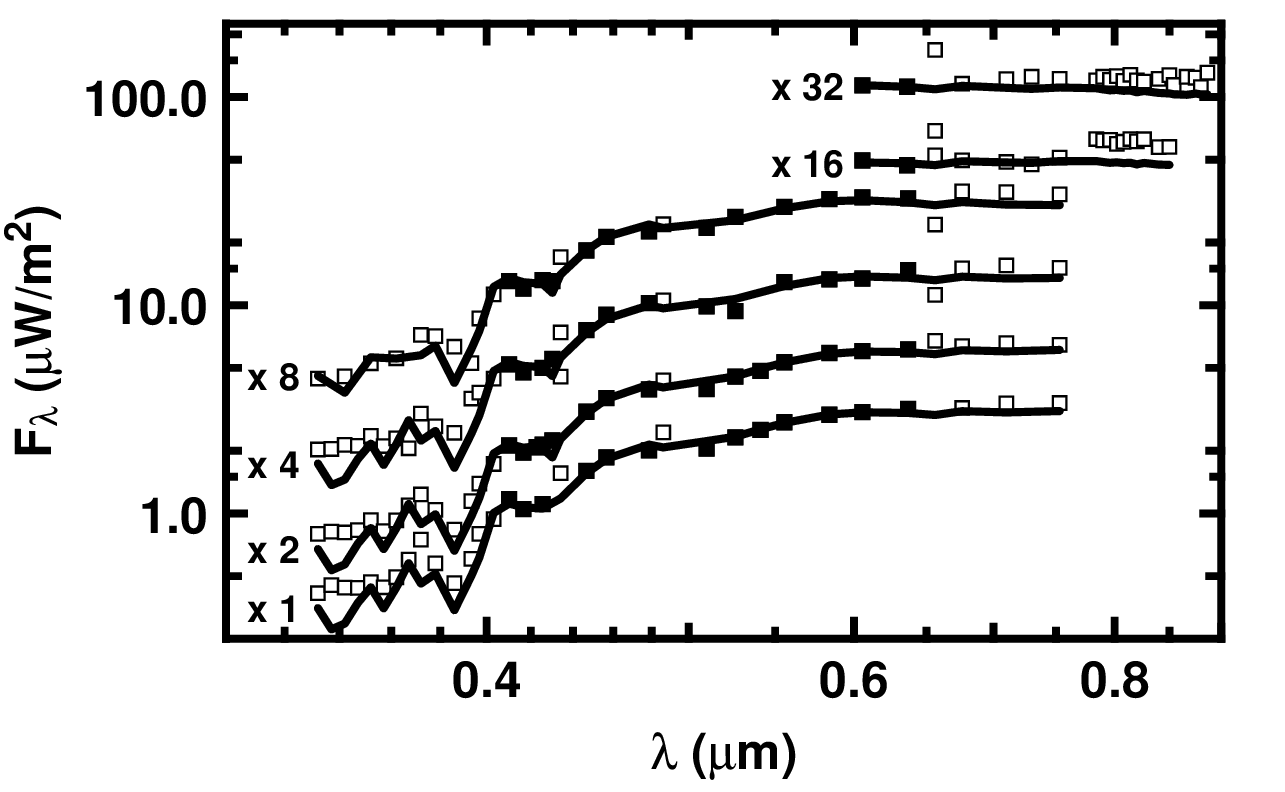}}
\caption
{\footnotesize{Fit to the photometry at six different epochs.  Black
squares are point that have been fitted; white squares represent other
wavelengths excluded from the fit.}\label{fig-ttau-4}}\end{figure*}

\noindent
\textbf{T~Tau~N}---
The stellar parameters ($T_{\rm eff}$ and $L_{\rm bol}$) of T~Tau~N
were obtained by fitting synthetic spectra (Lejeune et al.\ 1997) to
the optical part of the spectral energy distribution. In the absence
of recently published optical spectra with absolute flux calibration,
we decided to use narrow-band photometry taken at six different epochs
from 1965 to 1970 (Kuhi 1974). In order to eliminate the contamination
by the UV/blue (magnetospheric accretion) and red/IR (circumstellar
disk) excesses, we restricted the fit to the range 0.41--0.65\,$\mu$m.
Two points at 0.4340, 0.4861 $\mu$m display large variations between
epochs, they were also discarded as they are likely to be contaminated
by emission lines. As a consequence, 56 photometric measurements at 13
wavelengths and 6 epochs had to be fitted. (See Fig.\
\ref{fig-ttau-4}). We assumed that the star kept constant intrinsic
parameters over the 5 years of observation, but allowed the
circumstellar extinction to vary.  Such an hypothesis is supported by
long-term photometric observations (1986-2003) that show
color-magnitude diagrams of T~Tau elongated along the extinction
direction (Mel'Nikov \& Grankin 2005); Kuhi (1974) also measured
significant extinction variation in the period 1965-1970 using color
excesses.

\medskip

\begin{figure*}[!t]
\centerline{\includegraphics[height=1\textwidth,angle=-90]{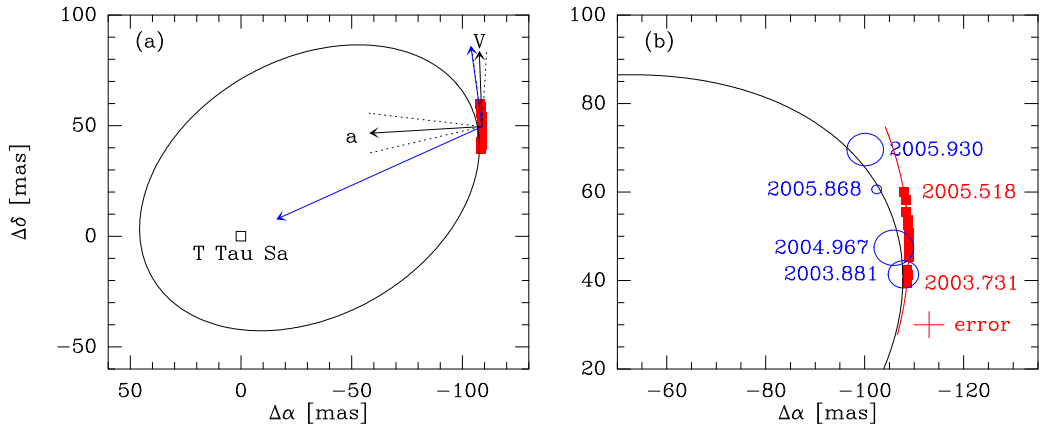}}
\caption
{\footnotesize{(a) VLBA positions (\textit{red squares}) registered to
T~Tau~Sa overimposed on the elliptical fit proposed by Duchêne et al.\
(2006). Also shown are the velocity and acceleration vectors for our
mean epoch deduced from our observations, as well as their
counterparts from the fit by Duchêne et al.\ (2006; \textit{blue
arrows}). The dotted black lines around the measured acceleration and
velocity show the error cone on the direction of each of these
vectors. (b) Zoom on the region corresponding to our observations. In
addition to the orbit and the VLBA positions, we show (in \textit{red
curve}) our best parabolic fit to our first 12 positions, as well as
several recent infrared positions (\textit{blue ellipses}). The
2003.881 position is from Duchêne et al.\ (2005), the 2004.967 and
2005.868 positions are from Duchêne et al.\ (2006), and the 2005.930
position is from Schaefer et al.\
(2006).}\label{fig-ttau-5}}\end{figure*}

\noindent
The non-linear fitting procedure used the Levenberg-Marquardt method
and the determination of errors was done using a Monte Carlo
simulation. The synthetic spectra were transformed into narrow-band
photometry by integration over the bandwidth of the measurements
(typically 0.05 $\mu$m). As the fitting procedure could not constrain
the metallicity, we assumed a solar one. Several fits using randomly
chosen initial guesses for $T_{\rm eff}$, $L_{\rm bol}$, and
extinctions were performed in order to ensure that a global minimum
$\chi^2$ was indeed reached. The errors reported by Kuhi (1974; 1.2\%)
had to be renormalized to 5.9\% in order to achieve a reduced $\chi^2$
of 1. This could result from an underestimation by the author or from
positive and negative contamination by spectral lines --indeed, Gahm
(1970) reports contamination as high as 20\% for RW Aur. The best
least-squares fit is represented in Fig.\ \ref{fig-ttau-5}, and yields
$T_{\rm eff}=5112_{-97}^{+99}$ K and $L_{\rm
bol}=5.11_{-0.66}^{+0.76}\,L_\odot$. The extinction varies between
1.02 and 1.34, within 1-$\sigma$ of the values determined by Kuhi
(1974) from color excesses. The effective temperature is consistent
with a K1 star as reported by Kuhi (1974).

\medskip

\noindent
In order to derive the age and mass of T~Tau~N, pre-main-sequence
isochrones by D'Antona \& Mazzitelli (1996) and Siess et al.\ (2000)
were used. The fitting procedure was identical to the previous one:
the age and mass were converted into effective temperature and
luminosity, which in turn were converted into narrow-band photometry
using the synthetic spectra. The derived parameters are shown in Tab.\
\ref{tab-ttau2}. The masses ($1.83^{+0.20}_{-0.16}$ and
$2.14^{+0.11}_{-0.10}$ \Msun) have overlapping error bars and are
consistent with values found in the literature (e.g.\ Duchêne et al.\
2006). The predicted ages, on the other hand, differ by a factor of
2. While the isochrones by D'Antona \& Mazzitelli (1996) give an age
in the commonly accepted range ($1.15^{+0.18}_{-0.16}$ Myr), a
somewhat larger value ($2.39^{+0.31}_{-0.27}$) is derived from Siess
et al.\ (2000). Note that the errors on the derived parameters are
entirely dominated by the modeling errors; the uncertainty on the
distance now represents a very small fraction of the error budget.

\begin{deluxetable}{lc}
\tabletypesize{\scriptsize}
\tablecolumns{2}
\tablewidth{0pc}
\tablecaption{\footnotesize{Parameters of T~Tau~N.}\label{tab-ttau2}}
\tablehead{
\multicolumn{1}{c}{~~~~~~~~~~Parameter~~~~~~~~~~} &
\multicolumn{1}{c}{~~~~~~~~~~Value~~~~~~~~~~} }
\startdata
Age [Myr] \dotfill                 & $2.39^{+0.31}_{-0.27}$ \tablenotemark{a}\\
\\[-0.3cm]
                                   & $1.15^{+0.18}_{-0.16}$ \tablenotemark{b}\\
\\[-0.3cm]
Mass [$M_\odot$] \dotfill          & $2.14^{+0.11}_{-0.10}$ \tablenotemark{a}\\
\\[-0.3cm]
                                   & $1.83^{+0.20}_{-0.16}$ \tablenotemark{b}\\
\\[-0.3cm]
$T_{\rm eff}$ [K] \dotfill         & $5112^{+99}_{-97}$\\
\\[-0.3cm]
$L_{\rm bol}$ [$L_\odot$] \dotfill & $5.11^{+0.76}_{-0.66}$\\
\\[-0.3cm]
$R_{\star}$ [$R_\odot$] \dotfill   & $2.89^{+0.24}_{-0.21}$\\
\\[-0.3cm]
${A_V}_{\rm MJD 39095.2}$ \dotfill & $1.34\pm0.17$\\
\\[-0.3cm]
${A_V}_{\rm MJD 39153.2}$ \dotfill & $1.37\pm0.17$\\
\\[-0.3cm]
${A_V}_{\rm MJD 39476.3}$ \dotfill & $1.20\pm0.17$\\
\\[-0.3cm]
${A_V}_{\rm MJD 40869.4}$ \dotfill & $1.02\pm0.17$\\
\\[-0.3cm]
${A_V}_{\rm MJD 39485.1}$ \dotfill & $1.36\pm0.19$\\
\\[-0.3cm]
${A_V}_{\rm MJD 39524.1}$ \dotfill & $1.16\pm0.19$\\
\enddata
\tablenotetext{a}{Siess et al.\ (2000)}
\tablenotetext{b}{D'Antona \& Mazzitelli (1997)}
\end{deluxetable}

\medskip

\noindent
\textbf{T~Tau~S}---
The two members of the T~Tau~S system have been studied in detail by
Duchêne and coworkers in a series of recent articles (Duchêne et al.\
2002, 2005, 2006). The most massive member of the system (T~Tau~Sa)
belongs to the mysterious class of ``infrared companions'', and is
presumably the precursor of an intermediate-mass star. T~Tau~Sb, on
the other hand is a very obscured, but otherwise normal, pre-main
sequence M1 star. The mass of both T~Tau~Sa and T~Tau~Sb were
estimated by Duchêne et al.\ (2006) using a fit to their orbital
paths. Those authors used the distance to T~Tauri that we deduced
without acceleration terms. Using the new distance determination
obtained in Sect.\ \ref{c5-absolute}, we can re-normalize those
masses. We obtain $M_{\rm Sa}=3.10\pm0.34$ \Msun, and $M_{\rm
Sb}=0.69\pm0.18$ \Msun. These values may need to be adjusted somewhat,
however, as the fit to the orbital path of the T~Tau~Sa/Sb system is
improved (see Sect.\ \ref{c5-relative}). Note finally, that the main
sources of errors on the masses are related to the orbital motion
modeling rather than to the uncertainties of the distance.

\subsection{HP~Tau system and HDE~283572}\label{c5-implications-hptau}

\begin{figure*}[!t]
\centerline{\includegraphics[height=1\textwidth,angle=0]{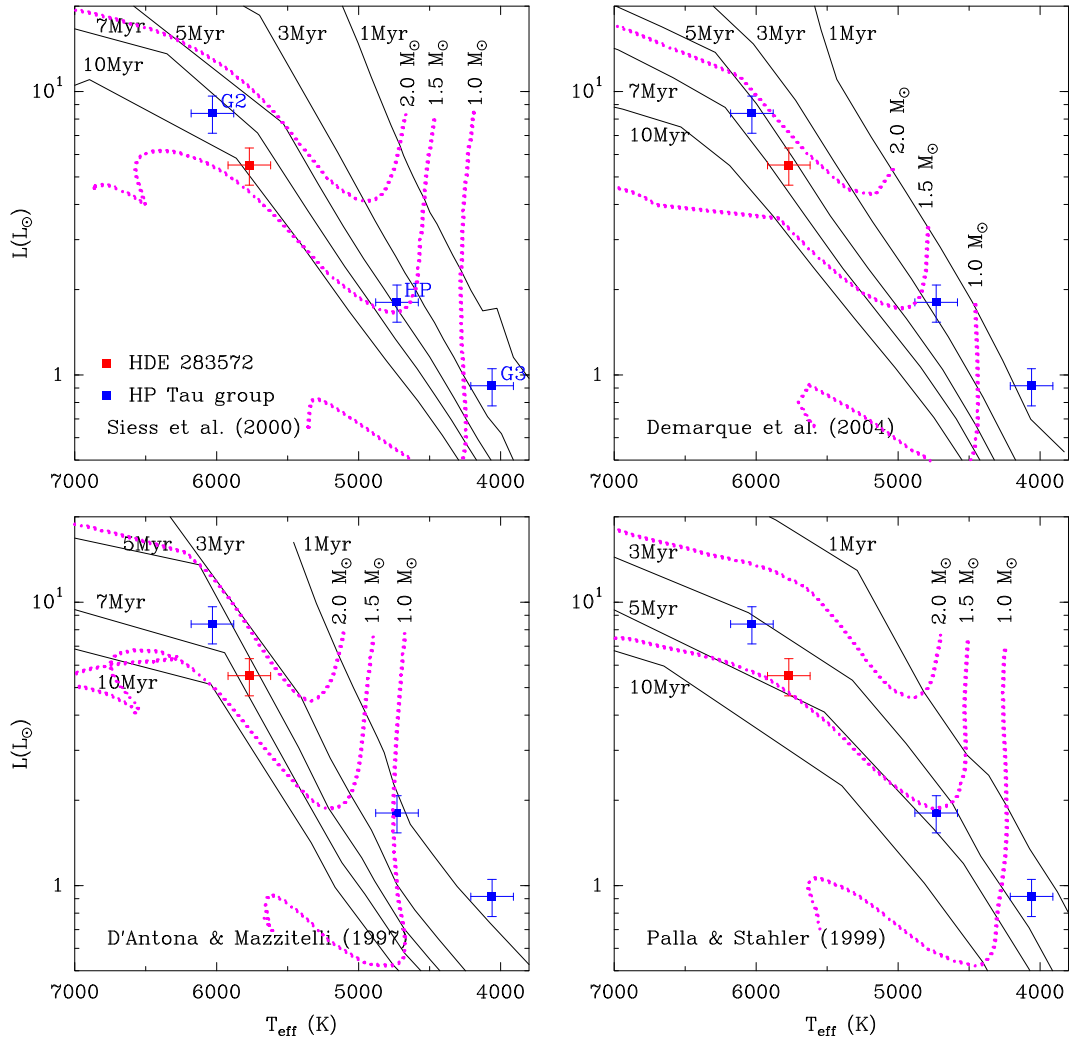}}
\caption
{\footnotesize{Positions of the three HP~Tau members (\textit{blue
symbols}) and of HDE~283572 (\textit{red symbol}) on an HR
diagram. From the coolest to the warmest, the three stars in the
HP~Tau group are HP~Tau/G3, HP~Tau, and HP~Tau/G2, as indicated in the
first panel. Isochrones (\textit{full back lines}) are shown at 1, 3,
5, 7, and 10 Myr for various models. For the same models, evolutionary
tracks for stars of 1.0, 1.5, and 2.0 \Msun\ are also shown as
\textit{dotted magenta lines}.}\label{fig-hptau-2}}\end{figure*}

As mentioned in Chapter \ref{chap-introduction}, HP~Tau/G2 is a member
of a compact group of four young stars, comprising HP~Tau itself,
HP~Tau/G1, G2, and G3. Given the small angular separations between
them, the members of this group are very likely to be physically
associated --indeed, HP~Tau/G2 and G3 are thought to form a bound
system. They are, therefore, very likely to be at the same distance
from the Sun. Using our accurate estimate of the distance to
HP~Tau/G2, we are now in a position to refine the determination of the
luminosities of all four stars. Little is known about HP~Tau/G1, but
the effective temperature and the bolometric luminosity (obtained
assuming $d=142$ pc) of the other three members are given in Briceño
et al.\ (2002). Those values (corrected to the new distance) allow us
to place the stars accurately on an HR diagram (Fig.\
\ref{fig-hptau-2}).

\medskip

\noindent
From their position on the HR diagram, one can (at least in principle)
derive the mass and age of the stars using theoretical pre-main
sequence evolutionary codes. Several such models are available, and we
will use four of them here: those of Siess et al.\ (2000), Demarque et
al.\ (2004; known as the Yonsei-Yale $Y^2$ models), D'Antona \&
Mazzitelli (1997), and Palla \& Stahler (1999; the models by Baraffe
et al.\ (1998) will not be used because they do not cover the mass
range of our stars). The isochrones for those four models at 1, 3, 5,
7, and 10 Myr are shown as solid black lines in Fig.\
\ref{fig-hptau-2}. Also shown are the evolutionary tracks (from the
same models) for stars of 1.0, 1.5, and 2.0 \Msun. The three HP~Tau
members are shown as blue symbols, and HDE~283572 is shown as a red
symbol.

\medskip

\noindent
A number of interesting points can be seen from Fig.\
\ref{fig-hptau-2}. First, there is reasonable agreement (within 40\%,
see below) between the masses predicted by different models. The best
case is that of HP~Tau/G2, for which the different models predict
masses consistent with each other at the 10\% level (between 1.7 and
1.9 \Msun). The situation for HP~Tau is somewhat less favorable, since
the models of Siess et al.\ (2000) or Palla \& Stahler (1999) predict
a mass of $\sim$ 1.5 \Msun, whereas those of D'Antona \& Mazzitelli
(1997) predicts a significantly smaller mass of $\sim$ 1.0
\Msun. Thus, there is a 35\% spread in the values predicted by
different models for the mass of that source. The least favorable
situation is for HP~Tau/G3. The mass of that source is about 0.8
\Msun\ according to the models of Siess et al.\ (2000), but slightly
less than 0.5 \Msun\ according to those of D'Antona \& Mazzitelli
(1997). This is a 40\% discrepancy. This tendency for pre-main
sequence evolutionary models to become more discrepant at lower masses
had been noticed before, and is discussed at length in Hillenbrand et
al.\ (2008). In the absence of dynamically measured masses, it is
impossible to assess which of the models used here provides the
``best'' answer.

\medskip

\noindent
Another interesting issue is related to the age predictions of the
different models. Since the different members of the HP~Tau group are
likely to be physically associated, they are expected to be nearly
coeval. This is particularly true of HP~Tau/G2 and HP~Tau/G3 which are
believed to form a loose binary system. Interestingly, most models
predict significantly different ages for the three sources (see Fig.\
\ref{fig-hptau-2}). The models by Siess et al.\ (2000) predict ages of
about 8 Myr and 3 Myr for HP~Tau/G2 and HP~Tau/G3, respectively. A
similar 5 Myr age difference is found for the models of Demarque et
al.\ (2004) and D'Antona \& Mazzitelli (1997): both predict ages
slightly smaller than 1 Myr for HP~Tau/G3, and somewhat larger than 5
Myr for HP~Tau/G2. It is possible that those differences could be
real, however, it should be noticed that the vast majority of low-mass
stars in Taurus (with spectral types M and late K) have ages smaller
than 3 Myr (Briceño et al.\ 2002). Moreover, mass-dependent systematic
effects in the age predictions made by evolutionary tracks have been
reported before. In particular, Hillenbrand et al.\ (2008) argued that
existing models could significantly over-predict the age of relatively
massive stars (M $\gtrsim$ 1.5 \Msun). HP~Tau/G2 is precisely such a
fairly massive star. So is HDE~283572, another young star in Taurus
with an accurate distance. The age estimate for that star based on the
models by Siess et al.\ (2000), Demarque et al.\ (2004) and D'Antona
\& Mazzitelli (1997) is 6--10 Myr (Fig.\ \ref{fig-hptau-2}), somewhat
larger than would be expected for Taurus. The only of the four models
considered here to predict similar ages for the three members of the
HP~Tau group is that of Palla \& Stahler (1999). Within the errors,
all three stars fall on the 3 Myr isochrone. Note that this value is
also consistent with the ages of lower mass stars in Taurus (Briceño
et al.\ 2002).

%
%

\newpage

\section{Relative astrometry}\label{c5-relative}

\subsection{T~Tau~S orbit}\label{c5-relative-ttaus}

The nature of the orbital motion between T~Tau~Sa and T~Tau~Sb has
been somewhat disputed in recent years. Using 20 years of VLA
observations, Loinard et al.\ (2003) concluded that the orbit between
T~Tau~Sa and T~Tau~Sb had been dramatically altered after a recent
periastron passage around 1996. Numerous near-infrared observations
obtained mostly between 2002 and now have been used by several authors
to constrain the orbital path. Perhaps the most complete study
published so far is that of Duchêne et al.\ (2006) who proposed an
orbit that could simultaneously reproduce the VLA observations used by
Loinard et al.\ (2003) and the near-infrared observations. The orbital
period proposed by these authors is 21 years, implying that the system
has completed a full revolution since the first VLA observations
published by Loinard et al.\ (2003). More recent near-infrared
observations, however, appear to be incompatible with the orbit
proposed by Duchêne et al.\ (2006). Instead, they suggest a
significantly longer orbital period of about 90 years (Köhler et al.\
2008). Such a longer orbital period, however, would be incompatible
with the measured VLA positions 20--25 years ago, and would again
require that something peculiar happened either to the radio source
associated with T~Tau~Sb or to the orbit of the T~Tau~Sa/Sb system.

\medskip

\noindent
In order to understand the relative motion between the members of the
T~Tau~S system, we need to express the motion of T~Tau~Sb relative to
the other members (T~Tau~N and T~Tau~Sa). Since only T~Tau~Sb is
detected in our VLBA observations, however, registering the positions
reported here to the other members of the system involves a number of
steps. The absolute position and proper motion of T~Tau~N has been
measured to great precision using over 20 years of VLA observations
(Loinard et al.\ 2003), so registering the position and motion of
T~Tau~Sb relative to T~Tau~N is fairly straightforward. Combining the
data used by Loinard et al.\ (2003) with several more recent VLA
observations, we obtained the following absolute position (at epoch
J2000.0) and proper motion for T~Tau~N:
\begin{eqnarray}
\alpha_{J2000.0}&=&\mbox{\dechms{04}{21}{59}{4321}}~\pm~\mbox{\mmsec{0}{0001}}
\nonumber \\%
\delta_{J2000.0}&=&\mbox{\decdms{19}{32}{06}{419}}~\pm~\mbox{\msec{0}{002}}
\nonumber \\%
\mu_\alpha\cos\delta&=&12.35~\pm~0.04~\mbox{mas yr$^{-1}$}\nonumber \\%
\mu_\delta&=&-12.80~\pm~0.06~\mbox{mas yr$^{-1}$.}\nonumber
\end{eqnarray}

\medskip

\noindent
Subtracting these values from the absolute positions and proper motion 
of T~Tau~Sb, we can obtain the positional offset between T~Tau~Sb and 
T~Tau~N, as well as their relative proper motion. For the median epoch 
of our observations, we obtain:
\begin{eqnarray}
\mu_\alpha\cos\delta{\rm (Sb/N)}&=&-8.33~\pm~0.07~{\rm mas~yr^{-1}}
\nonumber \\%
\mu_\delta{\rm (Sb/N)}&=&+11.62~\pm~0.11~{\rm mas~yr^{-1}}.\nonumber
\end{eqnarray}

\medskip

\noindent
The second step consists in registering the position and motion of
T~Tau~Sb to the center of mass of T~Tau~S using the parabolic fits
provided by Duchêne et al.\ (2006). Here, both the proper motion and
the acceleration must be taken into account. For the mean epoch of our
observations, we obtain:
\begin{eqnarray}
\mu_\alpha\cos\delta{\rm (Sb/CM)}&=&+0.3~\pm~0.9~{\rm mas~yr^{-1}}\nonumber \\%
\mu_\delta{\rm (Sb/CM)}&=&+9.3~\pm~0.8~{\rm mas~yr^{-1}}\nonumber \\%
a_\alpha\cos\delta{\rm (Sb/CM)}&=&+1.4~\pm~0.2~{\rm mas~yr^{-2}}\nonumber \\%
a_\delta{\rm (Sb/CM)}&=&-0.1~\pm~0.3~{\rm mas~yr^{-2}}.\nonumber 
\end{eqnarray}

\medskip

\noindent
The last correction to be made is the registration of the positions,
proper motions, and accelerations to T~Tau~Sa rather than to the
center of mass of T~Tau~S. This is obtained by simply multiplying the
values above by the ratio of the total mass of T~Tau~S (i.e.\ $M_{\rm
Sa}+M_{\rm Sb}$) to the mass of T~Tau~Sa. Using the masses given by
Duchêne et al.\ (2006), we obtain:
\begin{eqnarray}
\mu_\alpha\cos\delta{\rm (Sb/Sa)}&=&+0.4~\pm~1.1~{\rm mas~yr^{-1}}\nonumber \\%
\mu_\delta{\rm (Sb/Sa)}&=&+11.4~\pm~1.0~{\rm mas~yr^{-1}}\nonumber \\%
a_\alpha\cos\delta{\rm (Sb/Sa)}&=&+1.7~\pm~0.2~{\rm mas~yr^{-2}}\nonumber \\%
a_\delta{\rm (Sb/Sa)}&=&-0.1~\pm~0.3~{\rm mas~yr^{-2}}.\nonumber 
\end{eqnarray}

\begin{figure*}[!b]
\centerline{\includegraphics[width=0.7\textwidth,angle=270]{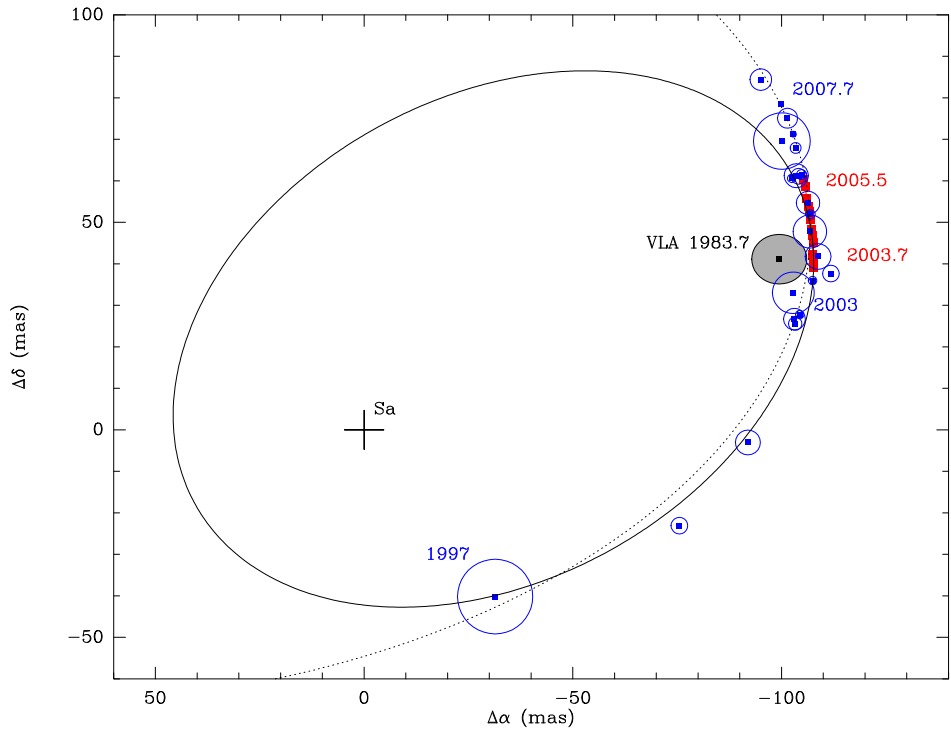}}
\caption{\footnotesize{
Relative position between T~Tau~Sb and T~Tau~Sa (located at the origin
of the system at position (0,0)). The \textit{blue symbols} correspond
to all the infrared observations tabulated by Köhler et al.\ (2008) as
well as Keck observations kindly provided by G.\ Schaefer. The
\textit{solid line} correspond to the fit published by Duchêne et al.\
(2006; with a period of about 21 yr) and the \textit{dotted line} to
the fit proposed by Köhler et al.\ (2008; with a period of about 90
yr). Our first 12 VLBA observations (\textit{red squares}) were
superimposed over the fits. The position of the VLA source in 1983.7
is shown as a \textit{grey symbol}.}\label{fig-ttau-6}}\end{figure*}

\medskip

\noindent 
These two vectors are shown in Fig.\ \ref{fig-ttau-5} together with
the VLBA positions registered to T~Tau~Sa, several infrared
observations and the elliptical fit obtained by Duchêne et al.\
(2006). The final error on the VLBA positions is the combination of
the original uncertainty on their measured absolute position, and of
the errors made at each of the steps described above. The final
uncertainty is about 3 mas in both right ascension and declination,
and is shown near the bottom right corner of Fig.\ \ref{fig-ttau-5}b.

\medskip

\noindent 
Given the uncertainties, the position of the VLBA source is generally
in good agreement with the infrared source position measured at
similar epochs. Indeed, the first 2 VLBA observations were obtained
almost exactly at the same time as the infrared image published by
Duchêne et al.\ (2005), and the positions match exactly. The position
of the VLBA source at the end of 2004 is also in agreement within
1$\sigma$ with the position of the infrared source at the same epoch
reported by Duchêne et al.\ (2006). The situation at the end of 2005,
however, is somewhat less clear. Extrapolating from the last VLBA
observation ($\sim$ 2005.5) to the end of 2005 gives a location that
would be in reasonable agreement with the position given by Schaefer
et al.\ (2006) but clearly not with the position obtained by Duchêne
et al.\ (2006). Note, indeed, that the two infrared positions are only
very marginally consistent with one another.

\medskip

\noindent 
Our VLBA observations suggest that T~Tau~Sb passed at the westernmost
point of its orbit around 2005.0, whereas according to the fit
proposed by Duchêne et al.\ (2006), this westernmost position was
reached slightly before 2004.0. As a consequence, the trajectory
described by the VLBA source is on average almost exactly north-south,
whereas according to the fit proposed by Duchêne et al.\ (2006),
T~Tau~Sb is already moving back toward the east (Fig.\
\ref{fig-ttau-5}). We note, however, that the fit proposed by
Duchêne et al.\ (2006, which gives an orbital period of $21.7\pm0.9$
yr) is very strongly constrained by their 2005.9 observation. Shaefer
et al.\ (2006), who measured a position at the end of 2005 somewhat
more to the north (in better agreement with our VLBA positions), argue
that they cannot discriminate between orbital periods of 20, 30 or 40
yr. Orbits with longer periods bend back toward the east somewhat
later (see Fig.\ 10 in Schaefer et al.\ 2006), and would be in better
agreement with our VLBA positions.

\medskip

\noindent 
Another element that favors a somewhat longer orbital period is the
acceleration measured here. According to the fit proposed by Duchêne
et al.\ (2006), the expected transverse proper motion and acceleration
are (G.\ Duchêne, private communication):
\begin{eqnarray}
\mu_\alpha\cos\delta{\rm (Sb/Sa)}&=&+1.7~\pm~0.2~{\rm mas~yr^{-1}}\nonumber \\%
\mu_\delta{\rm (Sb/Sa)}&=&+12.1~\pm~1.2~{\rm mas~yr^{-1}}.\nonumber \\%
a_\alpha\cos\delta{\rm (Sb/Sa)}&=&+3.1~\pm~0.5~{\rm mas~yr^{-2}}\nonumber \\%
a_\delta{\rm (Sb/Sa)}&=&-1.4~\pm~0.2~{\rm mas~yr^{-2}}.\nonumber
\end{eqnarray}

\begin{deluxetable}{crlrl}
\tabletypesize{\scriptsize}
\tablecolumns{5}
\tablewidth{0pc}
\tablecaption{\footnotesize{Separations between T~Tau~Sa and T~Tau~Sb from VLBA and near-infrared observations.}\label{tab-ttau1}}
\tablehead{
\colhead{Mean Date}               &
\colhead{$\Delta\alpha$}          &
\colhead{$\sigma_{\Delta\alpha}$} &
\colhead{$\Delta\delta$}          &
\colhead{$\sigma_{\Delta\delta}$} \\
\multicolumn{1}{c}{[years]}       &
\multicolumn{1}{c}{[mas]}         &
\multicolumn{1}{c}{}              &
\multicolumn{1}{c}{[mas]}         &
\multicolumn{1}{c}{}              }
\startdata
1999.956224 &   -88.26 &  0.47 & -10.60 &  0.75  \\%
\\[-0.2cm]
2003.732340 &  -107.19 &  0.05 &  40.05 &  0.08  \\%
\\[-0.2cm]
2003.882523 &  -107.20 &  0.05 &  41.89 &  0.08  \\%
\\[-0.2cm]
2004.041542 &  -106.96 &  0.06 &  43.21 &  0.13  \\%
\\[-0.2cm]
2004.238134 &  -107.11 &  0.05 &  46.13 &  0.09  \\%
\\[-0.2cm]
2004.369194 &  -106.86 &  0.05 &  47.72 &  0.09  \\%
\\[-0.2cm]
2004.522099 &  -106.68 &  0.05 &  49.38 &  0.09  \\%
\\[-0.2cm]
2004.713224 &  -106.46 &  0.05 &  51.72 &  0.08  \\%
\\[-0.2cm]
2004.860668 &  -106.16 &  0.05 &  53.53 &  0.08  \\%
\\[-0.2cm]
2004.994459 &  -105.94 &  0.05 &  54.90 &  0.09  \\%
\\[-0.2cm]
2005.150749 &  -105.52 &  0.05 &  56.79 &  0.08  \\%
\\[-0.2cm]
2005.355532 &  -105.21 &  0.09 &  59.50 &  0.19  \\%
\\[-0.2cm]
2005.519358 &  -104.71 &  0.05 &  61.36 &  0.09  \\%
\\[-0.2cm]
2008.412872 &   -90.88 &  0.15 &  90.96 &  0.19  \\%
\\[-0.2cm]
1997.780304 &   -31.40 &  7.89 & -40.18 &  8.33  \\%
\\[-0.2cm]
2000.139633 &   -75.55 &  2.08 & -23.09 &  2.70  \\%
\\[-0.2cm]
2000.887083 &   -91.95 &  2.99 &  -3.05 &  2.56  \\%
\\[-0.2cm]
2002.829587 &  -103.12 &  2.60 &  26.66 &  2.60  \\%
\\[-0.2cm]
2002.950055 &  -104.37 &  1.06 &  27.76 &  1.65  \\%
\\[-0.2cm]
2002.955530 &  -103.28 &  1.56 &  25.55 &  0.81  \\%
\\[-0.2cm]
2002.980172 &  -102.83 &  4.99 &  33.01 &  4.91  \\%
\\[-0.2cm]
2003.881607 &  -108.76 &  3.12 &  41.74 &  3.23  \\%
\\[-0.2cm]
2003.947317 &  -111.84 &  2.03 &  37.63 &  2.23  \\%
\\[-0.2cm]
2003.947317 &  -107.45 &  1.07 &  35.95 &  1.53  \\%
\\[-0.2cm]
2004.941841 &  -106.96 &  1.13 &  51.93 &  1.21  \\%
\\[-0.2cm]
2004.969220 &  -106.80 &  3.83 &  47.77 &  3.07  \\%
\\[-0.2cm]
2004.982909 &  -106.37 &  2.83 &  54.66 &  2.89  \\%
\\[-0.2cm]
2005.109516 &  -104.11 &  1.90 &  61.08 &  1.89  \\%
\\[-0.2cm]
2005.186178 &  -103.55 &  2.91 &  61.24 &  2.92  \\%
\\[-0.2cm]
2005.227246 &  -104.80 &  1.22 &  61.48 &  1.25  \\%
\\[-0.2cm]
2005.802207 &  -103.39 &  1.30 &  67.91 &  1.29  \\%
\\[-0.2cm]
2005.867917 &  -102.43 &  1.36 &  60.58 &  1.85  \\%
\\[-0.2cm]
2005.930889 &  -100.10 &  6.80 &  69.57 &  6.80  \\%
\\[-0.2cm]
2006.777566 &  -102.85 &  1.37 &  71.22 &  1.83  \\%
\\[-0.2cm]
2006.963744 &  -101.45 &  2.33 &  75.07 &  2.27  \\%
\\[-0.2cm]
2007.709118 &   -99.80 &  0.57 &  78.54 &  0.60  \\%
\\[-0.2cm]
2008.046544 &   -95.05 &  2.63 &  84.38 &  2.63  \\%
\enddata
\end{deluxetable}

\medskip

\noindent
Thus, while the expected and observed proper motions are in good
agreement, the expected acceleration is significantly larger that the
observed value (see also Fig.\ \ref{fig-ttau-5}). A smaller value of
the acceleration would be consistent with a somewhat longer orbital
period.

\medskip

\noindent
In order to summarize the last paragraphs, Fig.\ \ref{fig-ttau-6}
illustrates the problem: our VLBA positions (red squares) and the VLA
position around 1984 (grey symbol) are superimposed on the fits
proposed by Duchêne et al.\ (2006; solid line) and Köhler et al.\
(2008; dotted line). VLBA data appear to be in reasonable agreement
with all the published infrared positions obtained over the last few
years in the plot, except for the 2005.9 observation reported by
Duchêne et al.\ (2006). Thus, our observations are in better agreement
with the most recent infrared observations and with the fit proposed
by Köhler et al.\ (2008). The difficulty with such a long orbital
period is related to the older VLA position. Clearly, the observed VLA
position would be consistent with the 20--22 yr orbital period
proposed by Duchêne et al.\ (2006): between the old VLA observation
and the IR or VLBA data corresponding to 2003--2005, T~Tau~Sb would
have complete a full orbit. For a 90 yr orbital period, however, the
position expected for 1984 is located at about $\Delta\alpha\sim+100$
mas; $\Delta\delta\sim-70$ mas. This is clearly inconsistent with the
VLA position observed at that epoch.

\begin{figure*}[!b]
\centerline{\includegraphics[height=.7\textwidth,angle=270]{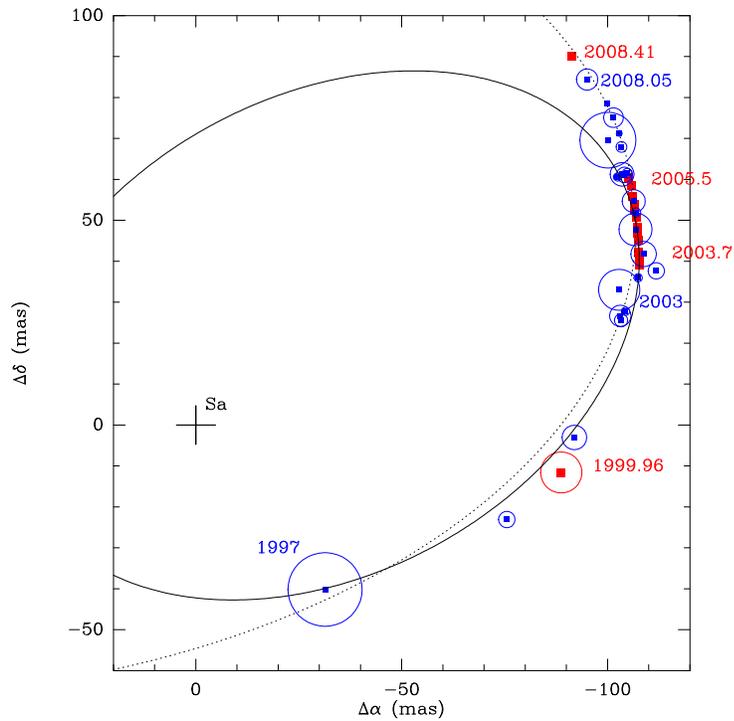}}
\caption{\footnotesize{
Relative position between T~Tau~Sb and T~Tau~Sa (located at the origin
of the system at position (0,0)). The \textit{blue symbols} correspond
to all the infrared observations tabulated by Köhler et al.\ (2008) as
well as Keck observations kindly provided by G.\ Schaefer. The
\textit{solid line} correspond to the fit published by Duchêne et al.\
(2006; with a period of about 21 yr) and the \textit{dotted line} to
the fit proposed by Köhler et al.\ (2008; with a period of about 90
yr). Our 14 VLBA observations (\textit{red squares}) were superimposed
over the fits. The series of VLBA observations between 2003 and 2005
(epochs 1 to 12 in Tab.\ \ref{tab-observaciones}) has been used to
measure the parallax to T~Tau~Sb. The point at 1999.96 is from VLBA
archive (observations by Smith et al.\ 2003; epoch 14 in Tab.\
\ref{tab-observaciones}). And the point at 2008.41 is the new VLBA
observation (epoch 13 in Tab.\ \ref{tab-observaciones}). Note that the
series of 14 VLBA observations agrees better with the latter
fit.}\label{fig-ttau-2}}\end{figure*}

\medskip

\noindent
A new VLBA observation were obtained for T~Tau~Sb on May 2008 (epoch
13 in Tab.\ \ref{tab-resultados}) to definitely allow us to
discriminate between the two orbits shown in Fig.\ \ref{fig-ttau-6},
or to provide crucial evidence for an orbit with an intermediate
period. In addition, there exists a good quality observation of
T~Tau~Sb in the VLBA archive from the end of 1999 (published by Smith
et al.\ 2003; epoch 14 in Tab.\ \ref{tab-resultados}). Our 14 VLBA
positions (red squares) were superimposed on the fits proposed by
Duchêne et al.\ (2006; solid line) and Köhler et al.\ (2008; dotted
line) in Fig.\ \ref{fig-ttau-2}, and it is clear that VLBA points are
in agreement with the fit proposed by Köhler et al.\ (2008).

\begin{figure*}[!b]
\centerline{\includegraphics[height=0.6\textwidth,angle=0]{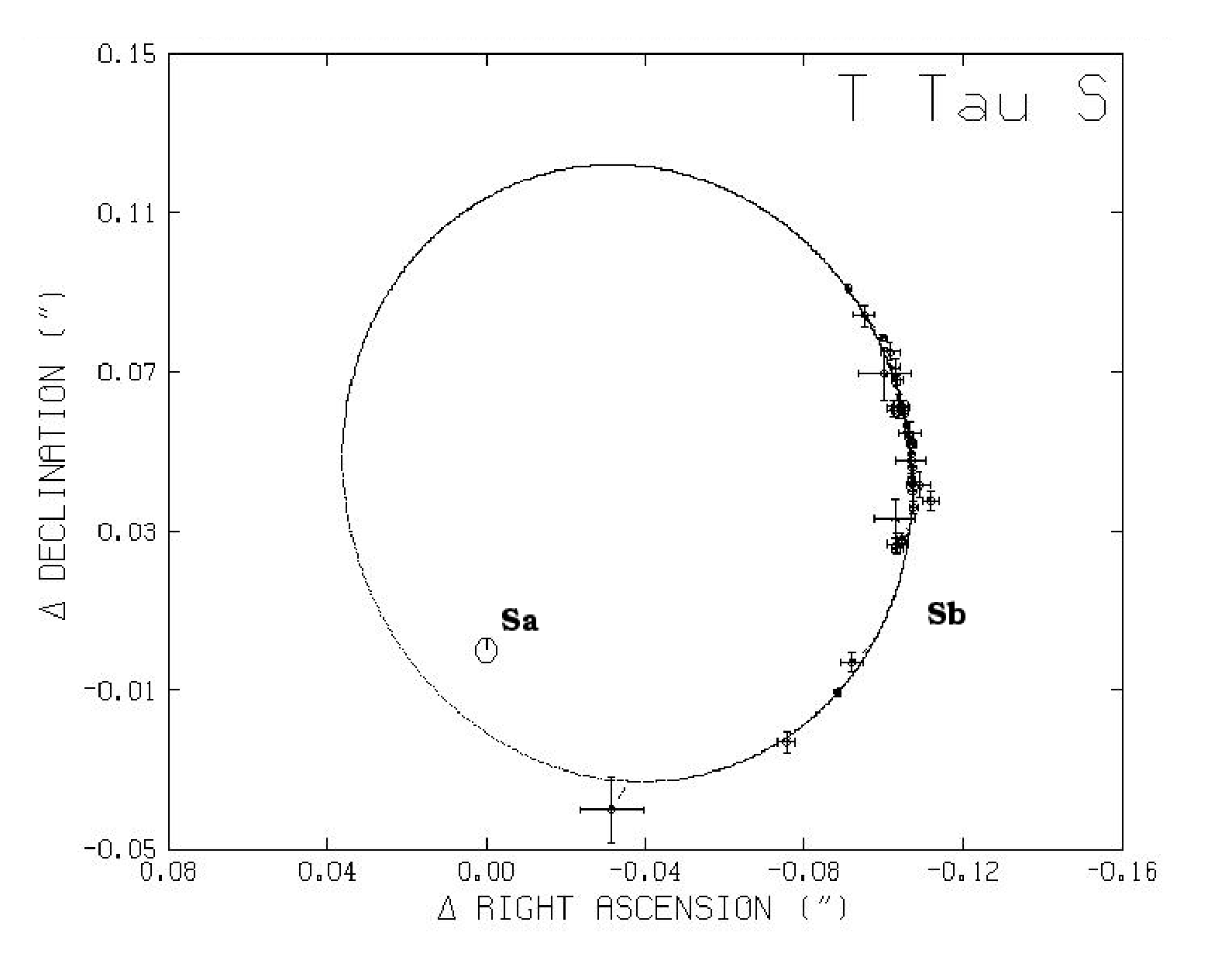}}
\caption{\footnotesize{
Relative position between T~Tau~Sb and T~Tau~Sa (located at the origin
of the system at position (0,0)). The crosses are near-infrared (all
the observations tabulated by Köhler et al.\ (2008) as well as Keck
observations kindly provided by G.\ Schaefer) and VLBA
observations. The solid line is our preliminary fit that corresponds
to a period of about 24 yr.}\label{fig-ttau-3}}\end{figure*}

\medskip

\noindent
However, to obtain an orbital fit of T~Tau~Sa/Sb system, we made use
of our 14 VLBA observations, and of all infrared observations
tabulated by Köhler et al.\ (2008) as well as Keck observations kindly
provided by G.\ Schaefer. The angular separations ($\Delta\alpha$,
$\Delta\delta$) between the primary (T~Tau~Sa) and the secondary
(T~Tau~Sb) are given in Tab.\ \ref{tab-ttau1} for both infrared and
radio observations. The preliminary orbital parameters derived with
those data are: period of $25.83\pm8$ yr, eccentricity of 0.78,
inclination of 53.14$^{\circ}$, and semi-major axis of 0.12
arcsec. Although the VLBA red points superimposed in Fig.\
\ref{fig-ttau-2} agrees better with the fit by Köhler et al.\ (2008),
an entirely new fit with near-infrared and VLBA data, gives a period
more like that from Duchêne et al.\ (2006). Our orbit model is shown
in Fig.\ \ref{fig-ttau-3}, but we should stress that the present fit
must still be considered preliminary. In order to give a definitive
orbit, we are now in the process of analyzing the 20 years of VLA data
and infrared observations from 1989 to 2009 in our data set.

\subsection{V773~Tau~A orbit}\label{c5-relative-v773tau}

We will concentrate on the 13 epochs when the source was double
(epochs 1, 2, 3, 4, 6, 7, 8, 10, 11, 13, 16, 18, 19), and analyze the
relative position of the two sub-components. As established by
Phillips et al.\ (1996) and Boden et al.\ (2007, hereafter B2007),
these double source observations reflect the relative astrometry of
the V773~Tau~A components and as such can be used to assess and update
the physical orbit obtained by B2007. There is, however, one
difficulty. The mechanism producing the radio emission is not directly
related to the mass of the star, so the primary is not necessarily the
brightest radio source in the system (in the present case, it is the
brightest in only about 60\% of the cases). This introduces a
degeneracy in the relative position angle between the stars, which we
removed using the preliminary orbital fit of B2007. The corresponding
angular separations ($\Delta\alpha$, $\Delta\delta$) between the
primary and the secondary are given in Tab.\ \ref{tab-fases} for the
13 epochs when the source is double. The uncertainties on
($\Delta\alpha$, $\Delta\delta$) quoted in Tab.\ \ref{tab-fases} are
based on the errors delivered by \textsf{JMFIT}, and are almost
certainly underestimated.

\begin{deluxetable}{ccrlrlc}
\tabletypesize{\scriptsize}
\tablecolumns{7}
\tablewidth{0pc}
\tablecaption{\footnotesize{Orbit phases and measured separations between two components of V773~Tau~A system.}\label{tab-fases}}
\tablehead{
\colhead{Epoch}                   &
\colhead{Mean Date}               &
\colhead{$\Delta\alpha$}          &
\colhead{$\sigma_{\Delta\alpha}$} &
\colhead{$\Delta\delta$}          &
\colhead{$\sigma_{\Delta\delta}$} &
\colhead{$\phi$}                  \\%
                                  &
\multicolumn{1}{c}{JD}            &
\multicolumn{1}{c}{(mas)}         &
                                  &
\multicolumn{1}{c}{(mas)}         &
                                  &
                                  }
\startdata
1  & 2453622.00 & $-$1.43  & 0.01     & $-$0.70  & 0.04     & 0.99 \\
\\[-0.2cm]
2  & 2453689.81 & 1.90     & 0.03     & 1.85     & 0.06     & 0.32 \\
\\[-0.2cm]
3  & 453756.63  & 2.79     & 0.04     & 0.78     & 0.12     & 0.63 \\
\\[-0.2cm]
4  & 2453827.44 & $-$1.31  & 0.01     & $-$0.71  & 0.03     & 0.01 \\
\\[-0.2cm]
5  & 2453899.24 & \dotfill & \dotfill & \dotfill & \dotfill & 0.42 \\
\\[-0.2cm]
6  & 2453984.01 & $-$1.34  & 0.02     & $-$0.35  & 0.05     & 0.08 \\
\\[-0.2cm]
7  & 2454336.05 & $-$1.96  & 0.03     & $-$1.34  & 0.06     & 0.97 \\
\\[-0.2cm]
8  & 2454342.03 & $-$1.21  & 0.02     & 0.08     & 0.03     & 0.08 \\
\\[-0.2cm]
9  & 2454349.01 & \dotfill & \dotfill & \dotfill & \dotfill & 0.22 \\
\\[-0.2cm]
10 & 2454354.99 & 2.61     & 0.11     & 1.75     & 0.18     & 0.33 \\
\\[-0.2cm]
11 & 2454359.98 & 2.92     & 0.03     & 1.69     & 0.05     & 0.43 \\
\\[-0.2cm]
12 & 2454364.97 & \dotfill & \dotfill & \dotfill & \dotfill & 0.53 \\
\\[-0.2cm]
13 & 2454370.95 & 2.82     & 0.06     & 0.43     & 0.11     & 0.65 \\
\\[-0.2cm]
14 & 2454376.93 & \dotfill & \dotfill & \dotfill & \dotfill & 0.77 \\
\\[-0.2cm]
15 & 2454382.92 & \dotfill & \dotfill & \dotfill & \dotfill & 0.88 \\
\\[-0.2cm]
16 & 2454390.89 & $-$1.40  & 0.02     & $-$0.60  & 0.05     & 0.04 \\
\\[-0.2cm]
17 & 2454396.88 & \dotfill & \dotfill & \dotfill & \dotfill & 0.16 \\
\\[-0.2cm]
18 & 2454400.87 & 0.77     & 0.03     & 1.36     & 0.05     & 0.24 \\
\\[-0.2cm]
19 & 2454421.81 & 2.73     & 0.01     & 0.57     & 0.03     & 0.64 \\
\enddata
\end{deluxetable}

\begin{figure*}[!b]
\centerline{\includegraphics[height=0.8\textwidth,angle=0]{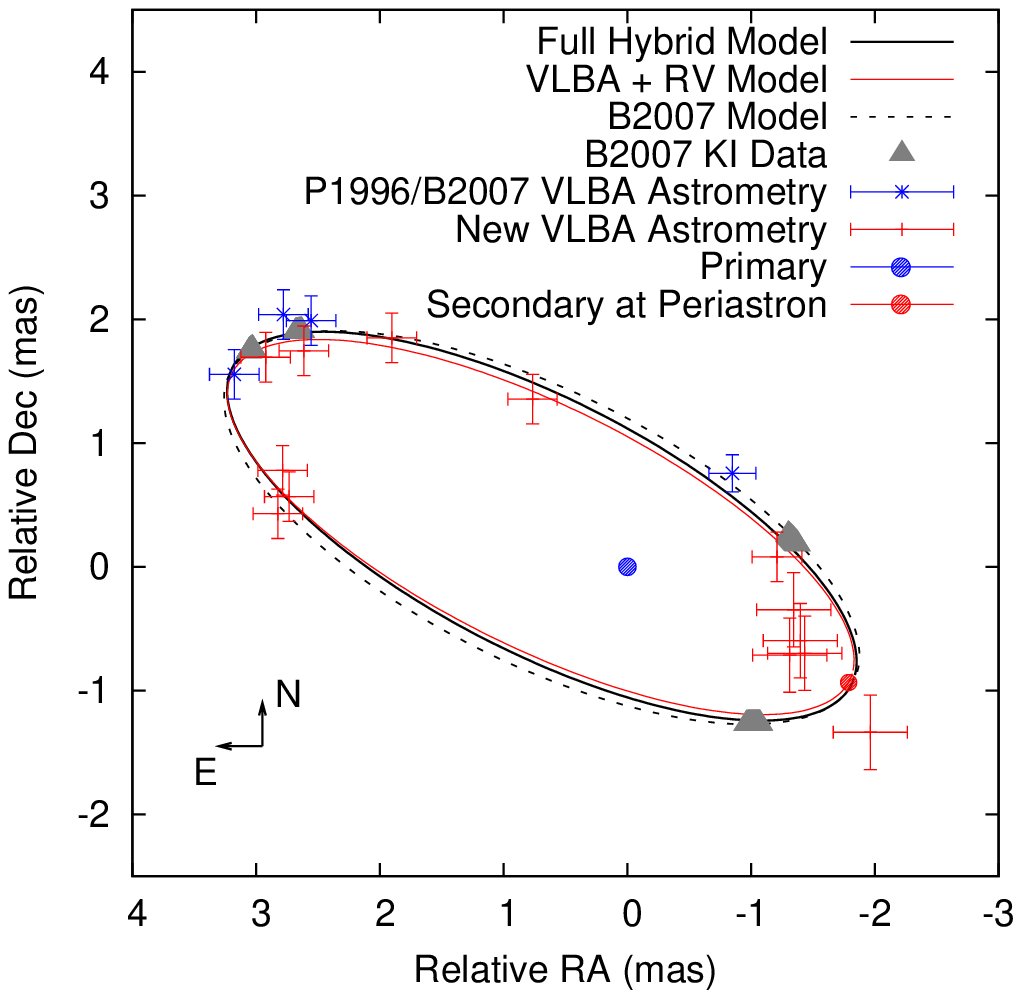}}
\caption
{\footnotesize{VLBA Relative astrometry and orbit models of
V773~Tau~A. Here VLBA relative astrometry from P1996, B2007 and this
work is shown with orbit model renderings for the B2007 orbit and two
models derived here (the ``VLBA + RV'' and ``Full Hybrid'' models
described in the text). We render the primary (Aa) component at the
origin and the secondary (Ab) at periastron. The apparent sizes of the
V773~Tau~A components are estimated by SED modeling (B2007) and
rendered to scale.}\label{fig-v773-2}}\end{figure*}

\medskip

\noindent
Figure \ref{fig-v773-2} shows all available relative astrometry
data on V773~Tau~A (from Phillips et al.\ 1993, B2007, and this work),
along with several orbit models including that from B2007 (dashed
line). Clearly there is general agreement between the old and new VLBA
relative astrometry on V773~Tau~A, and the B2007 orbit model. A close
inspection of the VLBA separations and the B2007 orbit model in Fig.\
\ref{fig-v773-2}, however, shows an interesting trend in the VLBA
astrometry. At most orbit phases the VLBA-derived separations and
B2007 orbit model are in good agreement. But in most observations near
periastron (secondary south-west of the primary) the VLBA separations
appear systematically smaller (i.e.\ secondary nearer the primary)
than predicted by the B2007 orbit. This suggests one of two
possibilities: there is a possible bias in the B2007 orbit solution,
or a possible bias in the VLBA-derived astrometry near periastron
(where there is known enhancement in the radio flaring; Massi et al.\
2002, 2006).

\begin{deluxetable}{lr@{$\pm$}lr@{$\pm$}l}
\tabletypesize{\scriptsize}
\tablecolumns{5}
\tablewidth{0pc}
\tablecaption{\footnotesize{Orbital parameters of V773~Tau.}
\label{tab-orbit}}
\tablehead{
\colhead{Orbital Parameter}         &
\multicolumn{2}{c}{B2007}           &
\multicolumn{2}{c}{``Full Hybrid''} }
\startdata
Period (days) \dotfill          & 51.1039  & 0.0021 & 51.1028  & 0.0019 \\
\textit{T$_o$} (MJD) \dotfill   & 53059.73 & 0.33   & 53059.68 & 0.30   \\
\textit{e} \dotfill             & 0.2717   & 0.0085 & 0.2732   & 0.0070 \\
\textit{K$_A$} (\kmps) \dotfill & 35.90    & 0.53   & 35.94    & 0.53   \\%
\textit{K$_B$} (\kmps) \dotfill & 41.5     & 1.4    & 41.5     & 1.4    \\%
$\gamma$ (\kmps) \dotfill       & 0.02     & 0.32   & 0.03     & 0.32   \\%
$\omega_A$ (deg) \dotfill       & 4.6      & 2.4    & 5.0      & 2.3    \\
$\Omega$ (deg) \dotfill         & 63.5     & 1.7    & 62.7     & 1.1    \\
\textit{i} (deg) \dotfill       & 66.0     & 2.4    & 67.8     & 1.6    \\
\textit{a} (mas) \dotfill       & 2.811    & 0.047  & 2.810    & 0.034  \\
\enddata
\end{deluxetable}

\medskip

\noindent
To investigate this issue we considered two different prescriptions
for integrating the new VLBA data into the orbit modeling. In the
first ``VLBA + RV'' model we considered the possibility that the KI
visibilities were the source of a possible orbit bias in B2007, and
used only the (old and new) VLBA-derived relative astrometry and
double-lined RV from B2007 to derive an orbit model. In the second
``Full Hybrid'' model we used all available data (VLBA astrometry, KI
visibilities, and RV), but assume that the VLBA separations near
periastron were biased by the enhanced flaring activity (e.g.\ if the
enhanced flares preferentially occur between the two stars), assigning
these points lower weight (specifically a factor of 50\% larger
1-sigma error per axis for orbit phases within 10\% of
periastron).\footnote{Our methods for orbit modeling with heterogeneous
RV and astrometric/visibility datasets are described in Boden et
al.~(2000) and are not repeated here. In all the orbit modeling we
have weighted all VLBA astrometry consistently with the weighting
derived in B2007, at 0.2 mas per axis 1-sigma, except as noted for
points near periastron in the ``Full Hybrid''.} We find that both new
orbit solutions are in excellent agreement with the original B2007
model, and renderings of both these orbit models are included in Fig.\
\ref{fig-v773-2}. It is clear that the KI visibility data is
reliable at its stated uncertainties, and the most plausible
hypothesis is that the VLBA relative astrometry near periastron
contains biases associated with the enhanced flaring activity
documented by Massi et al.\ (2002, 2006). This astrometric bias was
discussed in section \ref{c5-variability}. Tab.\ \ref{tab-orbit} gives
a direct comparison between the orbit models from B2007 and the ``Full
Hybrid'' model derived here; in all cases the orbital parameters
between the two models (which share a significant amount of underlying
data) are in excellent statistical agreement. For all subsequent
analysis we adopt this ``Full Hybrid'' model for the updated
V773~Tau~A orbit.

\medskip

\noindent
Given the agreement between the ``Full Hybrid'' model and the B2007
orbit model, the updated physical parameters are highly consistent
with B2007 estimates. Component dynamical masses resulting from this
orbit model $1.48\pm0.12$ \Msun\ and $1.28\pm0.07$ \Msun, for the
primary and the secondary respectively -- consistent with the B2007
estimates to 4\% and 0.5-sigma. Similarly, the system distance
estimate derived from the updated orbit model is $134.5\pm3.2$ pc,
again consistent with the value from B2007 to 1\% and 0.4-sigma. As a
final note, we should stress that the present fit must still be
considered preliminary as astrometric and RV observations to assess
the gravitational effect of the two other members of the system
(V773~Tau~B and C) are ongoing.

\chapter{Conclusions}\label{ch-conclusions}

\section{General conclusions}\label{c7-general}

\subsection{Taurus}

\textbf{Distances}---
For T~Tau~Sb, a resulting parallax of $6.82\pm0.03$ mas was obtained,
corresponding to a distance of $d=146.7\pm0.6$ pc. For Hubble~4 and
HDE~283572, obtained the parallaxes of $7.53\pm0.03$ mas
($d=132.8\pm0.5$ pc) and $7.78\pm0.04$ mas ($d=128.5\pm0.6$ pc),
respectively. HP~Tau/G2 yielded a parallax of $6.20\pm0.03$ mas,
corresponding to $d=161.2\pm0.9$ pc. Finally, for V773~Tau~A we
obtained a parallax of $7.57\pm0.20$ mas, corresponding to
$d=132.0\pm3.5$ pc. This increase the precision of the distance to
this star-forming region by 1-2 orders of magnitude.

\medskip

\noindent
\textbf{Proper motions}---
The proper motion of HDE~283572 was found to be linear and uniform,
but for Hubble~4 and HP~Tau/G2 we find significant residuals in
declination (but not in right ascension). It is unclear at the moment
whether these residuals are the consequence of an unseen companion, of
structure in the magnetospheres of the star, or of residual phase
errors in our calibration. V773~Tau~A was found to be a resolved
double radio source. The physical orbit of the system was constrained
by combining optical radial velocity measurements, Keck Interferometer
data, and our own VLBA data. This fit allowed us to refine the
determination of the masses of the two stars in the system, and to
deduce the position of the barycenter of the system at each epoch. A
fit assuming a uniformly accelerated proper motion appeared to provide
an adequate description of the trajectory of T~Tau~Sb. In this case,
our observations are not sufficient to constrain the orbit of the
system, but they provide information consistent with, and
complementary, existing near infrared data.

\medskip

\noindent
\textbf{Three-dimensional structure}---
The resulting mean distance to the Taurus complex is about 139 pc, in
excellent agreement with previous determinations. The total spatial
extent of Taurus on the sky is about $10^\circ$, corresponding to a
physical size of about 25 pc. The observations of the five stars
presented already provide some hints of what the three-dimensional
structure of Taurus might be. Hubble~4, HDE~283572, and V773~Tau~A
which were found to be at about 130 pc, are located in the same
portion of Taurus, near Lynds 1495. T~Tau~Sb is located in the
southern part of Taurus near Lynds 1551, its tangential velocity is
clearly different from that of Hubble~4, HDE~283572, and V773~Tau~A,
and it appears to be somewhat farther from us. Finally, HP~Tau/G2 is
located near the (Galactic) eastern edge of Taurus, and is the
farthest of the four sources considered here. Although additional
observations are needed to draw definite conclusions, our data,
therefore, suggest that the region around Lynds 1495 corresponds to
the near side of the Taurus complex at about 130 pc, while the eastern
side of Taurus corresponds to the far side at 160 pc. The region
around Lynds 1551 and T~Tau~Sb appears to be at an intermediate
distance of about 147 pc. Taurus has long been known to present a
filamentary structure. The two main filaments are roughly parallel to
one another, and have an axis ratio of about 7:1. Our observations
suggest that these filaments are oriented nearly along the line of
sight, i.e.\ roughly along the Galactic center--anticenter axis. This
peculiar orientation might indeed explain the low star-forming
efficiency of Taurus compared with other nearby star-forming regions.

\subsection{Ophiuchus}

\textbf{Distances}---
The resulting parallaxes of S1 and DoAr~21 are $8.55\pm0.50$ mas and
$8.20\pm0.37$ mas, respectively. This corresponds to
$d=116.9^{+7.2}_{-6.4}$ pc for S1 and $d=121.9^{+5.8}_{-5.3}$ pc for
DoAr~21. The resulting mean distance to the Ophiuchus complex is
$120.0^{+4.5}_{-4.2}$ pc, in excellent agreement with several recent
determination (e.g.\ Knude \& Hog 1999; de Geus et al.\ 1989; Lombardi
et al.\ 2008).

\medskip

\noindent
\textbf{Proper motions}---
Assuming the proper motion of S1 to be uniform appears to provide an
adequate fit, but with roughly periodic residuals (in both right
ascension and declination) with a period of about 0.7 years. This is
of the correct order of magnitude to be interpreted as the reflex
motion of S1 due to its known companion. DoAr~21 was found to behave
much like S1: a uniform proper motion provides a good fit but with
roughly periodic residuals (with a period of about 1.2 years in this
case), and the source was found to be double in at least one of our
images. We conclude that DoAr~21 is likely to belong to a binary
system.

\medskip

\noindent
\textbf{Three-dimensional structure}---
Ophiuchus is composed of a compact core, only about 2 pc across, and
filamentary structures (called ``streamers'') extending (in
projection) to about 10 pc. The Ophiuchus core is sufficiently compact
that we do not expect to resolve any structure along the line of
sight, and our observations show that it is at a distance of 120
pc. There could potentially be distance gradients of several parsecs
across the streamers. We note, however, that Schaefer et al.\ (2008)
determined the physical orbit of the binary system Haro 1-14C, and
deduced a distance of $111\pm19$ pc, in good agreement with our
determination. Haro 1-14C is located in the northern streamer
(associated with the darks clouds L1709/L1704), so the result of
Schaefer et al.\ (2008) suggests that that streamer is, if anything,
somewhat closer that the core.  This is, indeed, in agreement with
recent results of Lombardi et al.\ (2008).  On the other hand, Imai et
al.\ (2007) used the Japanese VLBI system (VERA) to determine the
parallax to the very young protostar IRAS~16293--2422 deeply embedded
in the southern Ophiuchus streamer (in L1689N). They obtain a distance
of $178^{+18}_{-37}$ pc, which would be more consistent with the older
value of 165 pc. Even including the streamers, Ophiuchus is only 10 pc
across in projection, so it is unlikely to be 60 pc deep. Thus, if the
results of Imai et al.\ (2007) are confirmed, they would indicate the
existence of several unrelated star-forming regions along the line of
sight.

\subsection{Multiplicity and variability}

At least 4 of the 7 sources in our sample (57\%) turn out to be tight
binary systems with separations between a few and a few tens of
milli-arcseconds. This represents a binarity fraction much larger than
that of main sequence stars for the same separation range. We argue
that a strong selection effect is likely to be at work. The systems
considered in this thesis were selected because they were known to be
non-thermal emitters previously detected with VLBI techniques. The
high binary rate may, therefore, indicate that tight binaries are more
likely to emit non-thermal radio emission than looser binaries or
single stars. This idea is reinforced by the observations of
V773~Tau~A where we confirm that the emitted flux depends strongly on
the separation between the two stars (the flux is strongest near the
periastron of the system, and weakest near apoastron). Non-thermal
emission is believed to be created during reconnection events in the
active magnetospheres of the stars. Reconnection events {\em between}
the stars (in addition to those within the individual magnetospheres)
in tight binary systems might naturally explain the higher radio flux
of such systems.

\section{Future prospects}\label{c7-future}

The results presented in this thesis have allowed us to refine the
determination of the distance to two important regions of nearby
low-mass star-formation and to start examine the three-dimensional
structure of Taurus and Ophiuchus complexes. They also raised a number
of issues that could be tackled with new data and suggested several
follow-up studies.

\medskip

\noindent
\textbf{Interacting binary V773~Tau~Aa/Ab}---
The radio flux of V773~Tau~A depends on the orbit phase, being highest
at periastron and weakest at apoastron. In most of our observations,
it is resolved into two components associated with the two stars in
the system. Our observations have shown, however, that near
periastron, the position of the radio sources is significantly
displaced from the position of the associated star. This is additional
evidence that the non-thermal emission in such systems is affected by
the presence of a close companion. In a recently accepted proposal, we
requested time to observe V773~Tau~A near periastron with the High
Sensitivity Array (a composite VLBI array comprised of the Very Large
Baseline Array, the Green Bank Telescope, and the Very Large Array
plus the Arecibo dish in Puerto Rico). These observations will allow
us to examine the spatial evolution of this interacting binary when it
is near periastron over a six hour period. This ought to shed light on
the origin of the variability of the source.

\medskip

\noindent
\textbf{Stellar structure and emission of pre-main sequence stars}---
We found interesting characteristics in the structure and emission of
the sources in both Taurus and Ophiuchus complexes. \textbf{(1)}
T~Tau~Sb was found to be somewhat extended toward the northeast
direction, and with a low brightness extension probably associated
with an accretion disk (Loinard et al.\ 2005). \textbf{(2)} Hubble~4
was found to have large declination residuals, that can be explained
if Hubble~4 had a companion, and the residuals reflect the
corresponding reflex motion (Torres et al.\ 2007), but the residuals
are relatively poorly constrained with the existing data, and
additional observations would help to confirm the periodicity in the
residuals. \textbf{(3)} For one epoch we detect that HP~Tau/G2
underwent a flaring event, reaching a flux 4 times higher than at
other epochs. One year after the flare, we found that the source is
extended with a deconvolved size in the north-south direction of about
2.5 mas. This increase in the source size might be due to variation in
the structure of the active magnetosphere (Torres et al.\
2009).

\medskip

\noindent
We are interested in studying with more detail, the structure and
emission mechanism of these stellar systems. Since we found that
V773~Tau~A system experienced large changes in structure and flux in
about one hour, we would like to observe some stars again and
re-analyze the data that have already been collected.

\medskip

\noindent
\textbf{Along the line of sight to Ophiuchus}---
To decide whether or not several unrelated star-forming regions exist
along the line of sight to Ophiuchus, we need more observations of
several new sources in the direction of the complex. VSSG~14 is a
source in line of sight to Oph-B subregion, and we have been observed
at 14 epochs from 2005 to 2007. Also, we have been obtained new
observations of the source Rox~39 that belongs to the region between
L1686 and L1689. We are in the process of reducing both VSSG~14 and
Rox~39 data. Additionally we would like to obtain observations in the
southern streamer of Ophiuchus, in particular of IRAS~16293--2422 to
check whether or not it is at nearly 180 pc, as reported by Imai et
al.\ (2007).

\medskip

\noindent
\textbf{Serpens and Perseus star-forming regions}---
Several other nearby star-forming regions have been studied in detail
at many wavelengths but have poorly determined distances. Non-thermal
sources are known to exist in these regions, so multi-epoch VLBA
observations would allow significant improvements in the determination
of their distances. New observations of Serpens are currently obtained
to measure the parallax to sources in these regions. Perseus will be
the subject of forthcoming proposals. These regions have been actively
studied in recently years, and our observations will provide distances
with a precision of about 3-5\%.

\medskip

\noindent
\textbf{A survey of radio emitting stars in Taurus and Ophiuchus}---
Our existing data already provide hints of the structure of both
Taurus and Ophiuchus. In addition, we have made a limited survey
(VLBA+VLA) of 17 sources in the Taurus star forming region recently,
that allowed us to detect several new sources. A large systematic VLBA
survey would almost certainly reveal many new non-thermal sources in
Taurus and Ophiuchus, that could then be observed to get their
parallax and proper motions. This, in turn, would allow us to
construct a high quality three-dimensional map of these two
interesting star-forming regions. We are planning to submit a large
VLBA proposal to initiate a large search for non-thermal radio sources
in Ophiuchus and Taurus. The sources that will be detected will be
observed at multiple epochs.


\begin{appendix}

\chapter
[Multiepoch VLBA Observations of T~Tauri South]{\normalsize Multiepoch
VLBA Observations of T~Tauri South}

\noindent
{\small Laurent Loinard, Amy J.\ Mioduszewski, Luis F.\ Rodríguez,
Rosa A.\ González-Lópezlira, Mónica Rodríguez \& Rosa M.\ Torres\\
\textit{The Astrophysical Journal, 619, L179, 2005}}

\addtocontents{toc}{
\noindent
Laurent Loinard, Amy J.\ Mioduszewski, Luis F.\ Rodríguez, Rosa A.\
González-Lópezlira, Mónica Rodríguez \& Rosa M. Torres\\
\textit{The Astrophysical Journal, 619, L179, 2005}
\protect\par}
\setcounter{page}{125}

\smallskip

\begin{quote}
\noindent
{\small \textbf{Abstract}. We present a series of seven observations
of the compact, nonthermal radio source associated with T~Tauri South
made with the Very Long Baseline Array (VLBA) over the course of 1
year. The emission is found to be composed of a compact structure most
certainly originating from the magnetosphere of an underlying pre-main
sequence star and a low brightness extension that may result from
reconnection flares at the star-disk interface. The accuracy of the
absolute astrometry offered by the VLBA allows us to make very precise
determinations of the trigonometric parallax and proper motion of
T~Tau South. The proper motion derived from our VLBA observations
agrees with that measured with the Very Large Array over a similar
period to better than 2 mas yr$^{-1}$, and it is fully consistent with
the infrared proper motion of T~Tau~Sb, the pre-main-sequence M star
with which the radio source has traditionally been associated. The
parallax, $\pi=7.07\pm0.14$ mas, corresponds to a distance of
$141.5^{+2.8}_{-2.7}$ pc.}
\end{quote}

\chapter
[VLBA Distance to Nearby Star-Forming Regions I] {\normalsize VLBA
Distance to Nearby Star-Forming Regions\\ I. The Distance to T~Tauri
with 0.4 $\%$ Accuracy}

\noindent
{\small Laurent Loinard, Rosa M.\ Torres, Amy J.\ Mioduszewski, Luis
F.\ Rodríguez, Rosa A.\ Gónzalez-Lópezlira, Régis Lachaume, Virgilio
Vázquez \& Erandy González\\
\textit{The Astrophysical Journal, 671, 546, 2007}}

\addtocontents{toc}{
\noindent
Laurent Loinard, Rosa M.\ Torres, Amy J.\ Mioduszewski, Luis F.\
Rodríguez, Rosa A.\ Gónzalez-Lópezlira, Régis Lachaume, Virgilio
Vázquez \& Erandy González\\
\textit{The Astrophysical Journal, 671, 546, 2007}
\protect\par}
\setcounter{page}{131}

\smallskip

\begin{quote}
\noindent
{\small \textbf{Abstract}. We present the results of a series of 12
3.6 cm radio continuum observations of T~Tau Sb, one of the companions
of the famous young stellar object T~Tauri. The data were collected
roughly every 2 months between 2003 September and 2005 July with the
Very Long Baseline Array (VLBA). Thanks to the remarkably accurate
astrometry delivered by the VLBA, the absolute position of T~Tau~Sb
could be measured with a precision typically better than about 100
$\mu$as at each of the 12 observed epochs. The trajectory of T~Tau~Sb
on the plane of the sky could therefore be traced very precisely and
was modeled as the superposition of the trigonometric parallax of the
source and an accelerated proper motion. The best fit yields a
distance to T~Tau~Sb of $147.6\pm0.6$ pc. The observed positions of
T~Tau~Sb are in good agreement with recent infrared measurements, but
they seem to favor a somewhat longer orbital period than that recently
reported by Duchêne and coworkers for the T~Tau~Sa/T~Tau~Sb system.}
\end{quote}

\chapter
[VLBA Distance to Nearby Star-Forming Regions II]{\normalsize VLBA
Distance to Nearby Star-Forming Regions\\ II. Hubble 4 and HDE 283572
in Taurus}

\noindent
{\small Rosa M.\ Torres, Laurent Loinard, Amy J.\ Mioduszewski \& Luis
F.\ Rodríguez\\
\textit{The Astrophysical Journal, 671, 1813, 2007}}

\addtocontents{toc}{
\noindent
Rosa M.\ Torres, Laurent Loinard, Amy J.\ Mioduszewski \& Luis F.\
Rodríguez\\
\textit{The Astrophysical Journal, 671, 181, 2007}
\protect\par}
\setcounter{page}{143}

\smallskip

\begin{quote}
\noindent
{\small \textbf{Abstract}.The nonthermal 3.6 cm radio continuum
emission from the naked T~Tauri stars Hubble~4 and HDE~283572 in
Taurus has been observed with the Very Long Baseline Array (VLBA) at
six epochs between 2004 September and 2005 December with a typical
separation between successive observations of 3 months. Thanks to the
remarkably accurate astrometry delivered by the VLBA, the trajectory
described by both stars on the plane of the sky could be traced very
precisely and modeled as the superposition of their trigonometric
parallax and uniform proper motion. The best fits yield distances to
Hubble~4 and HDE~283572 of $132.8\pm0.5$ and $128.5\pm0.6$ pc,
respectively. Combining these results with the other two existing VLBI
distance determinations in Taurus, we estimate the mean distance to
the Taurus association to be 137 pc with a dispersion (most probably
reflecting the depth of the complex) of about 20 pc.}
\end{quote}

\chapter
[A Preliminary Distance to the Core of Ophiuchus]{\normalsize A
Preliminary Distance to the Core of Ophiuchus,\\ with an Accuracy of 4
$\%$}

\noindent
{\small Laurent Loinard, Rosa M.\ Torres, Amy J.\ Mioduszewski \& Luis
F. Rodríguez\\
\textit{The Astrophysical Journal, 675, L29, 2008}}

\addtocontents{toc}{
\noindent
Laurent Loinard, Rosa M.\ Torres, Amy J.\ Mioduszewski \& Luis
F. Rodríguez\\
\textit{The Astrophysical Journal, 675, L29, 2008}
\protect\par}
\setcounter{page}{153}

\smallskip

\begin{quote}
\noindent
{\small \textbf{Abstract}. The nonthermal 3.6 cm radio continuum
emission from the young stars S1 and DoAr~21 in the core of Ophiuchus
has been observed with the VLBA at 6 and 7 epochs, respectively,
between June 2005 and August 2006. The typical separation between
successive observations was 2-3 months. Thanks to the remarkably
accurate astrometry delivered by the Very Long Baseline Array (VLBA),
the trajectory described by both stars on the plane of the sky could
be traced very precisely, and modeled as the superposition of their
trigonometric parallax and a uniform proper motion. The best fits
yield distances to S1 and DoAr~21 of $116.9^{+7.2}_{-6.4}$ and
$121.9^{+5.8}_{-5.3}$ pc, respectively. Combining these results, we
estimate the mean distance to the Ophiuchus core to be
$120.0^{+4.5}_{-4.2}$ pc, a value consistent with several recent
indirect determinations, but with a significantly improved accuracy of
4\%. Both S1 and DoAr~21 happen to be members of tight binary systems,
but our observations are not frequent enough to properly derive the
corresponding orbital parameters. This could be done with additional
data, however, and would result in a significantly improved accuracy
on the distance determination.}
\end{quote}

\chapter
[Tidal Forces as a Regulator of Star Formation in Taurus]{\normalsize
Tidal Forces as a Regulator of Star Formation in Taurus}

\noindent
{\small Javier Ballesteros-Paredes, Gilberto Gómez, Laurent Loinard,
Rosa M.\ Torres \& Bárbara Pichardo\\
\textit{Monthly Notices of the Royal Astronomical Society: Letters, 395, L81, 2009}}

\addtocontents{toc}{
\noindent
Javier Ballesteros-Paredes, Gilberto Gómez, Laurent Loinard, Rosa M.\
Torres \& Bárbara Pichardo\\
\textit{Monthly Notices of the Royal Astronomical Society: Letters, 395, L81, 2009}
\protect\par}
\setcounter{page}{159}

\smallskip

\begin{quote}
\noindent
{\small \textbf{Abstract}. Only a few molecular clouds in the solar
neighbourhood exhibit the formation of only low-mass
stars. Traditionally, these clouds have been assumed to be supported
against more vigorous collapse by magnetic fields. The existence of
strong magnetic fields in molecular clouds, however, poses serious
problems for the formation of stars and of the clouds themselves. In
this Letter, we review the three-dimensional structure and kinematics
of Taurus --the archetype of a region forming only low-mass stars-- as
well as its orientation within the Milky Way. We conclude that the
particularly low star formation efficiency in Taurus may naturally be
explained by tidal forces from the Galaxy, with no need for magnetic
regulation or stellar feedback.}
\end{quote}

\chapter
[VLBA Distance to Nearby Star-Forming Regions III]{\normalsize VLBA
Distance to Nearby Star-Forming Regions\\ III. HP~Tau/G2 and the 3D
Structure of Taurus Cloud}
\setcounter{page}{165}

\noindent
{\small Rosa M.\ Torres, Laurent Loinard, Amy J.\ Mioduszewski \& Luis
F.\ Rodríguez\\
\textit{The Astrophysical Journal, 698, 242, 2009}}

\addtocontents{toc}{
\noindent
Rosa M.\ Torres, Laurent Loinard, Amy J.\ Mioduszewski \& Luis F.\
Rodríguez\\
\textit{The Astrophysical Journal, 698, 242, 2009}
\protect\par}

\smallskip

\begin{quote}
\noindent
{\small \textbf{Abstract}. Using multiepoch Very Long Baseline Array
(VLBA) observations, we have measured the trigonometric parallax of
the weak-line T~Tauri star HP~Tau/G2 in Taurus. The best fit yields a
distance of $161.2\pm0.9$ pc, suggesting that the eastern portion of
Taurus (where HP~Tau/G2 is located) corresponds to the far side of the
complex. Previous VLBA observations have shown that T~Tau, to the
south of the complex, is at an intermediate distance of about 147 pc,
whereas the region around L1495 corresponds to the near side at
roughly 130 pc. Our observations of only four sources are still too
coarse to enable a reliable determination of the three-dimensional
structure of the entire Taurus star-forming complex. They do
demonstrate, however, that VLBA observations of multiple sources in a
given star-forming region have the potential not only to provide a
very accurate estimate of its mean distance, but also to reveal its
internal structure. The proper motion measurements obtained
simultaneously with the parallax allowed us to study the kinematics of
the young stars in Taurus. Combining the four observations available
so far, we estimate the peculiar velocity of Taurus to be about 10.6
km s$^{-1}$ almost completely in a direction parallel to the Galactic
plane. Using our improved distance measurement, we have refined the
determination of the position on the H-R diagram of HP~Tau/G2, and of
two other members of the HP~Tau group (HP~Tau itself and
HP~Tau/G3). Most pre-main sequence evolutionary models predict
significantly discrepant ages (by 5 Myr) for those three stars
--expected to be coeval. Only in the models of Palla \& Stahler do
they fall on a single isochrone (at 3 Myr).}
\end{quote}

\chapter
[VLBA Distance to Nearby Star-Forming Regions IV]{\normalsize VLBA
Distance to Nearby Star-Forming Regions\\ IV. Distance and Dynamical
Masses for V773~Tau~A}
\setcounter{page}{175}

\noindent
{\small Rosa M.\ Torres, Laurent Loinard, Andrew F.\ Boden, Amy J.\
Mioduszewski \& Luis F.\ Rodríguez\\
\textit{In preparation for The Astrophysical Journal}}

\addtocontents{toc}{
\noindent
Rosa M.\ Torres, Laurent Loinard, Andrew F.\ Boden, Amy J.\ Mioduszewski
\& Luis F.\ Rodríguez\\
\textit{In preparation for The Astrophysical Journal}
\protect\par}

\smallskip

\begin{quote}
\noindent
{\small \textbf{Abstract}. We present multi-epoch Very Long Baseline
Array (VLBA) observations of the 51-day binary component in the
quadruple young stellar system V773~Tau. Combined with previous
interferometric and radial velocity measurements, these new data
enable us to improve the characterization of the physical orbit of the
system.  In particular, we infer dynamical masses for the primary and
the secondary of 1.48 $\pm$ 0.12
\Msun, and 1.28 $\pm$ 0.07 \Msun, respectively, and a distance to the
system of 134.5 $\pm$ 3.2 pc.  Using the improved orbit, we can
calculate the absolute coordinates of the barycenter of the system
from the VLBA observations, and fit for its trigonometric parallax and
proper motion. The best fit yields a parallax of 7.7 $\pm$ 0.2 mas,
corresponding to a distance of 130.2$^{+3.5}_{-3.3}$ pc, in good
agreement with the estimate based on the orbital fit. In projection,
V773~Tau and two other young stars (Hubble 4 and HDE~283572) recently
observed with the VLBA are located near the dark cloud Lynds 1495, in
the central region of Taurus. These three stars appear to have similar
proper motions and trigonometric parallaxes, so we argue that the
weighted mean of their parallaxes provides a good estimate of the
distance to Lynds 1495. This weighted mean (7.62 $\pm$ 0.10 mas)
corresponds to $d$ = 131.2 $^{+1.8}_{-1.7}$ pc.}
\end{quote}

\chapter
[The Very Long Baseline Array]{The Very Long Baseline
Array}\label{chap-vlba}
\setcounter{page}{187}

\begin{quote}
\noindent
The Very Long Baseline Array (VLBA) is a system of ten radio
telescopes controlled remotely from the Domenici Science Operations
Center (DSOC) in Socorro, New Mexico (USA) by the National Radio
Astronomy Observatory (NRAO). The array works together as the world
largest dedicated, full-time astronomical instrument using the
technique of VLBI. From Mauna Kea on the Big Island of Hawaii to St.\
Croix in the U.S.\ Virgin Islands, the VLBA spans more than 5,000
miles. This Appendix attempts to summarize the instrument information
for astronomers who does not used the VLBA before. All the information
here has been adapted from the VLBA observational status summary
available through the VLBA astronomer page at
\textsf{http://www.vlba.nrao.edu/astro/}. It is included here for
completeness.
\end{quote}

\section{Antenna sites}\label{apendiceH-sites}

Tab.\ \ref{tab-antenas} gives the surveyed geographic locations of the
10 antennas comprising the VLBA. All locations are based on the WGS84
ellipsoid used by the GPS system, with Earth radius $a=6378.137$ km
and flattening $1/f=298.257223563$ (Napier 1995).

\medskip

\noindent
Several other radio telescopes often participate in VLBI observing in
conjunction with the VLBA. These include the Very Large Array (VLA),
either with up to 27 antennas added in phase (Y27) or with a single
antenna (Y1); the Green Bank Telescope (GBT); Arecibo (AR); Effelsberg
(Eb); the European VLBI Network (EVN); plus (occasionally) various
geodetic antennas or the NASA Deep Space Network.

\medskip

\noindent
A total of up to 100 hours per four-month trimester has been reserved
for a High Sensitivity Array (HSA) composed of the VLBA, VLA, GBT, AR,
and Eb. The VLA and GBT are NRAO facilities, while Arecibo is operated
by the National Astronomy and Ionosphere Center (NAIC), and Effelsberg
is operated by Max Planck Institut für Radioastronomie (MPIfR). Tab.\
\ref{tab-antenas} also lists the locations of the HSA telescopes.

\begin{deluxetable}{lrrr}
\tabletypesize{\scriptsize}
\tablecolumns{4}
\tablewidth{0pc}
\tablecaption{\footnotesize{Locations of VLBA and HSA telescopes.}
\label{tab-antenas}}
\tablehead{
\colhead{~~~~~~~~~~~~~~~~~~~~Location~~~~~~~~~~~~~~~~~~~~} &
\multicolumn{1}{c}{Latitud}            &
\multicolumn{1}{c}{Longitud}           &
\multicolumn{1}{c}{Elevation}                    }
\startdata
Saint Croix, Vinginia (SC)\dotfill     & \decdms{17}{45}{23}{68} &  \decdms{64}{35}{01}{07}  & 16   \\%
\\[-0.3cm]
Hancock, New Hampshire (HN)\dotfill    & \decdms{42}{56}{00}{99} &  \decdms{71}{59}{11}{69}  & 296  \\%
\\[-0.3cm]
North Liberty, Iowa (NL)\dotfill       & \decdms{41}{46}{17}{13} &  \decdms{91}{34}{26}{88}  & 222  \\%
\\[-0.3cm]
Fort Davis, Texas (FD)\dotfill	       & \decdms{30}{38}{06}{11} &  \decdms{103}{56}{41}{34} & 1606 \\%
\\[-0.3cm]
Los Alamos, New Mexico (LA)\dotfill    & \decdms{35}{46}{30}{45} &  \decdms{106}{14}{44}{15} & 1962 \\%
\\[-0.3cm]
Pie Town, New Mexico (PT)\dotfill      & \decdms{34}{18}{03}{61} &  \decdms{108}{07}{09}{06} & 2365 \\%
\\[-0.3cm]
Kitt Peak, Arizona (KP)\dotfill	       & \decdms{31}{57}{22}{70} &  \decdms{111}{36}{44}{72} & 1902 \\%
\\[-0.3cm]
Owens Valley, California (OV)\dotfill  & \decdms{37}{13}{53}{95} &  \decdms{118}{16}{37}{37} & 1196 \\%
\\[-0.3cm]
Brewster, Washington (BR)\dotfill      & \decdms{48}{07}{52}{42} &  \decdms{119}{40}{59}{80} & 250  \\%
\\[-0.3cm]
Mauna Kea, Hawaii (MK)\dotfill	       & \decdms{19}{48}{04}{97} &  \decdms{155}{27}{19}{81} & 3763 \\%
\\[-0.1cm]
Arecibo, Puerto Rico (AR)\dotfill      & \decdms{18}{20}{36}{60} &  \decdms{66}{45}{11}{10}  & 497  \\%
\\[-0.3cm]
Green Bank, West Virginia (GB)\dotfill & \decdms{38}{25}{59}{24} &  \decdms{79}{50}{23}{41}  & 807  \\%
\\[-0.3cm]
VLA, New Mexico (Y27)\dotfill          & \decdms{34}{04}{43}{75} &  \decdms{107}{37}{05}{91} & 2115 \\%
\\[-0.3cm]
Effelsberg, Germany (EB)\dotfill       & \decdms{50}{31}{30}{00} &  \decdms{$-$6}{53}{00}{30}& 319  \\%
\enddata
\end{deluxetable}

\section{Antennas}\label{apendiceH-antennas}

The main reflector of each VLBA antenna is a 25 m diameter dish which
is a shaped figure of revolution with a focal-length-to-diameter ratio
of 0.354. A 3.5 m diameter Cassegrain subreflector with a shaped
asymmetric figure is used at all frequencies above 1 GHz, while the
prime focus is used at lower frequencies. The antenna features a
wheel-and-track mount, with an advanced-design reflector support
structure. Elevation motion occurs at a rate of $30^\circ$ per minute
between a hardware limit of $2^\circ$ and a software limit of
$90^\circ$. Azimuth motion has a rate of 90$^\circ$ per minute between
limits of $-90^\circ$ and $450^\circ$. Antennas are stowed to avoid
operation in high winds, or in case of substantial snow or ice
accumulation.

\section{Performance parameters}\label{apendiceH-performance}

Tab.\ \ref{tab-frequency} gives the nominal frequency ranges for the
receiver/feed combinations available on all or most VLBA antennas
(Thompson 1995). Passband-limiting filters are described by Thompson
(1995). Measured frequency ranges are broader than nominal. Measured
frequency ranges may be especially important for avoiding radio
frequency interference (RFI), and for programs involving extragalactic
lines, rotation measures (Cotton 1995; Kemball 1999), and
multi-frequency synthesis (Conway \& Sault 1995; Sault \& Conway
1999).

\begin{deluxetable}{ccccccc}
\rotate
\tabletypesize{\scriptsize}
\tablecolumns{7}
\tablewidth{0pc}
\tablecaption{\footnotesize{Frequency ranges and typical performance parameters.}\label{tab-frequency}}
\tablehead{
\colhead{Receivers}                        &
\colhead{Nominal Frequency}                &
\colhead{Typical}                          &
\colhead{Center Frequency}                 &
\colhead{Typical}                          &
\colhead{Baseline Sensitivity}             &
\colhead{Image Sensitivity}                \\
\multicolumn{1}{c}{and Feeds}              &
\multicolumn{1}{c}{Range}                  &
\multicolumn{1}{c}{Zenith SEFD}            &
\multicolumn{1}{c}{for SEFD}               &
\multicolumn{1}{c}{Zenith Gain}            &
\multicolumn{1}{c}{$\Delta S^{256.2{\rm m}}$}        &
\multicolumn{1}{c}{$\Delta I^{256.8{\rm m}}_{\rm m}$}\\
\multicolumn{1}{c}{}                                 &
\multicolumn{1}{c}{[GHz]}                            &
\multicolumn{1}{c}{[Jy]}                             &
\multicolumn{1}{c}{[GHz]}                            &
\multicolumn{1}{c}{[K Jy$^{-1}$]}                    &
\multicolumn{1}{c}{[mJy]}                            &
\multicolumn{1}{c}{[$\mu{\rm Jy~beam}^{-1}$]}        }
\startdata
90 cm                                   &  0.312 - 0.342 &  2227   & 0.326  & 0.097  & 51.1\tablenotemark{a}                  & 350 \\%
50 cm                                   &  0.596 - 0.626 &  2216   & 0.611  & 0.088  & 101.1\tablenotemark{b}                 & 700\tablenotemark{b}\\%
21 cm\tablenotemark{c}                  &  1.35 - 1.75   &  296    & 1.438  & 0.096  & 3.3                                    & 32 \\%
18 cm\tablenotemark{c}                  &  1.35 - 1.75   &  303    & 1.658  & 0.100  & 3.7                                    & 36 \\%
13 cm\tablenotemark{d}                  &  2.15 - 2.35   &  322    & 2.275  & 0.093  & 3.6 	                              & 35 \\%
13 cm\tablenotemark{d}\tablenotemark{e} &  2.15 - 2.35   &  337    & 2.275  & 0.090  & 3.8                                    & 37 \\%
6 cm                                    &  4.6 - 5.1     &  312    & 4.999  & 0.130  & 3.5                                    & 34 \\%
4 cm                                    &  8.0 - 8.8     &  307    & 8.425  & 0.113  & 3.6                                    & 35 \\%
4 cm\tablenotemark{e}                   &  8.0 - 8.8     &  407    & 8.425  & 0.106  & 4.7                                    & 46 \\%
2 cm                                    &  12.0 - 15.4   &  550    & 15.369 & 0.104  & 6.2                                    & 60 \\%
1 cm\tablenotemark{f}                   &  21.7 - 24.1   &  502    & 22.236 & 0.107  & 5.9 	                              & 57 \\%
1 cm\tablenotemark{f}                   &  21.7 - 24.1   &  441    & 23.799 & 0.107  & 5.1                                    & 50 \\%
7 mm                                    &  41.0 - 45.0   &  1436   & 43.174 & 0.078  & 22.2\tablenotemark{a}\tablenotemark{g} & 151\\%
3 mm\tablenotemark{h}                   &  80.0 - 90.0   &  4000   & 86.2   & 0.025  & 57\tablenotemark{i}                    & 850\tablenotemark{j}\\%
\enddata
\tablenotetext{a}
{Assumes a fringe-fit interval of 1 minute.}
\tablenotetext{b}
{Assumes a fringe-fit interval of 1 minute and a data rate of 32 Mbps.}
\tablenotetext{c}
{Different settings of the same 20 cm receiver.}
\tablenotetext{d}
{Filters at NL, LA, and OV restrict frequencies to 2200-2400 MHz.}
\tablenotetext{e}
{With 13/4 cm dichroic.}
\tablenotetext{f}
{Different settings of the same 1 cm receiver. Continuum performance is
better at 23.8 GHz, away from the water line.}
\tablenotetext{g}
{Performance may be worse on some baselines due to poor subreflector orprimary
reflector shapes or poor atmospheric\\ conditions (almost universal at SC).}
\tablenotetext{h}
{``Average'' 3 mm antennas are assumed.}
\tablenotetext{i}
{Assumes a fringe-fit interval of 30 seconds and a recording rate of 512 Mbps.}
\tablenotetext{j}
{Assumes 4 hours of integration with 7 antennas recording at a rate of
512 Mbps.}
\end{deluxetable}

\medskip

\noindent
Also appearing in Tab.\ \ref{tab-frequency} are parameters
characterizing the performance of a typical VLBA antenna for the
various receiver/feed combinations. Columns [3] and [5] give typical
VLBA system equivalent flux densities ($SEFD\/$s) at zenith and
opacity-corrected gains at zenith, respectively. These were obtained
from averages of right circularly polarized (RCP) and left circularly
polarized (LCP) values from 10 antennas, measured at the frequencies
in column [4] by VLBA operations personnel during regular pointing
observations. In 2007, Germany's Max Planck Insitut für
Radioastronomie funded a program to enhance the 1 cm sensitivity of
the VLBA by installing modern low noise amplifiers to replace the
original VLBA hardware. This program, implemented by NRAO, was
completed in early 2008 and achieved its goal of reducing the zenith
$SEFD\/$s by more than 30\%.

\medskip

\noindent
The typical zenith $SEFD\/$s can be used to estimate $rms$ noise
levels on a baseline between 2 VLBA antennas ($\Delta S$ for a single
polarization) and in a VLBA image ($\Delta I_{\rm m}$ for a single
polarization). Characteristic values for $\Delta S^{\rm 256, 2 m}$
assuming a fringe-fit interval of $\tau_{\rm ff} = 2$ minutes and for
$\Delta I^{\rm 256, 8 h}_{\rm m}$ assuming a total integration time on
source of $t_{\rm int} = 8$ hours also appear in Tab.\
\ref{tab-frequency}. The tabulated baseline sensitivities for 90 cm,
50 cm, and 7 mm assume a fringe-fit interval of 1 minute, since 2
minutes is unrealistically long. All the baseline and image
sensitivities in the table, except for 50 cm and 3 mm, assume an
aggregate recording bit rate equal to the typical value of 256
Mbps. This rate is commonly achieved by recording a total bandwidth
$\Delta\nu$ of 64 MHz (usually 32 MHz per polarization) with 2-bit
(4-level) sampling.

\medskip

\noindent
Opacity-corrected zenith gains are needed for current techniques for
amplitude calibration. These zenith gains vary from antenna to
antenna, and are monitored by VLBA operations and communicated to
users. The typical values appearing in Tab.\ \ref{tab-frequency} are
meant to be illustrative only.

\section{VLBA signal path}\label{apendiceH-path}

This section describes the devices in the signal path at a VLBA
antenna site. Devices from 1 to 6 and 8 to 11 are located at the
antenna; all others are in the site control building (Napier 1995;
Thompson 1995; Rogers 1995). In collaboration with the South African
KAT group, MIT Haystack Observatory, and the CASPER group at UC
Berkeley, NRAO currently is developing a digital back end that will
enable data to be delivered to recording systems at a rate of 4
Gbps. The new back end systems will be implemented in 2010.

\begin{enumerate}

\item \textit{Antenna and subreflector}--- These concentrate the radio
  frequency (RF) radiation. Antenna pointing and subreflector position
  are controlled by commands from the site computer based on the
  current observing schedule and/or provided by the array operators or
  by the site technicians.

\item \textit{Feed}--- The feed collects the RF radiation. All feeds
  and receivers are available at any time, and are selected by
  subreflector motion controlled by the computer. The shaped
  subreflector illuminates all feeds above 1 GHz; these feeds are
  located on a ring at the Cassegrain focus, and changes from one feed
  to another (hence changes in observing band) take only a few
  seconds. In addition, a permanently installed dichroic enables
  simultaneous 2.3/8.4 GHz observations. The 330 and 610 MHz feeds are
  crossed dipoles mounted on the subreflector near prime
  focus. Therefore, it is possible to make simultaneous 330/610 MHz
  observations.

\item \textit{Polarizer}--- This device converts circular
  polarizations to linear for subsequent transmission. For receivers
  above 1 GHz, the polarizer is at cryogenic temperatures.

\item \textit{Pulse cal}--- This system injects calibration tones
  based on a string of pulses at intervals of 1.0 or 0.2
  microseconds. Pulses thus are generated at frequency intervals of 1
  MHz or 5 MHz.

\item \textit{Noise cal}--- This device injects switched, well
  calibrated, broadband noise for system temperature
  measurements. Synchronous detection occurs in the intermediate
  frequency (IF) distributors and base band converters. Switching is
  done at 80 Hz.

\item \textit{Receiver}--- The receiver amplifies the signal. Most
  VLBA receivers are HFETs (Heterostructure Field Effect Transistors)
  at a physical temperature of 15 K, but the 90 cm and 50 cm receivers
  are GaAsFETs (Gallium Arsenide Field Effect Transistors) at room
  temperature. Each receiver has 2 channels, one for RCP and one for
  LCP. The 1 cm, 7 mm, and 3 mm receivers also perform the first
  frequency down conversion.

\item \textit{Maser}--- The maser is a very stable frequency standard
  with two output signals, one at 100 MHz and one at 5 MHz. The 100
  MHz output is the reference for the front end synthesizers and the
  pulse cal system. The 5 MHz output is the reference for the base
  band converters, the formatter, and the antenna timing.

\item \textit{Local oscillator transmitter and receiver}--- The local
  oscillator (LO) transmitter and receiver multiplies the 100 MHz from
  the maser to 500 MHz and sends it to the antenna vertex room. A
  round trip phase measuring scheme monitors the length of the cable
  used to transmit the signal so that phase corrections can be made
  for temperature and pointing induced variations.

\item \textit{Front end synthesizer}--- The front end synthesizer
  generates the reference signals used to convert the receiver output
  from RF to IF. The lock points are at $(n \times 500) \pm 100$ MHz,
  where $n$ is an integer. The synthesizer output frequency is between
  2.1 and 15.9 GHz. There are 3 such synthesizers, each of which is
  locked to the maser. One synthesizer is used for most wavelengths,
  but two are used at 1 cm, at 7 mm, at 3 mm, and for the wide band
  mode at 4 cm.

\item \textit{IF converter}--- The IF converter mixes the receiver
  output signals with the first LO generated by a front end
  synthesizer. Two signals between 500 and 1000 MHz are output by each
  IF converter, one for RCP and one for LCP. The same LO signal is
  used for mixing with both polarizations in most cases. However, the
  4 cm IF converter has a special mode that allows both output signals
  to be connected to the RCP output of the receiver and to use
  separate LO signals, thereby allowing the use of spanned bandwidths
  exceeding 500 MHz. Also, the 90 cm and 50 cm signals are combined
  and transmitted on the same IFs. The 50 cm signals are not frequency
  converted, while the 90 cm signals are upconverted to 827 MHz before
  output.

\item \textit{IF cables}--- There are four of these, labeled A, B, C,
  and D. Each IF converter normally sends its output signals to A and
  C, or else to B and D, although switching is available for other
  possibilities if needed. By convention, the RCP signals are sent to
  A or B while the LCP signals are sent to C or D. Normally only 2
  cables will be in use at a time. Certain dual frequency modes,
  especially 13 cm and 4 cm, can use all four cables.

\item \textit{IF distributors}--- The IF distributors make 8 copies of
  each IF, one for each base band converter. They also can optionally
  switch in 20 db of attenuation for solar observations. There are two
  IF distributors, each handling two IFs. Power detectors allow the
  determination of total and switched power in the full IF bandwidth
  for system temperature determinations and for power level setting.

\item \textit{Base band converters}--- The base band converters (BBCs)
  mix the IF signals to base band and provide the final analog
  filtering. Each of 8 BBCs generates a reference signal between 500
  and 1000 MHz at any multiple of 10 kHz. Each BBC can select as input
  any of the four IFs. Each BBC provides the upper and lower sidebands
  as separate outputs, allowing for a total of 16 ``BB channels'',
  where one BB channel is one sideband from one BBC. Allowed
  bandwidths per BBC are 0.0625, 0.125, 0.25, 0.5, 1, 2, 4, 8, and 16
  MHz. Thus the 16 possible BB channels can cover an aggregate
  bandwidth up to 256 MHz. The BBC signals are adjusted in
  amplitude. With automatic leveling turned on, the power in the
  signals sent to the samplers is kept nearly constant, which is
  important for the 2-bit (4-level) sampling mode. The BBCs contain
  synchronous detectors that measure both total power and switched
  power in each sideband for system temperature determination.

\item \textit{Samplers}--- Samplers convert the analog BBC outputs to
  digital form. There are two samplers, each of which handles signals
  from 4 BBCs. Either 1-bit (2-level) or 2-bit (4-level) sampling may
  be selected. A single sample rate applies to all BB channels; rates
  available are 32, 16, 8, 4, or 2 Msamples per second on each
  channel.

\item \textit{Formatter}--- The formatter selects the desired bit
  streams from the samplers, adds time tags, and supports various
  other functions required to record efficiently on the VLBA original
  tape-based data acquisition system. Although no longer required for
  disk-based recording, these functions are still used in the
  transitional Mark 5A system. Auxiliary detection of up to 16
  pulse-calibration tones and state counts also is supported in the
  formatter.

\item \textit{Recorders}--- The VLBA records on Mark 5A recording
  systems, also in use at the VLA and the GBT. Each unit records on
  two removable modules, sequentially in most cases, but in parallel
  at the 512-Mbps data rate that is the highest currently
  supported. Each module comprises eight commercial disk drives. The
  current VLBA complement of modules is based primarily on 250 or 300
  Gbyte disks, for a total of 2-2.4 Terabytes of recording
  capacity. Presently, a few modules with eight 500-750-Gbyte disks
  (4-6 Terabyte total capacity) are available. Thus, a single module
  lasts for approximately 17-52 hours if recorded continuously at 256
  Mbps, or commensurately shorter periods for recording at 512 Mbps.

\item \textit{Site computer}--- A VME site computer running VxWorks
  controls all site equipment based on commands in the current
  observing schedule or provided by the array operators or by the site
  technicians. All systems are set as requested in the current
  schedule for each new observation.

\item \textit{Monitor and control bus}--- This carries commands from
  the site computer to all site hardware and returns data from the
  site hardware to the computer.

\item \textit{GPS receiver}--- This device acquires time from the
  Global Positioning System (GPS). GPS time is usually used to monitor
  the site clock, providing critical information for data
  correlation. GPS time is occasionally used to set the site clock if
  it is disrupted for some reason. Five of the stations have
  co-located geodetic GPS receivers that are part of the International
  Global Navigation Satellite Systems network.

\end{enumerate}

\section{Recording format}\label{apendiceH-recording}

The VLBA records data on Mark 5A disk-based systems in VLBA
format. Although the VLBA cannot record Mark 4 format as such, there
is a high degree of compatibility between Mark 4 and VLBA formats. In
general, disks in either VLBA or Mark 4 formats can be played back,
for the same observation if necessary, on any VLBA or Mark 4
correlator.

\medskip

\noindent
As part of the VLBA sensitivity upgrade, the NRAO have a goal to
convert the VLBA to Mark 5C recording systems, capable of recording
data at 4 Gbps, by 2010. The data then will be ``format-free'' data
recorded as standard disk files, though there will be a compatibility
mode possible to record in formats useful for earlier versions of Mark
5 hardware correlators.

\section{Correlator}\label{apendiceH-correlator}

As part of the VLBA sensitivity upgrade, NRAO currently is integrating
the DiFX software correlator into the operational environment of the
VLBA, and performing tests to validate its results by comparison with
those of the original VLBA correlator. The DiFX system was developed
at Swinburne University in Melbourne, Australia (Deller et al.\
2007). DiFX will become the operational VLBA correlator in 2010.

\medskip

\noindent
The VLBA correlator, located at the DSOC, accommodates the full range
of scientific investigations for which the array was designed. The
correlator supports wideband continuum, high-resolution spectroscopy,
bandwidth synthesis, polarimetric, and gated observations.

\medskip

\noindent
The correlator is designed to process all observations involving VLBA
stations. With its 20-station capacity and sub-arraying capabilities,
it is designed to correlate an extended array combining the VLBA with
as many as 10 other stations. At present, the VLBA correlator has
available 17 data inputs from Mark 5A recorders, so the rare
observations requiring correlation of more than 17 stations including
the VLBA require multiple correlator passes.

\medskip

\noindent
Each station input comprises 8 parallel ``channels'', which operate at
a fixed rate of 32 Msamples per second, for either 1- or 2-bit
samples. Observations at lower sample rates generally can be processed
with a speed-up factor of 2 (for 16 Msamples per second) or 4 (for 8
Msamples per second or less) relative to observe time. Special modes
are invoked automatically to enhance sensitivity when fewer than 8
channels are observed, or when correlating narrowband or oversampled
data. The correlator accepts input data recorded in VLBA or Mark 4
longitudinal format, or on Mark 5A disk modules, and plays these data
back on tape or disk drives similar to the station recorders.

\medskip

\noindent
Each input channel can be resolved into 1024, 512, 256, 128, 64, or 32
``spectral points'', subject to a limit of 2048 points per baseline
across all channels. The correlator cannot process maximally (16-fold)
oversampled data at the highest spectral resolution, which effectively
prohibits 1024-channel resolution at the narrowest bandwidth of 62.5
kHz. Adjacent, oppositely polarized channels can be paired to produce
all four Stokes parameters; in this case correlator constraints impose
a maximum spectral resolution of 128 points per polarization state.

\medskip

\noindent
The correlator forms cross-spectral power measurements on all relevant
baselines in a given sub-array, including individual antenna
``self-spectra''. These can be integrated over any integral multiple
of the basic integration cycle, 131.072 milliseconds ($2^{17}$
microsec). Adjacent spectral points may be averaged while integrating
to reduce spectral resolution.

\medskip

\noindent
Correlator output is written in a ``FITS Binary Table'' format, and
includes editing flags plus amplitude, weather, and pulse calibration
data logged at VLBA antennas at observe time (Flatters 1998; Ulvestad
1999). All results are archived on digital-audio-tape (DAT)
cassettes. The output data rate is limited to 1.0 Mbytes per second
(MB/s), which must be shared among all simultaneous correlator
sub-arrays. Data are copied from the archive for distribution to users
on a variety of media, with DAT and Exabyte currently given primary
support.

\medskip

\noindent
Operation of the correlator is governed primarily by information
obtained from the VLBA control system monitor data or from foreign
stations log files. Supervision of the correlation process is the
responsibility of VLBA operations personnel, and requires no
participation by the observers.

\section{Angular resolution}\label{apendiceH-angular}

Tab.\ \ref{tab-maximum} gives the maximum lengths rounded to the
nearest km ($B^{\rm km}_{\rm max}$) for each of the 45 VLBA internal
baselines as well as the baselines to other HSA telescopes. A measure
of the corresponding resolution ($\theta_{\rm HPBW}$) in
milliarcseconds (mas) is
\begin{equation}\begin{split}\label{ec-theta}
\theta_{\rm HPBW} \sim 2063\times
\frac{\lambda^{\rm cm}}{B^{\rm km}_{\rm max}} \ \ {\rm mas}
\end{split}\end{equation}
where $\lambda^{\rm cm}$ is the receiver wavelength in cm (Wrobel
1995). A uniformly weighted image made from a long $(u,v)$ plane track
will have a synthesized beam with a slightly narrower minor axis
FWHM. At the center frequencies appearing in Tab.\ \ref{tab-frequency}
and for the longest VLBA baseline, $\theta_{\rm HPBW}$ is 22, 12, 5,
4.3, 3.2, 1.4, 0.85, 0.47, and 0.32 mas for receivers named 90, 50,
21, 18, 13, 6, 4, 2, and 1 cm, plus 0.17 mas at 7 mm. The longest
VLBA-only baseline at 3 mm is currently the one between MK and NL,
which is about 30\% shorter than the longest baseline at other
wavelengths.

\begin{deluxetable}{ccccccccccccccc}
\rotate
\tabletypesize{\scriptsize}
\tablecolumns{15}
\tablewidth{0pc}
\tablecaption{\footnotesize{Maximum VLBI Baseline Lengths in km ($B^{\rm km}_{\rm max}$).}\label{tab-maximum}}
\tablehead{
\colhead{}    &
\colhead{SC}  &
\colhead{HN}  &
\colhead{NL}  &
\colhead{FD}  &
\colhead{LA}  &
\colhead{PT}  &
\colhead{KP}  &
\colhead{OV}  &
\colhead{BR}  &
\colhead{MK}  &
\colhead{EB}  &
\colhead{AR}  &
\colhead{GB}  &
\colhead{Y27} }
\startdata
SC  & ...  & 2853 & 3645 & 4143 & 4458 & 4579 & 4839 & 5460 & 5767 & 8611  & 6822  & 238  & 2708 & 4532 \\%
\\[-0.3cm]
HN  & 2853 & ...  & 1611 & 3105 & 3006 & 3226 & 3623 & 3885 & 3657 & 7502  & 5602  & 2748 & 829  & 3198 \\%
\\[-0.3cm]
NL  & 3645 & 1611 & ...  & 1654 & 1432 & 1663 & 2075 & 2328 & 2300 & 6156  & 6734  & 3461 & 1064 & 1640 \\%
\\[-0.3cm]
FD  & 4143 & 3105 & 1654 & ...  & 608  & 564  & 744  & 1508 & 2345 & 5134  & 8084  & 3922 & 2354 & 515  \\%
\\[-0.3cm]
LA  & 4458 & 3006 & 1432 & 608  & ...  & 236  & 652  & 1088 & 1757 & 4970  & 7831  & 4246 & 2344 & 226  \\%
\\[-0.3cm]
PT  & 4579 & 3226 & 1663 & 564  & 236  & ...  & 417  & 973  & 1806 & 4795  & 8014  & 4365 & 2551 & 52   \\%
\\[-0.3cm]
KP  & 4839 & 3623 & 2075 & 744  & 652  & 417  & ...  & 845  & 1913 & 4466  & 8321  & 4623 & 2939 & 441  \\%
\\[-0.3cm]
OV  & 5460 & 3885 & 2328 & 1508 & 1088 & 973  & 845  & ...  & 1214 & 4015  & 8203  & 5255 & 3323 & 1025 \\%
\\[-0.3cm]
BR  & 5767 & 3657 & 2300 & 2345 & 1757 & 1806 & 1913 & 1214 & ...  & 4398  & 7441  & 5585 & 3326 & 1849 \\%
\\[-0.3cm]
MK  & 8611 & 7502 & 6156 & 5134 & 4970 & 4795 & 4466 & 4015 & 4398 & ...   & 10328 & 8434 & 7028 & 4835 \\%
\\[-0.1cm]
EB  & 6822 & 5602 & 6734 & 8084 & 7831 & 8014 & 8321 & 8203 & 7441 & 10328 & ...   & 6911 & 6335 & 8008 \\%
\\[-0.3cm]
AR  & 238  & 2748 & 3461 & 3922 & 4246 & 4365 & 4623 & 5255 & 5585 & 8434  & 6911  & ...  & 2545 & 4317 \\%
\\[-0.3cm]
GB  & 2708 & 829  & 1064 & 2354 & 2344 & 2551 & 2939 & 3323 & 3326 & 7028  & 6335  & 2545 & ...  & 2516 \\%
\\[-0.3cm]
Y27 & 4532 & 3198 & 1640 & 515  & 226  & 52   & 441  & 1025 & 1849 & 4835  & 8008  & 4317 & 2516 & ...  \\%
\enddata
\end{deluxetable}

\section{Time resolution}\label{apendiceH-time}

Time resolution is set by the VLBI correlator accumulation time. At
the VLBA correlator it is about 2 seconds for most programs, although
a minimum accumulation time of 131 milliseconds is available. The
combination of time and spectral resolution for an observation must
result in a correlator output rate of less than 1.0 Megabyte per
second (MB/s). The limits on time resolution will become far more
flexible when the DiFX correlator becomes operational in 2010.

\section{Baseline sensitivity}\label{apendiceH-baseline}

Adequate baseline sensitivity is necessary for VLBI fringe
fitting. Typical baseline sensitivities are listed in Tab.\
\ref{tab-frequency}. Alternatively, the following formula can be used
in conjunction with the typical zenith $SEFD\/$s for VLBA antennas
given in Tab.\ \ref{tab-frequency} to calculate the $rms$ thermal
noise ($\Delta S$) in the visibility amplitude of a
single-polarization baseline between two identical antennas (Walker
1995a; Wrobel \& Walker 1999):
\begin{equation}\begin{split}\label{ec-ds1}
\Delta S &= \frac{1}{\eta_{\rm s}}\times
\frac{SEFD}{\sqrt{2\times\Delta\nu\times\tau_{\rm ff}}} \ \ {\rm Jy.}
\end{split}\end{equation}

\medskip

\noindent
In the last equation $\eta_{\rm s}\le 1$ accounts for the VLBI system
inefficiency (e.g., quantization in the data recording and correlator
approximations). For the VLBA correlator $\eta_{\rm s}\approx 0.5$ for
1-bit sampling and $\eta_{\rm s}\approx 0.7$ for 2-bit sampling. For
non-identical antennas 1 and 2, Eq.\ \ref{ec-ds1} is modified to the
following:
\begin{equation}\begin{split}\label{ec-ds2}
\Delta S &= \frac{1}{\eta_{\rm s}}\times
\frac{\sqrt{SEFD_{\rm 1}SEFD_{\rm 2}}}{2\times\Delta\nu\times\tau_{\rm ff}}
\ {\rm Jy.}
\end{split}\end{equation}

\medskip

\noindent
The bandwidth in Hz is $\Delta\nu$. For a continuum target, use the BB
channel width or the full recorded bandwidth, depending on
fringe-fitting mode, and for a line target, use the BB channel width
divided by the number of spectral points per BB channel. $\tau_{\rm
ff}$ is the fringe-fit interval in seconds, which should be less than
or about equal to the coherence time $\tau_{\rm coh}$. Eqs.\
\ref{ec-ds1} and \ref{ec-ds2} hold in the weak source limit. About the
same noise can be obtained with either 1-bit (2-level) or 2-bit
(4-level) quantization at a constant overall bit rate; cutting the
bandwidth in half to go from 1-bit to 2-bit sampling is approximately
compensated by a change in $\eta_{\rm s}$ that is very nearly equal to
$\sqrt{2}$ (Moran \& Dhawan 1995). The actual coherence time
appropriate for a given VLBA program can be estimated using observed
fringe amplitude data on an appropriately strong and compact source.

\section{Image sensitivity}\label{apendiceH-image}

Typical image sensitivities for the VLBA are listed in Tab.\
\ref{tab-frequency}. Alternatively, the following formula may be
used in conjunction with the typical zenith $SEFD\/$s for VLBA
antennas given in Tab.\ \ref{tab-frequency} (or a different $SEFD\/$
for lower elevations or poor weather) to calculate the $rms$ thermal
noise ($\Delta I_{\rm m}$) expected in a single-polarization image,
assuming natural weighting (Wrobel 1995; Wrobel \& Walker 1999):
\begin{equation}\begin{split}\label{ec-dim}
\Delta I_{\rm m} &= \frac{1}{\eta_{s}}\times
\frac{SEFD}{N\times(N-1)\times\Delta_{\nu}\times t_{\rm int}}
\ \ {\rm Jy \ beam}^{-1} ,
\end{split}\end{equation}
where $N$ is the number of VLBA antennas available; $\Delta\nu$ is the
bandwidth [Hz]; and $t_{\rm int}$ is the total integration time on
source [s]. The expression for image noise becomes rather more
complicated for a set of non-identical antennas such as the HSA, and
may depend quite strongly on the data weighting that is chosen in
imaging. The best strategy is to estimate image sensitivity using the
EVN sensitivity calculator at
\textsf{http://www.evlbi.org/cgi-bin/EVNcalc}.

\medskip

\noindent
If simultaneous dual polarization data are available with the above
value of $\Delta I_{\rm m}$ per polarization, then for an image of
Stokes $I$, $Q$, $U$, or $V$,
\begin{equation}\begin{split}\label{ec-di}
\Delta I = \Delta Q = \Delta U = \Delta V = \frac{\Delta I_{\rm m}}{\sqrt{2}} .
\end{split}\end{equation}
For a polarized intensity image of $P = \sqrt{Q^2 + U^2}$,
\begin{equation}\begin{split}\label{ec-dp}
\Delta P = 0.655\times\Delta Q = 0.655\times\Delta U .
\end{split}\end{equation}

\medskip

\noindent
It is sometimes useful to express $\Delta I_{\rm m}$ in terms of an
$rms$ brightness temperature in Kelvins ($\Delta T_{\rm b}$) measured
within the synthesized beam. An approximate formula for a
single-polarization image is
\begin{equation}\begin{split}\label{ec-dtb}
\Delta T_{b} \sim 320\times\Delta I_{\rm m}\times(B^{\rm km}_{\rm max})^2
\ {\rm K} .
\end{split}\end{equation}
%

\section{Calibration transfer}\label{apendiceH-calibration}

Data necessary to perform accurate calibration for the VLBA are
supplied as part of the correlator output files, and will appear
within the Astronomical Image Processing System (\aips) as extension
tables attached to the FITS files. These tables include GC (gain), TY
(system temperature), and WX (weather) tables for amplitude
calibration, PC (pulse-cal) tables for system phase calibration, and
FG (flag) tables for editing. For non-VLBA antennas, some or all of
these tables may be missing, since relevant monitor data are not
available at the time of correlation. For example, for the HSA, GC and
TY information are available for most antennas, except that
calibration of the phased VLA requires additional information about
the flux density of at least one source. Flag (FG) tables for non-VLBA
antennas are absent or only partially complete, lacking information
about antenna off-source times. In such cases, the ``flag'' input file
that is output by the \textsf{SCHED} software may be very useful for
flagging data when antennas are not on source; this file appears to be
quite good at predicting the on-source times for the GBT, Arecibo, and
Effelsberg, but presently underestimates by about 10 seconds the time
it takes for Y27 to change source.

\section{Amplitude calibration}\label{apendiceH-amplitude}

Traditional calibration of VLBI fringe amplitudes for continuum
sources requires knowing the on-source system temperature in Jy
($SEFD\/$; Moran \& Dhawan 1995). System temperatures in K ($T_{\rm
sys}$) are measured ``frequently'' in each BB channel during
observations with VLBA antennas; ``frequently'' means at least once
per source/frequency combination or once every user-specified interval
(default is 2 minutes), whichever is shorter. These $T_{\rm sys}$
values are required by fringe amplitude calibration programs such as
\textsf{ANTAB/APCAL} in \aips ~or \textsf{CAL} in the Caltech VLBI
Analysis Programs. Such programs can be used to convert from $T_{\rm
sys}$ to $SEFD\/$ by dividing by the VLBA antenna zenith gains in K
Jy$^{-1}$ provided by VLBA operations, based upon regular monitoring
of all receiver and feed combinations. $T_{\rm sys}$ and gain values
for VLBA antennas are delivered in TY and GC tables, respectively.

\medskip

\noindent
An additional loss of sensitivity may occur for data taken with 2-bit
(4-level) quantization, due to non-optimal setting of the voltage
thresholds for the samplers. This usually is a relatively minor, but
important, adjustment to the amplitude calibration. In the VLBA, for
instance, the system design leads to a systematic (5\% to 10\%)
calibration offset of the samplers between even and odd BB channels;
for dual polarization observations, this may lead to a systematic
offset between RR and LL correlations that must be accounted for in
the calibration. The combination of the antenna and sampler
calibrations may be found and applied in \aips ~using the procedure
\textsf{VLBACALA}.

\medskip

\noindent
Post-observing amplitude adjustments might be necessary for an antenna
position dependent gain (the ``gain curve'') and for the atmospheric
opacity above an antenna, particularly at high frequencies (Moran \&
Dhawan 1995). The GC table described above contains gain curves for
VLBA antennas. Opacity adjustments can be made with \aips ~task
\textsf{APCAL} if weather data are available in a WX table.

\medskip

\noindent
Although experience with VLBA calibration shows that it probably
yields fringe amplitudes accurate to 5\% or less at the standard
frequencies in the 1-10 GHz range, it is recommended that users
observe a few amplitude calibration check sources during their VLBA
program. Such sources can be used (1) to assess the relative gains of
VLBA antennas plus gain differences among base band channels at each
antenna; (2) to test for non-closing amplitude and phase errors; and
(3) to check the correlation coefficient adjustments, provided
contemporaneous source flux densities are available independent of the
VLBA observations. These calibrations are particularly important if
non-VLBA antennas are included in an observation, since their a priori
gains and/or measured system temperatures may be much less accurate
than for the well-monitored VLBA antennas. The VLBA gains are measured
at the center frequencies appearing in Tab.\ \ref{tab-frequency};
users observing at other frequencies may be able to improve their
amplitude calibration by including brief observations, usually of
their amplitude check sources, at the appropriate
frequencies. Amplitude check sources should be point-like on inner
VLBA baselines. Sources may be selected from the VLBI surveys
available through
\textsf{http://www.vlba.nrao.edu/astro/obsprep/sourcelist/}.

\section{Phase calibration and imaging}\label{apendiceH-phase}

\begin{enumerate}

\item
 \textit{Fringe finders}--- VLBI fringe phases are much more difficult
 to deal with than fringe amplitudes. If the a priori correlator model
 assumed for VLBI correlation is particularly poor, then the fringe
 phase can wind so rapidly in both time (the fringe rate) and in
 frequency (the delay) that no fringes will be found within the finite
 fringe rate and delay windows examined during correlation. Reasons
 for a poor a priori correlator model include source position and
 antenna location errors, atmospheric (tropospheric and ionospheric)
 propagation effects, and the behavior of the independent clocks at
 each antenna. Users observing sources with poorly known positions
 should plan to refine the positions first on another instrument. To
 allow accurate location of any previously unknown antennas and to
 allow NRAO staff to conduct periodic monitoring of clock drifts, each
 user must include at least two ``fringe finder'' sources which are
 strong, compact, and have accurately known positions. Typically, a
 fringe finder should be observed for 5 minutes every 1-3 hours.

\item
 \textit{The pulse cal system}--- VLBA observers using more than 1 BBC
 will want to sum over the BBCs to reduce noise levels. This should
 not be done with the raw signals delivered by the BBCs: the
 independent local oscillators in each BBC introduce an unknown phase
 offset from one BBC to the next, so such a summation of the raw
 signals would be incoherent. A so-called ``phase cal'' or ``pulse
 cal'' system (Thompson 1995) is available at VLBA antennas to
 overcome this problem. This system, in conjunction with the LO cable
 length measuring system, is also used to measure changes in the
 delays through the cables and electronics which must be removed for
 accurate geodetic and astrometric observations. The pulse cal system
 consists of a pulse generator and a sine-wave detector. The interval
 between the pulses can be either 0.2 or 1 microsecond. They are
 injected into the signal path at the receivers and serve to define
 the delay reference point for astrometry. The weak pulses appear in
 the spectrum as a ``comb'' of very narrow, weak spectral lines at
 intervals of 1 MHz (or, optionally, 5 MHz). The detector, located at
 the VLBA antennas, measures the phase of one or more of these lines,
 and their relative offsets can be used to correct the phases of data
 from different BBCs. The VLBA pulse cal data are logged as a function
 of time and delivered in a PC table. The \aips ~software can be used
 to load and apply these data. However, some VLBA observers may still
 want to use a strong compact source to do a ``manual'' pulse cal if
 necessary (Diamond 1995). For example, spectral line users will not
 want the pulse cal ``comb'' in their spectra, so they should ensure
 that their observing schedules both disable the pulse cal generators
 and include observations suitable for a ``manual'' pulse cal. Manual
 pulse calibration also is likely to be necessary for any non-VLBA
 antennas included in an observation, because they may have no tone
 generators, or else may not have detectors located at the antenna. In
 addition, it is necessary at 3 mm, where the VLBA antennas have no
 pulse calibration tones.

\item
 \textit{Fringe fitting}--- After correlation and application of the
 pulse calibration, the phases on a VLBA target source still can
 exhibit high residual fringe rates and delays. Before imaging, these
 residuals should be removed to permit data averaging in time and, for
 a continuum source, in frequency. The process of finding these
 residuals is referred to as fringe fitting. Before fringe fitting, it
 is recommended to edit the data based on the a priori edit
 information provided for VLBA antennas. Such editing data are
 delivered in the FG table. The old baseline-based fringe search
 methods have been replaced by more powerful global fringe search
 techniques (Cotton 1995; Diamond 1995). Global fringe fitting is
 simply a generalization of the phase self-calibration technique, as
 during a global fringe fit the difference between model phases and
 measured phases are minimized by solving for the antenna-based
 instrumental phase, its time slope (the fringe rate), and its
 frequency slope (the delay) for each antenna. Global fringe fitting
 in \aips ~is done with the program
\textsf{FRING} or associated procedures. If the VLBA target source is
a spectral line source or is too weak to fringe fit on itself, then
residual fringe rates and delays can be found on an adjacent strong
continuum source and applied to the VLBA target source.

\item
 \textit{Editing}--- After fringe-fitting and averaging, VLBA
 visibility amplitudes should be inspected and obviously discrepant
 points removed (Diamond 1995; Walker 1995b). Usually such editing is
 done interactively using tasks in \aips ~or the Caltech program
 \textsf{DIFMAP} (Shepherd 1997). Note that VLBA correlator output
 data also will include a flag (FG) table derived from monitor data
 output, containing information such as off-source flags for the
 antennas during slews to another source.

\item
 \textit{Self-calibration, imaging, and deconvolution}--- Even after
 global fringe fitting, averaging, and editing, the phases on a VLBA
 target source can still vary rapidly with time. Most of these
 variations are due to inadequate removal of antenna-based atmospheric
 phases, but some variations also can be caused by an inadequate model
 of the source structure during fringe fitting. If the VLBA target
 source is sufficiently strong and if absolute positional information
 is not needed, then it is possible to reduce these phase fluctuations
 by looping through cycles of Fourier transform imaging and
 deconvolution, combined with phase self-calibration in a time
 interval shorter than that used for the fringe fit (Cornwell 1995;
 Walker 1995b; Cornwell \& Fomalont 1999). Fourier transform imaging
 is straightforward (Briggs et al.\ 1999), and done with \aips ~task
 \textsf{IMAGR} or the Caltech program \textsf{DIFMAP} (Shepherd
 1997). The resulting VLBI images are deconvolved to rid them of
 substantial sidelobes arising from relatively sparse sampling of the
 $(u,v)$ plane (Cornwell et al.\ 1999). Such deconvolution is
 achieved with \aips ~tasks based on the CLEAN or Maximum Entropy
 methods or with the Caltech program \textsf{DIFMAP}.

 Phase self-calibration just involves minimizing the difference
 between observed phases and model phases based on a trial image, by
 solving for antenna-based instrumental phases (Cornwell 1995; Walker
 1995b; Cornwell \& Fomalont 1999). After removal of these
 antenna-based phases, the improved visibilities are used to generate
 an improved set of model phases, usually based on a new deconvolved
 trial image. This process is iterated several times until the phase
 variations are substantially reduced. The method is then generalized
 to allow estimation and removal of complex instrumental antenna
 gains, leading to further image improvement. Both phase and complex
 self-calibration are accomplished with the \aips ~task \textsf{CALIB}
 and with program \textsf{DIFMAP} in the Caltech VLBI Analysis
 Programs. Self-calibration should only be done if the VLBA target
 source is detected with sufficient signal-to-noise in the
 self-calibration time interval (otherwise, fake sources can be
 generated) and if absolute positional information is not needed.

 The useful field of view in VLBI images can be limited by finite
 bandwidth, integration time, and non-coplanar baselines (Wrobel 1995;
 Cotton 1999; Bridle \& Schwab 1999; Perley 1999).

\item
 \textit{Phase referencing}--- If the VLBA target source is not
 sufficiently strong for self-calibration or if absolute positional
 information is needed but geodetic techniques are not used, then VLBA
 phase referenced observations must be employed (Beasley \& Conway
 1995). Currently, more than half of all VLBA observations employ
 phase referencing. A VLBA phase reference source should be observed
 frequently and be within a few degrees of the VLBA target region,
 otherwise differential atmospheric (tropospheric and ionospheric)
 propagation effects will prevent accurate phase transfer. VLBA users
 can draw candidate phase calibrators from the source catalog in use
 at the VLBA correlator, distributed with the NRAO program
 \textsf{SCHED}; easy searching for the nearest calibrators is
 available on-line through the VLBA Calibrator Survey (Beasley et
 al. 2002) at \textsf{http://www.vlba.nrao.edu/astro/calib/}. Most of
 these candidate phase calibrators have positional uncertainties below
 1 mas.

 Calibration of atmospheric effects for either imaging or astrometric
 experiments can be improved by the use of multiple phase calibrators
 that enable multi-parameter solutions for phase effects in the
 atmosphere. For further information, see the \aips ~Memos 110 (task
 \textsf{DELZN}, Mioduszewski 2004) and 111 (task \textsf{ATMCA},
 Fomalont \& Kogan 2005), available from
 \textsf{http://www.aips.nrao.edu/aipsdoc.html}

 Ionospheric corrections can even be of significant benefit for
 frequencies as high as 5 GHz or 8 GHz. These corrections may be made
 with the \aips ~task \textsf{TECOR}, or the procedure
 \textsf{VLBATECR}. In addition, it is strongly recommended that the
 most accurate Earth-Orientation values be applied to the calibration,
 since correlation may have taken place before final values were
 available; this may be done with \aips ~task \textsf{CLCOR} or with
 the \aips ~procedure \textsf{VLBAEOPS}.

 The rapid motion of VLBA antennas often can lead to very short time
 intervals for the slew between target source and phase reference
 source. Some data may be associated with the wrong source, leading to
 visibility points of very low amplitude at the beginnings of
 scans. Application of the \aips ~program \textsf{QUACK} using the
 `TAIL' option will fix this problem.

\end{enumerate}

\section{Observing}\label{apendiceH-observing}

Each VLBA program is run remotely from the DSOC by VLBA operations. No
observing assistance by a VLBA user is expected, although VLBA
operations should be able to reach the observer by telephone during
the program. As the program progresses, the array operator monitors
the state of the antennas and tape recording systems, mainly using a
compact yet comprehensive display program. Various logging,
calibration, and flagging data are automatically recorded by the
monitor and control system running on the station computer at each
VLBA site. If necessary, the array operator can request local
assistance from a site technician at each VLBA antenna. Recorded media
are automatically shipped from each VLBA antenna to the correlator
specified by the observer.

\section{Post-processing software}\label{apendiceH-post}

\aips ~is a set of programs for the analysis of continuum and line VLBI
observations involving one or more BB channel. These programs are
available for a wide range of computer operating systems, including
various flavors of Linux and the Mac-OS/X operating system. Extensive
on-line internal documentation can be accessed within \aips. An entire
chapter in the \aips ~\cook ~(NRAO staff, 2007) provides useful
``how-to'' guidance for those reducing VLBI data. Appendix C of the
\aips ~\cook ~provides a step-by-step guide to calibrating many types
of VLBA data sets in \aips. A new ``frozen'' version of \aips ~is
produced each year, and a newer version (currently \textsf{31DEC08})
is updated and made available throughout the calendar year. New
capabilities such as improved astrometric calibration and simpler
data-reduction procedures are implemented frequently in recently
version of \aips.

\section{Data archive}\label{apendiceH-data}

An archive of all output from the VLBA correlator is maintained at the
DSOC. The user who proposed the observations will retain a proprietary
right to the data for a fixed interval of 12 months following the end
of correlation of the last observations requested in the original
proposal or a direct extension of that proposal. Thereafter, archived
data will be available to any user on request. An on-line data archive
has been developed, and data beginning from 1998 currently are on line
(\textsf{https://archive.nrao.edu/archive/archiveproject.jsp}). The
most recent data are available either as multiple correlator output
files or as large FITS files, sometimes with default calibrations
attached.

\medskip

\noindent
Data are distributed to users on a variety of media, with DDS3 and
Exabyte currently given primary support. Distribution also is possible
via ftp. For the initial distribution to the user proposing the
observations, this will occur automatically, soon after correlation is
complete, provided a medium has been specified. Distributed data will
conform to the new FITS binary table standard for interferometry data
interchange (Flatters 1998), which is read by \aips ~task
\textsf{FITLD}.

\end{appendix}


\end{document}